\documentclass[a4paper,11pt]{article}
\pdfoutput=1

\usepackage{jcappub}
 
\usepackage{graphicx,longtable,mathrsfs,color,array}
\usepackage{hyperref}
\usepackage{bm}
\usepackage[usenames,dvipsnames, table]{xcolor} 
\usepackage{amssymb,amsmath,mathtools,mathrsfs,slashed} 
\usepackage{soul}
\usepackage{epsfig,subfigure,placeins,float} 
\usepackage{tikz}
\usepackage{booktabs,longtable,ctable,multirow} 
\usepackage{exscale,relsize} 
\usepackage[normalem]{ulem} 
\usepackage{enumerate}
\usepackage{enumitem}
\usepackage{comment}
\usepackage{multirow}
\usepackage{colortbl}
\usepackage{rotating}
\usepackage{imakeidx}
\usepackage{makecell}
\usepackage{hhline} 
\usepackage{placeins}
\usetikzlibrary{positioning}
\usetikzlibrary{arrows}
\newcommand{\bn}{{\bf n}}
\newcommand*\bell{\ensuremath{\boldsymbol\ell}}
\newcommand{\msun}{{\, \rm M}_\odot}


\title{Testing gravitational wave polarizations with LISA}

\author{
Shingo Akama$^{a}$,
Maxence Corman$^{b}$\footnote{Corresponding author: \href{mailto:maxence.corman@aei.mpg.de }{maxence.corman@aei.mpg.de}},
Paola C. M. Delgado$^{c}$\footnote{Corresponding author: \href{mailto:delgado@fzu.cz}{delgado@fzu.cz }},
Alice Garoffolo$^{d}$, 
Macarena Lagos$^{e}$\footnote{Project coordinator: \href{mailto:macarena.lagos.u@unab.cl}{macarena.lagos.u@unab.cl}},
Alberto Mangiagli$^{b}$\footnote{Project coordinator: \href{mailto:alberto.mangiagli@aei.mpg.de}{alberto.mangiagli@aei.mpg.de}},
Sylvain Marsat$^{f}$,
Manuel Piarulli$^{f}$,
Gianmassimo Tasinato$^{h}$,
Jann Zosso$^{i}$

\vspace{0.5cm}

Giuseppe Gaetano Luciano$^{j}$,
Nils A. Nilsson$^{k}$, 
Leandros Perivolaropoulos$^{m}$,
Kristen Schumacher Aloh$^{nor}$,
Benjamin Sutton$^{p}$,
Roxane Theriault$^{a}$,
Amresh Verma$^{q}$,
Yiqi Xie$^{rt}$,
Mian Zhu$^{sa}$
}
\affiliation[a]{Faculty of Physics, Astronomy and Applied Computer Science, Jagiellonian University, 30-348 Krakow, Poland
}
\affiliation[b]{%
  Max Planck Institute for Gravitational Physics (Albert Einstein Institute), Am M\"uhlenberg 1, DE-14476 Potsdam, Germany
}
\affiliation[c]{%
  CEICO, FZU – Institute of Physics of the Czech Academy of Sciences, Na Slovance 1999/2, 182 00 Prague, Czech Republic %
}
\affiliation[d]{%
 Center for Particle Cosmology, Department of Physics and Astronomy, University of Pennsylvania 209 South 33rd Street, Philadelphia, Pennsylvania 19104, USA
}
\affiliation[e]{%
  Institute of Astrophysics, Department of Physics and Astronomy, Universidad Andrés Bello, Santiago, Chile %
}
\affiliation[f]{%
  Université de Toulouse, CNRS/IN2P3, L2IT, Toulouse, France %
}

\affiliation[h]{%
  Physics Department, Swansea University, SA2 8PP, UK, and\\
  Dipartimento di Fisica e Astronomia, Universit\`a di Bologna,\\ INFN, Sezione di Bologna, viale B. Pichat 6/2, 40127 Bologna, Italy
}
\affiliation[i]{%
 Center of Gravity, Niels Bohr Institute, Blegdamsvej 17, 2100 Copenhagen, Denmark
}
\affiliation[j]{%
Departamento de Qu\'{\i}mica, F\'{\i}sica y Ciencias Ambientales y del Suelo,
Escuela Polit\'{e}cnica Superior -- Lleida, Universidad de Lleida, Av. Jaume
II, 69, 25001 Lleida, Spain %
}
\affiliation[k]{%
Cosmology, Gravity and Astroparticle Physics Group, Center for Theoretical Physics of the Universe, Institute for Basic Science, Daejeon 34126, Korea, and LTE, Observatoire de Paris, Université PSL, CNRS, LNE, Sorbonne Universit\'e, 61 avenue de l’Observatoire, 75 014 Paris, France %
}

\affiliation[m]{%
  Department of Physics, University of Ioannina, 45110 Ioannina, Greece
}
\affiliation[n]{%
  TAPIR, California Institute of Technology, Pasadena, California 91125, USA%
}
\affiliation[o]{%
  Department of Physics, The University of Texas at Austin, 2515 Speedway, Austin, Texas 78712, USA%
}
\affiliation[p]{%
  Theoretical Particle Physics and Cosmology Group, Physics Department,
King’s College London, University of London, \\ Strand, London WC2R 2LS, United Kingdom %
}
\affiliation[q]{%
  Physics Department, Ariel University, Ariel 40700, Israel %
}
\affiliation[r]{%
  Illinois Center for Advanced Studies of the Universe, Department of Physics, University of Illinois at Urbana-Champaign, Urbana, Illinois 61801, USA %
}
\affiliation[s]{%
  College of Physics, Sichuan University, Chengdu 610065, China %
}
\affiliation[t]{%
  Canadian Institute for Theoretical Astrophysics, University of Toronto, Toronto, Ontario M5S 3H8, Canada %
}

\author[]{\\ \vspace{.2cm}
\centering \texttt{(For the LISA Cosmology Working Group)}
\vspace{.2cm}
}

\abstract{
In this paper we quantify the ability of the Laser Interferometer Space Antenna (LISA) to test the presence of non-tensorial polarizations as well as modifications to the tensor ones in gravitational waves emitted from massive black hole binaries. We employ the Parametrized Post-Einsteinian (PPE) formalism to model deviations from General Relativity (GR) for tensor, vector, and scalar polarizations. Our PPE parametrization is inspired by post-Newtonian waveforms from four modified gravity theories: Horndeski, Einstein-\ae ther, Rosen's bimetric, and Lightman-Lee. We consistently implement these modifications across the inspiral, merger, and ringdown phases, ensuring proper waveform alignment and tapering. Subsequently, we perform Fisher forecasts to derive expected constraints on deviations from General Relativity and map these constraints to the parameter spaces of the four gravity theories.  For tensor polarizations, LISA achieves constraints on amplitude modifications ranging between $\sim 10^{-4}-10^{-2}$ precision level, depending on the frequency evolution of the modifications, for systems with $10^5-10^7 \msun$ at $z = 1$. We find that LISA can distinguish breathing and longitudinal scalar polarizations only for relatively light binaries with $M \lesssim 10^4  \msun$, beyond which these modes become degenerate in the detector response. Importantly, constraints on vector polarizations are approximately 2-3 times more precise than for scalar polarizations. For both vector and scalar modes, amplitude measurements reach precisions ranging between $\sim 10^{-8}-10^{-2}$, depending on the frequency evolution of the modifications, for systems with $10^5-10^7 \msun$ at $z = 1$. These results demonstrate LISA's potential to probe gravity in the strong-field regime via gravitational wave polarizations.
}

\begin{document}

\begin{figure}
\begin{flushright}
{\includegraphics[width = 0.2 \textwidth]{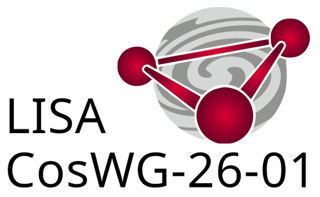}}\\[5mm]
\end{flushright}
\end{figure}

\maketitle
\flushbottom

\section{Introduction}\label{sec:introduction}
While the theory of General Relativity (GR) has been exquisitely confirmed around the Solar System by constraining deviations to less than the $10^{-4}$ level~\cite{Will:2014kxa}, its predictive power reaches its limits at high energies,
where GR is known to be incomplete. Gravitational waves (GWs) are a key observable that provides direct information on the high-energy regime of gravity, since the signals detectable to date come from extreme astrophysical compact objects with strong gravitational fields 
(see~\cite{Will:2014kxa,Yunes:2013dva,Yunes:2016jcc,Ezquiaga:2018btd,Yunes:2025xwp} for comprehensive reviews on tests of gravity with GWs).

GR predicts that GWs carry two tensor polarization modes, corresponding to the transverse, traceless perturbations of spacetime. However, most modified theories of gravity proposed in the literature predict the existence of additional polarization modes beyond these tensor ones, including scalar and vector polarizations. The earliest theoretical studies of extra polarizations date back to the 1970s \cite{Eardley:1973zuo,Eardley:1973br}. Testing for these extra polarizations provides a powerful model-independent probe of the GW generation mechanism and can help discriminate between GR and alternative theories of gravity. 

Current GW observations from ground-based detectors such as LIGO \cite{LIGOScientific:2014pky}, Virgo \cite{VIRGO:2014yos} and KAGRA \cite{Somiya:2011np} face significant challenges in constraining these additional polarization modes. The relatively low signal-to-noise ratios (SNRs) of detected events, combined with the requirement for a network of multiple detectors to simultaneously observe a given event, have limited the ability to place stringent bounds on non-tensorial polarizations \cite{LIGOScientific:2017ous,LIGOScientific:2017ycc,LIGOScientific:2018czr,LIGOScientific:2018dkp,LIGOScientific:2019fpa,LIGOScientific:2020tif,LIGOScientific:2021sio}.

The Laser Interferometer Space Antenna (LISA) promises to improve our ability to test GW polarization content. 
Quantifying rigorously this statement, for GWs  emitted by massive black hole binaries (MBHBs), is the focus of this work. 
LISA will observe signals from massive binary black hole mergers that last from weeks to months, accumulating large SNRs \cite{LISA:2024hlh,Caprini:2025mfr}. Crucially, LISA's orbital motion around the Sun will naturally modulate the detector's response to different polarization modes over the course of the observation, providing a built-in mechanism to disentangle the various polarization components. This combination of long observation times, high SNRs, and time-varying detector response will enable LISA to place unprecedented constraints on scalar and vector GW polarizations, offering a unique opportunity to test the nature of gravity in the strong-field, dynamical regime.

The possibility of constraining GW polarization content relies on the substantial theoretical work that has gone into modeling GW signals in modified gravity theories, with important implications for several types of GW detectors.
Parameterized frameworks for additional polarizations during the inspiral and merger phases have been developed in~\cite{Yunes:2009ke,Mishra:2010tp,Li:2011cg,Healy:2011ef,Chatziioannou:2012rf,Arun:2012hf,Isi:2017fbj,Takeda:2018uai,Berti:2018cxi,Berti:2018vdi,Xie:2025voe}, while ringdown modifications and tests of the no-hair theorem have been studied in~\cite{Dreyer:2003bv, Berti:2005ys, Gossan:2011ha,Meidam:2014jpa,Tattersall:2017erk, Glampedakis:2017dvb,Bhattacharyya:2017tyc,Franciolini:2018uyq, Evstafyeva:2022rve, Maselli:2023khq, Pitte:2024zbi, Berti:2025hly}. These theoretical frameworks provide the foundation for interpreting observations in terms of specific modified gravity theories.

Direct searches for non-GR polarizations from individual transient GW events have been extensively investigated for current ground-based interferometers \cite{Nishizawa:2009bf, Hayama:2012au, Hagihara:2018azu, Hagihara:2019ihn, Pang:2020pfz, Chatziioannou:2021mij, Wong:2021cmp, Omiya:2023rhj, Hu:2023soi, Yunes:2024lzm}. These techniques have already yielded weak observational constraints on deviations from GR through the search of extra polarizations \cite{LIGOScientific:2017ous, LIGOScientific:2017ycc, LIGOScientific:2018czr, LIGOScientific:2018dkp, LIGOScientific:2019fpa, LIGOScientific:2020tif, LIGOScientific:2021sio}. Forecasts for future terrestrial detectors are presented in \cite{Amalberti:2021kzh, Abac:2025saz}. 
In the milli-Hertz band, initial analyses have been carried out for space-based observatories, studying the response and sensitivity to different polarizations, e.g.~\cite{Tinto_2010,Blaut:2012zz, Philippoz:2017ywb,Liang:2019pry, Zhang:2020khm,Liu:2020mab, Wang:2021owg}, while \cite{Zhang:2021fha} has analyzed the expected SNR of monochromatic extra polarizations with LISA, in the context of double white dwarf systems.
%
%
Joint analyses combining ground- and space-based detectors have also been proposed~\cite{Philippoz:2018xrm}, including cross-correlation studies between LISA, Taiji, and TianQin \cite{Omiya:2020fvw, Hu:2022byd, Hu:2024toa, Wang:2025avl, Jiang:2025abg}.

Complementary constraints on extra polarizations come from stochastic GW backgrounds. At nanohertz frequencies, pulsar timing arrays (PTAs) can probe GW polarizations through the stochastic background \cite{Lee_2008, Anholm:2008wy, Chamberlin:2011ev, Gair:2015hra, Cornish:2017oic, Niu:2018oox, Boitier:2020xfx, Liang:2023ary, Cannizzaro:2023mgc, Bernardo:2023zna, Hu:2025rlz, Mihaylov:2018uqm, Tasinato:2023zcg}, with combined analyses discussed in \cite{Qin:2018yhy,Inomata:2024kzr}, and these techniques have yielded weak constraints on extra polarizations from PTAs \cite{NANOGrav:2021ini, Wu:2021kmd, NANOGrav:2023ygs}. Prospects for future PTA facilities such as the Square Kilometer Array (SKA) are summarized in \cite{Janssen:2014dka}. Beyond PTAs, the stochastic GW background in other frequency bands provides complementary tests through analyses of its polarization content \cite{Callister:2017ocg, OBeirne:2018slh, JimenezCruz:2025wqa}.

Besides MBHBs, which are the primary target of LISA, other sources provide promising avenues to test directly or indirectly the presence of additional degrees of freedom in gravity. Scalarization might produce scalar fields around compact objects \cite{PhysRevD.50.7304, Sotiriou:2013qea, Cardoso:2013fwa, Silva:2017uqg, Doneva:2022ewd} which can affect the evolution of extreme mass-ratio inspirals (EMRIs) \cite{1996PThPh..96..713O, Cardoso:2011xi,Yunes:2011aa, Pani:2011xj,  Cardoso:2018zhm, Maselli:2020zgv, Jiang:2021htl, Barsanti:2022vvl, Cardoso:2023dwz} since the scalar field can modify the tensor polarizations detected in LISA. Neutron star mergers in modified gravity theories are also studied in \cite{Barausse:2012da,Sampson:2014qqa, Shibata:2013pra, Lam:2024azd}. In addition, indirect observations from binary-pulsar also provide indirect constraints on the presence of scalar and vector fields \cite{Wex:2014nva,Seymour:2020yle}.

Beyond direct searches for extra GW polarizations through individual sources or stochastic backgrounds, several complementary approaches exist to test for deviations from GR. Within the LISA Consortium, dedicated studies have analyzed various signatures of modified gravity in GW observations; see \cite{Gair:2012nm,Yagi:2013du} for comprehensive reviews on tests of gravity with space-based detectors.
One important class of such tests involves GW propagation effects: additional signatures of modified gravity can be probed through their impact on GW cosmological propagation, including polarization mixing, modified dispersion relations, and amplitude damping. Some of these effects have been parametrized in \cite{Nishizawa:2017nef, Belgacem:2017ihm, Ezquiaga:2021ler, Jenks:2023pmk} and incorporated into LISA Working Group analyses \cite{LISACosmologyWorkingGroup:2019mwx, Barausse:2020rsu, LISACosmologyWorkingGroup:2022jok, LISA:2022kgy, LISA:2024hlh}, providing constraints on the parameter space of modified gravity theories that are complementary to the detection of extra GW polarizations.
Further cosmological tests of GR arise from gravitational lensing of GWs \cite{Mukherjee:2019wcg, Mukherjee:2019wfw, Garoffolo:2019mna, Goyal:2020bkm, Ezquiaga:2020dao, Tasinato:2021wol, Dalang:2021qhu, Finke:2021znb, Balaudo:2022znx, Ezquiaga:2022nak, Menadeo:2025hgf, Tizfahm:2024rcj}, which can reveal modifications to propagation in strong gravitational fields.

Finally, multi-messenger observations also offer valuable constraints: cross-correlations between GW signals and electromagnetic observations provide additional avenues to probe general deviations from GR \cite{LIGOScientific:2018dkp, Belgacem:2018lbp, Lagos:2019kds, Mastrogiovanni:2020gua, Baker:2020apq, Mukherjee:2020mha, Chen:2023wpj, Lagos:2024boe}, supported by improved modeling of black hole populations \cite{Mancarella:2021ecn} and by studies of scalar-field clustering \cite{Garoffolo:2020vtd}. 

\smallskip

While the studies cited above have made important contributions to understanding detector responses to different polarization modes and to testing modified gravity through various complementary channels, a comprehensive forecast for the detectability and parameter estimation of extra polarizations from MBHB mergers with LISA is missing from the literature. The aim of this paper is to fill this gap by providing quantitative forecasts for LISA's ability to constrain non-tensorial polarization content in GWs from MBHB systems, including a well-motivated frequency evolution of extra polarizations from inspiral to merger.

Our analysis begins with a detailed review of the six extra polarizations in metric theories of gravity, their associated spacetime distortions and spin weights. 
Importantly, we clarify the distinction between extra degrees of freedom versus extra GW polarizations, since some modified gravity theories with extra degrees of freedom may not excite any non-tensorial polarization. 
Then, we perform a review of LISA's Time-Delay Interferometry (TDI) response to all polarizations. A key finding of our analysis is that LISA's million of kilometers arm lengths and sensitivity to $\sim 10^{-2}-10^{-1}$Hz frequencies ensure that  it operates   in a regime where the breathing and longitudinal scalar modes produce measurably different responses in the TDI observables \cite{Tinto_2010}, enabling independent constraints on each mode. This represents a significant advance over current ground-based detectors, which cannot differentiate between these two scalar modes because the gravitational wavelengths they observe are much larger than the detector arm lengths.

We employ the Parametrized Post-Einsteinian (PPE) formalism \cite{Yunes:2009ke,Cornish:2011ys,Sampson:2013lpa,Perkins:2020tra,Mezzasoma:2022pjb}, which provides a theory-agnostic framework for describing frequency-dependent deviations from GR in the phase and amplitude of gravitational waveforms coming from metric theories of gravity, across all possible polarization modes--two tensor, two vector, and two scalars. This approach allows us to test for generic departures from GR without committing to a specific alternative theory, while still maintaining the ability to interpret our results in the context of concrete theoretical models. Nonetheless, while the PPE formalism provides a broad parametric language for deviations from GR, it is not assumption-free. Several implicit conditions must hold for the framework to yield meaningful constraints: (i) the GW-emitting systems in the modified theory must admit binary solutions sufficiently similar to those in GR, so that the inspiral dynamics and associated radiation can be described as a perturbative deformation of the GR waveform; (ii) the dominant modifications can be captured by a small number of leading-order post-Newtonian (PN) corrections; and (iii) the framework is restricted, by construction, to metric theories of gravity. These assumptions constrain the class of theories that can be encompassed and tested with this formalism and need to be considered in order to avoid biases from incorrect waveform models \cite{Gupta:2024gun}.

Our PPE parameterization is inspired by the PN inspiral waveforms predicted in four well-studied modified gravity theories: Horndeski gravity \cite{Higashino:2022izi}, Einstein-\ae ther theory \cite{Zhang:2019iim}, Rosen's bimetric theory \cite{Rosen:1974ua}, and Lightman-Lee theory \cite{Lightman:1973kun} (see
\cite{Chatziioannou:2012rf} for a useful, modern formulation of the latter
two theories). Each of these theories predicts distinctive signatures in the various polarization modes of GWs, making them ideal testbeds for exploring the range of possible deviations from GR. By constructing a PPE model that captures the essential features of these theories, we can assess LISA's constraining power in a manner that is both theoretically motivated and broadly applicable. For instance, in these theories we find the phase modifications of the extra polarization modes to be directly linked to phase modifications in the tensor mode, reducing the dimensionality of the possible parameter space. As a result, our PPE model contains a total of 8 free parameters to be tested: one describing the phase of all polarizations, one for the amplitude modification of tensor polarizations, and two for the amplitude of vector, breathing, and longitudinal polarizations in two dominant angular harmonics $(\ell=|m|=1)$ and $(\ell=|m|=2)$. 

Using Fisher matrix forecasts with the \texttt{lisabeta} code \cite{Marsat:2020rtl}, we estimate the precision with which LISA will be able to measure the 8 PPE parameters of modified gravity, for individual massive black hole binaries. We then map these projected constraints back onto the parameter spaces of the four specific modified gravity theories considered here, providing quantitative predictions for how well LISA will be able to test or rule out these alternatives to GR.

Our Fisher forecast analysis reveals that LISA will achieve constraints on the amplitude of extra polarizations with uncertainties ranging between $\sigma\sim 10^{-8}-10^{-2}$ for sources at $z=1$ and total mass between $10^5-10^7 \msun$, with a large dependence on the assumed frequency evolution of these polarizations. We also find about a factor of $2-3$ stronger constraints on vector with respect to  scalar polarizations (for the same given inspiral frequency scaling for vector and scalar modes) across most of the parameter space, owing to enhanced detector response to vector modes. In addition, LISA is expected to enable discrimination between breathing and longitudinal scalar polarizations for binaries with total mass $M \lesssim  10^4 \msun$.

This paper is organized as follows. In Section~\ref{sec:revpol} we review definitions of the six GW polarizations in metric theories of gravity. In Section~\ref{sec:Detecting_extra_polarizations}
we discuss how LISA's TDI approach can distinguish different polarizations. In Section~\ref{sec:ppe_formalism} we present the PPE formalism and we discuss the numerical approach adopted in the Fisher  forecast. In Section~\ref{sec:ModGrav} we describe the inspiral waveforms of extra polarizations of four modified gravity theories, and map them to the PPE formalism. In Section~\ref{sec:results_for_extra_pol} we present the Fisher  constraints on the PPE parameters. In Section \ref{sec:discu_and_concl} we conclude summarizing our results and discussing their future impact.

\paragraph{Notation and conventions}

Throughout this paper we will be setting the speed of light to unity, $c=1$ as well as Planck constant $\hbar=1$. We assume a fiducial flat $\Lambda$CDM cosmology with $h = 0.6774$, $\Omega_m = 0.3075$ and $\Omega_{\Lambda} = 0.6925$ \cite{Planck:2015fie}. In addition, we collect regularly used acronyms in Table \ref{tab:acronyms}, and the definitions and descriptions of the main GR parameters used in this paper in Table \ref{tab:Notations}.

\begin{table}[h!]
\begin{center}
\begin{tabular}{|l l|}
 \hline
Acronym & Meaning \\
 \hline
 BH & Black hole \\
 BHB & Black hole binary\\
 MBHB & Massive black hole binary\\
 GR & General relativity \\
 GW & Gravitational wave \\
 DoF & Degree of freedom \\
 SVT &  Scalar-vector-tensor\\
 FTI & Flexible-Theory-Independent \\
 IMR & Inspiral-merger-ringdown\\
 LISA & Laser Interferometer Space Antenna \\
 LVK &  LIGO-Virgo-KAGRA\\
 PN & Post-Newtonian \\
  PPN & Parametrized post-Newtonian \\
 PPE &  Parametrized post-Einsteinian\\
  SPA &  Stationary phase approximation\\
  TT &  Transverse-traceless \\
  TDI & Time-delay Interferometry \\
 PSD & Power spectral density \\
 SNR & Signal-to-noise ratio \\
 \hline
\end{tabular}
\end{center}
\caption{Summary of acronyms regularly used in this paper. \label{tab:acronyms}}
\end{table}

\begin{table}[h!]
    \centering
    \begin{tabular}{|c | c|c|}
    \hline
        Parameter  & Definition (if needed) & Description \\ \hline
        $z$ & & Redshift  \\ \hline
        $m_1$ & & Primary source-frame BH mass \\ \hline
        $m_2$ & & Secondary source-frame BH mass \\ \hline
        $M$ &$ M \equiv m_1 + m_2$ & Source-frame total mass  \\ \hline
         $M_z$ &$ M_z \equiv (1+z)M$ & Redshifted source-frame total mass  \\ \hline
        ${\cal M}_z$ & ${\cal M}_z \equiv (1 + z) \, \frac{(m_1 m_2)^\frac35}{M^\frac15} \equiv \eta^{3/5}M_z $ & Redshifted chirp mass   \\ \hline
         $\eta$ &$\eta \equiv \frac{m_1 m_2}{M^2}$  & Symmetric mass ratio \\
        \hline
         $d_L$ & & Luminosity distance  \\ \hline
         $t_c$ & & Time to coalescence \\ \hline
        $\chi_1$ & & Primary BH spin magnitude \\ \hline
        $\chi_2$ & & Secondary BH spin magnitude\\ \hline
        $\chi_{\rm PN}$ & \begin{tabular}{@{}c@{}} $\chi_{\rm PN} \equiv \eta/113 \, [(113q + 75)\chi_1$ + \\ $(113/q + 75)\chi_2]$ \end{tabular}  & PN spin parameter \\ \hline
        $\chi_m$ & $\chi_{\rm m} \equiv q \chi_1/ (1+q)  - \chi_2 / (1+q) $ & Mass-weighted spin difference \\ \hline
        $\iota$ & & Inclination \\ \hline
        $\beta_S$ & & Latitude of the source \\ \hline
        $\lambda_S$ & & Longitude of the source \\ \hline
        $\phi_c$ & & Phase at coalescence \\ \hline
        $\psi$ & & Polarization phase \\ \hline
        $G_N$ & & Newtonian gravitational constant \\
    \hline
    \end{tabular}
    \caption{Definitions and descriptions of the main GR parameters used in the paper to characterize waveforms and perform Fisher  forecasts. } 
    \label{tab:Notations}
\end{table}

\section{\label{sec:revpol}Review on GW polarizations}
\label{Sec:ReviewGWpol}

It is known that in GR the GW field in the far-away wave zone\footnote{Following  the definition in~\cite{poisson_will_2014}, the wave-zone is the three-dimensional spatial region where the length of the position wave vector $R$ is larger than the characteristic wavelength $\lambda_c$
of the GW field. Furthermore, the \emph{far-away} wave zone represents an asymptotically spatially flat region, where the terms decaying faster than $1/R$ have become negligible in the GW tensor $h_{\alpha\beta}$.} has two transverse polarizations, namely the \textit{plus} and \textit{cross} modes. 
In the case of a wave that propagates along the $z$-direction, such polarizations correspond to the non-vanishing $(x,y)$ components of the metric tensor $h^{xx}=-h^{yy}=h_+$ and $h^{xy} = h^{yx}=h_\times$, in cartesian coordinates. Within GR, these two dynamical degrees of freedom (DoFs) also govern the GW detector response, since matter is minimally coupled to the metric. More precisely, these two polarization modes govern the geodesic deviation equation between two time-like geodesics. Indeed, in GR, the electric part of the leading-order asymptotic Riemann tensor is entirely determined by these two dynamical transverse polarizations \cite{poisson_will_2014}. 

\bigskip
On the other hand, in extended theories of gravity, there can exist additional dynamical DoFs. In fact, the existence of additional dynamical DoFs can be viewed as a unique signature beyond GR, due to Lovelock's theorem \cite{poisson_will_2014,papantonopoulos2014modifications,Will:2018bme}. However, not all of these gravitational DoFs might directly influence the detector response to the associated asymptotic radiation. In this work, we will define GW polarization through the detector response, and hence extra DoFs in modified gravity will not always be interpreted as additional GW polarizations.
It is therefore necessary to distinguish between the concepts of \textit{dynamical degrees of freedom}, defined as the number of independent, dynamical, gauge-invariant, perturbative solutions to a differential equation of motion and the \textit{gravitational polarizations} that govern the detector response to asymptotic radiation (it should be noted that these dynamical DoFs can still affect the tensorial waveforms). Most notably, this is so within the framework of so called \textit{metric theories} of gravity \cite{Dicke:1964pna,misner2017gravitation,poisson_will_2014,papantonopoulos2014modifications,Will:2014kxa,Will:2018bme,YunesColemanMiller:2021lky,Zosso:2024xgy}. More precisely, metric theories of gravity only admit up to six modes of gravitational polarization \cite{Eardley:1973zuo,poisson_will_2014,Will:2018bme}. This statement crucially relies on the definition of metric theories of gravity.

\bigskip
Based on the Einstein equivalence principle, metric theories encompass all theories that comply with the principles of universal and minimal coupling through the existence of a single physical metric $g_{\mu\nu}$, that is the only gravitational field that directly couples to the matter sector described through a collection of fields $\Psi_m$. In this paper, we will only be considering such minimally coupled theories (note this means we are working in the so-called Jordan frame). More precisely, assuming the existence of a covariant and local action of a metric theory, additional gravitational fields $\Psi$, identified through their non-minimal coupling to the physical metric, do not directly appear in the matter action of the theory, where matter fields are minimally coupled to the physical metric only. The general action of a metric theory therefore takes the schematic form
    \begin{equation}\label{eq:ActionMetricTheory}
       S=\int d^4x\,\sqrt{-g} \left(\frac{1}{2\kappa}\mathcal{L}_\text{G}[g,\Psi]+\mathcal{L}^\text{min}_\text{m}[g,\Psi_\text{m}]\right)\,,
    \end{equation}
where $\kappa=8\pi G_N $, with a matter Lagrangian $\mathcal{L}^{\text{min}}_\text{m}$, where matter fields $\Psi_m$ are minimally coupled to the metric $g$ only, and a gravitational Lagrangian $\mathcal{L}_\text{G}$ depending on the metric, as well as a possible set of additional non-minimal fields $\Psi$.
The principles of universal and minimal coupling ensure that all non-gravitational laws of physics are locally recovered within the normal coordinates of the physical metric, therefore satisfying  Einstein's weak equivalence principle. Moreover, together with general covariance of the action, such a split between minimally coupled matter fields and additional gravitational fields ensures the existence of a well-defined and locally conserved energy-momentum tensor of matter fields~\cite{poisson_will_2014,papantonopoulos2014modifications,carroll2019spacetime}. Furthermore, universal and minimal coupling of matter to a single physical metric is what allows one to unambiguously talk about the metric as a property of spacetime, rather than an additional field over spacetime~\cite{Will:2018bme}.

In particular, the fact that within metric theories only the physical metric tensor directly couples  to matter fields also allows for a simple definition of the GW detector response, entirely based on the geodesic deviation equation, as the latter does not depend on the equations of motion of the theory. Therefore, as in GR, the geodesic deviation equation between two timelike geodesics of any metric theory of gravity is determined through the electric part of the Riemann tensor associated to the physical metric, which has only up to six independent dynamical modes~\cite{Will:2018bme}. These six potential modes of gravitational polarizations are directly associated to the six degrees of freedom within the physical metric. A simple proof of these statements can be given by restricting to the spacetime metric in the far-away wave zone, and assuming that the field equations reduce to wave equations in this zone. 

In this section, we work in the far-away wave zone, and assume asymptotic flatness, that is, we assume an asymptotic Minkowski background spacetime with leading perturbations of order $1/r$. Later on, we will indeed include the cosmological propagation effects of GR (namely, redshifting of frequency and an amplitude decay inversely proportional to the luminosity distances). Nonetheless, we will always neglect modified propagation effects that would occur in theories beyond GR.\footnote{Examples of such cosmological propagation effects are birefringence triggered by gravitational parity violations, changes in GW amplitude, or propagation speed modifications associated to the spontaneous Lorentz breaking of cosmological solutions beyond GR. See e.g.~\cite{Ezquiaga:2018btd, Belgacem:2019pkk}.} 
Furthermore, we will assume for simplicity that also in the asymptotically flat regime any difference in the propagation speed of different polarizations is negligible, such that all the polarizations of GWs arrive at the detector at the same time. We therefore in particular require that any mass of dynamical degrees of freedom and any scale of hard Lorentz breaking have a negligible effect on the propagation speeds. We note that this assumption may be violated in environments where the propagating degrees of freedom acquire an effective mass through coupling to surrounding matter (e.g.\ \cite{Cardoso:2013fwa}), which would in principle induce differential arrival times between polarization modes. However, we expect this effect to be negligible for the systems and distances considered here: GWs from MBHBs travel in near-vacuum for the vast majority of their propagation path, and any medium-induced modification of the dispersion relation would only be relevant in the immediate vicinity of the source or observer, contributing negligibly to the total phase accumulated over cosmological distances. Environmental effects arising from interactions between the binary and the accretion disk or dynamical friction modifying the orbital evolution are a separate and potentially more significant concern, but are beyond the scope of this work.

This section is organized as follows. In Section \ref{subsec:geodesic_equation} we review the number of degrees of freedom that affect the geodesic equation of a point particle, which determines the maximum number of polarizations that GWs can have in metric theories. In Section \ref{subsec:Diffeomorphism_Invariance} we discuss a gauge-invariant approach to count again the number of polarizations that GWs can have.  In Section \ref{subsec:Polarizations} we show the decomposition of the six polarizations into tensor, vector and scalars.
In Section \ref{subsec:spin_weight} we analyze the spin weight of each polarization, associating tensor polarizations to spin 2, vectors to spin 1, and scalars to spin 0.
In Section \ref{subsec:spin_weighted_spherical_decomposition} we perform general angular harmonic decompositions of the six polarizations, based on their spin weight, and impose amplitude conditions for the non-precessing binary systems considered in this paper. 
In Section \ref{subsec:waveform_frequency_domain} we present the Fourier decomposition of the angular-harmonic amplitudes of all polarizations.

\subsection{\label{subsec:geodesic_equation}Geodesic equation} 

In order to define the six GW polarizations that can be present in metric theories of gravity, we consider the generic response of a typical GW detector. Assuming an idealized detector built  of two test masses that are moving freely and slowly, the corresponding detector response is governed by the geodesic deviation equation. Concretely, denoting the space-like separation vector between the two timelike geodesics traced out by the masses by $s^\alpha$ and assuming the relative distance to be small compared with the radiation’s characteristic wavelength\footnote{Note that for LISA this approximation does not hold at the high-frequency end of its sensitivity and the corresponding detector response needs to be generalized as we show in Section\ \ref{subsec:LISA_Antenna_Response} below. For simplicity and without loss of generality, we will nevertheless define the GW polarizations under this assumption.}, the evolution of the spatial part of $s^\alpha$ obeys the following approximate form of the equation of geodesic deviation~\cite{poisson_will_2014} 
\begin{equation}
\label{geoeq}
    \frac{d^2s_j}{dt^2}=-R_{0j0k}s^k\,,
\end{equation}
where the $t$ is the time of a local rest-frame and $R_{0j0k}$ are the electric components of the Riemann tensor associated to the physical metric, with Latin indexes indicating spatial components. It is important to note that Eq.~\eqref{geoeq} holds in any metric theory of gravity: hence, it does not depend on the equations of motion of the theory but crucially relies on the assumption that only the physical metric $g$ directly couples to matter. In addition, Eq.\ (\ref{geoeq}) is valid up to first order in the separation between the test masses, compared to the scale of curvature variation (given by the GW wavelength $\lambda$). As a result, this is an approximate expression valid for low frequencies, when $s/\lambda \ll 1$.

By expanding the Riemann tensor up to first order in the metric perturbation we obtain
\begin{align}\label{eq:RiemannFirstOrder}
     R_{0j0k} &=-\frac{1}{2}\left(\ddot{h}_{jk}+\partial_{j}\partial_kh_{00}-\partial_{j}\dot{h}_{0k}-\partial_{k}\dot{h}_{0j},
    \right) 
\end{align}
where an over-dot stands for a partial time derivative, i.e. $\dot{\,} = \partial_0$. We continue the characterization by  fixing a particularly useful frame. Namely, in a local  patch of spacetime, it is always possible to choose a synchronous frame \cite{WaldBook,landau_classical_2003,carroll2019spacetime,carroll2019spacetime}, in which the perturbations of the physical metric $h_{\mu\nu}$ are chosen to be purely spatial $h_{00}=h_{0i}=0$. This frame, also called ``Gaussian normal coordinates'', exists for any spacetime region that is small enough so that a congruence of timelike geodesics do not cross,  is the most convenient choice of coordinate system  to describe the response of an interferometer based on time-travel distance changes between freely-falling test masses.

Through Eqs.~\eqref{eq:RiemannFirstOrder} it is easily seen that, in this particular gauge, the response matrix of the geodesic deviation equation simply corresponds to the metric perturbations 
\begin{align}\label{eq:RiemannSynchronous}
     R_{0j0k} &=-\frac{1}{2}\ddot{h}_{jk},
\end{align}
where the 6 independent components of $h_{ij}$ ($10$ initial - 4 gauge choices) are labelled as follows
\begin{equation}\label{eq:MetricSixPolarizations}
    h_{ij} = \begin{pmatrix}
h_b+h_+ & h_\times & h_{v1}\\
h_\times & h_b-h_+ & h_{v2}\\
h_{v1} & h_{v2} & h_l\\
\end{pmatrix}\,.
\end{equation}

\subsection{\label{subsec:Diffeomorphism_Invariance}Diffeomorphism invariance}
In the previous section, we derived the geodesic equation under a specific gauge choice. We now turn to the question of gauge invariance of GW polarizations, and examine their explicit relation to the metric perturbations \( h_{\mu\nu} \) given by:
\begin{equation}\label{eq:MetricPert}
    g_{\mu\nu}=\eta_{\mu\nu}+h_{\mu\nu}\,.
\end{equation}
From now on, we assume an asymptotically flat spacetime representing a localized source. This setup is standard in the study of GW experiments. We adopt source-centered asymptotic rest frame coordinates \( \{t, x, y, z\} \). Working within the framework of linearized perturbations around the asymptotically flat background, the metric perturbation transforms under the associated linear gauge transformations as
\begin{equation}\label{eq:gaugeTransf}
    h_{\mu\nu} \rightarrow h_{\mu\nu} - \mathcal{L}_\xi\,\eta_{\mu\nu}=h_{\mu\nu}-2 \partial_{(\mu} \xi_{\nu)}\,,
\end{equation}
with $\mathcal{L}_\xi$ the Lie derivative and the small vector $\xi_{\nu}$ characterizing the four gauge degrees of freedom. 
Following \cite{Flanagan:2005yc,Heisenberg:2024cjk},  the perturbed physical metric in the asymptotically flat region $h_{\mu\nu}$ can be decomposed within a scalar-vector-tensor (SVT) decomposition of the following form
\begin{subequations}\label{eq:SVTmetricDecomp}
\begin{align}
    h_{00}&=2A\,,\\
    h_{0i}&=B^\text{T}_i + \partial_i B\,,\\
    h_{ij}&=\frac{1}{3} \delta_{ij} D + E^\text{TT}_{ij} + \partial_{(i}E^\text{T}_{j)} + \left(\partial_i\partial_j -\frac{1}{3} \delta_{ij} \partial^2\right) E \,,
\end{align}
\end{subequations}
where $D \equiv \delta^{ij}h_{ij}$ defines the trace. This is the most general decomposition of $h_{\mu\nu}$ into its irreducible representations under $3d$ rotations with a subsequent Helmholtz decomposition of the resulting spatial vectors and tensors, with the properties $\partial^i B^\text{T}_i = 0$ and $\partial^i E^\text{T}_i = 0$ for the transverse vectors, and $\partial^i  E_{ij}^\text{TT} = \delta^{ij} E_{ij}^\text{TT}=0$ for the transverse-traceless (TT) component.
These components transform under Eq.~\eqref{eq:gaugeTransf} as
\begin{align}\label{eqn:GaugeInvTransf}
    &A  \rightarrow A-\dot{\xi_0} \,, \qquad\qquad B \rightarrow B-\xi_0-\dot{\xi}\,, \qquad D \rightarrow D-2 \nabla^2 \xi \,, \qquad E  \rightarrow E -2 \xi \,, \\
    &B^\text{T}_i  \rightarrow B^\text{T}_i-\dot{\xi}^\text{T}_i \,, \,\qquad  E^\text{T}_i  \rightarrow E^\text{T}_i-2 \xi^\text{T}_i, \\
    &E_{i j}^\text{TT}  \rightarrow E_{i j}^\text{TT},
\end{align}
where $\xi_0$ is the zeroth component of the generator, while the spatial part has been split according to $\xi_i = \xi^\text{T}_i + \partial_i \xi$, using the Helmholtz decomposition again.
Given these transformation laws, it is clear that the following quantities 
\begin{subequations}\label{eqn:GaugeInvQuantity}
\begin{align}
    \Phi & \equiv A-\dot{B}+\frac{1}{2} \ddot{E} \,,\\
    \Theta & \equiv \frac{1}{3}\left(D-\nabla^2 E\right) \,,\\
    \Xi_i & \equiv B^\text{T}_i-\frac{1}{2} \dot{E}^\text{T}_i \,,
\end{align}
\end{subequations}
and $E_{ij}^\text{TT}$ are invariant under the linear diffeomorphism~\eqref{eq:gaugeTransf}. 
Interestingly, the linearized Riemann tensor can be written as
\begin{align}\label{eq:Gauge invariant Riemann first}
    R_{0j0k} 
    %
    &=-\frac{1}{2} \ddot{E}_{ij}^\text{TT} - \frac{\delta_{ij} }{2}  \ddot{\Theta} + \partial_{(i} \dot{\Xi}_{j)} -\partial_i \partial_j \Phi\,,
\end{align}
showing that the electric components of the Riemann tensor, which enter the geodesic deviation equation, are gauge invariant under linear diffeomorphism.
Moreover, as we will show later, this decomposition of the Riemann tensor will  provide a straightforward recipe to identify the number of GW polarization states within a given metric theory of gravity.

In order to relate the gauge invariant metric perturbations to the GW polarizations we require an additional step. Recall we work with
an asymptotic Minkowski background metric  $\eta_{\mu\nu}$ in the far-away wave zone, around which we consider first-order perturbations characterized by $\mathcal{O}(1/r)$ terms, where $r$ denotes the corresponding source-centered radial coordinate. Therefore, any gauge invariant perturbation mode that contributes non-trivially at leading order Eq.~\eqref{eq:Gauge invariant Riemann first} is part of the asymptotic radiation, whose phases $e^{-i(\omega t-k\,\bm n \cdot \bm x)}$ by definition travel at the phase velocity 
\begin{equation}\label{eq:PhaseVelocityDef}
    V\equiv \omega/k\,,
\end{equation}
such that each mode depends on time through the particular combination $t - r/V$. Here, $\bm{n}$ is the unit vector pointing in the direction of propagation, $n_i = \partial_i r$ defined in Eq.~\eqref{N} below. Because of this universal behavior, spatial and time derivatives are related to each other,
    \begin{equation}\label{eq:identity derivatives}
    \partial_{i} F = -\, \frac{n_i}{V}\, \dot{F} \,,
    \end{equation} 
for any generic function $F(t - r / V)$. This finally allows us to rewrite Eq.~\eqref{eq:Gauge invariant Riemann first} as
\begin{align}
    R_{0i0j} = -\frac{1}{2} \partial_0 \partial_0 \left[E_{ij}^\text{TT} + \frac{2}{V} n_{(i} \Xi_{j)} + \frac{2}{V^2} n_i n_j \Phi+ \delta_{ij} \Theta \right]\,.\label{eqn:Apolarization}
\end{align}
We have therefore confirmed that the six GW polarizations in a generic metric theory of gravity can be written in terms of manifestly gauge invariant quantities and therefore represent physical observables.
In this case, the linearized Riemann tensor assumes the following form
\begin{equation}\label{eq:RiemannTensorWaveZone}
    R_{0j0k}=-\frac{1}{2}\ddot{S}_{jk}\,,
\end{equation}
where the six independent components in the symmetric spatial tensor $S_{jk}$ represent the 6 polarization modes of the GW. Comparing with Eq.~\eqref{eq:RiemannSynchronous}, we see that $S_{ij} = h_{ij}$ in synchronous gauge.
Using Eq.~\eqref{eq:RiemannTensorWaveZone} the geodesic deviation equation Eq.~\eqref{geoeq} is easily integrated to give to leading order
\begin{align}\label{eq:integratedGeodesicDeviation}
    \Delta s_i (t) = \frac{1}{2} S_{ij}(t) s^j (t_0) \,,
\end{align}
where
\begin{equation}
    \Delta s_i (t)\equiv s_i (t)-s_i (t_0)\,,
\end{equation}
for some initial reference time $t_0$ before the presence of any GW.

\subsection{\label{subsec:Polarizations}Polarization basis}
We introduce a basis for rank-$2$ tensors.
In a spatial orthonormal basis of the outgoing radiation $(\bm{n},\bm{\theta},\bm{\phi})$, with $\bm{n}$ the radial direction of propagation, the six polarization basis tensors in Eq.~\eqref{eq:Sijexpand} explicitly read 
\begin{subequations}\label{eq:PolTensors}
\begin{align}
e^+_{ij}&\equiv\theta_i\theta_j-\phi_i\phi_j\,,& e^\times_{ij}&\equiv\theta_i\phi_j+\phi_i\theta_j\,,& e^b_{ij}&\equiv \theta_i\theta_j+\phi_i\phi_j\,,\\
e^{v1}_{ij}&\equiv n_i\theta_j+\theta_in_j\,,& e^{v2}_{ij}&\equiv n_i\phi_j+\phi_in_j\,,& e^l_{ij} &\equiv n_in_j\,.
\end{align}
\end{subequations}
Introducing the polar angles $\{\theta, \phi\}$ of the source-centered coordinate system, a standard choice for the orthonormal basis is\footnote{With these choices of the transverse space, the vector $\bm{\theta}\, (\bm{\phi})$ points in the direction of increasing colatitude (longitude) on the surface of a sphere as the notation suggests.}
\begin{eqnarray}
\label{N}
\bm{n}&\equiv&\left[\sin\theta\cos\phi,\,\sin\theta\sin\phi,\,\cos\theta\right],\\[2mm]
\label{theta}
\bm{\theta}&\equiv&\left[\cos\theta\cos\phi,\,\cos\theta\sin\phi,\,-\sin\theta\right],\\[2mm]
\label{varphi}
\bm{\phi}&\equiv&\left[-\sin\phi,\cos\phi,0\right].
\end{eqnarray}
Notice that $(\bm{\theta},\bm{\phi})$ axes are defined up to a rotation around $\bm{n}$. This rotational freedom will be discussed later to determine the spin weight of the GW polarizations. 

\begin{figure}[h!]
\centering
\includegraphics[width = 0.75\textwidth]{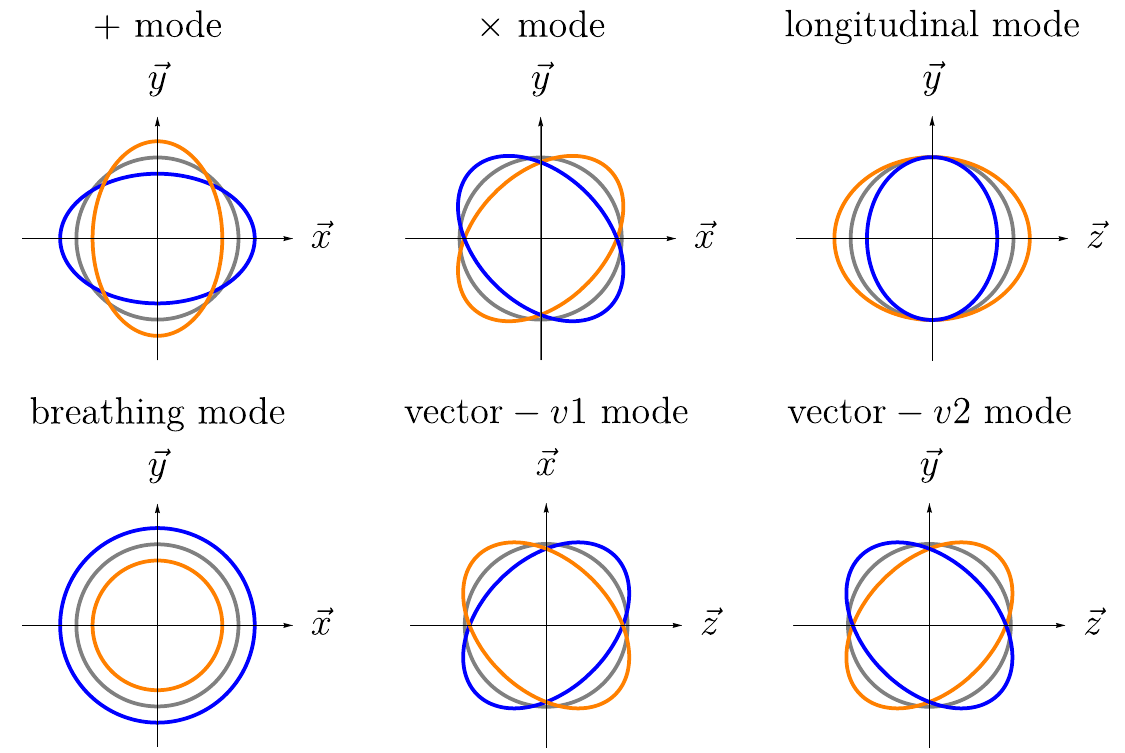}
 \caption{Illustration of how the polarizations of a GW traveling towards $\hat{z}$ affect a ring of tests particles. The gray circle represents the unperturbed ring, while the blue and orange ellipses represent the perturbed ring at different moments in time, oscillating between orange and blue shapes. Figure adapted from the pioneering work 
 \cite{Eardley:1973br} (see also  
 \cite{Will:2018bme}).}
 \label{fig:6polarizations}
\end{figure}

In the rank-2 tensor basis constructed from $(\bm{n},\bm{\theta},\bm{\phi})$, these polarization tensors take the form 
\begin{equation}\label{pol_tensors}
\begin{aligned}
&{\left[e^{+}\right]_{i j}=\left(\begin{array}{ccc}
1 & 0 & 0 \\
0 & -1 & 0 \\
0 & 0 & 0
\end{array}\right), \qquad \left[e^{v1}\right]_{i j}=\left(\begin{array}{lll}
0 & 0 & 1 \\
0 & 0 & 0 \\
1 & 0 & 0
\end{array}\right), \qquad \left[e^{l}\right]_{i j}=\left(\begin{array}{lll}
0 & 0 & 0 \\
0 & 0 & 0 \\
0 & 0 & 1
\end{array}\right), 
\quad} \\
& {\left[e^{\times}\right]_{i j}=\left(\begin{array}{lll}
0 & 1 & 0 \\
1 & 0 & 0 \\
0 & 0 & 0
\end{array}\right),
\qquad \quad \left[e^{v2}\right]_{i j}=\left(\begin{array}{lll}
0 & 0 & 0 \\
0 & 0 & 1 \\
0 & 1 & 0
\end{array}\right), \qquad \left[e^{b}\right]_{i j}=\left(\begin{array}{lll}
1 & 0 & 0 \\
0 & 1 & 0 \\
0 & 0 & 0
\end{array}\right) .}
\end{aligned}
\end{equation}
We can expand the symmetric spatial tensor $S_{ij}$ defined in Eq.~\eqref{eq:RiemannTensorWaveZone} in the  complete orthogonal basis above
\begin{align}\label{eq:Sijexpand}
S_{ij}= \sum_{p} e^{p}_{ij} S_p = e^+_{ij}\,S_++e^\times_{ij}\,S_\times+e^{v1}_{ij}\,
S_{v1}+e^{v2}_{ij}\,S_{v2}+e^b_{ij}\,S_b+e^l_{ij}\,S_l\,,
\end{align}
where $p = \{ +, \times, v1, v2,b, l\}$, defining six independent GW polarizations that can govern a general GW detector response.  This way $S_{jk}$ takes the following matrix form
\begin{equation}\label{eq:Sinz}
S_{jk}=  \begin{pmatrix}
S_b+S_+ & S_\times & S_{v1}\\
S_\times & S_b-S_+ & S_{v2}\\
S_{v1} & S_{v2} & S_l\\
\end{pmatrix}\,,
\end{equation}
as anticipated in Eq.\eqref{eq:MetricSixPolarizations}. The effects of the different polarizations on a ring of test particles can be visualized in Figure \ref{fig:6polarizations}. 

As we will understand in more detail below, these include the two familiar tensor polarizations $S_+$ and $S_\times$ from GR, as well as two additional vector polarizations $S_{v1}$ and $S_{v2}$ and two scalar polarizations $S_b$, known as the \textit{breathing} mode, and $S_l$ the \textit{longitudinal} scalar mode.

Comparing Eq.~\eqref{eqn:Apolarization} with the definitions of the six GW polarizations in Eqs.~\eqref{eq:RiemannTensorWaveZone} and \eqref{eq:Sijexpand}, one can naturally identify the following relations:
\begin{subequations}\label{eq:GWgenPoldef}
\begin{align}
S_+&\equiv\frac{1}{2}e_+^{ij}\,E^{TT}_{ij}\,,& S_\times&\equiv\frac{1}{2}e_\times^{ij}\,E^{TT}_{ij}\,,& S_b&\equiv \Theta\,,\\
S_{v1}&\equiv \frac{\theta^i}{V}\,\Xi^T_{i}\,,& S_{v2}&\equiv \frac{\phi^i}{V}\,\Xi^T_{i}\,,& S_l&\equiv \,\Theta + \frac{2}{V^2} \Phi\,,
\end{align}
\end{subequations}
where indices are lowered and raised with the background Minkowski metric. Since we are in a gauge where the perturbations are only spatial then, in practice, their indices are raised and lowered simply with a Kronecker delta $\delta_{ij}$.

While a metric theory of gravity might have up to six GW polarization modes, the actual number of polarizations that are present within a given theory depends on the equations of motion for each theory. In GR, for instance, it is the linearized Einstein equations that determine that only the TT component $E^{TT}_{ij}$ of the metric perturbation propagates. On the other hand, in a scalar-tensor theory, it might be the scalar equations of motion that determine that the additional scalar degree of freedom in the theory excites an additional scalar GR polarization. At this point, however, it is important to stress the difference between the concepts of dynamical degrees of freedom of a theory, whose number is a priori not limited
and also depends on its matter content, and the GW polarizations defined here, of which any metric theory can only have up to six. Indeed, while additional dynamical DoFs of a metric theory are necessary to excite additional GW polarizations, it can very well be that such additional DoFs do not induce any additional GW polarization depending on their coupling to the physical metric, as we show in an explicit example of Horndeski theory in Appendix~\ref{App:Horndeski plus}.

\subsection{\label{subsec:spin_weight}Spin weight}

Figure~\ref{fig:6polarizations} reveals the corresponding scalar, vector and tensor nature of the different GW polarizations: while the breathing and longitudinal scalar polarizations are invariant under rotations about the direction of propagation, the two vector modes are invariant under rotations by $2\pi$, while $+$ and $\times$ tensor modes are invariant under rotations of $\pi$ only. 

More formally, these statements can be understood by noting that for a given direction of propagation $\bm{n}$ we have a freedom in defining the transverse basis in Eqs.~\eqref{theta} and \eqref{varphi} characterized by a rotation of an additional angle $\psi$ about the axis of propagation. This freedom can conveniently be described by introducing the following complex basis of the transverse space
\begin{equation}\label{eq:Def m}
    m_i\equiv \frac{1}{\sqrt{2}}(\theta_i+i\phi_i)\,,
\end{equation}
alongside its complex conjugate
\begin{equation}\label{eq:Def mbar}
    \bar m_i= \frac{1}{\sqrt{2}}(\theta_i-i\phi_i)\,.
\end{equation}
These vectors have so called definite spin-weight $s=1$ and $s=-1$, respectively, as determined by their behavior under counterclockwise rotations about the direction of propagation
\begin{equation}\label{eq:SpinweightTransformation m}
    m'_i\rightarrow e^{i\psi}m_i\,, \quad  \bar{m}'_i\rightarrow e^{-i\psi}\bar{m}_i\, .
\end{equation}
With the aid of these vectors one can construct complex scalar quantities $S_{s}$ of the detector response matrix $S_{ij}$ with definite spin-weight $s$ in the following way. For instance, one can isolate the $+$ and $\times$ polarizations by defining the complex function $S_{{-2}}$ of definite spin-weight $s=-2$ through
\begin{equation}
     S_{{-2}}\equiv S_{ij}\bar{m}_i\bar{m}_j=\frac{1}{2}S_{ij}(e^+_{ij}-i\, e^\times_{ij})=S_+-iS_\times\,.
\end{equation}
Similarly, for vector and scalar perturbations we can define
\begin{align}
    &  S_{{-1}}\equiv \frac{1}{\sqrt{2}} S_{ij}n_{(i} \bar{m}_{j)}=S_{v1}-iS_{v2}\,,\\
    & S_{0,b}\equiv S_{ij}m_i\bar{m}_j=S_b\,,\\
    &S_{0,l}\equiv S_{ij}n_in_j=S_l\,.
\end{align}
In this form, the tensorial nature of the different polarizations is evident. Moreover, the definition of these quantities of definite spin-weight will prove very useful in the following, as they allow for an expansion in spin-weighted-spherical harmonics that explicitly takes into account the freedom in rotating the transverse plane of the GWs (see Section~\ref{subsec:Extra_Polarizations} below).

We have so far reviewed that the six GW polarizations in a generic metric theory of gravity can be written in terms of manifestly gauge invariant quantities and therefore represent physical observables. Nonetheless, global transformations can also be performed and those could mix the different polarization states. As shown in \cite{Will:2018bme}, a global rotation around the propagation direction or a boost along the propagation direction $n_i$ will not mix polarizations. However, rotations or boosts along directions other than $n_i$ will indeed mix the scalar, vector, and tensor polarizations, since the metric is being projected onto an unconventional frame where the new $n_{i}'$ does not represent the physical propagation direction. Here, we will fix these freedoms by always assuming that we are at the source rest-frame and $n_i$ is indeed the propagation direction (as assumed for all the modified gravity models that will be discussed in Section\ \ref{sec:ModGrav}). It may be possible to argue that the relative LISA-source peculiar motions always introduce frames where the source is effectively boosted, nonetheless since these relative velocities are non-relativistic, the mixing between polarizations are expected to be negligible, yet an exact quantification of these mixing effects remains to be performed.

\subsection{\label{subsec:spin_weighted_spherical_decomposition}Spin-weighted spherical decomposition}

Working again in synchronous gauge ($S_{ij} = h_{ij}$), we expand the GW on the polarization basis $h_{ij} = \sum_p e^{p}_{ij} h_p$, where $p$ is the polarization index taking values $p = \{ +, \times, v1, v2, b, l\}$.
As shown in the previous section, the quantities $ h_{+} \pm i  h_{\times}$,\, $h_{v1} \pm i  h_{v2}$, $h_{b}$, $h_{l}$ carry spin-weight of $|s| = 2,\, 1,\, 0,\,0 $, respectively. Hence, it is convenient to decompose them on a basis of spin-weighted spherical harmonics and work at the level of the multipole components.
We first use the spin weight of each polarization to define the following angular decomposition in time domain: 
\begin{align}
    h_{+}(t,\iota,\phi_c)-ih_{\times}(t,\iota,\phi_c)&=\sum_{\ell = 0}^{+\infty} \sum_{m = -\ell}^{\ell}  h_{T}^{(\ell,m)}(t)Y_{-2}^{(\ell, m)}(\iota,\phi_c), \label{tensor_lm}\\
     h_{v_1}(t,\iota,\phi_c)-i h_{v_2}(t,\iota,\phi_c)&=\sum_{\ell = 0}^{+\infty} \sum_{m = -\ell}^{\ell}   h_{V}^{(\ell,m)}(t)Y_{-1}^{(\ell, m)}(\iota,\phi_c),\\
  h_{b,l}(t,\iota,\phi_c)&=\Re \sum_{\ell = 0}^{+\infty} \sum_{m = -\ell}^{\ell}   h_{B,L}^{(\ell,m)}(t)Y^{(\ell, m)}(\iota,\phi_c)\label{scalar_lm},
\end{align}
where $Y_{s}^{(\ell,m)}$ are the spin-weighted spherical harmonics (SWSH), which are functions of $\iota$ 
(the inclination angle) and $\phi_c$ (the coalescence phase) for non-precessing binary systems.  
Note that $h_{T,V,B,L}^{(\ell,m)}(t)$ are complex quantities, whereas $h_{+,\times, v1,v2, b, l}(t)$  are real. 

Furthermore, for any source satisfying equatorial (planar) symmetry, such as non-precessing binaries, the metric is expected to satisfy the same reflection symmetries as long as the additional fields of modified gravity also preserve parity. In particular, this means that the expressions as functions of $(\pi-\iota,\phi_c )$ must also be a solution to the equations of motion. In order for that to happen, the following conditions must be satisfied (see \cite{Faye:2012we} for a more detailed discussion for tensor modes):
\begin{align}
    h_T^{(\ell,-m)}=(-1)^\ell h_T^{(\ell,m)*}\,, \label{eq:T_symmetry}\\
    h_V^{(\ell,-m)}=(-1)^{\ell-1} h_V^{(\ell,m)*}\,,\\
 h_{B,L}^{(\ell,-m)}=(-1)^\ell h_{B,L}^{(\ell,m)*}\, .\label{eq:S_symmetry}
\end{align}
Next, we can use these conditions to re-express the polarizations only in terms of $|m|$:
\begin{align}
    & h_+(t)=\sum_{\ell, |m|} h_T^{(\ell, |m|)}Y_+^{(\ell,|m|)}+\text{c.c.}\,,\quad 
    h_\times(t)=\sum_{\ell, |m|} h_T^{(\ell, |m|)}Y_\times^{(\ell,|m|)}+\text{c.c.}\,,\\
    & h_{v1}(t)=\sum_{\ell, |m|} h_V^{(\ell, |m|)}Y_{v1}^{(\ell,|m|)}+\text{c.c.}\,,\quad 
    h_{v2}(t)=\sum_{\ell, |m|} h_V^{(\ell, |m|)} Y_{v2}^{(\ell,|m|)}+\text{c.c.}\,,\\
     & h_{b,l}(t)=\sum_{\ell, |m|} h_{B,L}^{(\ell, |m|)}Y_s^{(\ell,|m|)}+\text{c.c.}\,,
\end{align}
where c.c.\ stands for complex conjugate and we have defined 

\begin{subequations}\label{eq:SWSHPols}
\begin{align}
    &Y_+^{(\ell,|m|)}=\frac{1}{2}\left(Y_{-2}^{(\ell,|m|)}+(-1)^\ell Y_{-2}^{(\ell,-|m|)*} \right)\,,\\
    &Y_\times^{(\ell,|m|)}=\frac{i}{2}\left(Y_{-2}^{(\ell,|m|)}-(-1)^\ell Y_{-2}^{(\ell,-|m|)*} \right)\,,\\
    & Y_{v1}^{(\ell,|m|)}=\frac{1}{2}\left(Y_{-1}^{(\ell,|m|)}+(-1)^{\ell-1} Y_{-1}^{(\ell,-|m|)*} 
    \right)\,,\\
    & Y_{v2}^{(\ell,|m|)}=\frac{i}{2}\left(Y_{-1}^{(\ell,|m|)}-(-1)^{\ell-1} Y_{-1}^{(\ell,-|m|)*} 
    \right)\,,\\
   & Y_s^{(\ell,|m|)}= \frac{1}{2}\left(Y^{(\ell,|m|)}+(-1)^{\ell} Y^{(\ell,-|m|)*} 
    \right)\,.
\end{align}
\end{subequations}
These expressions assume a specific choice of the polarization angle $\psi$, as mentioned in Section \ref{subsec:spin_weight}. For a general polarization angle, see Appendix \ref{app:rotations}. In addition, notice that these expressions and the planar symmetry conditions hold for modes with $m\not=0$, which is enough for practical purposes since $m=0$ modes will not be included in the frequency-domain waveforms used in this paper. In general, the $m=0$ modes are primarily associated to the phenomenon of gravitational displacement memory, describing a non-oscillatory component within the asymptotic radiation that leaves behind a permanent shift in proper distances after the passage of the radiation. Given that LISA is expected to provide the first single-event detection of gravitational memory \cite{Inchauspe:2024ibs}, it would be interesting to include these modes in future considerations.

\subsection{\label{subsec:waveform_frequency_domain}Waveform in frequency domain}
In GW data analysis, one often works with the Fourier transform of the signal since detectors have colored noise.  
Each polarization is then expressed as  \footnote{We highlight that we follow the conventions of \cite{Marsat:2020rtl} but other papers assume $-f$ \cite{Piarulli:2025rvr,Garcia-Quiros:2020qpx}. }: 

\begin{equation}\label{eq:WaveFormFourier}
    \tilde{h}_p(f) = \int_{-\infty}^{\infty}  h_p(t) e^{2\pi i f t} dt \quad \text{or, equivalently,} \quad h_p(t) =
    \int_{-\infty}^{\infty}  \tilde{h}_p(f) e^{-2\pi i f t} df .
\end{equation}
Note that since each polarization $p=+,\times, v1, v2, b, l$ is described by a real function in time domain, one will have that $ \tilde{h}_p(f)= \tilde{h}_p(-f)^*$ and can then re-write the Fourier domain expression as an integral over positive frequencies only:
\begin{eqnarray}
   h_p(t) =  \int_{0}^{\infty}  \tilde{h}_p(f) e^{-2\pi i f t} df + \text{c.c.}  \label{Eq:FourierInt}
\end{eqnarray}
In order to obtain the waveform in Fourier space, one has to compute Eq.~\eqref{eq:WaveFormFourier}. This is usually achieved under the Stationary Phase Approximation (SPA) \cite{Droz:1999qx}, which assumes that the GW is composed by a slowly varying real amplitude $ A_p(t)$, times a rapidly oscillating phase $\Phi(t)$. Given that $h_p(t)$ is real, it will be written as $h_p(t) = A_p(t) \, (e^{i \Phi(t)} + e^{-i \Phi(t)})$. This means that the Fourier integral is of the form
\begin{equation}
    \tilde{h}_p(f) = \int_{-\infty}^{\infty}  A_p(t) \left[  e^{2\pi i f t - i \Phi(t)}  + e^{2\pi i f t + i \Phi(t)} \right] dt  \label{eq:spa}.
\end{equation}
Solving this integral through the SPA, selects only the first of the two terms in the square bracket when $f>0$, since the second one does not have a stationary point, i.e.\ a value of $t$ such that $2\pi f  + \dot{\Phi}(t) =0$. Terms without stationary points contribute subdominantly to the Fourier integral hence they can be neglected~\cite{Chatziioannou:2012rf}. 

After obtaining $ \tilde{h}_p(f)$ one can then use Eqs.\ (\ref{tensor_lm})-(\ref{scalar_lm}) to obtain the 
frequency-domain expressions for the harmonics coefficients of the decomposition over spin-weighted spherical harmonics as:
\begin{align}
  \tilde{h}_+(f)-i \tilde{h}_\times(f) &= \sum_{\ell, m}\tilde{h}_T^{(\ell,m)}(f)Y^{(\ell,m)}_{-2}, \\
   \tilde{h}_{v1}(f)-i \tilde{h}_{v2}(f) &= \sum_{\ell, m} \tilde{h}_V^{(\ell,m)}(f) Y^{(\ell,m)}_{-1}, \\
  \tilde{h}_{b,l}(f)&= \sum_{\ell, m} \tilde{h}_{B,L}^{(\ell,m)}(f)Y^{(\ell,m)},
\end{align}
and from now on we will focus on modeling the functions $\tilde{h}_T^{(\ell,m)}(f)$, $\tilde{h}_V^{(\ell,m)}(f)$ and $\tilde{h}_{B,L}^{(\ell,m)}(f)$ in modified gravity. Note that, due to Eqs.\ \eqref{eq:T_symmetry}-\eqref{eq:S_symmetry} we can focus solely on modeling $m>0$ modes as the $m<0$ modes can be derived from them. As discussed in \cite{Marsat:2020rtl}, one can furthermore choose a Fourier convention where  the functions $\tilde{h}_P^{(\ell,|m|)}(f)$ have support only in $f>0$, whereas the functions $\tilde{h}_P^{(\ell,-|m|)}(f)$ are inferred to have support for $f<0$.


\section{\label{sec:Detecting_extra_polarizations}Detector response to GW polarizations}

\begin{figure}[h!]
\centering
\includegraphics[width = 0.6\textwidth]{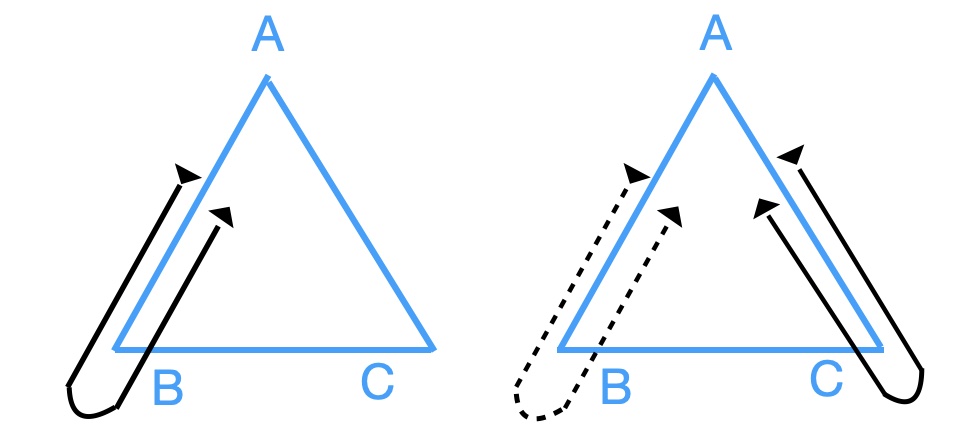}
 \caption{
 LISA consists of three spacecraft, labeled $A$, $B$, and $C$, arranged in an equilateral triangle with constant arm length \( L \).
\textbf{Left:} The solid line illustrates a single round-trip light path from spacecraft $A$ to $B$, as described in Section \ref{subsec:LISA_Antenna_Response}.
\textbf{Right:} The solid and dashed lines represent the two round-trip light paths that form the Michelson interferometer signal, discussed in Section \ref{subsec:aet_channels}.}
 \label{fig:channels}
\end{figure}

The space-based interferometer LISA offers several advantages over its ground-based counterparts in the search for possible  GW extra polarizations. In particular, LISA’s orbital motion relative to the Earth provides characteristic modulation effects (see e.g.\ \cite{Yunes:2009ke,Takeda:2019gwk}) and better sky localization, while its response functions exhibit a distinct dependence on high-frequency regimes that can help discriminate between different polarization states. In this section, we briefly review LISA’s response to GWs  and demonstrate its sensitivity to distinct polarization modes.
This section is organized as follows. 
In Section \ref{subsec:LISA_Antenna_Response}, we establish the theoretical framework and rigorously derive how GWs affect the geodesic of a photon propagating between two spacecrafts, $A$ and $B$. In Section \ref{subsec:aet_channels}, we apply this formalism to the LISA configuration, analyzing two photon round trips between three detectors, $A$, $B$, and $C$, positioned at the vertices of an equilateral triangle. A schematic representation of the system is shown in Figure \ref{fig:channels}.

\subsection{\label{subsec:LISA_Antenna_Response}One-way Doppler shift}

The one-way Doppler response to a GW characterized by six polarizations is derived in \cite{Tinto_2010} and reproduced here for convenience. 
With LISA in mind, we want to obtain the leading order effect of GWs on the frequency of light traveling between two spacecrafts, and establish a linear relation between the resulting Doppler response and the GW signal. In doing so, we will assume that the distance between the spacecrafts is small in comparison with the distance to the source, such that the spherical wavefront of GWs can be approximated by a superposition of plane-waves with fixed direction $\hat k$. 
Moreover, we assume that the each mode in the plane-wave expansion satisfies a fixed dispersion relation $\omega=V(k) \, k$, where by definition $V$ is the phase velocity of the wave, that may depend on $k$. 

For simplicity we will consider that every polarization mode travel at the same speed. 
Without loss of generality, let us consider a GW propagating in the $z$-direction in a coordinate system centered at the spacecraft $A$, and a second spacecraft $B$ located at a coordinate distance $L$ (see Figure~ \ref{fig:channels}). The perturbed spacetime takes the form
\begin{equation}
    ds^2=-dt^2+\left[\delta_{ij} +h_{ij}\left(t-\frac{z}{V}\right)\right]dx^i dx^j\,, 
    \label{eq:metric pert}
\end{equation}
with \(|h_{ij}| \ll1\) and where \(\{t,x,y,z\}\) are the background Minkowski coordinates, and the GW $h_{ij}$ can be expanded on the polarization basis  in Eq.~\eqref{pol_tensors}.

Observe that by definition the coordinate distance $L$ between the spacecrafts remains constant at all times within the synchronous gauge, as the coordinates correspond to comoving coordinates. 
While the classic treatment of the effect of GWs on two freely falling test-masses is based on a study of the geodesic deviation equation, such an analysis receives corrections at the high-frequency end of the LISA sensitivity spectrum. This is because the geodesic deviation equation is only valid up to corrections of the order $L/\lambda$, where $\lambda$ is the typical wavelength of the radiation \cite{Maggiore:2007ulw}. Thus, in order to correctly capture the high-frequency spectrum of LISA with wavelengths shorter than the LISA arms, another, more general treatment is required. A rather straightforward method of directly obtaining a Doppler response due to the presence of radiation goes back to W.L. Burke \cite{Burke:1975zz} (see also \cite{Estabrook:1975jtn,Tinto_2010}) that we will follow below.

Consider a photon emitted at spacecraft $A$ in the direction of spacecraft $B$. Its 4-momentum in terms of the metric perturbation at spacecraft $A$ (denoted by $h^{(A)}_{\rho\xi}$) can be written in the form
\begin{equation}\label{eq:Momentum0}
    P^{(A)}_{\rho}=\nu_A\left(-\delta^t_{\rho}+\ell^{AB}_\rho+\frac{1}{2}h^{(A)}_{\rho\xi}\ell_{AB}^{\xi}\right),
\end{equation}
where $\nu_A$ is the photon's frequency at emission and \( {\ell}^{AB}_\rho \) describes the spatial vector at $A$ pointing in the direction of spacecraft $B$, such that $\ell^{AB}_{t}=0$ and $\ell^{AB}_{i}\ell^{AB}_{j}\delta^{ij}=1$. In terms of the azimuthal and polar angles \((\theta,\phi)\) of the coordinate system centered at $A$, this spatial vector takes the form
\begin{equation}
\ell_{\rho}^{AB}=\sin\theta \cos\phi \ \delta^x_{\rho} + \sin\theta \sin\phi \ \delta^y_{\rho} + \cos\theta \ \delta^z_{\rho}.
    \label{eq:nvector}
\end{equation}
In the following, we will denote this spatial vector as
\begin{equation}
\ell_{\rho}^{AB}\equiv \hat \ell_\rho\,,
\end{equation}
and also define its upper index counterpart as $\hat\ell^\rho=\eta^{\rho\sigma}\hat\ell_\sigma$.
The last term in Eq.~\eqref{eq:Momentum0} is proportional to the GW at some particular emission time $t$, and consists of the perturbative addition necessary to retain the null property $P_\rho P^\rho=0$ of the 4-momentum within the perturbed spacetime \eqref{eq:metric pert}. Since we choose the spatial origin to be at spacecraft $A$, we have that
\begin{equation}\label{eq:h0}
    h^{(A)}_{\rho\xi}=h_{\rho\xi}(t)\,,
\end{equation}
in terms of the metric perturbation defined in Eq.~\eqref{eq:metric pert}.
Moreover, note that an observer at spacecraft $A$ with unit tangent vector $O^\rho=-\delta_t^{\rho}$ indeed measures a frequency of $O^\rho P_\rho=\nu_A$.

Similarly, the 4-momentum of the photon as it arrives at spacecraft $B$ can again be written as 
\begin{equation}\label{eq:momentumP1}
    P^{(B)}_{\rho}=\nu_B\left(-\delta^t_{\rho}+\ell^{BA}_{\rho}+\frac{1}{2}h^{(B)}_{\rho\xi}\ell_{BA}^{\xi}\right),
\end{equation}
where the GW signal arrives at a later time, as
\begin{equation}\label{eq:h1}
    h^{(B)}_{\rho\xi}=h_{\rho\xi}\left(t+T/2-\frac{\hat{k}\cdot\hat{\ell}}{V}\,T/2\right)\,,
\end{equation}
with $T \equiv 2L$ denoting the round-trip
light travel time between spacecraft $A$ and $B$. At leading order, the GW signal at $B$ is received at coordinate time $t + T/2$ and position $z = \hat{k} \cdot \hat{\ell}\, T/2$.
Meanwhile, the frequency $\nu_B$ and direction vector $\ell_{BA}$ at $B$, satisfying $\ell^{BA}_i \ell^{BA}_j\delta^{ij} = 1$, are formally obtained by parallel transporting the four-momentum $P^{(A)}_{\rho}$ from $A$ to $B$ along the photon worldline, defined as an autoparallel trajectory
\begin{equation}\label{eq:Autoparallels}
    (\nabla_\xi P_{\rho})P^\xi=0\,.
\end{equation}
A convenient shortcut in computation is however provided by noting that the spacetime given by Eq.~\eqref{eq:metric pert} admits three Killing vectors, which expressed in a linear combination read
\begin{equation}
K^{\rho}=a_1\delta_x^{\rho}+a_2\delta_y^{\rho}+a_3\left(\frac{\delta_t^{\rho}}{V}+\delta_z^{\rho}\right),
    \label{eq:Kvector}
\end{equation}
where \(a_1, a_2, a_3\) are constants. Indeed, it is readily verified that this vector satisfies the Killing equation
\begin{equation}\label{eq:KillingEq}
    \nabla_\xi K_{\rho}+ \nabla_\rho K_{\xi}=0\,,
\end{equation}
for the perturbative plane-wave radiation that satisfy
\begin{equation}
    \partial_t h_{ij}\left(t-\frac{z}{V}\right)=-V\partial_z h_{ij}\left(t-\frac{z}{V}\right)\,.
\end{equation}
Note that by comparing Eqs.~\eqref{eq:nvector} and \eqref{eq:Kvector} and setting $(a_1,a_2,a_3)=C\hat{\ell}$, where $C$ is an arbitrary constant, Eq.~\eqref{eq:Kvector} can also be written as
\begin{equation}
    K^{\rho}=C\left[\hat\ell^{\rho}+\frac{\hat{k}\cdot\hat{\ell}}{V}\,\delta_t^{\rho}\right]\,.
\end{equation}
Thus, Eqs.\ \eqref{eq:Autoparallels} and \eqref{eq:KillingEq} imply that
\begin{equation}
    \nabla_\xi(P_{\rho}K^\rho)P^\xi=\partial_\xi(P_{\rho}K^\rho)P^\xi=0\,.
\end{equation}
Since this must hold for any $P^\xi$, we have that $P_{\rho}K^{\rho}=constant$ everywhere along the photon's worldline, resulting in particular in the relationship
\begin{equation}\label{eq:Killing Equation Parallel Transport}
P^{(A)}_{\rho}K_{(A)}^{\rho}=P^{(B)}_{\rho}K_{(B)}^{\rho}\,.
\end{equation}

Notice that in writing $\ell^{BA}_{\rho}$ defined at $B$ in Eq.~\eqref{eq:momentumP1} we have formally allowed for a non trivial shift in the direction vector defined at the origin as it is parallel transported to the location of spacecraft $B$. Clearly, the leading order of such a modification $\ell^{BA}_{\rho}=\hat\ell_{\rho}+\delta\ell_\rho$ needs to be of the order of the perturbation field $h_{\rho\xi}$. However, since the Killing vectors in \eqref{eq:Kvector} retain their form throughout our spacetime patch, the Killing equations $\delta\ell_{\rho}K^{\rho}=0$ imply that $\delta\ell_{x}=\delta\ell_{y}=0$ and  $\delta\ell_{t}+\delta\ell_{z}=0$. On the other hand, the definition of the direction vectors requires $\ell^{BA}_{i}\ell^{BA}_{j}\delta^{ij}=1=\hat\ell_i\hat \ell_j\delta^{ij}+2\delta\ell_{i}\hat\ell_j\delta^{ij}+\mathcal{O}(h^2)$ and therefore to leading order also $\delta\ell_z=0$. Thus, the parallel transport of the direction vector is in fact trivial and we can write $\ell^{BA}_{\rho}=\hat\ell_\rho$.

From \eqref{eq:Killing Equation Parallel Transport} it then follows that the ratio between the frequencies at spacecraft $A$ and $B$ can be written as
\begin{equation}
\frac{\nu_B}{\nu_A}=\frac{1-\hat{k} \cdot \hat{\ell}/V+\frac{1}{2} \ell^\rho h^{(A)}_{\rho \xi} \ell^{\xi}}{1-\hat{k} \cdot \hat{\ell}/V+\frac{1}{2} \ell^\rho h_{\rho \xi}^{(B)} \ell^{\xi}}.
\end{equation}
Expanding this expression to first order in the perturbation, while using equations \eqref{eq:h0} and \eqref{eq:h1}, we obtain the result for the one-way Doppler response
\begin{equation}\label{eq:DopplerResponse1}
    \Delta\nu(t) \equiv \frac{\nu_B-\nu_A}{\nu_A} =\frac{\ell^i \ell^j h_{ij}(t)-\ell^i \ell^jh_{ij}(t+[1-\hat{k} \cdot \hat{\ell}/V\,]L) }{2(1-\hat{k} \cdot \hat{\ell}/V)}\,.
\end{equation}
Note that this expression is well-defined, since even for a non-transverse wave traveling along a detector arm $\hat{k} \cdot \hat{\ell}=V$ the numerator still vanishes together with the denominator. The actual limit results in
\begin{equation}
    \lim_{\hat{k} \cdot \hat{\ell}\,\rightarrow V}\Delta\nu(t)=-L\, \ell^i \ell^j \dot h_{ij}(t)\,.
\end{equation}
Thus, while the Doppler response for a wave traveling along the detector arm vanishes for a transverse GW as expected, the longitudinal components still contribute.
Additionally, note that for GWs traveling at the speed of light, $V=1$, this coincides with the result in~\cite{Tinto_2010}.

\subsection{\label{subsec:aet_channels}Detector response function: A, E and T channels}
In the context of time delay interferometry, another key quantity is the phase shift, $\Delta \varphi(t)$ of the laser beam at the vertex, which measure the change in the proper distance along the arm of the detector. This is related to the one-way Doppler response through 
\begin{align}
    \Delta \varphi(t) = \int^t_0 d t' \Delta \nu(t') \,.
\end{align}
By comparing the phase difference between the light that makes a round trip between two vertices, and the one stored in the initial laser cavity, one could obtain the measurement of a GW. However, this single-arm interferometer is not feasible due to the presence of laser noise and the intrinsic fluctuations of the detectors' arms. 
To remove the noise, it is best to consider combinations of the photons signals that travel along different arms of the detector~\cite{Cornish:2001bb,Vallisneri:2005ji,Smith:2019wny,Armstrong_1999,Tinto:2020fcc}. 
Therefore, the actual phase difference observed at a LISA interferometer vertex, $\Phi$, can be expressed as the sum of the gravitational response, represented by the interferometric phase shift $\Delta\varphi$, and the noise contribution, ${\cal N}$,
\begin{equation}
\Phi_{A_{BC}}(t)=\Delta \varphi_{A_{BC}}(t)+{\cal N}_{A_{BC}}(t)\,,
\end{equation}
where the subscript ${A_{BC}}$ denotes the signal detected in the interferometer formed by arms $AB$ and $AC$. See Fig~\ref{fig:channels}.

Following~\cite{Smith:2019wny}, we can relate the signal part of the phase shift to the GW as 
\begin{equation}
\Delta\varphi_{A_{BC}}(t) = \int_{-\infty}^\infty df \int d^2 \bn \sum_{p} {h}_p(f,\bn)e^{i 2\pi f t} F^p_{A_{BC}}(\bn,f;t)\,,\label{eq:int_phase}
\end{equation}
with 
 $p$ indicating the six polarization 
indexes $p = ( +, \times, v1, v2, b, l)$
and $h_p(f,\bn)$ the Fourier component of the GW, such that $h^*_p (f, \bn)=h_p (-f, \bn)$,  and
\cite{Cornish:2001bb}
\begin{align}
F^p_{A_{BC}}(\bn,f;t) &=  \frac{1}{2} e^{-i 2\pi f\bn \cdot \vec x_A(t)}e_{ab}^p(\bn) \left[ \mathcal{F}^{ab}(\hat{\ell}_{AB}(t) \cdot \bn,f) -\mathcal{F}^{ab}(\hat{\ell}_{AC}(t) \cdot \bn,f)\right] \label{eq:gain1} \\
\mathcal{F}^{ab}(\bell \cdot \bn,f) &=
\frac{W(f)\bell^a \bell^b}{2} \Big\{ {\rm sinc}\left[\frac{f}{2 f_*}(1-\bell\cdot\bn)\right] e^{-i\frac{f}{2f_*}(3+\bell\cdot\bn)} 
\nonumber\\
& \hskip2.5cm
+ {\rm sinc}\left[\frac{f}{2 f_*}(1+\bell\cdot\bn)\right] e^{-i\frac{f}{2f_*}(1+\bell\cdot\bn)} \Big\}\,.
\label{eq:gain2}
\end{align}
We use $\hat{\ell}_{AB}$ to indicate a unit vector
pointing from vertex $A$ to vertex $B$, 
as in Section~\ref{subsec:LISA_Antenna_Response}, while the quantity $f_* = 1/(2\pi L)$ denotes the characteristic frequency associated with the interferometer arm length $L$. The unit vector $\boldsymbol{n}$ specifies the direction of GW propagation. 
For $W = 1$, the above expressions fully describe the round-trip paths $ABA$ and $ACA$, corresponding to the Michelson interferometer configuration, which will be the focus of our analysis in this section (see Figure~\ref{fig:channels}, right panel).
To further suppress noise sources, time-delay interferometry (TDI) incorporates longer paths, in which the accumulated phase follows the same functional form but includes a time-delay transformation $t \rightarrow t - 2L$. In the Fourier domain, this delay introduces a frequency-dependent phase shift. By choosing the transfer function $W(f) = 1 - e^{-2if/f_*}$, one ensures that the contributions from full round trips in the TDI signal are properly accounted for. For a comprehensive review of time-delay interferometry, see e.g.~\cite{Tinto:2020fcc}. 
 %
Notice that the time dependence of Eq.~\eqref{eq:int_phase} is all contained in the instrument response functions.

\begin{figure}[h!]
\centering
\includegraphics[width = 1.\textwidth]{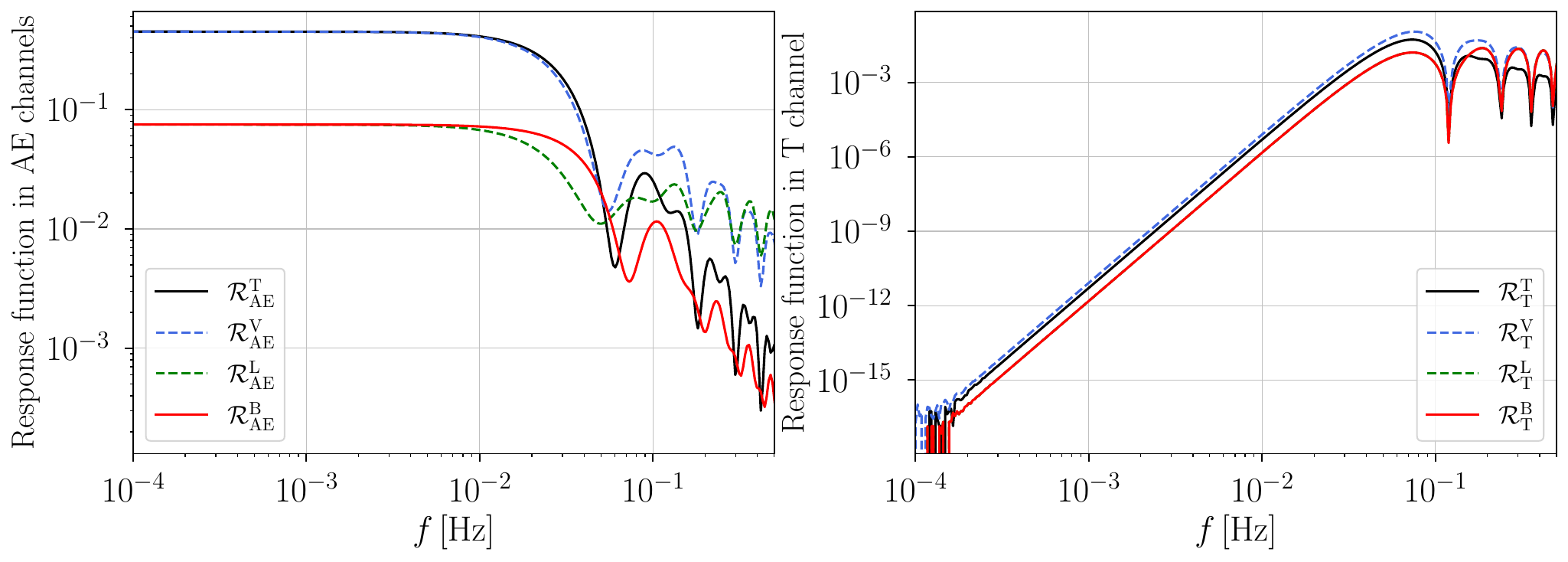}
 \caption{Combination of sky-averaged response functions in the AE channels (left) and T channel (right) for different polarizations.}
 \label{fig:responsefcomp}
\end{figure}

To understand which combination of interferometers gives the cleanest information, we analyze the correlation between the responses at different vertices of the detector constellation. 
Assuming that the signal and the noise are uncorrelated, the total response consists of the sum of the GW signal and noise,  
\begin{eqnarray}\label{2point}
\langle \Phi_{A_{BC}}(t_1) \, \Phi_{X_{YZ}}(t_2) \rangle& =& \frac{1}{2} \int_{-\infty}^\infty df 
\Big[ \mathcal{R}^{\rm tens}_{A_{BC},X_{YZ}}(f;t_1,t_2) {I}_T(f) + \mathcal{R}^{\rm vec}_{A_{BC},X_{YZ}}(f;t_1,t_2) {I}_V(f) 
\nonumber
\\
&&+
 \mathcal{R}^{\rm br}_{A_{BC},X_{YZ}}(f;t_1,t_2) {I}_b(f) + \mathcal{R}^{\rm long}_{A_{BC},X_{YZ}}(f;t_1,t_2) {I}_l(f) 
 + N_{A_{BC},X_{YZ}}(f) \Big]\,,
 \nonumber
\\
\end{eqnarray}
where  $N_{A_{BC},X_{YZ}}(f)$ is the correlated noise spectrum between the arm combinations of the detector. We  combine together the contributions of the two tensor and two vector degrees of freedom, and we  introduce the isotropic intensities of tensor, vectors and two scalar sectors as
\begin{eqnarray}\label{eq:correlator}
\sum_{p=+,\times }\langle h_p^* (f,  \bn) h_{p} (f',  \bn')\rangle&=&
\,\delta(f-f')\,\frac{\delta^{(2)} (\bn-\bn')}{4 \pi}\,I_T(f)\,,
\\
\sum_{\lambda=v1,v2 }\langle h_\lambda^* (f,  \bn) h_{\lambda} (f',  \bn')\rangle&=&
\,\delta(f-f')\,\frac{\delta^{(2)} (\bn-\bn')}{4 \pi}\,I_V(f)\,,
\\
\langle h_b^* (f,  \bn) h_{b} (f',  \bn')\rangle&=&
\,\delta(f-f')\,\frac{\delta^{(2)} (\bn-\bn')}{8 \pi}\,I_b(f)\,,
\\
\langle h_l^* (f,  \bn) h_{b} (f',  \bn')\rangle&=&
\,\delta(f-f')\,\frac{\delta^{(2)} (\bn-\bn')}{8 \pi}\,I_l(f)\,,
\end{eqnarray}
In Eq.~\eqref{2point}, the intensity response  functions are given by
\begin{eqnarray}
\mathcal{R}^{\rm tens}_{A_{BC},X_{YZ}} &=&
 \int \frac{d^2 {\bf n}}{4\pi}e^{2 \pi i f (t_1-t_2)}\left[F^+_{A_{BC}}(\bn,f;t_1)F^{*+}_{X_{YZ}}(\bn,f;t_2)+F^\times_{A_{BC}}(\bn,f;t_1)F^{*\times}_{X_{YZ}}(\bn,f;t_2)\right],
\label{eqn:rint1}
\nonumber
\\
\mathcal{R}^{\rm vec}_{A_{BC},X_{YZ}} &=&
 \int \frac{d^2 {\bf n}}{4\pi}e^{2 \pi i f (t_1-t_2)}\left[F^{v_1}_{A_{BC}}(\bn,f;t_1)F^{*v_1}_{X_{YZ}}(\bn,f;t_2)+F^{v_2}_{A_{BC}}(\bn,f;t_1)F^{*v_2}_{X_{YZ}},(\bn,f;t_2)\right],
\label{eqn:rint2}
\nonumber
\\
\mathcal{R}^{\rm br}_{A_{BC},X_{YZ}} &=&
 \int \frac{d^2 \bn}{4\pi}e^{2 \pi i f (t_1-t_2)}
 F^{b}_{A_{BC}}(\bn,f;t_1)F^{*b}_{X_{YZ}}(\bn,f;t_2),
\label{eqn:rint3}
\nonumber
\\
\mathcal{R}^{\rm long}_{A_{BC},X_{YZ}} &=&
 \int \frac{d^2 \bn}{4\pi}e^{2 \pi i f (t_1-t_2)}
 \,
 F^{l}_{A_{BC}}(\bn,f;t_1)F^{*l}_{X_{YZ}}(\bn,f;t_2),
\label{eqn:rint4}
\end{eqnarray}
which are functions of $(f;t_1, t_2)$. With these expressions, we can look for combinations of different interferometers which decorrelate the noise contribution. These are called $A, E, T$ orthogonal channels.
Consider the space spanned by the arm combinations $(I,J)=(A_{BC}$, $B_{CA}$, $C_{AB})$. 
  The correlation matrix $\langle \Phi_I (t_i)\Phi_J (t_j)\rangle$  is symmetric under the interchange of $I\leftrightarrow J$. Its structure can be schematically written as 
 \begin{equation}
\langle \Phi_I (t_i) \Phi_J (t_j)\rangle=\left( 
\begin{matrix}
C_1 & C_2 & C_2  \\
C_2 & C_1 & C_2 \\
C_2 & C_2 & C_1 
\end{matrix}
\right)
\,,
\label{eq:genmac}
\end{equation}
since LISA is an equilateral triangle (to first approximation). The orthogonal channels are obtained by diagonalizing the previous
matrix. The eigenvectors of Eq.~\eqref{eq:genmac} are
\begin{eqnarray}
\Phi_A&=&
\frac{1}{\sqrt{6}} \left(\Phi_{A_{BC}}-2  \Phi_{B_{CA}} +\Phi_{C_{AB}} \right)\,,
\\
\Phi_E&=&
\frac{1}{\sqrt{2}} \left(\Phi_{A_{BC}}-\Phi_{C_{AB}} \right)\,,
\\
\Phi_T&=&
\frac{1}{\sqrt{3}} \left(\Phi_{A_{BC}}+  \Phi_{B_{CA}} +\Phi_{C_{AB}} \right)\,,
\end{eqnarray}
while its eigenvalues result
\begin{eqnarray}
C_A&=&C_E\,=\,C_1-C_2\,,
\label{eq:reschA}
\\
C_T&=&C_1+2C_2\,.
\label{eq:reschT}
\end{eqnarray}
The structure of the previous eigenvalues indicate how to build the orthogonal
response functions for each channel and each sector.  Given Eq.\ \eqref{eq:reschA},
the detector response in the $A$, $E$ channel is identical. We show the results
in Figure\ \ref{fig:responsefcomp}. 
Similar results,
also using a somehow different notation, have been found in \cite{Blaut:2012zz,Liang:2019pry,Zhang:2019oet}.

For simplicity, in plotting those curves we do not consider the motion of the interferometer, and 
we choose the spacecraft positions as  \cite{Smith:2019wny}
\begin{eqnarray}
\vec x_A&=&[0,\,0,\,0],
\\
\vec x_B&=&L\,[1/2,\,\sqrt{3}/{2},\,0],
\\
\vec x_C&=&L\,[-1/2,\,\sqrt{3}/{2},\,0],
\end{eqnarray}
with $L\,=\,2.5 \times 10^6$ km. 
Of course, since we integrate over all GW directions, there is no lack of generality in the previous choice.
Notice that at low frequencies the sensitivity to
tensors and vectors is the same. The same property
holds for the two scalar polarizations.
Also, 
notice that at low frequencies the $T$ channel is insensitive to 
 signal in all sectors.

\smallskip

A notable feature of the $A$ and $E$ channel response functions is that, in the low-frequency limit, they exhibit identical behavior for both the tensor–vector pair and the scalar longitudinal–breathing pair of GW polarizations. For the tensor–vector case, this degeneracy arises as a consequence of integrating over all sky directions. In contrast, for the scalar longitudinal–breathing pair, the degeneracy persists even without sky averaging (see e.g.~\cite{Liang:2019pry}).
It is important to emphasize, however, that the different polarization modes become distinguishable at higher frequencies, where the response functions develop distinct frequency-dependent features.
To make this distinction explicit, we present---for the first time---analytic expansions of the response functions in the low-frequency regime. Introducing the dimensionless frequency variable \( x = f / f_\star \), we perform a Taylor expansion and obtain:
\begin{eqnarray}
{\cal R}_{\rm AE}^{\rm T}&=&\frac{9}{20}-\frac{169}{1120} x^2+\frac{85}{4032} x^4-
\frac{178273}{106444800}\,x^6,
\\
{\cal R}_{\rm AE}^{\rm V}
&=&\frac{9}{20}-\frac{6 }{35} \,x^2+\frac{55 }{2016}\,x^4-\frac{15863 }{6652800}\,x^6,
\\
{\cal R}_{\rm AE}^{\rm B}
&=&\frac{3}{40}-\frac{23 }{2240}\,x^2+\frac{x^4}{4032}+\frac{211}{53222400}\, x^6,
\\
{\cal R}_{\rm AE}^{\rm L}
&=&\frac{3}{40}-\frac{13 }{448}\,x^2+\frac{149}{28800}\, x^4-\frac{2287 }{4656960}\,x^6
\,,
\end{eqnarray}
 and 
 \begin{eqnarray}
 {\cal R}_{\rm T}^{\rm T}&=&
\frac{x^6}{4032},
\hskip 0.7cm \hskip0.7cm
{\cal R}_{\rm T}^{\rm V}
\,=\, \frac{x^6}{2520}, \hskip 0.7cm\hskip0.7cm
{\cal R}_{\rm T}^{\rm B}
\,=\,{\cal R}_{\rm T}^{\rm L}\,=\,\frac{x^6}{13440}\,.
 \end{eqnarray}
 Hence, we observe that in the $(A, E)$ channels the response to different polarizations is degenerate in the limit $f/f_\star \to 0$, with distinctions only emerging at quadratic order in $f^2/f_\star^2$. In contrast, in the $T$-channel the responses to all polarizations first enter at order $f^6/f_\star^6$. These differences in the LISA response at intermediate frequencies may therefore help in separating polarization modes, in contrast to ground-based detectors, which operate exclusively in the small-frequency regime and thus cannot exploit such frequency-dependent structure.
In Section~\ref{subsubsec:long_mode}, we analyze the small-frequency degeneracy of the LISA response for the scalar longitudinal-breathing pair in more detail and quantify its dependence on the total binary mass. Owing to the frequency content of the associated waveforms, we find that this degeneracy is unavoidable for high-mass systems, whereas binaries with total masses $M \lesssim 10^4 \msun$ can break the degeneracy and allow these scalar modes to be distinguished.

For the purpose of obtaining analytical results, in this section we focused on an idealized configuration of the LISA detector: equal arm lengths, static geometry, and sky-averaged response. We have explicitly verified that our analytical expressions are recovered in the appropriate limit of the \texttt{lisabeta}. However \texttt{lisabeta} 
allows for a more sophisticated numerical treatment and, in the following sections, we will make use of this capability to go beyond the idealized approximations implemented here.

\section{\label{sec:ppe_formalism}Parametrized formalism}
GW signals from a general metric theory of gravity can be formulated through the parametrized post-Einsteinian (PPE) method in a model-independent way \cite{Yunes:2009ke, Chatziioannou:2012rf} during the inspiral. This is an analytic formalism in which, through a limited number of parameters, the parameterized waveform captures deviations from GR in the GW's amplitude and phase, in a perturbative way. These parameters can then be mapped to specific modified gravity theories. 

We assume that waveform deviations are generated during the emission of the signal, and ignore possible additional modifications during cosmological propagation (i.e.\ the only propagation effects are those of redshift and an amplitude decay with luminosity distance). This is motivated in gravity theories that modify GR in the 
high-energy regime only, while reducing to GR in the low-energy limit where deviations are suppressed. 
Nonetheless, our approach could be extended in the future to also include propagation effects \cite{Mirshekari:2011yq, Belgacem:2019pkk, Ezquiaga:2018btd, LIGOScientific:2021sio}. We also note that a number of theories studied in the literature are equipped with screening mechanisms, which suppress modifications on high energies or short scales, in order to satisfy solar system and binary pulsar current constraints: see e.g. \cite{Jain:2010ka,Babichev:2013usa,Joyce:2014kja,Burrage:2017qrf} for reviews. 
For some modified gravity theories, deviations in the gravitational waveform emitted by binary black holes should be suppressed, given that the energies involved are much higher than those characteristic of solar system dynamics. However, this suppression is not firmly established, as most screening studies are conducted within static, spherically symmetric setups rather than more general or dynamical frameworks. Nonetheless, studies in dynamical systems \cite{Dar:2018dra,Bezares:2021dma,Shibata:2022gec,deRham:2024xxb,Cayuso:2024ppe} have revealed that screening effects can suppress scalar dipole emission in certain classes of scalar-tensor theories. It remains unclear whether such effects suppress modifications to GW waveforms in the four theories considered in this paper. In fact, the effectiveness of screening is highly dependent on the specific gravitational interactions considered, and there indeed exist theories in which the spherically symmetric solutions closely resemble those of GR, while the GW emission or cosmological time evolution remains significantly altered or unscreened~\cite{Kimura:2011dc, Babichev:2011iz, Tattersall:2017erk}. Moreover, screening can also affect the detection process itself by suppressing the coupling between the detector and additional polarizations, potentially rendering them unobservable~\cite{Garoffolo:2021fxj}. 
In this work, we adopt an agnostic perspective, ignoring the potential effect of screening mechanisms, and allow for general deviations in the emitted waveform, parameterized using the PPE formalism. 

This section is organized as follows. In Section~\ref{subsec:tensor_polarizations} we give an overview of the PPE formalism for tensor polarizations, for the quadrupole and higher angular modes. In Section~\ref{subsec:Extra_Polarizations} we extend the PPE formalism to additional polarizations, motivated by the four modified gravity theories considered in this paper. In Section~\ref{subsec:mapping_to_gravity_theories} we map the PPE parameters to specific gravity theories. Finally, in Section~\ref{subsec:numerical_implementation} we describe our numerical implementation.

\subsection{\label{subsec:tensor_polarizations}Tensor polarizations}
Originally, the PPE formalism was introduced to parameterize modifications of the plus ($+$) and cross ($\times$) polarizations, neglecting the possibility of additional polarizations,  in the dominant $(\ell=2,|m|=2)$ angular harmonic in the frequency domain as \cite{Yunes:2009ke}
\begin{eqnarray}
    \tilde{h}_{T}^{(2,2)}(f)=  \tilde{h}_{\rm GR}^{(2,2)}(f) \; (1 + \alpha u^{a}_2)\; e^{i\beta u^{b}_2}\;, \label{Eq:PPE_original}
\end{eqnarray}
where $\tilde{h}_{\rm GR}^{(2,2)}(f)$ is the GR waveform of the corresponding harmonic and $u_2$ is related to the dimensionless orbital velocity in the $|m|=2$ angular harmonic. More generally, we can define $u_m$ as
\begin{equation}
    u_{m}=\left(\frac{2\pi G_N \mathcal{M}_z f}{|m|}\right)^{1/3} \,.  
    \label{eq:ul-PPE-extra}
\end{equation} 
Here, the constant $G_N$ refers to the Newtonian gravitational constant in GR, ${\cal M}_z = (1 + z) \times\frac{(m_1 m_2)^{3/5}}{(m_1 + m_2)^{1/5}}$ denotes the redshifted chirp mass,  $z$ is the redshift of the source, $m_1$ ($m_2$) are the primary (secondary) source-frame masses and $f$ is the observed frequency. 
In Eq.~\eqref{Eq:PPE_original} the magnitudes of the deviations from GR in amplitude and phase are parametrized by $(\alpha,\beta)$, while the constants $(a,b)$ describe their scaling with frequency, encapsulated in $u_2$. While $(\alpha,\beta)$ parametrize deviations in amplitude and phase in GR and are therefore typically treated as free parameters, the constants $(a,b)$ can be derived within beyond-GR theories (see Table~1 and Table~2 in \cite{Tahura:2018zuq}).

A generalization of the PPE formalism to a general $(\ell,|m|)$ angular harmonic of the tensor modes has been proposed in \cite{Mezzasoma:2022pjb}, where deviations from GR in the tensor polarizations can be more generally expressed as
\begin{eqnarray}
    \tilde{h}_{T}^{(\ell,|m|)}(f)=  \tilde{h}_{\rm GR}^{(\ell,|m|)}(f) (1 + \alpha_{\ell m} u_m^{a_{\ell m}})e^{i(|m|/2)\beta u_m^{b}}\,,
    \label{eq:hlm_extra_pol}
\end{eqnarray}
where $\tilde{h}_{\rm GR}^{(\ell,|m|)}$ is the GR waveform for the $(\ell,|m|)$ harmonic and  $u_m$ is reported in Eq.~\eqref{eq:ul-PPE-extra}. Within this generalization, the amplitude parameter $\alpha_{\ell m}$ and the constant $a_{\ell m}$ generally depend on $\ell$ and $m$, whereas the corresponding phase parameter $\beta$ and constant $b$ are expected to be the same for all angular harmonics.

The GR waveform can be expressed in terms of an amplitude and phase as $\tilde{h}_{\rm GR}^{(\ell,|m|)}=A^{(\ell, |m|)}e^{-i\Psi_{GR}^{(m)}}$. Although all the $m$ components contribute to a certain multipole $\ell$ value, the dominant contribution in a PN series of the amplitude comes from those such that $\ell = |m|$. 
Keeping only leading terms in the amplitude, leads to the well known {\it restricted} amplitude. On the other hand, the phase depends only on the azimuthal number, $|m|$~\cite{Blanchet:2023bwj}. Therefore, we consider

\begin{align}
    A^{(\ell=|m|,|m|)}&\equiv A^{(m)} = \frac{2\pi}{(3|m|)^{1/2}}\frac{ G^2_N \mathcal{M}_z^2}{d_L} \eta^{(2-|m|)/5} u_m^{(2|m|-11)/2}, \label{Eq:h_PN_aell} \\
    \Psi^{(m)}_{GR} &= - 2 \pi ft_c + \frac{\pi}{4} - \frac{3 |m|}{256 u_{m}^5} \sum_{n = 0}^7 u_{m}^{n/3} \left( c_n^{PN} + l_n^{PN} \ln u_{m} \right)\,, \label{Eq:Psi_GR}
\end{align}
where $d_L$ is the luminosity distance, $t_c$ is the time to coalescence, $\eta = m_1 m_2/M^2$ is the symmetric mass ratio, and $M = m_1 + m_2$ is the total source-frame mass. Additionally, $c_n^{PN}$ and $l_n^{PN}$ are known as PN coefficients and are obtained through a perturbative method of the order of $\mathcal{O}(v^8)$ (or 3.5 PN order) at the characteristic velocity $v$ of the binary \cite{Buonanno_2009,Blanchet:2013haa}.

Notice that any $m=0$ contribution in the time domain waveform is not transported over in Fourier space in the SPA. This is because the phase $\Phi$ in Eq.~\eqref{eq:spa} vanishes in those cases (since the phase is proportional to $|m|$ as shown in the third term in Eq.~\eqref{Eq:Psi_GR}). Thus, the phase will not contain stationary points, and any $(\ell, m=0)$ mode is neglected. As a consequence, we will generally consider the lowest possible angular harmonic to be $\ell = |m|=1$.

\subsection{\label{subsec:Extra_Polarizations}All polarizations}

Even though in certain theories of gravity, such as Einstein-dilaton–Gauss-Bonnet \cite{Kanti:1995vq} (a special subclass of Horndeski gravity), additional polarizations are not excited at leading PN order, many other theories (including other subclasses of Horndeski) allow additional polarizations to be excited at the same or even lower PN order than the tensor modes. In such cases, a natural strategy to test GR is to search for the presence of extra polarizations, rather than restricting to deviations in the tensor sector alone. Indeed, in this work we are interested in the detection prospect of LISA for observing additional polarizations, not only modifications to the tensor modes signal. As a consequence, we want to generalize the PPE formalism to include additional polarizations. Following the work in \cite{Chatziioannou:2012rf}, we include the lowest angular harmonic contributions to each polarization mode, namely $\ell=|m|=1$ and $\ell=|m|=2$ (see Section \ref{subsec:waveform_frequency_domain}), and we will only add the leading-order PN contribution to the amplitude and phase. 
Therefore, we can express the PPE parametrized tensor, scalar and vector waveforms as
\begin{align}
    \tilde{h}_{T}^{(\ell,|m|)}(f) &=  A^{(m)} (1 + \alpha_{\ell m} u_m^{a_{\ell m}})  e^{-i\Psi^{(m)}_{GR}} \,  e^{i(|m|/2)\beta u_m^{b}} \,,\label{Eq:PPE_Tlm} \\
    \tilde{h}_{P}^{(2,2)}(f) & = A^{(2)} \,  \alpha_{P2} u_2^{a_{P}} \, e^{-i\Psi^{(2)}_{GR}} \,  e^{i 2\beta_P u_2^{b}}\, , \label{Eq:PPE_p22}\\
    \tilde{h}_{P}^{(1,1)}(f) & = A^{(1)} \,  \alpha_{P1} u_1^{a_{P}} \,  e^{-i\Psi^{(1)}_{GR}} \, e^{i \beta_P u_1^{b}}\, , \label{Eq:PPE_p11}
\end{align}
where Eq.~\eqref{Eq:PPE_Tlm} corresponds to Eq.~\eqref{eq:hlm_extra_pol} but we wrote explicitly the contribution of the GR amplitude $A^{(m)}$ and phase $\Psi^{(m)}_{GR}$. In Eqs.~\eqref{Eq:PPE_p22}-\eqref{Eq:PPE_p11} $P$ can be $\{V,B,L\}$ for the vector, breathing, and longitudinal polarizations, respectively. Note that this parametrization is based on the work in \cite{Chatziioannou:2012rf}, yet our convention for the phase is opposite in sign compared to theirs. Furthermore, we have introduced an additional amplitude prefactor in the definition of $A^{(m)}$ in Eq.\ \eqref{Eq:h_PN_aell}, so that our definition of $a_{p}$ is shifted and the definitions of $\alpha_{P1}$ and $\alpha_{P2}$ are rescaled, when compared to \cite{Chatziioannou:2012rf}.

The extra polarizations can appear in the $(\ell=|m|=1)$ or  $(\ell=|m|=2)$ angular harmonics and are defined as deformations of a GR tensor waveform, with deviation parameters $(\alpha_{P1}, \alpha_{P2}, \beta_{P})$. In addition, the modified frequency evolution for the extra polarizations is parametrized with $(a_P,b)$ in  Eqs.~\eqref{Eq:PPE_p22}-\eqref{Eq:PPE_p11}, where $a_P$ is assumed to be the same for both angular harmonics $(\ell=|m|=1)$ and  $(\ell=|m|=2)$ of extra polarizations, as found in the gravity theories analyzed in Section \ref{sec:ModGrav}, but not necessarily the same as the tensor parameter $a$. However, $b$ is found to be the same in the extra polarizations as in the tensor polarization in Eq.~\eqref{Eq:PPE_Tlm}. Moreover, in all the cases considered here we find that $2\beta_P=\beta$. This arises because the phase evolution of all polarizations is primarily determined by the binary's orbital dynamics (but note that this relation may not always hold in beyond-GR theories that we do not consider in this work), and because we are ignoring propagation effects or assuming all polarizations to propagate at similar speeds. 
Consequently, from now we do not treat $\beta_P$ as an independent parameter and use a single parameter $\beta$ to describe the phase modifications in the scalar, vector, or tensor sectors. Under this assumption, the presence of additional polarizations with $\beta \neq 0$  necessarily implies a phase modification in the tensor mode.
The converse, however, does not hold: the tensor phase may be modified even
in the absence of extra polarizations, provided 
 $\alpha_{P1}=\alpha_{P2}=0$. In other words, any phase modification in the scalar or vector sectors will also manifest in the tensor polarization, but not necessarily the other way around.
Still related to this point, from Eqs.~\eqref{Eq:PPE_p22}–\eqref{Eq:PPE_p11} it is clear that meaningful constraints on phase modifications in the scalar and vector sectors require nonzero amplitude parameters $\alpha_{P1}$ and $\alpha_{P2}$ and the values of $\alpha_{P1}$ and $\alpha_{P2}$ themselves will affect the constraints on $\beta$. We detail how we choose $\alpha_{P1}$ and $\alpha_{P2}$ for these cases in Section \ref{subsec:gr_injections}.

According to~\cite{Chatziioannou:2012rf}, the form of Eqs.\ \eqref{Eq:PPE_Tlm}-\eqref{Eq:PPE_p11} can be understood by symmetry and dimensional reasoning. The time-domain waveform of the $(\ell,|m|=\ell)$ harmonic should scale as
\begin{align}
    h_p^{(\ell=|m|,|m|)}(t) \sim \frac{\mu}{d_L} v^{|m|} e^{-im\Phi}\, , \label{Eq:h_PN_TD}
\end{align}
where $\Phi$ is the time-domain orbital phase and $v$ is the dimensionless binary velocity $v=r\omega_{\rm orb}$ with $\omega_{\rm orb}$ being the orbital velocity and $r$ the binary radius.
The two-body reduced mass $\mu=m_1m_2/M = \eta M$ is introduced to cancel the dimension of $d_L$ in geometrical units, and is chosen because it is a symmetric combination of the component masses and vanishes when either of the component masses vanishes. 
The frequency-domain waveform can be sketched by applying the SPA to Eq.\ \eqref{Eq:h_PN_TD}. Taking account of power-law modifications to the binary binding energy and the power of radiation, one eventually arrives at
\begin{align}
    \tilde{h}_p^{(\ell,|m|=\ell)}(f)\sim \frac{\mathcal{M}_z^2}{|m|^{1/2}d_L} \eta^{(2-|m|)/5} u_m^{(2|m|-11)/2} e^{-i\Psi^{(m)}_{\rm GR}} (1+\delta A^{(m)}) e^{i\delta \Psi^{(m)}}, \label{Eq:h_PN_full_SPA}
\end{align}
at the leading-PN order, where $(\delta A^{(m)},\delta \Psi^{(m)})$ are small power-law deformations.  In GR, the $|m|=1$ mode is typically suppressed for nearly equal-mass binaries, and the amplitude takes $|m|=2$ as the leading-PN mode. In modified gravity, however, if the $|m|=1$ mode is allowed to scale as Eq.\ \eqref{Eq:h_PN_full_SPA}, it will become non-negligible as its amplitude PN order will be lower than the $|m|=2$ mode. Therefore, one should at least consider PPE modifications introduced to both $\ell=|m|=1$ and $\ell=|m|=2$ harmonics. Then, Eqs.\ \eqref{Eq:PPE_p22} and \eqref{Eq:PPE_p11} can be separated out from the GR contribution in Eq.\ \eqref{Eq:h_PN_full_SPA}, with $A^{(m)}$ in Eq.\ \eqref{Eq:h_PN_aell} correctly accounting for the amplitude scaling with $(d_L,\mathcal{M}_z,\eta,u_m)$.

Before we continue, we would like to discuss some of the assumptions we made. First, in our approach we include both amplitude and phase modifications (though GW detectors typically measure phases with higher precision than amplitudes \cite{Chatziioannou:2012rf}). However, since calculations for the parameters ($\alpha_{\ell m}, a_{\ell m}$) for different modified gravity theories have not been performed systematically, we only include amplitude modifications for the $(\ell=|m|=2)$ tensor angular harmonic and neglect those of higher harmonics.
In other words, we assume  $\alpha_{\ell m}\ne 0$ only for $\ell=|m|=2$ and Eq.~\eqref{Eq:PPE_Tlm} can be split as
\begin{align}
    \tilde{h}_{T}^{(2,2)}(f) &=  A^{(2)} \, (1 + \alpha_{22} u_2^{a_{22}})\,  e^{-i\Psi^{(2)}_{GR}} \,  e^{i\beta u_2^{b}} \,,\label{Eq:PPE_T22_assumptions} \quad {\rm if} \,\, \ell=|m|=2 \\
    \tilde{h}_{T}^{(\ell,|m|)}(f) &=  A^{(m)}  e^{-i\Psi^{(m)}_{GR}} \,  e^{i(|m|/2)\beta u_m^{b}} \,\label{Eq:PPE_Tlm_assumptions} \quad \rm otherwise
\end{align}
and, from now on, we rename 
\begin{align}
        \alpha_{22} &= \alpha_T\, , \\
        a_{22} &= a\, ,
\end{align}
for consistency with the extra-polarization amplitude modifications. 

Second, our PPE model contains a total of 13 free parameters, namely $\{\alpha_T, \alpha_{P2}, \alpha_{P1}, a, $ $ a_{P1}, a_{P2}, \beta, b\}$, for tensor and extra polarizations, which are the focus of this paper. In practice, various modified gravity theories predict the same values for the frequency scaling at leading order $(a,a_{P1},a_{P2},b)$ so those will be fixed, and we will perform Fisher  forecasts for the amplitude and phase parameters $(\alpha_T,\alpha_{P2}, \alpha_{P1},\beta)$, which can then be translated into constraints on specific modified gravity theories. In general, the amplitude parameters  $(\alpha_T,\alpha_{P2}, \alpha_{P1})$ could be complex, while the phase parameter $\beta$ is real. A complex phase of the parameters $(\alpha_T, \alpha_{P2}, \alpha_{P1})$ will induce global phase shifts in the corresponding polarization, so we will simplify the analysis by assuming them to be real, i.e.\ $(\alpha_T, \alpha_{P2}, \alpha_{P1})\in \mathbb{R}$, but a generalization could be considered in the future. Related to this point, \cite{Islam:2019dmk} reports that the phase $\phi$ of $\alpha_T=|\alpha_T|e^{i\phi_{\alpha_T}}$ is poorly determined in the context of the LIGO-Virgo-KAGRA (LVK). We decided to verify this sentence in the context of LISA, focusing only on  the case of complex $\alpha_T$. The results of this test are presented in Section \ref{subsec:complex_alpha}, where we show that the phase of $\alpha_T$ can be constrained only in a few optimistic scenarios.

Third, in modified gravity theories, a PN calculation can be performed at leading order to obtain predictions for the extra polarizations in GWs, when assuming small deviations from GR. For self-consistency, we will thus generally assume that the amplitude parameters $(\alpha_T,\alpha_{P1},\alpha_{P2})$ have an upper bound given by:
\begin{align}
& {\rm max} \left|\alpha_T u_2^{a}\right|<1, \quad \text{for the tensor part} \label{Eq:limit_alpha_tensor}\\
 & {\rm{max}} \left| \frac{\alpha_{P m} u^{a_P}_{m}A^{(m)}}{A^{(2)}}\right| < 1  \quad \rm \text{for $m=1$ or $m=2$} \label{Eq:limit_alpha_extrapol} .
\end{align}
In the second line we have assumed that the amplitude of the extra polarizations is small compared to the dominant $(\ell=|m|=2)$ tensor mode in GR. Since this expression  may generally depend on frequency (through $u_m$), we impose this bound at the observed frequency that maximizes the left-hand side. 
We note that the PPE formalism only includes the leader-order modifications to the phase and amplitude of the waveforms. However, in most known alternative theories of gravity, deviations from GR extend over all PN orders. 
We restrict our analysis to the leading-order PN corrections for several reasons, while acknowledging that this choice carries important caveats. First, in most of the modified gravity theories considered here, higher-order PN corrections to the extra polarization waveforms have not yet been computed analytically; the leading-order term is often the only one available in the literature. Second, including multiple PN coefficients simultaneously introduces significant parameter degeneracies: varying several PN orders at once leads to strongly correlated posteriors that are largely uninformative unless one employs dimensionality-reduction techniques such as principal component analysis, as suggested in Ref.~\cite{Saleem:2021nsb}.
However, it is known that restricting to the leading PN order can be insufficient for a complete test of GR. Higher-order corrections can shift or even dominate the signal for certain theory parameters or binary configurations (e.g.\ \cite{Cardoso:2023dwz}), potentially biasing parameter recovery and leading to false GR violations or, conversely, masking genuine deviations. This concern has been comprehensively discussed in \cite{Gupta:2024gun}. More broadly, the use of truncated waveform templates has long been recognized as a potential source of systematic bias in tests of GR \cite{Yunes:2009ke}, and the PPE framework itself only captures deviations that admit a PN series representation \cite{Xie:2024ubm}.

For the purposes of this work, we regard leading-order PN terms as a well-motivated first step that is consistent with the current state of analytical waveform modeling in modified gravity. Nonetheless, LISA's high SNR and long observation baselines may ultimately make it sensitive to sub-leading PN effects, and a full multi-order PPE analysis or a fully theory-specific Bayesian inference will be necessary to confirm any detected deviation. Extending this analysis to higher PN orders represents a natural and important direction for future work.

\subsection{\label{subsec:mapping_to_gravity_theories}Mapping to gravity theories}

To date, the calculation of extra polarization waveforms in modified gravity has only been done for a handful of theories, including Horndeski \cite{Higashino:2022izi}, Einstein-\ae ther \cite{Zhang:2019iim}, Rosen's theory \cite{Rosen:1974ua}, and Lightman-Lee Theory \cite{Lightman:1973kun}. Nonetheless, the PPE framework is in principle applicable to any metric gravity theory that admits a PN expansion for its GW emission, but the calculation of the relevant PN coefficients for new theories is technically demanding and remains an active area of research. Extending the present analysis to other theoretically better-motivated and observationally less constrained theories constitutes an important direction for future work. We thus focus on the four aforementioned theories and perform a mapping of their parameters to the PPE model (see  Section \ref{sec:ModGrav}). In this section, we summarize the results on the PPE mapping. 

In Horndeski gravity, the PPE parameters depend on 7 modified gravity parameters, including a modified gravitational constant $G_*$, the scalar field mass $m_s$, the black hole hair difference $\Delta\hat\alpha$, a mass-weighted scalar hair $\Gamma$, a parameter $g_4$ arising solely from non-minimal coupling, and two other parameters $\delta_0$ and $\kappa_4$ capturing both non-canonical scalar and modified gravity effects. 
More details on this theory and its PPE mapping can be found in Section \ref{sec:ST_theory}. In Einstein-\ae ther, the PPE parameters depend on 6 modified gravity parameters, including the black hole sensitivities $s_{1,2}$, and four coupling constants $c_{\theta}$, $c_\sigma$, $c_\omega$ and $c_a$. These 6 parameters get combined in specific ways to lead to the quantities $\epsilon_x$, $\kappa_3$, and $a_{bL}^{\text{\ae}}$  found in the tables below. The specific details can be found in Section \ref{sec:EA_theory}. In Rosen's theory, the PPE parameters depend on only 2 modified gravity parameters, the black hole sensitivities $s_1$ and $s_2$, which get combined in specific ways to produce $k_{\rm R}$ and ${\cal G}$ in the tables below. More details can be found in Section \ref{sec:Rosen_theory}. Finally, in Lightman-Lee theory, the PPE parameters only depend on 1 modified gravity parameter, ${\cal G}$ describing the black holes difference in sensitivities. Details can be found in Section~ \ref{sec:LL_theory}.

\begin{table*}[h!]
\centering
\begin{tabular}{|c|c|c|c|c|c|c|c|c|c|c|}
\hline
Theories &  \multicolumn{2}{c}{$a$} & \multicolumn{2}{|c}{$b$} & \multicolumn{2}{|c}{$a_{B}$}  &  \multicolumn{2}{|c}{$a_{L}$} & \multicolumn{2}{|c|}{$a_{V}$}  \\ \hline
Horndeski (Section \ref{sec:ST_theory}) & \cellcolor{blue!5} -2 &  \cellcolor{green!5} 0 &  \cellcolor{blue!3} -7 &  \cellcolor{green!5} -5 & \multicolumn{2}{c}{$0$} & \multicolumn{2}{|c}{-6} & \multicolumn{2}{|c|}{$-$}  \\ \hline
Einstein-\ae ther (Section \ref{sec:EA_theory}) &  \cellcolor{blue!5} -2 & \cellcolor{green!5} 0 &  \cellcolor{blue!3} -7 &  \cellcolor{green!5} -5 & \cellcolor{blue!5} -2 & \cellcolor{green!5} 0 & \cellcolor{blue!5} -2 & \cellcolor{green!5} 0 & \cellcolor{blue!5} -2 &\cellcolor{green!5} 0  \\ \hline
Rosen's theory (Section \ref{sec:Rosen_theory}) & \multicolumn{2}{c|}{$0$ }&  \cellcolor{blue!5} -7 &\cellcolor{green!5} -5 & \multicolumn{2}{c}{0} & \multicolumn{2}{|c}{0}  & \multicolumn{2}{|c|}{0}   \\ \hline
Lightman-Lee Theory (Section \ref{sec:LL_theory}) & \multicolumn{2}{c|}{$0$ }&  \cellcolor{blue!5} -7 & \cellcolor{green!5}-5 &  \multicolumn{2}{c}{0} & \multicolumn{2}{|c}{0}  & \multicolumn{2}{|c|}{0}   \\ 
\hline
\end{tabular}
\caption{Mapping of the frequency scaling parameters $(a,b,a_B,a_L, a_{V})$. Two possible values are obtained for some theories, depending on the modified gravity parameters dominating. In the case of Horndeski two cases of $(a=-2,b=-7)$ and $(a=0,b=-5)$ are possible, as indicated by the cell colors. In Rosen's and Lightman-Lee theories, the value $a=0$ is always obtained, regardless of the value of $b$.}
\label{table:PPE_mapping_ab}
\end{table*}

Table \ref{table:PPE_mapping_ab} collects the mapping on frequency scaling for the tensor and extra polarizations in the four modified gravity theories we consider in this study. Depending on which parameters dominate in the theory, different frequency scalings will determine the leading deviation from GR. For this reason, we indicate two possible scenarios in blue and green colors in Table \ref{table:PPE_mapping_ab}. As an example, in Horndeski, vector modes do not propagate (and hence there is no valid $a_V$ quoted), while the tensor and scalar polarizations allow two scenarios in the combinations $(a=-2,b=-7,a_B=0,a_L=-6)$ or $(a=0,b=-5,a_B=0,a_L=-6)$. This Table can be considered to be a generalization of Table~1 and Table~2 in \cite{Tahura:2018zuq} that report only the value of $a$ and $b$ for the tensor polarization, which highlights the remaining theoretical work to be done in the future, in order to make a more comprehensive analysis of extra polarizations in more gravity theories. 
Furthermore, notice that previous works mostly focused on the case when $b=-7$ (corresponding to a dipole emission dominating the modified gravity effects), but here we generalize the results to include regions of the parameter space of the modified gravity theories in which the dipole contribution may be subdominant compared to quadrupole contributions, and thus include also the case of $b=-5$. Recalling our definition of phase modification in Eqs.~\eqref{Eq:PPE_T22_assumptions}-\eqref{Eq:PPE_Tlm_assumptions}, it is clear that a value of $b=-7$ ($b=-5$) modifies the contribution of -1PN (0PN) of the standard phase in GR.

Table \ref{table:PPE_mapping_beta} shows the mapping of the phase parameter $\beta$, in the four modified gravity theories. Recall that in all cases we obtain 
$\beta=2\beta_P$, and thus the quoted $\beta$ parameter determines the phase of all polarizations. Here, we show in blue and green the value of $\beta$ corresponding to the blue and green scenarios of frequency scalings in Table \ref{table:PPE_mapping_ab}.

\begin{table*}[h!]
\centering
\begin{tabular}{|c|c|c|}
\hline
Theories & \multicolumn{2}{c|}{$\beta$}  \\ \hline
Horndeski (Section \ref{sec:ST_theory}) & \cellcolor{blue!5} $ - \frac{5 \kappa_4}{1792}\biggl(\frac{G_*}{G_N}\biggr)^{-\frac73} (\Delta\hat \alpha)^2 \eta^\frac25 $ & \cellcolor{green!5}   Eq.~\eqref{eq: ST-beta-first-branch}    \\ \hline
Einstein-\ae ther (Section \ref{sec:EA_theory}) &\cellcolor{blue!5}  $- \frac{3 }{224} \eta^\frac25 \epsilon_x [(1-s_1)(1-s_2)]^\frac23$ & \cellcolor{green!5}  $ \frac{3}{64} \left[ \frac{[(1-s_1)(1-s_2)]^{-\frac23}}{(2-c_a)\kappa_3}\right]^{(1)}$   \\ \hline
Rosen's theory (Section \ref{sec:Rosen_theory}) &  \cellcolor{blue!5} $-\frac{50}{16464} \kappa_{\rm R}^{-\frac23} {\cal G}^2 \eta^{\frac25}$ & \cellcolor{green!5}  \fcolorbox{red}{green!5}{$  -\frac{3}{256} \left[ 1 + \frac87 \kappa_{\rm R}^{-\frac56} \right]$ }     \\ \hline
Lightman-Lee Theory (Section \ref{sec:LL_theory}) &  \cellcolor{blue!5} $\frac{625}{16464} {\cal G}^2 \eta^{2/5}$ & \cellcolor{green!5} \fcolorbox{red}{green!5}{$ -\frac{3}{224} - \frac{3}{256}$ }  \\ 
\hline
\end{tabular}
\caption{Mapping of $\beta$ to modified gravity theories. The cell colors indicate the value of $\beta$ parameters associated to the corresponding $(a,b)$ values in Table \ref{table:PPE_mapping_ab}. Red boxes represent cases that do not admit GR limits (cf.\ with the corresponding sections for more details).} 
\label{table:PPE_mapping_beta}
\end{table*}

Notice that Horndeski and Einstein-\ae ther theories are smoothly connected to GR in the sense that $\beta=0$ is a possible choice of parameter, whereas non-zero values of $\beta$ are always predicted in Rosen's and Lightman-Lee theories. We use red boxes to highlight scenarios where a smooth GR limit cannot be taken, and hence the theories predict definite non-vanishing values for the PPE parameters.  In Section \ref{sec:constrain-specific-theories} we show that numerical results from GR injections can be directly translated to constraints into Horndeski and Einstein-\ae ther parameters, but for  Rosen and  Lightman-Lee we perform tailored non-GR injections to estimate constraints on their parameters.

In Table \ref{table:PPE_mapping_alpha2} we show the mapping for the tensor amplitude PPE parameter $\alpha_T$ for the GW polarizations in the four modified gravity theories. 
\begin{table*}[h!]
\centering
\begin{tabular}{|c|c|c|}
\hline
Theories &   \multicolumn{2}{c|}{$\alpha_T$} \\ \hline
Horndeski (Section \ref{sec:ST_theory}) & \cellcolor{blue!5}  $  -\frac{5 \kappa_4}{48} \biggl(\frac{G_*}{G_N}\biggr)^{\frac16}(\Delta\hat{\alpha})^2 \eta^\frac25 $ & \cellcolor{green!5} $ -1+\biggl(\frac{G_*}{G_N}\biggr)^{\frac56}\left(1+\frac{1}{3}\delta_0-\frac{\kappa_4}{12}\Gamma^2\right) $   \\ \hline
Einstein-\ae ther (Section \ref{sec:EA_theory})& \cellcolor{blue!5}  $  -\frac{\eta^{2/5} \epsilon_x}{2  [(1-s_1)(1-s_2)]^{2/3}  } $   &\cellcolor{green!5} $\left[ \sqrt{\frac{(2-c_a)}{2 \kappa_3}} [(1-s_1)(1-s_2)]^{\frac13} \right]^{(1)}$\\ \hline
Rosen's theory (Section \ref{sec:Rosen_theory}) & \multicolumn{2}{c|}{\fcolorbox{red}{white}{$-1 + i \sqrt{\frac27} \kappa_{\rm R}^{-3/4}$}}  \\ \hline
Lightman-Lee Theory (Section \ref{sec:LL_theory}) & \multicolumn{2}{c|}{\fcolorbox{red}{white}{$-1 - i \sqrt{\frac27}$ }} \\ 
\hline
\end{tabular}
\caption{Mapping of $\alpha_T$ to modified gravity theories, determining the $\ell=|m|=2$ angular harmonic waveforms. In practice, in the tests performed we assume $
\alpha_T$ to be real, and hence results are mapped to $|\alpha_T|$. }
\label{table:PPE_mapping_alpha2}
\end{table*}
From this table we can see explicitly that some theories do predict complex values for amplitude parameters. Nonetheless, we will only perform Fisher  forecasts assuming real values, which can thus be mapped to the modulus of the amplitude parameters quoted in these tables.

Finally, in Tables \ref{table:PPE_mapping_alphaX2} and  \ref{table:PPE_mapping_alpha1} we show the mapping of the extra polarization amplitude PPE parameters $(\alpha_{p2},\alpha_{p1})$, respectively. In some cases, the expressions are too long to quote, so we refer to the corresponding equations in Section \ref{sec:ModGrav} where the complete expressions can be found. 

\begin{table}[h!]
\centering
\resizebox{\textwidth}{!}{%
\begin{tabular}{|c|c|c|c|c|c|c|}
\hline
Theories &    \multicolumn{2}{c|}{$\alpha_{B2}$}  & \multicolumn{2}{c|} {$ \alpha_{L2}$} & \multicolumn{2}{c|}{$ \alpha_{V2}$} \\ \hline
Horndeski (Section \ref{sec:ST_theory})  
&\multicolumn{2}{c|}{$-\frac{1}{\sqrt{6}}\kappa_4g_4\Gamma\left(G_*/G_N\right)^{5/6}$} &\multicolumn{2}{c|} {$\alpha_{B2}(G_Nm_s\mathcal{M}_z)^2/4$} &\multicolumn{2}{c|} {--}   \\ \hline
Einstein-\ae ther (Section \ref{sec:EA_theory}) & \cellcolor{blue!5} Eq.~\eqref{eq:EAalphaB2a2} & \cellcolor{green!5}Eq.~\eqref{eq:EAalphaB2a0}  & \multicolumn{2}{c|}{$a^{\text{\ae}}_{bL} \alpha_{B2}$} & \cellcolor{blue!5} Eq.~\eqref{eq:EAalphaV2a2} & \cellcolor{green!5}Eq.~\eqref{eq:EAalphaV2a0}  \\ \hline
Rosen's theory (Section \ref{sec:Rosen_theory})  &\multicolumn{2}{c|}{\fcolorbox{red}{white}{ $-\frac{i}{\sqrt{21}} \kappa_{\rm R}^{-3/4}$ }}& \multicolumn{2}{c|}{\fcolorbox{red}{white}{$-\frac{2 i}{\sqrt{21}} \kappa_{\rm R}^{-3/4}$}} & \multicolumn{2}{c|}{\fcolorbox{red}{white}{$- \sqrt{\frac27} \kappa_{\rm R}^{-3/4}$ }  }  \\ \hline
Lightman-Lee Theory (Section \ref{sec:LL_theory})& \multicolumn{2}{c|}{\fcolorbox{red}{white}{$ -i\sqrt{\frac{3}{7}}$}} & \multicolumn{2}{c|}{\fcolorbox{red}{white}{$\frac{2 i}{\sqrt{21}}$} }& \multicolumn{2}{c|}{\fcolorbox{red}{white}{$\sqrt{\frac{2}{7}}$ }}   \\
\hline
\end{tabular}
} 
\caption{Mapping of $(\alpha_{B2}, \alpha_{L2},\alpha_{V2})$ to modified gravity theories, determining the $\ell=|m|=2$ angular harmonic waveforms. In the case of Horndeski and Einstein-\ae ther theories, the expressions quoted for $\alpha_{L2}$ are valid for both cases of green and blue frequency scalings. }
\label{table:PPE_mapping_alphaX2}
\end{table}

\begin{table}[h!]
\centering
\resizebox{\textwidth}{!}{
\begin{tabular}{|c|c|c|c|c|c|c|}
\hline
Theories &  \multicolumn{2}{c|}{ $\alpha_{B1}$}  & \multicolumn{2}{c|}{$ \alpha_{L1}$} & \multicolumn{2}{c|}{$ \alpha_{V1}$} \\ \hline
Horndeski (Section \ref{sec:ST_theory}) 
& \multicolumn{2}{c|}{ $-\frac{1}{2}\sqrt{\frac{10}{3}}\kappa_4g_4(\Delta\hat\alpha)\biggl(\frac{G_*}{G_N}\biggr)^{1/2}$} 
& \multicolumn{2}{c|} {$\alpha_{B1}(G_N m_s \mathcal{M}_z)^2$} 
& \multicolumn{2}{c|}{--} 
\\ \hline
Einstein-\ae ther (Section \ref{sec:EA_theory}) 
& \cellcolor{blue!5} Eq.~\eqref{eq:EAalphaB1a2} 
& \cellcolor{green!5} Eq.~\eqref{eq:EAalphaB1a0} 
& \multicolumn{2}{c|}{$a^{\text{\ae}}_{bL} \alpha_{B1}$}
& \cellcolor{blue!5} Eq.~\eqref{eq:EAalphaV1a2} 
& \cellcolor{green!5} Eq.~\eqref{eq:EAalphaV1a0} 
\\ \hline
Rosen's theory (Section \ref{sec:Rosen_theory}) 
& \multicolumn{2}{c|}{ $\frac{4 i }{3} \sqrt{\frac{5}{21}} \mathcal{G} \kappa_{\rm R}^{-7/12}$ } 
& \multicolumn{2}{c|}{ $\frac{4 i }{3} \sqrt{\frac{5}{21}} \mathcal{G} \kappa_{\rm R}^{-7/12}$ } 
& \multicolumn{2}{c|}{ $ \frac{4}{3} \sqrt{\frac{10}{21}} \mathcal{G} \kappa_{\rm R}^{-7/12}$ } 
\\ \hline
Lightman-Lee Theory (Section \ref{sec:LL_theory}) 
& \multicolumn{2}{c|}{ $-  \frac{25 i}{6} \sqrt{\frac{5}{21}} \mathcal{G}$ } 
& \multicolumn{2}{c|}{ $-  \frac{10 i}{3} \sqrt{\frac{5}{21}} \mathcal{G}$ } 
& \multicolumn{2}{c|}{ $\frac{10}{3} \sqrt{\frac{10}{21}} \mathcal{G}$ } 
\\
\hline
\end{tabular}
}
\caption{Mapping of $(\alpha_{B1}, \alpha_{L1}, \alpha_{V1})$ to modified gravity theories, determining the $\ell = |m| = 1$ angular harmonic waveforms.}
\label{table:PPE_mapping_alpha1}
\end{table}

\subsection{\label{subsec:numerical_implementation}Numerical implementation}
To generate the GW signal,
we adopt the state-of-the-art phenomenological model \texttt{IMRPhenomXHM} \cite{Garcia-Quiros:2020qpx}, implemented in the \texttt{lisabeta} code \cite{Marsat:2020rtl}, which describes the inspiral, merger, and ringdown of quasi-circular, non-precessing black hole binaries. This model includes the tensor polarizations predicted by GR for the dominant quadrupole mode $(\ell = |m| = 2)$, as well as higher-order angular harmonics 
$(\ell, |m|) = (2, 1), (3, 3), (3, 2), (4, 4), (4, 3)$.
However since GR and therefore \texttt{IMRPhenomXHM}
does not predict any emission in the $(\ell = |m| = 1)$ angular harmonic, we model the amplitude of the extra polarizations using the leading-order analytical PN expression for $A^{(m)}$ 
defined in Eq.~\eqref{Eq:h_PN_aell}. For the amplitude of the tensor polarization, as well as the phase of both the tensor and extra polarizations, 
we instead use those provided by the waveform model itself. 
In the case of cirular and aligned spins, the GW waveform can be described by 11 parameters: the rest-frame primary and secondary masses, $m_1$ and $m_2$, the spin magnitudes along the binary angular momentum, $\chi_1$ and $\chi_2$, two angles to describe the sky latitude and longitude, $\beta_S$ and $\lambda_S$, the luminosity distance $d_L$, the inclination $\iota$, the phase at coalescence  $\phi$ , the time at coalescence $t_c$, and the polarization angle $\psi$.

We compute the optimal SNR as \cite{Moore:2014lga}
\begin{equation}
    {\rm SNR^2} = 4 \int_{f_{\rm min}}^{f_{\rm max}} \frac{|\tilde{h}(f)|^2 }{S_n(f)} df,
\end{equation}
where $\tilde{h}(f)$ is the frequency domain GW waveform summed over all angular harmonics and $S_n(f)$ is the noise power spectral density  (PSD). To model LISA noise PSD, we adopt the estimate ``SciRDv1'' as in \cite{Babak:2021mhe}. We set $f_{\rm min} = 10^{-5} \rm \, Hz$ and $f_{\rm max} = 0.5 \rm \, Hz$. We assume 5 years of overall mission duration, with 80\% duty cycle.  We add the confusion background from unresolved galactic binaries to the LISA noise PSD, according to the fits presented in \cite{2021PhRvD.104d3019K}. The amplitude of the background is taken for four years of mission duration. The SNR is calculated for each TDI channel A, E and T, with a total obtained as a squared sum. 

To perform the parameter estimation,  we adopt the Fisher matrix formalism \cite{PhysRevD.77.042001}, where the Fisher matrix is defined as 
\begin{equation}
    \Gamma^{ab} \equiv \left( \frac{\partial \tilde{h}}{\partial \theta^a} \bigg| \frac{\partial \tilde{h}}{\partial \theta^b}    \right),
\end{equation}
where $\partial \tilde{h}/\partial \theta^a$ is the partial derivative of the GW waveform respect to the parameter $\theta^a$ evaluated at the injected value and $(\cdot|\cdot)$ is the standard inner product between two complex quantities, defined as
\begin{equation}
    (a|b) \equiv 2 \int_{0}^{\infty} \frac{\tilde{a}^{*}(f)\tilde{b}(f) + \tilde{a}(f)\tilde{b}^{*}(f)}{S_n(f)} df \,.
\end{equation}
For each parameter, we compute the partial derivative $\partial \tilde{h}/\partial \theta^a$ numerically as \cite{10.5555/1403886}
\begin{equation}
    \frac{\partial \tilde{h}}{\partial \theta^a} \approx \frac{\tilde{h}(\theta^a + \delta \theta^a) - \tilde{h}(\theta^a)}{\delta \theta^a},
\end{equation}
where $\delta \theta^a$ is the infinitesimal step for the corresponding parameter. The values of $\delta \theta^a$ must be chosen appropriately in order to ensure that $\partial \tilde{h}/\partial \theta^a$ is well defined: values that are too large no longer approximate a derivative, while values that are too small may suffer from numerical precision issues. We report in Appendix~\ref{sec:Fisher steps} the values adopted for $\delta \theta^a$ for all parameters, together with the tests performed to determine them. The covariance matrix is defined as the inverse of the Fisher matrix, $\Sigma = \Gamma^{-1}$. The expected marginalized statistical error on a parameter $\theta^a$ is computed from the corresponding diagonal element of the covariance matrix as
\begin{equation}
\sigma_{\theta^a} \equiv \sqrt{ \langle (\delta \theta^a)^2 \rangle } = \sqrt{\Sigma^{aa}},
\end{equation}
while the off-diagonal elements represent the correlations between parameters.
As mentioned earlier, we fix $(a,a_{P1},a_{P2},b)$, and therefore do not
test our ability to determine at which PN order a modification is present.
In addition, for simplicity we assume that only one deviation term is present, either in the amplitude or in the phase. Hence, we compute the Fisher matrix for the standard GR parameters plus one modification at a time, for a total of 12 parameters at the same time. Such an approach does not let us explore the correlation between various modified gravity parameters. These cases can be the subject of a later study. Nevertheless, as a preliminary investigation, in Section
\ref{subsec:amp-phase-modifications} we report the results for few cases where we compute the Fisher matrix assuming simultaneous modifications to the amplitude and phase  of the tensor polarization --  i.e.\ a Fisher analysis with 13 parameters, consisting of the standard 11 GR parameters plus $\alpha$ and $\beta$ -- finding no correlation between these two parameters.  
For the 11 parameters describing the binary in GR, we compute the Fisher matrix with respect to the quantities $[{\cal M}_z, q, \chi_{\rm PN},\chi_{\rm m}, t_c, d_L, \iota, \phi, \lambda_S, \beta_S, \psi]$ (see Table~\ref{tab:Notations}) \footnote{We also tested another parametrization of the Fisher  as $[M_z, q, \chi_1,\chi_2, t_c, d_L, \iota, \phi, \lambda_S, \beta_S, \psi]$, where $M_z$ is the redshifted total mass,  without finding any difference.}.

We point out that in the case of $b=-5$, the parameter $\beta$ in the tensor part is degenerate with the redshifted chirp mass $\mathcal{M}_z$. In fact, following Eq.~\eqref{Eq:PPE_T22_assumptions}, the total phase of the tensor polarization, including the lowest 0PN order, for the $\ell=|m|=2$ mode can be expressed as
\begin{align}
    \Psi_{tot}^{(2)} & = -\Psi_{GR}^{(2)} + \beta u_2^{-5} = \nonumber \\ 
    & =  2 \pi ft_c - \frac{\pi}{4} + \frac{3}{128} \frac{1}{\mathcal{M}_z^{5/3}}(\pi f)^{-5/3} + \frac{\beta}{ \mathcal{M}_z^{5/3}} (\pi f)^{-5/3} \nonumber  \\
    & = 2 \pi ft_c - \frac{\pi}{4} + \frac{\frac{3}{128} + \beta}{\mathcal{M}_z^{5/3}} (\pi f)^{-5/3}.
\end{align}
In other words, any change in $\beta$ can be compensated by an appropriate change in the chirp mass $\mathcal{M}_z$. This degeneracy is well-known in the context of the FTI approach \cite{Sanger:2024axs,Piarulli:2025rvr} and --  when comparing our Fisher approach to the full Bayesian approach in \cite{Piarulli:2025rvr} -- we find that our Fisher analysis underestimates the uncertainties on $\beta$. In order to reduce this degeneracy, we introduce a novel parameter $\beta'$ defined as 
\begin{equation}
    \beta' = \beta -\frac{3}{128} + \left( \frac{3}{128} + \beta_{inj} \right) \left( \frac{\mathcal{M}_z}{\mathcal{M}_{z, inj}} \right)^{5/3},
    \label{eq:beta_prime}
\end{equation}
where $\mathcal{M}_{z, inj}$ is the true redshifted chirp mass assuming that GR is the correct theory. We then calculate the Fisher  matrix with respect to this $\beta'$ parameter.
In practice, the parameter $\beta'$ allows us to compute the Fisher matrix along the line of degeneracy itself passing through the point 
$(\beta_{inj}, \mathcal{M}_{inj})$, with angular coefficient $(\frac{3}{128} + \beta)/\mathcal{M}_z^{5/3}$ and assuming that $\beta_{inj} \ne 0$. Once we obtain the uncertainties on $\beta'$, we can translate them back to uncertainties on $\beta$ by simply inverting Eq.~\eqref{eq:beta_prime}. We adopt this reparameterization for all the phase modifications with $b = -5$, but we report the results as a function of $\beta$ for clarity.

For specific gravity theories, it is possible to use a PN expansion to obtain analytical expressions for the modified emission of GWs during the early inspiral (as in Section \ref{sec:ModGrav}), which inspired the PPE approach used here. However, it has been shown that restricting the modified signal to the inspiral where only the quadrupole mode dominates can make the likelihood of the extrinsic parameters very degenerate (see Figure 9 and 10 in \cite{Marsat:2020rtl} for an example). Conversely, including the merger and ringdown as well as higher harmonics has proven to be crucial in breaking degeneracies and improving the localization of the source \cite{Arun:2007qv,Arun:2007hu,Trias:2008pu,Arun:2008zn,Marsat:2020rtl,Pratten:2022kug,Pitte:2023ltw} -- although predictions in modified gravity during the merger and ringdown require numerical, or  more refined  analytical techniques.

Numerical relativity has produced first proof-of-principle beyond GR
gravitational waveforms in scalar-tensor theories e.g.\ \cite{Shibata:2013pra,Barausse:2012da}, Einstein-Maxwell-Dilaton theories \cite{Hirschmann:2017psw}, cubic Horndeski theories \cite{Figueras:2021abd}, higher derivative theories
\cite{Cayuso:2023aht}, effective field theories for dark energy \cite{terHaar:2020xxb,Bezares:2021dma},
Einstein-scalar-Gauss-Bonnet (EsGB) gravity \cite{East:2020hgw, East:2021bqk,East:2022rqi,Franchini:2022ukz,Corman:2022xqg, AresteSalo:2023mmd,AresteSalo:2025sxc,Corman:2025wun,Lara:2025kzj}
and dynamical Chern-Simons gravity \cite{Okounkova:2017yby,Okounkova:2019dfo,Okounkova:2019zjf}.
The main challenge is that many modified gravity theories are not well-posed mathematical problems, and cannot be uniquely solved.
Therefore, most beyond GR simulations of compact objects have so far mainly focused:\\
(i) on a very limited number of theories that preserve the mathematical structure of the
field equations in GR e.g.\ subclass 
of scalar-tensor theories that lead to neutron stars developing a scalar charge
\cite{Damour:1993hw,Damour:1996ke}
or Einstein-Maxwell-Dilaton
models where electrically charged black holes develop a scalar charge
\cite{Garfinkle:1990qj} \\
(ii) on treating the modifications to GR only perturbatively to by-pass the intrinsic pathologies of the underlying equations of
motion, as in dynamical Chern-Simons  \cite{Okounkova:2017yby,Okounkova:2019dfo,Okounkova:2019zjf}, \\
(iii) on reformulating the field equations, addressing directly potential pathologies by
the introduction of additional, regularizing fields \cite{Allwright:2018rut,Cayuso:2017iqc,Cayuso:2020lca},\\
(iv) on a subclass of Horndeski gravity where it was recently shown that the equations of motion
possess a well-posed initial value problem. This subclass, referred to as ``four derivative scalar-tensor
theory'' can be shown to lead to a
well-posed initial value problem in a modified generalized harmonic (MGH) formulation
\cite{Kovacs:2020pns,Kovacs:2020ywu},
as well as modified conformal and covariant Z4 (CCZ4) formulation \cite{AresteSalo:2022hua,AresteSalo:2023mmd}
so long as the coupling parameter that determines the beyond-GR corrections
is much smaller than all other length scales in
the problem. Otherwise, the theory also displays pathologies---in particular
the complete breakdown of hyperbolicity resulting in an ill-posed problem
(e.g.\ \cite{Ripley:2019hxt,Bernard:2019fjb}).\\

Ignoring (ii) which was shown to exhibit secular effects, and thus cannot faithfully track the solutions when the corrections to GR are non-negligible \cite{Corman:2024cdr}, these works suggest that at least for four derivative scalar-tensor theories (studied in \cite{East:2020hgw,Corman:2022xqg,AresteSalo:2022hua,AresteSalo:2023mmd,AresteSalo:2025sxc,Corman:2025wun,Lara:2025kzj}) and higher-derivative theories \cite{Cayuso:2023aht}, 
modifying GR results in a phase shift of the gravitational waveform which
depends on the sign and magnitude of modification to GR, however the peak amplitude of the GW at merger depends only very weakly on the modified gravity couplings, at least for values consistent with current GR bounds or such that we are in regime of validity of gravity theory as an Effective Field Theory. The effect of modifying gravity on the frequency and decay rate of the quasinormal modes has been too small to reliably quantify from numerical
data but has been studied using perturbative methods (see Section 3.1.3 in \cite{Berti:2025hly} for a summary of results obtained so far) and has been shown to be very small.

Based on the results from numerical simulations showing that any modifications is suppressed close to the merger and, following the Flexible-Theory-Independent (FTI) approach in the context of LVK \cite{Mehta:2022pcn,Sanger:2024axs} and LISA \cite{Piarulli:2025rvr}, 
we will thus perform our analysis including the merger and ringdown portions of the signal, tapering down the GR modifications in these parts of the GW signal (i.e.\ assuming that the modified gravity effects become negligible after the inspiral). 
While the inspiral only assumes the sources to be point-like objects, the merger and ringdown phases assume that the binary components are black holes that merge into a Kerr remnant described by GR. We note that this assumption is internally consistent for the three theories in our analysis that admit black hole solutions (Horndeski, Einstein-\ae ther, and Lightman-Lee), though even in those cases the post-merger dynamics in the modified theory are unknown and may differ from GR at some level. For Rosen's bimetric theory, which does not admit black hole solutions (see Section~\ref{sec:Rosen_theory}), this part of the waveform model is not self-consistent with the theory: if the binary components are black hole mimickers rather than GR black holes, neither the merger dynamics nor the ringdown spectrum would necessarily resemble those of a Kerr black hole. The inspiral-phase constraints derived for Rosen's theory are therefore our most robust result for that theory; the IMR constraints in Section \ref{sec:constrain-specific-theories} should be interpreted with this limitation in mind. Developing self-consistent merger-ringdown models for non-GR compact object binaries remains an open challenge for the field.

During the merger, the modifications will be smoothly tapered off to zero using a Planck windowing function defined as \cite{McKechan:2010kp} 
\begin{equation}
W(f;\Delta f, f_{\rm tape}) = \left[1+ \exp{\left(- \frac{\Delta f}{f-(f_{\rm tape}-\Delta f)} -\frac{\Delta f}{f -f_{\rm tape}} \right)}\right]^{-1}
\label{eq:window_func},
\end{equation}
which smoothly transitions between one and zero around
$f_{\rm tape}$ over the range of $\Delta f$. Following \cite{Mehta:2022pcn,Sanger:2024axs,Piarulli:2025rvr}, we apply this window function to the second derivative of the phase with respect to frequency, as this quantity is directly linked to the orbital evolution of the system and naturally provides a time–to-frequency correspondence once the integration constants, that fix the phase and time shifts, are specified (see Appendix A of \cite{Piarulli:2025rvr} for further details). At an early stage of this work, we also explored an alternative alignment formalism, detailed in Appendix~\ref{sec:alignment}, but we found that it produced results inconsistent with those obtained using the FTI approach. For this reason, we ultimately decided to adopt the FTI method. We refer the interested reader to Appendix~\ref{sec:alignment} for more details. 

The tapering frequency depends on the mode: for the dominant harmonic, we set $f_{\rm tape} = f^{\rm peak}_{22}$, where $f^{\rm peak}_{22}$ is the peak frequency of the $(\ell,m)=(2,2)$ mode. For other harmonics, we set $f_{\rm tape} = \gamma \cdot\frac{|m|}{2} f^{\rm peak}_{22}$ where $\gamma$ is a constant of order unity. Similarly, we fix $\Delta f$ as a fraction of $f_{\rm tape}$, $\Delta f \equiv \delta \cdot f_{\rm tape}$, for each harmonic. The choice of $(\gamma,\delta)$ is completely arbitrary. Earlier analyses on LVK events \cite{LIGOScientific:2018dkp,LIGOScientific:2020tif,LIGOScientific:2020zkf,LIGOScientific:2020stg,LIGOScientific:2021sio} used $\gamma = 0.35$; however, following a more recent study
\cite{Sanger:2024axs}, we set $\gamma = 1.0$. 
In addition, we performed tests on $\delta$ and found it to not significantly affect the results, and we therefore fix its value to $\delta = 0.2$. With this choice, the Planck window suppresses the non-GR modifications in the interval $[0.9 f_{\rm tape}, 1.1 f_{\rm tape}]$ for each harmonic. However, we note that in this configuration non-GR modifications remain active beyond $f^{\rm peak}_{22}$, where the signal is dominated by the merger and GR modifications are expected to be suppressed. \cite{Piarulli:2025rvr}  adopted a different strategy, setting $\gamma=0.35$ and the end of the Planck window at $f^{\rm peak}_{22}$, thereby restricting non-GR modifications to the inspiral portion of the signal. The choice of cutoff frequency is non-trivial but we found no significant differences between our results and those of \cite{Piarulli:2025rvr}.

\begin{figure}[h!]
\centering
\includegraphics[width = \textwidth]{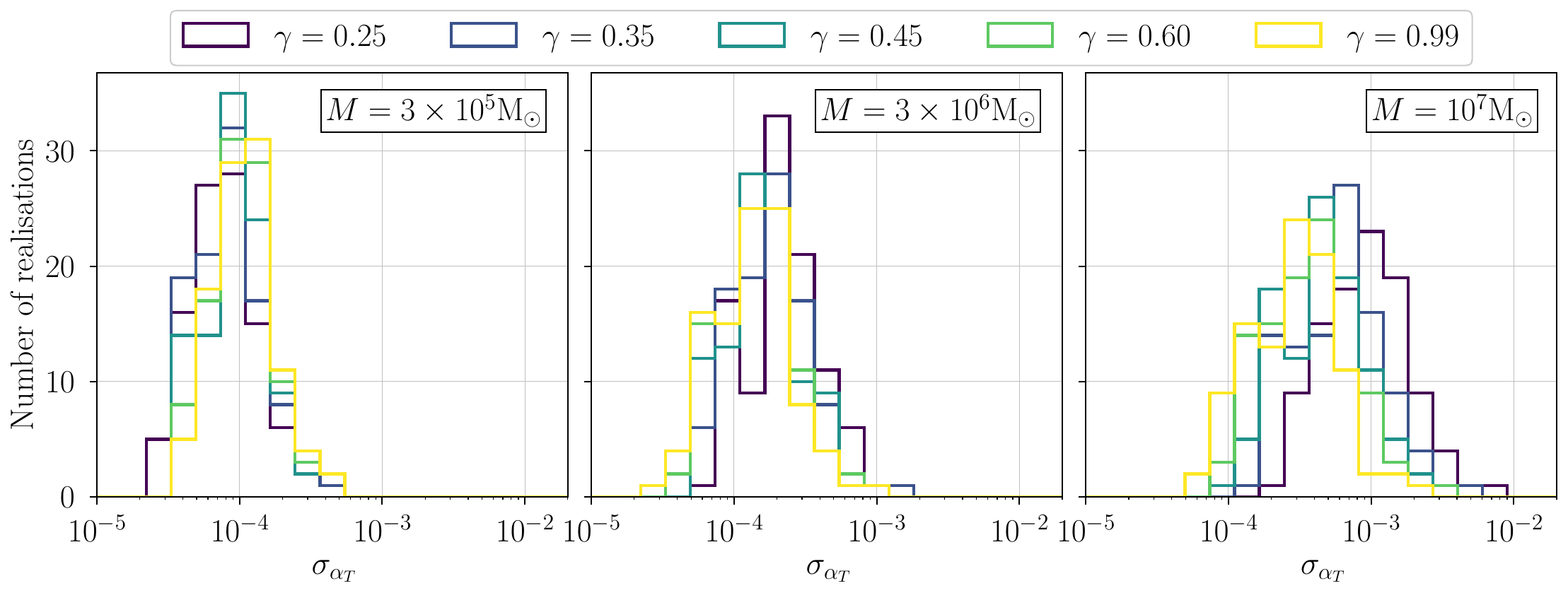}
 \caption{Distribution of the absolute errors on $\alpha_T$ for different termination frequency (as reported in the legend) and different total mass $M$ and $z=1$. We assume a reference value of $\alpha_T=0$ and $a=-2$. Higher masses are more affected. For the rest of the paper, we adopt $\gamma=0.99$. }
 \label{fig:fcut_tgr_comparison} 
\end{figure}

In Figure~\ref{fig:fcut_tgr_comparison}  we show how the constraints on $\alpha_T$ change for three systems with $M=3\times 10^5 \msun$, $3\times 10^6 \msun$ and $10^7 \msun$ at $z=1$ as a function of $\gamma$. To produce this plot, we randomized all the binary parameters over 100 realizations, with the exception of the total mass, the redshift and the value of $\alpha_T$ that is always assumed to be zero. The histograms represent the distribution of absolute errors on $\alpha_T$ from the Fisher  analysis. It is clear that the choice of $\gamma$ affects primarily the case with $M = 10^{7} \msun$, because for heavy-systems we observe only a small portion of the inspiral compared to the $M = 3\times 10^{5} \msun$ case. As a consequence, the value of $\gamma$ influences more the amount of non-GR modification that is observable, with tighter constraints obtained for $\gamma = 0.99$. In agreement with previous studies \cite{Sanger:2024axs, Piarulli:2025rvr}, we find similar behavior for all the other amplitude and phase modifications considered in this work.

A more conservative approach would require to limit the analysis only to the inspiral portion of the signal. In this way, it would not be necessary to taper the modifications that might introduce some spurious effects in the analysis. Since most of the previous work on tests of GR \cite{Mehta:2022pcn, Sanger:2024axs,Piarulli:2025rvr}  includes the merger and ringdown portion of the signal, we follow a similar approach and the results reported in Section~\ref{subsec:gr_injections} have been obtained including the entire signal. However, in Section \ref{sec:inspiral-only-injections}, we present the results for GR injections in a few cases where we consider only the inspiral portion of the signal.

Finally, we also add a phase correction for each $(\ell,m)$ harmonic so that the phase of each mode and its derivative vanish at a reference frequency, defined as 
\begin{equation}
    f^{\rm ref}_{\ell m} \equiv \frac{|m|f^{\rm peak}_{22}}{2} .
    \label{eq:fref_lm}
\end{equation}
This correction ensures that the
alignment between the GR waveform and the modified
GR waveform remains the same in  time domain. We discuss more details on how we perform the alignment in Appendix~\ref{sec:alignment}.

\section{\label{sec:ModGrav}Modified gravity waveforms}

As discussed in the previous section, in this work we will focus on four particular metric theories of gravity, which have a known PN expansion for the extra polarizations beyond GR. We will  present each of these theories, define the relevant parameters determining the frequency domain waveforms, and perform an explicit mapping to the PPE parameters.

The early inspiral waveforms derived in this section are obtained by treating the binary components as point-like compact objects, following the standard approach. No assumption on the internal structure of the objects is made at this stage beyond their masses and the theory-dependent sensitivity parameters, which encode how the gravitational binding energy of each body responds to the external scalar or vector fields present in the theory \cite{Will:2014kxa}. This generality means that the inspiral-phase PPE constraints derived here apply in principle to any sufficiently compact binary (BHs, neutron stars, exotic compact objects, or black hole mimickers) as long as the PN approximation is valid.

In considering additional gravitational polarizations, it is important to keep in mind that the presence of additional propagating degrees of freedom in a metric theory does not guarantee the existence of extra polarizations within the gravitational radiation, since the former crucially depend on the couplings between the gravitational fields (recall the discussion in Section~\ref{sec:revpol}). Therefore, the additional vector and scalar waveforms of each particular theory below can only be directly constrained if the parameters of the theory that govern the coupling between extra gravitational fields and the physical metric allow for the presence of polarizations beyond the tensor ones. In Appendix~\ref{sApp:HorndeskiPolarizations} we provide an explicit example of such a dependence of the direct detectability of gravitational polarizations on the parameters of a theory.

\subsection{Horndeski gravity\label{sec:ST_theory}}
As it is widely known, Horndeski theory represents the most general scalar-tensor metric theory of gravity with at most second order equations of motion constructed out of a physical metric tensor $g_{\mu\nu}$ and a real-valued scalar field $\phi$  \cite{Horndeski:1974wa,Nicolis:2008in,Deffayet:2009wt,Deffayet:2009mn,Heisenberg:2018vsk,Kobayashi:2019hrl}. More precisely, its equations of motion include only up to two derivative operators per field, which ensures that the theory is free from Ostrogradsky instabilities \cite{Ostrogradsky:1850fid}.  Horndeski gravity encompasses a vast class of scalar-tensor theories such as $f(R)$ gravity~\cite{Bergmann:1968aj,Ruzma:1969JETP,Buchdahl:1970MN,Sotiriou:2008rp,DeFelice:2010aj}, Brans-Dicke (BD) gravity~\cite{Brans:1961sx,Dicke:1961gz}, and scalar Gauss-Bonnet (sGB) gravity~\cite{Zwiebach:1985uq,Gross:1986mw} (see also Appendix~\ref{App:Horndeski plus}), that has been widely applied to observational tests of gravity (see, e.g.\ Refs.~\cite{Koyama:2015vza,Quartin:2023tpl}).
Consistent covariant generalizations of Horndeski theories
exist though, called DHOST \cite{Langlois:2015cwa,Crisostomi:2016czh,BenAchour:2016fzp}, allowing for higher-than-two
derivatives on the field equations of motion by exploiting constraint relations. We do not discuss   DHOST theories in this work.

Horndeski gravity \cite{Horndeski:1974wa,Nicolis:2008in,Deffayet:2009wt,Deffayet:2009mn,Kobayashi:2019hrl} can propagate  additional scalar degrees of freedom. As shown in Refs.~\cite{Hou:2017bqj,Heisenberg:2024cjk}, depending on the parameter space of the theory, the theory
 can either excite none, only one, or two additional scalar polarizations within the gravitational radiation. More precisely, if the scalar DoF is massive it may excite both the transverse breathing and longitudinal polarizations, while the massless case will at most excite the transverse breathing mode. As can be expected, due to a lack of any vector DoFs, also the vector polarizations are absent. In Appendix~\ref{App:Horndeski plus} we offer a concise derivation of these statements together with a definition of the full Horndeski Lagrangian. 

Based on the considerations in Appendix~\ref{App:Horndeski plus}, in this work, we will concentrate on a particular subset of Horndeski gravity, which is most relevant for the study of additional polarizations
\begin{equation}
\label{eq:Horndeskisubclass}
    S^H_\text{G} = \int d^4x\sqrt{-g}\biggl[G_2(\phi,X) - G_3(\phi,X) \Box \phi + G_4(\phi) R\biggr] ~,
\end{equation}
where $X:=-\nabla_\mu\phi\nabla^\mu\phi$, and $G_i(\phi,X)$ are arbitrary functions of $\phi$ and $X$. 
Observe that this subclass is precisely the theory, which in a cosmological context would satisfy the observational luminality bound from the GW170817 event \cite{LIGOScientific:2017zic} (see, e.g., \cite{Langlois:2017dyl}). 
The gravitational waveforms in frequency domain from this subclass have been previously studied in Ref.~\cite{Higashino:2022izi} in the context of a binary system in the approximation of a quasi-circular orbit composed of two point-like particles. These are the frequency-domain waveforms that we will present below.

As discussed in Section~\ref{sec:revpol}, for simplicity and in order to be consistent with available data analysis techniques, we will only consider the parameter spaces in which all modes in a theory propagate at similar speeds. Since in Horndeski gravity on asymptotically flat spacetime the tensor modes always travel at the speed of light, this implies that we will restrict ourselves to a parameter space, in which the scalar modes propagate with a speed close to unity. This is assured by assuming that the effective mass of scalar mode remains much smaller than the GW frequency. 

Moreover, obviously any additional scalar polarization will only be present in GW data if the corresponding scalar mode is actually excited in the process of emission. As discussed, in this work we will however remain agnostic about the precise emission mechanism and simply assume that all dynamical degrees of freedom of a given theory are excited. This is justified by the fact that the dynamical processes of radiation emission in compact binary coalescences are not well understood and impossible to fully describe analytically. Thus, even though the subclass of Horndeski theory in Eq.~\eqref{eq:Horndeskisubclass} does not admit any asymptotically flat BH solutions with a scalar hair in the analytically treatable spherically symmetric and static case~\cite{Sotiriou:2011dz,Graham:2014mda,Faraoni:2017ock,Minamitsuji:2022vbi}, the binary system might still dynamically lead to the formation of relevant scalar charge that sources an emission of the scalar mode.

\subsubsection{Frequency-domain waveforms}
The two tensor and two scalar modes propagate in the above class of scalar-tensor theories. The gravitational waveforms have been studied in Ref.~\cite{Higashino:2022izi} (see also Refs.~\cite{Liu:2018sia,Liu:2020moh,Liu:2022qcx}). Compared to the literature, we will rotate the polarization bases by $\pi/2$ around the propagation direction for the purpose of PPE mapping in the present paper, which changes the signs of the tensor amplitudes in Horndeski gravity. See Appendix~\ref{app:rotations} for more details. The waveforms in frequency domain are of the form (see Refs.~\cite{Liu:2018sia,Higashino:2022izi} and Appendix~\ref{sec:app-deerivation-ST} for the derivation of the waveforms), \footnote{The luminosity distance appearing in the GW amplitudes can differ from the electromagnetic luminosity distance, depending on the time variation of the background scalar field and modified gravity parameters, see e.g., Ref.~\cite{Liu:2022qcx}.
However, we do not consider such cosmological propagation effects, and hence all waveforms are expressed in terms of the electromagnetic luminosity distance.}
\begin{align}
\tilde h^{(2,2)}_{T,{\rm ST}}&=A^{(2)}e^{-i\Psi^{(2)}_{\rm GR}}\biggl(\frac{G_*}{G_N}\biggr)^{5/6}\biggl[1+\frac{1}{3}\delta_0-\frac{\kappa_4}{12}\Gamma^2-\frac{5}{48}\kappa_4(\Delta\hat\alpha)^2\biggl(\frac{G_*}{G_N}\biggr)^{-2/3}\eta^{2/5}u_2^{-2}\biggr]e^{i\delta\Psi^{(2)}_{T,\rm ST}},\\
\tilde h^{(2,2)}_{B,{\rm ST}}&=A^{(2)}e^{-i\Psi^{(2)}_{\rm GR}}\biggl[-\frac{1}{\sqrt{6}}\kappa_4g_4\Gamma\biggl(\frac{G_*}{G_N}\biggr)^{5/6}\biggr]e^{i\delta\Psi^{(2)}_{B,\rm ST}},\\
\tilde h^{(2,2)}_{L,{\rm ST}}&=A^{(2)}e^{-i\Psi^{(2)}_{\rm GR}}\biggl[-\frac{m_s^2}{4(\pi f)^2}\frac{1}{\sqrt{6}}\kappa_4g_4\Gamma\biggl(\frac{G_*}{G_N}\biggr)^{5/6}\biggr]e^{i\delta\Psi^{(2)}_{L,\rm ST}},
\end{align}
for the quadrupole modes $\ell=2$, and 
\begin{align}
\tilde h^{(1,1)}_{B,{\rm ST}}&=A^{(1)}e^{-i\Psi^{(1)}_{\rm GR}}\biggl[-\frac{1}{2}\sqrt{\frac{10}{3}}\kappa_4g_4(\Delta\hat\alpha)\biggl(\frac{G_*}{G_N}\biggr)^{1/2}\biggr]e^{i\delta\Psi^{(1)}_{B,\rm ST}},\\
\tilde h^{(1,1)}_{L,{\rm ST}}&=A^{(1)}e^{-i\Psi^{(1)}_{\rm GR}}\biggl[-\frac{m_s^2}{(2\pi f)^2}\frac{1}{2}\sqrt{\frac{10}{3}}\kappa_4g_4(\Delta\hat\alpha)\biggl(\frac{G_*}{G_N}\biggr)^{1/2}\biggr]e^{i\delta\Psi^{(1)}_{L,\rm ST}},
\end{align}
for the dipole modes $\ell=1$
where $\delta\Psi^{(2)}_{T,\rm ST}, \delta\Psi^{(m)}_{B,\rm ST}$, and $\delta\Psi^{(m)}_{L,\rm ST}$ stand for the deviation from the GR phase defined by\footnote{The mass-dependent term comes from the propagation in flat space. Since we only take into account the redshift and do not include any propagation effect in an expanding universe, the mass-dependent term is kept in its form in flat space. A consistent treatment of that term would require a full calculation of waveforms on an expanding background.}
\begin{align}
\delta\Psi^{(2)}_{T,\rm ST}&:=-\frac{3}{128}\biggl\{1-\biggl(\frac{G_*}{G_N}\biggr)^{-5/3}\biggl[1-\frac{2}{3}\delta_0-\frac{\kappa_4}{6}\Gamma^2-\frac{5\kappa_4}{42}(\Delta\hat\alpha)^2\left(\frac{G_*}{G_N}\right)^{-2/3}\eta^{2/5}u_2^{-2}\biggr]\biggr\}u_2^{-5}, \label{eq:phase-correction-Horndeski}\\
\delta\Psi^{(2)}_{B,\rm ST}&=\delta\Psi^{(2)}_{L,\rm ST}\notag\\
&:=-\frac{m_s^2 d_L}{4\pi f}-\frac{3}{128}\biggl\{1-\biggl(\frac{G_*}{G_N}\biggr)^{-5/3}\biggl[1-\frac{2}{3}\delta_0-\frac{\kappa_4}{6}\Gamma^2-\frac{5\kappa_4}{42}(\Delta\hat\alpha)^2\left(\frac{G_*}{G_N}\right)^{-2/3}\eta^{2/5}u_2^{-2}\biggr]\biggr\}u_2^{-5},\\
\delta\Psi^{(1)}_{B,\rm ST}&=\delta\Psi^{(1)}_{L,\rm ST}\notag\\
&:=-\frac{m_s^2 d_L}{4\pi f}-\frac{3}{256}\biggl\{1-\biggl(\frac{G_*}{G_N}\biggr)^{-5/3}\biggl[1-\frac{2}{3}\delta_0-\frac{\kappa_4}{6}\Gamma^2-\frac{5\kappa_4}{42}(\Delta\hat\alpha)^2\left(\frac{G_*}{G_N}\right)^{-2/3}\eta^{2/5}u_1^{-2}\biggr]\biggr\}u_1^{-5}.
\end{align}
Modified gravity parameters in the waveforms are summarized in Table~\ref{tab:Notations-ST}.
Note that the above waveforms of extra polarizations are non-vanishing only in the presence of non-vanishing scalar hairs and non-minimal coupling (i.e., only for $\hat\alpha_{(1,2)}\neq0$ and $g_4\neq0$). Here, the phases of extra polarizations have a mass-dependent term that vanishes if the group velocity of all polarizations is identical. The PPE parametrization in Eq.\ \eqref{Eq:PPE_Tlm}-\eqref{Eq:PPE_p11} with $\beta=2\beta_P$ effectively assumes all polarization modes to propagate at the same speed, thus  hereafter we ignore the mass-dependent phase term.

\begin{table}[h!]
    \centering
    \renewcommand{\arraystretch}{1.3}
    \begin{tabular}{|c|c|c|c|}
    \hline
       Quantity   & Definition & Description & \begin{tabular} {@{}c@{}} GR value \\ (canonical \\
        scalar) \end{tabular} \\ \hline
        $G_*$  & $G_*\equiv(16\pi \bar G_4)^{-1}$ &  modified Newtonian constant  & $G_N$\\ \hline
        $g_4$ & $g_4\equiv\kappa^{-1/2}({\rm d}\ln G_4/{\rm d}\phi)|_{\varphi=0}$ & perturbation coefficient of $G_4$ & $0$\\ \hline
         $m_s^2$ & \begin{tabular} {@{}c@{}}
         $m_s^2\equiv-\bar G_{2\phi\phi}$ \\
        $\times[\bar G_{2X}-2\bar G_{3\phi}+3(\bar G_{4\phi})^2/\bar G_4]^{-1}$ \end{tabular}
         & effective mass squared &  \\ \hline
        $\hat\alpha_{(1,2)}$ & \begin{tabular} {@{}c@{}}
        $\hat\alpha_{(1,2)}\equiv$ \\
        $\kappa^{-1/2}[{\rm d}\ln m_{(1,2)}/{\rm d}\phi]|_{\varphi=0}$\\
        $-g_4/2$ \end{tabular} & \begin{tabular} {@{}c@{}} primary ($1$) and secondary $(2)$ \\ BH scalar hair \end{tabular} & $0$\\ \hline
        $\Delta\hat\alpha$ & $\Delta\hat\alpha\equiv\hat\alpha_1-\hat\alpha_2$ & BH scalar hair difference & $0$\\ \hline
        $\Gamma$ & $\Gamma\equiv -2(m_2\hat\alpha_1+m_1\hat\alpha_2)/M$&  mass-weighted scalar hair & $0$\\ \hline
        $\kappa_4$ & \begin{tabular} {@{}c@{}}
         $\kappa_4\equiv \kappa \bar G_4$ \\
        $\times[\bar G_{2X}-2\bar G_{3\phi}+3(\bar G_{4\phi})^2/\bar G_4]^{-1}$ \end{tabular}
        & \begin{tabular} {@{}c@{}} parameter capturing \\ both non-canonical scalar \\
        and modified gravity effects \end{tabular} & $1/2$\\ \hline
        $\delta_0$ & $\delta_0\equiv 4\kappa_4\hat\alpha_1\hat\alpha_2$ & \begin{tabular} {@{}c@{}} parameter capturing \\ both non-canonical scalar \\
        and modified gravity effects \end{tabular} & $0$\\
    \hline
    \end{tabular}
    \caption{Modified gravity parameters for the GW waveforms in Horndeski gravity. In this table, we denoted a scalar perturbation by $\varphi=\phi-\bar\phi$ with a background scalar field $\bar\phi$. A quantity with a bar represents evaluation at the background, $\varphi=0$. As has been shown in Ref.~\cite{Higashino:2022izi}, the parameters $\hat\alpha_{(1,2)}$ are directly related to the scalar charges, and thus we call $\hat\alpha_1$ ($\hat\alpha_2$) the primary (secondary) BH scalar hair.
    }
    \label{tab:Notations-ST}
\end{table}

\subsubsection{Mapping to PPE parametrization}
Let us summarize the PPE parameters for the waveforms. By mapping the frequency-domain waveforms into our PPE parametrization in Eqs.~\eqref{Eq:PPE_Tlm}-\eqref{Eq:PPE_p11}, 
we obtain the following expressions, 
\begin{align}
\alpha_T u_2^a&=-1+\biggl(\frac{G_*}{G_N}\biggr)^{5/6}\biggl[1+\frac{1}{3}\delta_0-\frac{\kappa_4}{12}\Gamma^2-\frac{5\kappa_4}{48}(\Delta\hat\alpha)^2\biggl(\frac{G_*}{G_N}\biggr)^{-2/3}\eta^{2/5}u_2^{-2}\biggr], \label{eq: ST-alpha-u2-a}\\
\alpha_{B2}u_2^{a_B}&=-\frac{1}{\sqrt{6}}\kappa_4g_4\Gamma\biggl(\frac{G_*}{G_N}\biggr)^{5/6},\\
\alpha_{L2}u_2^{a_L}&=-\frac{m_s^2}{4(\pi f)^2}\frac{1}{\sqrt{6}}\kappa_4g_4\Gamma\biggl(\frac{G_*}{G_N}\biggr)^{5/6},\\
\alpha_{B1}u_1^{a_B}&=-\frac{1}{2}\sqrt{\frac{10}{3}}\kappa_4g_4(\Delta\hat\alpha)\biggl(\frac{G_*}{G_N}\biggr)^{1/2},\\
\alpha_{L1}u_1^{a_L}&=-\frac{m_s^2}{(2\pi f)^2}\frac{1}{2}\sqrt{\frac{10}{3}}\kappa_4g_4(\Delta\hat\alpha)\biggl(\frac{G_*}{G_N}\biggr)^{1/2},
\end{align}
and
\begin{align}
\beta u_2^b&=2\beta_B u_2^b=2\beta_L u_2^b=-\frac{3}{128}\biggl\{1-\biggl(\frac{G_*}{G_N}\biggr)^{-5/3}\biggl[1-\frac{2}{3}\delta_0-\frac{\kappa_4}{6}\Gamma^2\notag\\
&\quad\quad\quad\quad\quad\quad\quad\quad\quad\ -\frac{5\kappa_4}{42}(\Delta\hat\alpha)^2\left(\frac{G_*}{G_N}\right)^{-2/3}\eta^{2/5}u_2^{-2}\biggr]\biggr\}u_2^{-5}\label{eq: ST-beta-u2-b}\\
\beta_B u_1^b&=\beta_L u_1^b=-\frac{3}{256}\biggl\{1-\biggl(\frac{G_*}{G_N}\biggr)^{-5/3}\biggl[1-\frac{2}{3}\delta_0-\frac{\kappa_4}{6}\Gamma^2-\frac{5\kappa_4}{42}(\Delta\hat\alpha)^2\left(\frac{G_*}{G_N}\right)^{-2/3}\eta^{2/5}u_1^{-2}\biggr]\biggr\}u_1^{-5}.\label{eq: ST-beta1-u1-b}
\end{align}
From Eqs.~\eqref{eq: ST-alpha-u2-a},~\eqref{eq: ST-beta-u2-b}, and~\eqref{eq: ST-beta1-u1-b}, one can find that the current class of scalar-tensor theories identifies two branches in which modifications in the waveforms appear at different PN order, which are characterized by\footnote{In principle, it would be possible to consider another branch in which the mass-dependent term in the phase that we have ignored dominates the others.
However, if one detunes model parameters in such a way that terms with a lower power of frequency can be dominant in the above phase, there might be a possibility that such a term is comparable to would-be subdominant terms such as higher-PN corrections. We thus consider the above two branches only, but it would be interesting to consider another branch, investigate whether other corrections are always smaller than the mass-dependent term even after detuning parameters, and study the impact of the mass-dependent term on the SNR.}
\begin{align}
(1).\quad &\ \biggl|\biggl(\frac{G_*}{G_N}\biggr)^{-5/6}-1\biggr|, \biggl|\biggl(\frac{G_*}{G_N}\biggr)^{5/3}-1\biggr|, |\delta_0|, |\kappa_4\Gamma^2|\gg|\kappa_4(\Delta\hat\alpha)^2|\biggl(\frac{G_*}{G_N}\biggr)^{-2/3}\eta^{2/5}u_m^{-2},\\
(2).\quad &\ \biggl|\biggl(\frac{G_*}{G_N}\biggr)^{-5/6}-1\biggr|, \biggl|\biggl(\frac{G_*}{G_N}\biggr)^{5/3}-1\biggr|, |\delta_0|, |\kappa_4\Gamma^2|\ll|\kappa_4(\Delta\hat\alpha)^2|\biggl(\frac{G_*}{G_N}\biggr)^{-2/3}\eta^{2/5}u_m^{-2}.
\end{align}
Here, $\delta_0$ and $\kappa_4\Gamma^2$ are of order $\kappa_4\hat\alpha_{(1,2)}^2$, while $\kappa_4(\Delta\hat\alpha)^2\left(\frac{G_*}{G_N}\right)^{-2/3}\eta^{2/5}u_m^{-2}$ is of order $\kappa_4(\Delta\hat\alpha)^2(1/v^2)$, and thus we generally have $\delta_0, \kappa_4\Gamma^2\ll\kappa_4(\Delta\hat\alpha)^2\left(\frac{G_*}{G_N}\right)^{-2/3}\eta^{2/5}u_m^{-2}$ under the PN approximation.
However, this hierarchy can be reversed for the case of $\Delta\hat\alpha\approx0$, which motivates us to consider the above two cases.

For the first branch, we have 
\begin{align}
\alpha_T&=-1+\biggl(\frac{G_*}{G_N}\biggr)^{5/6}\left(1+\frac{1}{3}\delta_0-\frac{\kappa_4}{12}\Gamma^2\right)\ {\rm with}\ a=0, \label{eq: ST-alpha-first-branch}\\
\alpha_{B2}&=-\frac{1}{\sqrt{6}}\kappa_4g_4\Gamma\biggl(\frac{G_*}{G_N}\biggr)^{5/6}\ {\rm with}\ a_B=0, \label{eq: ST-alphab2-first-branch}\\
\alpha_{L2}&=-\frac{1}{4}(G_Nm_s\mathcal{M}_z)^2\frac{1}{\sqrt{6}}\kappa_4g_4\Gamma\biggl(\frac{G_*}{G_N}\biggr)^{5/6}\ {\rm with}\ a_L=-6,\label{eq: ST-alphal2-first-branch}\\
\alpha_{B1}&=-\frac{1}{2}\sqrt{\frac{10}{3}}\kappa_4g_4(\Delta\hat\alpha)\biggl(\frac{G_*}{G_N}\biggr)^{1/2},\label{eq: ST-alphab1-first-branch}\\
\alpha_{L1}&=-(G_Nm_s\mathcal{M}_z)^2\frac{1}{2}\sqrt{\frac{10}{3}}\kappa_4g_4(\Delta\hat\alpha)\biggl(\frac{G_*}{G_N}\biggr)^{1/2},\label{eq: ST-alphal1-first-branch}
\end{align}
and  
\begin{align}
\beta=2\beta_B=2\beta_L&=-\frac{3}{128}\biggl[1-\biggl(\frac{G_*}{G_N}\biggr)^{-5/3}\left(1-\frac{2}{3}\delta_0-\frac{\kappa_4}{6}\Gamma^2\right)\biggr]\ {\rm with}\ b=-5. \label{eq: ST-beta-first-branch}
\end{align}

For the second branch, we have
\begin{align}
\alpha_T&=-\frac{5\kappa_4}{48}(\Delta\hat\alpha)^2\biggl(\frac{G_*}{G_N}\biggr)^{1/6}\eta^{2/5}\ {\rm with}\ a=-2, \label{eq: ST-alpha-second-branch}\\
\alpha_{B2}&=-\frac{1}{\sqrt{6}}\kappa_4g_4\Gamma\biggl(\frac{G_*}{G_N}\biggr)^{5/6}\ {\rm with}\ a_B=0, \label{eq: ST-alphab2-second-branch}\\
\alpha_{L2}&=-\frac{1}{4}(G_Nm_s\mathcal{M}_z)^2\frac{1}{\sqrt{6}}\kappa_4g_4\Gamma\biggl(\frac{G_*}{G_N}\biggr)^{5/6}\ {\rm with}\ a_L=-6, \label{eq: ST-alphal2-second-branch}\\
\alpha_{B1}&=-\frac{1}{2}\sqrt{\frac{10}{3}}\kappa_4g_4(\Delta\hat\alpha)\biggl(\frac{G_*}{G_N}\biggr)^{1/2}, \label{eq: ST-alphab1-second-branch}\\
\alpha_{L1}&=-(G_Nm_s\mathcal{M}_z)^2\frac{1}{2}\sqrt{\frac{10}{3}}\kappa_4g_4(\Delta\hat\alpha)\biggl(\frac{G_*}{G_N}\biggr)^{1/2},\label{eq: ST-alphal1-second-branch}
\end{align}
and 
\begin{align}
\beta=2\beta_B=2\beta_L&=-\frac{5\kappa_4}{1792}\biggl(\frac{G_*}{G_N}\biggr)^{-\frac73}(\Delta\hat\alpha)^2\eta^{2/5}\ {\rm with}\ b=-7.\label{eq: ST-beta-second-branch}
\end{align}
Note that the two subcases share the same expression of the PPE parameters $\alpha_P$ for the extra polarizations.

\subsection{Einstein-æther Theory}\label{sec:EA_theory}
Einstein-æther theory (${\text{\AE}}$)~\cite{Eling:2004dk,Khodadi:2020gns,Zhang:2019iim,Oost:2018tcv,Yagi:2013ava} is a prominent example of a locally Lorentz violating vector-tensor theory~\cite{BeltranJimenez:2008zzi,Tasinato:2014eka,Heisenberg:2014rta,Kimura:2016rzw,Heisenberg:2018mxx}: theories where an additional vector field, the {\it æther} vector field, is coupled to the metric.
The imposed presence of a non vanishing vector field introduces a preferred frame and, hence, break local Lorentz invariance allowing to test the validity of the Lorentz symmetry in the gravitational sector, one of the pillars of GR. Note that crucially, in \AE gravity, as a metric theory, Lorentz symmetry is only violated in the gravity sector, such that it is an interesting theory to consider even in the light of the existing tight constraints on Lorentz symmetry in the matter sector. In \AE theory, the æther vector field is constrained to have a unit norm, thus breaking Lorentz symmetry explicitly. In contrast to theories with spontaneous Lorentz symmetry breaking, such as Generalized Proca gravity \cite{Heisenberg:2014rta,BeltranJimenez:2016rff}, the constraints introduced through the explicit breaking permit for the presence of a larger variety of viable vector-metric coupling operators that do not change the number of propagating degrees of freedom.
Through these operators, the æther field interacts with the metric and modifies the gravitational dynamics, leading to unique predictions for the behavior of GWs, the dynamics of compact objects~\cite{Jacobson:2000xp, Foster:2006az,Foster:2005dk,Eling:2006df} and cosmology~\cite{Carroll:2004ai}.
In general, vector-tensor theories are also motivated by 
\begin{itemize}
    \item {\it Quantum Gravity and High-Energy Physics}: At high energies, near the Planck scale, the classical description of spacetime provided by GR is expected to break down. Vector-tensor theories offer a framework to explore potential modifications to the spacetime structure that could arise from quantum gravitational effects. These theories can provide insights into how Lorentz symmetry might be violated at high energies, leading to new physics beyond GR \cite{Jacobson:2008aj, Elliott:2005va}.
    \item {\it Dark Energy and Dark Matter}: Vector fields in vector-tensor theories can play a role in cosmology by contributing to the dynamics of dark energy and dark matter. For instance, vector fields could act as candidates for dark energy, driving the accelerated expansion of the universe. They might also interact with dark matter, influencing its distribution and behavior on cosmic scales \cite{Carroll:2004ai}.
    \item {\it Gravitational Wave Polarizations}: In GR, GWs have only two polarization modes: the plus and cross modes. However, vector-tensor theories predict additional polarization modes due to the presence of the vector field. These extra modes can affect the propagation and detection of GWs, providing a potential observational signature of Lorentz symmetry breaking \cite{Yagi:2013ava}.
\end{itemize}

The total action of \AE theory is given by $S = S_{\text{æ}} + S_{\text{mat}}$, where the gravitational sector reads
by~\cite{Schumacher:2023cxh,Jacobson:2000xp,Jacobson:2007veq}:
\begin{equation}
S = -\frac{1}{16\pi G_{\text{æ}}} \int \sqrt{-g} \, d^4x \left[ R + \lambda (U^\mu U_\mu - 1) + \frac{1}{3} c_\theta \theta^2 + c_\sigma \sigma^{\mu\nu} \sigma_{\mu\nu} + c_\omega \omega^{\mu\nu} \omega_{\mu\nu} + c_a A^\mu A_\mu \right]\,,
\end{equation}
where $G_{\text{æ}}$ is the "bare" gravitational constant related to Newton's constant through $G_{\text{æ}} = \left( 1 - \frac{c_a}{2} \right) G_N $ and $\lambda$ is the Lagrange multiplier enforcing the unit norm condition of the æther's four-velocity $U^\mu$. 
The quantities $\theta$, $\sigma_{\mu\nu}$, $\omega_{\mu\nu}$, and $A^\mu$ are respectively the expansion, shear, vorticity (or twist), and acceleration of the æther's four-velocity $U^\mu$. 
The action displays four dimensionless coupling constants: $c_\theta$, $c_\sigma$, $c_\omega$ and $c_a$ which regulate the strength of the interactions between the æther field and the metric \cite{Jacobson:2007veq, Foster:2006az}. These have to satisfy observational bounds and theoretical stability constraints~\cite{Oost:2018tcv,Yagi:2013ava,Schumacher:2023cxh}.
In particular, to avoid gradient instabilities and ghost modes, the squared speeds of the GW polarizations 
\begin{align}
c_T^2 &= \frac{1}{1 - c_\sigma}, \qquad 
c_V^2 = \frac{c_\sigma + c_\omega - c_\sigma c_\omega}{2c_a (1 - c_\sigma)}, \qquad 
c_S^2 = \frac{(c_\theta + 2c_\sigma)(1 - c_a/2)}{3c_a (1 - c_\sigma)(1 + c_\theta/2)}.\label{eq:AEScalarSpeed}
\end{align}
must all be $\geq 0$.  Additionally, the energy densities of the different modes
\begin{equation}
    E_T = \frac{k^2 |A|^2}{8\pi G_{\text{\ae}}} \,, \qquad E_V = \frac{k^2 |A|^2}{8\pi G_{\text{\ae}}}  \frac{c_\sigma + c_\omega (1 - c_\sigma)}{1 - c_\sigma}\,, \qquad E_S = \frac{k^2 |A|^2}{8\pi G_{\text{\ae}}}  c_a (2 - c_a)
\end{equation}
 must be positive to avoid negative energy propagation.

\subsubsection{Frequency domain waveforms}
In \AE theory, all the six polarization modes are excited, with the unique signature that the two scalar modes are proportional to each other~\cite{Schumacher:2023cxh,Zhang:2017srh} as $\tilde{h}^{\text{\ae}}_{L} = a^{\text{\ae}}_{BL}\tilde{h}^{\text{\ae}}_{B} (f)$ with $a^{\text{\ae}}_{BL} = 1 + 2 \beta^{\text{\ae}}_2$ for both angular harmonics $(2,2)$ and $(1,1)$ (see Table~\ref{tab:AEparameters} for the definitions of the parameters), with propagation speed given in Eq.~\eqref{eq:AEScalarSpeed}.  
The waveforms in frequency domain are given by~\cite{Schumacher:2023cxh,Zhang:2017srh}
\begin{align}
    \tilde{h}^{(2,2)}_{T}(f) &= A^{(2)}  e^{ - i   \Psi^{(2)}_{GR}}  \left[ A^{(2)}_\text{\ae} e^{i \delta \Psi^{(2)}_{\text{\ae}} } \right]   e^{-i 2\pi f d_L (1 - c_T^{-1})}, \label{eqn:hAET2} \\
    \tilde{h}^{(2,2)}_{V} (f) &=  A^{(2)} e^{ - i   \Psi^{(2)}_{GR}}\left[\frac{ A^{(2)}_{\text{\ae}}  e^{i \delta \Psi^{(2)}_{\text{\ae}}  }}{2}  \left[  \frac{ -i \beta^{\text{\ae}}_1}{c_\sigma + c_\omega - c_\sigma c_\omega} \frac{1}{c_V} \left( \mathcal{S} - \frac{c_{\sigma}}{1 - c_{\sigma}} \right)  \right]  \right]      e^{-i2 \pi f d_L\left(1 - c_{V}^{-1} \right)}, \label{eqn:hAEV2} \\
    \tilde{h}^{(2,2)}_{B} (f) &=  A^{(2)}   e^{ - i   \Psi^{(2)}_{GR}} \left[ \frac{A^{(2)}_{\text{\ae}} e^{i \delta \Psi^{(2)}_{\text{\ae}} } }{\sqrt{6}}   \left[ \frac{-1}{2 - c_{a}} \left(3c_{a} (Z - 1)-\frac{ 2 \mathcal{S}}{c_S^2} \right)  \right] \right]    e^{-i2 \pi f d_L\left(1 - c_{S}^{-1} \right)}, \label{eqn:hAEB2}
\end{align}
for the quadrupole, and the dipole $\ell = 1$ modes
\begin{align}
    \tilde{h}^{(1,1)}_{V}(f) &= A^{(1)} e^{ - i   \Psi^{(1)}_{GR}}  \left[ A^{(1)}_{\text{\ae}} e^{i \delta \Psi^{(1)}_{\text{\ae}}  }  \left[ \frac{-\beta^{\text{\ae}}_1}{c_\sigma + c_\omega - c_\sigma c_\omega}  \right] \right] e^{-i2 \pi f d_L\left(1 - c_{V}^{-1} \right)} ,\label{eqn:hAEV1}\\
    \tilde{h}^{(1,1)}_{B} (f) &= A^{(1)} e^{ - i   \Psi^{(1)}_{GR}} \left[  A^{(1)}_{\text{\ae}} e^{i \delta \Psi^{(1)}_{\text{\ae}} }    \left[ \frac{ \sqrt{2} i   }{(2 - c_{a})c_S} \right] \right]   e^{-i2 \pi f d_L\left(1 - c_{S}^{-1} \right)}, \label{eqn:hAEB1}
\end{align}
where $A^{(m)}$ and $\Psi^{(m)}_{GR}$ are the GR amplitude and phase defined in Eqs.~\eqref{Eq:Psi_GR} and~\eqref{Eq:h_PN_aell}  (note the different convention in the phase of coalescence $\phi_c = - \Phi_c$ of~\cite{Schumacher:2023cxh}). 
In the expressions above, we have introduced the following phase and amplitude modifications:
\begin{align}
    \delta \Psi^{(2)}_{\text{\ae}} (f) &= \frac{3 }{64 } \frac{(1-s_1)(1-s_2)}{(2-c_a) \kappa_3} \left( G_N \pi \bar{\cal M} f\right)^{-\frac53} \left[1 -    \frac{4 \, \eta^\frac25 \epsilon_x}{7 \left( G_N \pi \bar{\cal M} f\right)^{\frac23}} \right] + 2 \pi f t_c -\frac{\pi}{4}+ \Psi^{(2)}_{\rm GR} \,,\label{eq:Psi2AE} \\
    \delta \Psi^{(1)}_{\text{\ae}} (f) &= \frac{3 }{128 } \frac{(1-s_1)(1-s_2)}{(2-c_a) \kappa_3} \left(2 G_N \pi \bar{\cal M} f\right)^{-\frac53} \left[1 - \frac{4 \,  \eta^\frac25 \epsilon_x }{7 \left( 2 G_N \pi \bar{\cal M} f\right)^{\frac23}}\right]+ 2 \pi f t_c -\frac{\pi}{4} + \Psi^{(1)}_{\rm GR} \,,\label{eq:Psi1AE} \\
    A^{(2)}_{\text{\ae}} 
    &= \sqrt{\frac{(2-c_a)}{2 \kappa_3}} [(1-s_1)(1-s_2)]^{\frac13} \left[ 1 - \frac{\eta^{\frac25} \epsilon_x}{2  [(1-s_1)(1-s_2)]^{\frac23} u^2_2 }  \right] \,,\label{eq:A2AE} \\ 
    A^{(1)}_{\text{\ae}} 
    &=  \sqrt{\frac{5}{96}} \sqrt{\frac{(2-c_a)}{2\kappa_3}} \Delta s  \left[ 1 - \frac{\eta^{\frac25} \epsilon_x}{2 [(1-s_1)(1-s_2)]^{\frac23} u^2_1 }   \right] \,, \label{eq:A1AE}
\end{align}
where $s_i$ are the sensitivities of the compact objects and all the quantities are defined in Table~\ref{tab:AEparameters}.
Note that here we have included an additional $8 \sqrt{\pi/5}$ factor in the amplitude of the $\ell = 2$ modes, and $4 \sqrt{\pi/3}$ for the $\ell = 1$, coming from the spin-weighted spherical harmonics, compared to~\cite{Schumacher:2023cxh}\footnote{We correct a factor of $2$ missing in the expression of in $A^{(1)}_{\text{\ae}}$ of~\cite{Schumacher:2023cxh}, which does not affect the results of that paper.}. 
Note that the modified frequency evolution of the waveform is computed up to the $-1$PN order compared to the leading GR one, namely $\propto u^{-5}_m$. Therefore, in Eqs.~\eqref{eq:Psi2AE},~\eqref{eq:Psi1AE} we keep $\Psi^{(m)}_{\rm GR}$ up to the same PN order, i.e $\Psi^{(m)}_{\rm GR} = - 2 \pi ft_c + \frac{\pi}{4} - 3 m /(256 u_{m}^5)$.

We expand the phase and amplitude close to GR values $s_1, s_2, c_a \approx 0$ and $\kappa_3 \approx 1$,
\begin{align}
    \delta \Psi^{(2)}_{\text{\ae}} (f) &=  \left[\frac{3 }{64 } \frac{[(1-s_1)(1-s_2)]^{-\frac23}}{(2-c_a) \kappa_3} \right]^{(1)} u^{-5}_2  +   \left[-    \frac{3 \, \eta^\frac25 \epsilon_x}{224 [ (1-s_1)(1-s_2)]^{-\frac23}} \right] u^{-7}_2\label{eq:Psi2AEApprox} \\
    \delta \Psi^{(1)}_{\text{\ae}} (f) &=  \left[\frac{3 }{128} \frac{[(1-s_1)(1-s_2)]^{-\frac23}}{(2-c_a) \kappa_3} \right]^{(1)} u^{-5}_1  +   \left[-    \frac{3 \, \eta^\frac25 \epsilon_x}{448 [ (1-s_1)(1-s_2)]^{-\frac23}} \right] u^{-7}_1\label{eq:Psi1AEApprox} \\
    A^{(2)}_{\text{\ae}} &=  1 +  \left[ \sqrt{\frac{(2-c_a)}{2 \kappa_3}} [(1-s_1)(1-s_2)]^{\frac13} \right]^{(1)} -  \left[\frac{\eta^{\frac25} \epsilon_x}{2  [(1-s_1)(1-s_2)]^{\frac23}  }  \right] u^{-2}_2 \label{eq:A2AEApprox} \\ 
    A^{(1)}_{\text{\ae}} &=   \sqrt{\frac{5}{96}} \sqrt{\frac{(2-c_a)}{2\kappa_3}} \Delta s  \left[ 1 - \frac{\eta^{\frac25} \epsilon_x}{2 [(1-s_1)(1-s_2)]^{\frac23} u^2_1 }   \right] \label{eq:A1AEApprox}
\end{align}
with the understanding that each of the parenthesis above has to be kept at first order in the expansion of the parameters around the GR values. The apex $(1)$ above the square brackets above stands for retaining only the first order term in a Taylor expansion of the quantity in the brackets around the GR values.

\begin{table}[h!]
    \centering
    \renewcommand{\arraystretch}{1.4}
    \begin{tabular}{|c|c|c|}
    \hline
    Quantity & Definition & $c_T^2 = c_V^2 = c_S^2 = 1$ \\ 
    \hline
    $a^{\text{\ae}}_{BL}$ & $1 + 2\beta^{\text{\ae}}_2$ & \\ \hline
    $\bar{{\cal M}}$ & $(1-s_1)(1-s_2){\cal M}_z$ & \\ \hline
    $\mu_A$ & $\displaystyle \frac{m_A}{m_1 + m_2}$ & \\ \hline
    $\beta^{\text{\ae}}_1$ & $\displaystyle -\frac{2 c_\sigma}{c_V}$ & 0 \\ \hline
    $\beta^{\text{\ae}}_2$ & $\displaystyle \frac{c_a - c_\sigma}{2 c_a (1 - c_\sigma) c_S^2}$ & $\tfrac{1}{2}$ \\ \hline
    $Z$ & $\displaystyle \frac{(\alpha^{\text{\ae}}_1 - 2\alpha^{\text{\ae}}_2)(1 - c_\sigma)}{3(2c_\sigma - c_a)}$ & $\tfrac{4}{3}$ \\ \hline
    $\alpha^{\text{\ae}}_1$ & $\displaystyle 4\,\frac{c_\omega (c_a - 2 c_\sigma) + c_a c_\sigma}{c_\omega(c_\sigma - 1) - c_\sigma}$ & $-4 c_a$ \\ \hline
    $\alpha^{\text{\ae}}_2$ & $\displaystyle \frac{\alpha^{\text{\ae}}_1}{2} + \frac{3 (c_a - 2 c_\sigma)(c_\theta + c_a)}{(2 - c_a)(c_\theta + 2 c_\sigma)}$ & 0 \\ \hline
    $\Delta s$ & $s_1 - s_2$ & \\ \hline
    $\epsilon_x$ & $\displaystyle \frac{5\Delta s^2}{32 \kappa_3}  \mathcal{C}$ & $\displaystyle \frac{5 \Delta s^2}{4} \frac{1}{c_a [(12 -c_a) - 4 {\cal S} + (10-3c_a) {\cal S}^2]}$\\ \hline
    $\kappa_3$ & $\displaystyle \mathcal{A}_1 + \mathcal{A}_2 \mathcal{S} + \mathcal{A}_3 \mathcal{S}^2$ & $\displaystyle \frac{(12 -c_a) - 4 {\cal S} + (10-3c_a) {\cal S}^2}{6(2-c_a)}$ \\ \hline
    ${\cal S}$ & $s_1 \mu_1 + s_2 \mu_2$ & \\ \hline
    $\mathcal{A}_1$ & $\displaystyle \frac{1}{c_T} + \frac{2 c_{a}c_{\sigma}^2}{(c_\sigma + c_\omega - c_\sigma c_\omega)^2 c_V} + \frac{3 c_{a} (Z - 1)^2}{2(2 - c_{a}) c_S}$ & $\displaystyle \frac{12 - 5c_a}{6(2-c_a)}$\\ \hline
    $\mathcal{A}_2$ & $\displaystyle - \frac{2c_{\sigma}}{(c_\sigma + c_\omega - c_\sigma c_\omega) c_V^3} - \frac{2(Z - 1)}{(2 - c_{a})c_S^3}$ & $\displaystyle - \frac{2}{3(2-c_a)}$\\ \hline
    $\mathcal{A}_3$ & $\displaystyle \frac{1}{2c_{a} c_V^5} + \frac{2}{3c_{a}(2 - c_{a})c_S^5}$ & $\displaystyle \frac{10 - 3c_a}{6 c_a (2 - c_a)}$\\ \hline
    $\mathcal{C}$ & $\displaystyle \frac{4}{3 c_{a} c_V^3} + \frac{4}{3 c_{a} (2 - c_{a})c_S^3}$ & $\displaystyle \frac{4}{3 c_a} \frac{3-c_a}{2-c_a}$\\
    \hline
    \end{tabular}
    \caption{Parameters and definitions in Einstein-\ae ther theory. The third column lists simplified expressions assuming $c_T^2 = c_V^2 = c_S^2 = 1$, and when empty is because the expressions remain the same.}
    \label{tab:AEparameters}
\end{table}

\subsubsection{Mapping to PPE parametrization} 
We now proceed to map the frequency-domain waveforms into the PPE parameterization, i.e.  
Specifically, we aim to recast Eqs.~\eqref{eqn:hAET2}-\eqref{eqn:hAEB1}  in the form of Eqs.~\eqref{Eq:PPE_Tlm},~\eqref{Eq:PPE_p22} and~\eqref{Eq:PPE_p11}. 
We note that, even though the mapping equations are written assuming equal propagation speeds $c_T^2 = c_V^2 = c_S^2$, while Eqs.~\eqref{eqn:hAET2}-\eqref{eqn:hAEB1} do not necessarily assume so, here we provide the general expressions for the PPE parameters for generality. 
Their specific values when all polarizations travel at the same speed can be found by reducing the parameter space suitably. Interestingly, when $c_T^2 = c_V^2 = c_S^2 = 1$, the amplitude of the vector modes vanishes, as it can be seen from Table~\ref{tab:AEparameters}.

The mapping of the phase leads to $\beta=2\beta_P$ with 
\begin{align}
      |m| \beta_P \, u_m^b &= \delta \Psi^{(m)}_{\text{\ae}}
\end{align}
for both angular harmonics $\ell=|m|=1$ and  $\ell=|m|=2$, while the amplitudes are mapped as follows:
\begin{align}
    (1 +\alpha_T u_2^a ) &= A^{(2)}_{\text{\ae}} \,,\\
    \alpha_{B2} \, u_2^{a_{B}} &=  \frac{A^{(2)}_{\text{\ae}} }{ \sqrt{6}}  \left[ \frac{-1}{2 - c_{a}} \left(3c_{a} (Z - 1)-\frac{ 2 \mathcal{S}}{c_S^2} \right)  \right]  \,, \qquad \alpha_{B1} \, u_1^{a_{B}} = A^{(1)}_{\text{\ae}}     \left[ \frac{\sqrt{2} i}{(2 - c_{a})c_S} \right]   \,, \\
     \alpha_{V2} \, u_2^{a_{V}} &=  \frac{A^{(2)}_{\text{\ae}}}{2} \left[  \frac{-i\beta^{\text{\ae}}_1}{c_\sigma + c_\omega - c_\sigma c_\omega} \frac{1}{c_V} \left( \mathcal{S} - \frac{c_{\sigma}}{1 - c_{\sigma}} \right)  \right] \,, \quad \alpha_{ V1} \, u_1^{a_{V}} = A^{(1)}_{\text{\ae}} \left[ \frac{ - \beta^{\text{\ae}}_1}{c_\sigma + c_\omega - c_\sigma c_\omega} \right]\,.
\end{align}
The next goal is to extract the specific values of the PPE parameters. Looking at Eqs.~\eqref{eq:Psi1AEApprox}, ~\eqref{eq:A2AEApprox} and~\eqref{eq:A1AEApprox} we learn that this is possible only if we separately consider the cases where the binary evolution is driven by the quadrupole or the dipole emission. Indeed, while the coefficients of $u_m$ in those equations are supposed to be small in this parametrized approach, the different frequency evolution can change the relative amplitude of the corrections.

\paragraph{Dipole.}

First, we consider the case where the modified phase and amplitude are dominated by the dipole emission. This means considering the $-1$PN terms  in Eqs.~\eqref{eq:Psi1AEApprox},~\eqref{eq:A2AEApprox} and~\eqref{eq:A1AEApprox}, i.e.\ those $\propto u^{-5}_m$ in the phase and $\propto u^{0}_m$ in the amplitude.
In this case we obtain 
\begin{align}
    b = -5 \,, \qquad \beta_P = \left[\frac{3 }{128 } \frac{[(1-s_1)(1-s_2)]^{-\frac23}}{(2-c_a) \kappa_3} \right]^{(1)}\,,
\end{align}
with also $\beta = 2 \beta_P$ and 
\begin{align}
    a&= a_B = a_V = 0 \\
    \alpha_T &= \left[ \sqrt{\frac{(2-c_a)}{2 \kappa_3}} [(1-s_1)(1-s_2)]^{\frac13} \right]^{(1)} \label{eq:EAalphaTa0}\\
    \alpha_{B2} &= \frac{1}{\sqrt{6}} \left[ \sqrt{\frac{(2-c_a)}{2 \kappa_3}} [(1-s_1)(1-s_2)]^{\frac13} \right]^{(1)} \left[ \frac{-1}{2 - c_{a}} \left(3c_{a} (Z - 1)-\frac{ 2 \mathcal{S}}{c_S^2} \right)  \right]  \label{eq:EAalphaB2a0}\\
    \alpha_{V2} &= \frac{1}{2} \left[ \sqrt{\frac{(2-c_a)}{2 \kappa_3}} [(1-s_1)(1-s_2)]^{\frac13} \right]^{(1)} \left[  \frac{-i\beta^{\text{\ae}}_1}{c_\sigma + c_\omega - c_\sigma c_\omega} \frac{1}{c_V} \left( \mathcal{S} - \frac{c_{\sigma}}{1 - c_{\sigma}} \right)  \right] \label{eq:EAalphaV2a0} \\
    \alpha_{B1} &= \sqrt{\frac{5}{96}} \sqrt{\frac{(2-c_a)}{2\kappa_3}} \Delta s  \left[ \frac{\sqrt{2} i}{(2 - c_{a})c_S} \right] \label{eq:EAalphaB1a0}\\
    \alpha_{V1} &=\sqrt{\frac{5}{96}} \sqrt{\frac{(2-c_a)}{2\kappa_3}} \Delta s  \left[ \frac{  - \beta^{\text{\ae}}_1}{c_\sigma + c_\omega - c_\sigma c_\omega} \right]   \label{eq:EAalphaV1a0}\,.
\end{align}

\paragraph{Quadrupole.} 
Second, we consider the case where the modified phase and amplitude evolutions are driven by  quadrupole emission. This means considering terms $\propto u^{-5}_m$ in the phase and $\propto u^{0}_m$ in the amplitude in Eqs.~\eqref{eq:Psi1AEApprox},~\eqref{eq:A2AEApprox} and~\eqref{eq:A1AEApprox}.
In this case we obtain
\begin{align}
     b = -7\,, \qquad \beta_P = -    \frac{3 \, \eta^\frac25 \epsilon_x}{448 [ (1-s_1)(1-s_2)]^{-\frac23}}\,,
\end{align}
and 
\begin{align}
    a&= a_B = a_V = -2 \\
    \alpha_T &= -  \left[\frac{\eta^{\frac25} \epsilon_x}{2  [(1-s_1)(1-s_2)]^{\frac23}  }  \right] \label{eq:EAalphaTa2} \\
    \alpha_{B2} &= -\frac{1}{\sqrt{6}}  \left[\frac{\eta^{\frac25} \epsilon_x}{2  [(1-s_1)(1-s_2)]^{\frac23}  }  \right] \left[ \frac{-1}{2 - c_{a}} \left(3c_{a} (Z - 1)-\frac{ 2 \mathcal{S}}{c_S^2} \right)  \right]  \label{eq:EAalphaB2a2}\\
    \alpha_{V2} &= - \frac{1}{2} \left[\frac{\eta^{\frac25} \epsilon_x}{2  [(1-s_1)(1-s_2)]^{\frac23}  }  \right] \left[  \frac{- i\beta^{\text{\ae}}_1}{c_\sigma + c_\omega - c_\sigma c_\omega} \frac{1}{c_V} \left( \mathcal{S} - \frac{c_{\sigma}}{1 - c_{\sigma}} \right)  \right]  \label{eq:EAalphaV2a2}\\
    \alpha_{B1} &= -\sqrt{\frac{5}{96}} \sqrt{\frac{(2-c_a)}{2\kappa_3}} \Delta s   \frac{\eta^{\frac25} \epsilon_x}{2 [(1-s_1)(1-s_2)]^{\frac23}}  \left[ \frac{ \sqrt{2} i}{(2 - c_{a})c_S} \right] \label{eq:EAalphaB1a2}\\
    \alpha_{V1} &=  - \sqrt{\frac{5}{96}} \sqrt{\frac{(2-c_a)}{2\kappa_3}} \Delta s   \frac{\eta^{\frac25} \epsilon_x}{2 [(1-s_1)(1-s_2)]^{\frac23} } \left[ \frac{ - \beta^{\text{\ae}}_1}{c_\sigma + c_\omega - c_\sigma c_\omega} \right]    \label{eq:EAalphaV1a2}\,.
\end{align}

\subsection{Massive Gravity: Rosen theory}\label{sec:Rosen_theory}
In this section we map to the PPE parametrization Rosen's theory~\cite{Rosen:1974ua,RosenPhysRevD.3.2317}. 
This is an example of bimetric theory with flat, non-dynamical prior geometry, $\eta_{\mu\nu}$, and a dynamical rank-$2$ tensorial gravitational field, $g_{\mu\nu}$, whose action takes the form
\begin{equation}
    S = \frac{1}{32 \pi G_N} \int d^4 x \, \sqrt{-\eta} \, \eta^{\mu\nu} g^{\alpha \beta} g^{\gamma \delta} \bar \nabla_\mu g_{\alpha [ \gamma} \bar \nabla_{|\nu| }g_{\beta]\delta}\,,
\end{equation}
where $\bar \nabla_\mu$ are the derivative of the flat metric. Rosen's theory is designed to reproduce the post-Newtonian limit  of GR except for the parametrized post-Newtonian (PPN) parameter $\alpha_2$, which has been constrained through observations to be less than $4 \times 10^{-7}$~\cite{Will:2005va}.
On the contrary, the strong field regime and the radiative sector display substantial differences with respect to GR:  Rosen theory does not admit black hole solutions; the upper mass limit for neutron stars is higher;  cosmological models are different from the standard case, and its the polarization content is richer~\cite{Will:1977zz,Rosen:1974ua}. Although Rosen theory does not admit black hole solutions, the theory does support highly compact stellar configurations. The authors of \cite{1985ApJ...291..417H} showed that stable compact objects in bimetric gravity can reach central densities orders of magnitude higher than the GR maximum, which would act as black hole mimickers, i.e.\ horizonless bodies that can be as compact as or more compact than neutron stars. The inspiral-phase PPE waveforms derived in this section apply to such objects, since the PN point-particle treatment is agnostic about internal structure.

The absence of black holes in Rosen theory rules it out by observations. In addition, the theory's GW predictions have been shown to be inconsistent with timing observations of the Hulse-Taylor binary pulsar \cite{Will:1977zz}, since the orbital period undergoes through rapid changes driven by the dipole emission of gravitational radiation allowed in the theory. We thus include Rosen theory here not as a physically viable candidate for the true theory of gravity, but as a well-studied theoretical laboratory for exploring the PPE formalism with complete extra polarization content. It remains one of the few theories for which closed-form PN waveforms including all six polarization modes have been derived, making it a valuable testbed for understanding the PPE parameter mapping and the detectability of extra polarizations. The constraints derived in Section \ref{sec:constrain-specific-theories} for Rosen's theory should therefore be read as illustrating the method, rather than as physical bounds on a viable theory.
 
In the Eardly's~\cite{Eardley:1973zuo}  $E(2)$ classification, Rosen's theory belongs to the $II_6$ class, making some of the six polarization observer-dependent. 

\subsubsection{Frequency domain waveforms}
The GW's solution for Rosen theory have been computed in, e.g.~\cite{Will1977ApJ...214..826W}, and reported in~\cite{Chatziioannou:2012rf}. 
These are given by
\begin{align}
    \tilde{h}^{(2,2)}_{T}(f) &= A^{(2)}  e^{- i \Psi_{GR}^{(2)} } \left(i \sqrt{\frac27}  k_{\text R}^{-3/4} \right) e^{- i \delta \Psi^{(2)}_{\text R} } \\
    \tilde{h}^{(2,2)}_{V}(f)  &= A^{(2)} e^{- i \Psi_{GR}^{(2)} }  \left(  -\sqrt{\frac27}  k_{\text R}^{-3/4} \right)  e^{- i  \delta \Psi^{(2)}_{\text R}} \\
    \tilde{h}^{(2,2)}_{B} (f) &=  A^{(2)}e^{- i \Psi_{GR}^{(2)}} \left( -\frac{  i }{\sqrt{21}}    k_{\text R}^{-3/4}  \right) e^{- i  \delta \Psi^{(2)}_{\text R}} \\
    \tilde{h}^{(2,2)}_{L}(f) &= A^{(2)}e^{- i \Psi_{GR}^{(2)} }  \left(-\frac{  2 i }{\sqrt{21}}    k_{\text R}^{-3/4} \right)  e^{- i  \delta \Psi^{(2)}_{\text R}} \\
\end{align}
for $\ell =2$, and for $\ell = 1$ we have 
\begin{align}
    \tilde{h}^{(1,1)}_{V}(f)&= A^{(1)} e^{- i \Psi_{GR}^{(1)} } \left( \frac43 \sqrt{\frac{10}{21}}    {\cal G} k_{\text R}^{-7/12} \right) e^{- i  \delta \Psi^{(1)}_{\text R} } \\
    \tilde{h}^{(1,1)}_{B} (f) &=   A^{(1)}  e^{- i \Psi_{GR}^{(1)} } \left( \frac{4 i}{3} \sqrt{\frac{5}{21}}{\cal G} k_{\text R}^{-7/12} \right) e^{- i  \delta \Psi^{(1)}_{\text R} } \\
    \tilde{h}^{(1,1)}_{L}(f) &=   A^{(1)}  e^{- i \Psi_{GR}^{(1)} } \left(\frac{4 i}{3} \sqrt{\frac{5}{21}}{\cal G} k_{\text R}^{-7/12} \right)  e^{- i  \delta \Psi^{(1)}_{\text R} } 
\end{align}
with $A^{(m)}$ and $\Psi_{GR}^{(m)}$ the GR amplitude and phase as in Eqs.~\eqref{Eq:h_PN_aell} and~\eqref{Eq:Psi_GR}, and the parameters ${\cal G}$, which is the difference  of the self-gravitating binding energy per unit mass of the binary constituents, and $k_{\text R}$ are defined as 
\begin{equation}
    {\cal G} \equiv \frac{s_1}{m_1} - \frac{s_2 }{m_2 } \,, \qquad k_{\text R} \equiv 1 - (4 s_1 s_2)/3\,,
\end{equation}
where $s_i$ are the sensitivities of the two bodies.
Their values in General Relativity are ${\cal G} = 0$ and $k_{\text R} = 1 $~\cite{Will:1977zz,Will1977ApJ...214..826W}. 
The modification to the phase $\delta \Psi^{(m)}_{\text R}$ are given by 
\begin{equation}
       \delta \Psi_{\text R}^{(m)} =  \frac{3 |m| }{224 u_m^5} k_{\text R}^{-5/6} + \frac{25 |m|}{8232} \, \frac{k_{\text R}^{-2/3} {\cal G}^2 \eta^{2/5}}{u_m^7} +  \frac{3 |m|}{256 u^5_m}\,.
\end{equation}
Similarly to Einstein-\ae ther theory, the phase modifications are known in Rosen's theory only up to the $-1$PN correction compared to the leading order. Hence, we also truncate at $u^{-5}_m$ order.

\subsubsection{Mapping to PPE parametrization} 
Now we want to match the PPE parameterizations, namely Eqs.~\eqref{Eq:PPE_Tlm},~\eqref{Eq:PPE_p22} and~\eqref{Eq:PPE_p11}. In order to do so, the first step is to linearize the expressions above considering that deviations from GR are small. This is particularly tricky for Rosen theory, as it can be seen from the previous expressions, the $\ell = 2$ vector and scalar modes do not vanish for ${\cal G}= 0$ and $k_{\text R} = 1$, and also the phase does not converge to the GR one.
Comparing the expressions we have that $a = a_B = a_L = a_V = 0$ and 
\begin{align}
    \alpha_T &= -1 + i \sqrt{\frac27}  k_{\text R}^{-3/4} \,, \\ 
    \alpha_{V2} &= - \sqrt{\frac27}   k_{\text R}^{-3/4}  \,, \qquad \alpha_{V1} =  \frac43 \sqrt{\frac{10}{21}}    {\cal G} k_{\text R}^{-7/12}   \,,  \\
    \alpha_{B2} &= -\frac{ i }{\sqrt{21}}    k_{\text R}^{-3/4} \,, \qquad \alpha_{B1} = \frac{4 i}{3} \sqrt{\frac{5}{21}}{\cal G} k_{\text R}^{-7/12} \,, \\
    \alpha_{L2} &= -\frac{2 i }{\sqrt{21}}    k_{\text R}^{-3/4}\,, \qquad \alpha_{L1} =\frac{4 i}{3} \sqrt{\frac{5}{21}}{\cal G} k_{\text R}^{-7/12} \,.
\end{align}
The matching of the phase gives the usual $2\beta_P u_2^b =  - \delta \Psi^{(2)}_{\rm R}$, and the tensor phase modification can be found from $\beta = 2 \beta_P$. However, to extract the specific values of the PPE parameters, we have to consider separately the cases where the dynamics of the binary is driven by the dipole or the quadrupole contribution. 

\paragraph{Dipole}
As before, in this case we consider the $-1$PN contributions. This gives

\begin{equation}
    b = -5\,, \qquad \beta_P = - \frac{6 }{512} \left[ 1 + \frac87  k_{\text R}^{-5/6} \right]\,,
\end{equation}

\paragraph{Quadrupole}
In this case we consider the $0$PN order,
while the phase factors are  
\begin{equation}
    b = -7\, \,, \qquad \beta_P = - \frac{50}{16464}k_{\text R}^{-2/3} {\cal G}^2 \eta^{2/5} \,.
\end{equation}

\subsection{Lightman-Lee theory}\label{sec:LL_theory}

The Lightman--Lee theory~\cite{Lightman:1973kun} is a fully conservative metric theory of gravity formulated within the PPN framework. It introduces an auxiliary flat background metric ${\bar \eta}_{\mu\nu}$ 
and a dynamical tensor $B_{\mu\nu}$
in addition to the physical metric $g_{\mu\nu}$, while matter couples minimally to the latter, ensuring geodesic motion and respect of the weak equivalence principle. Its
action
is
\begin{equation}
 S\,=\,-\frac{1}{16 \pi\,G_N}
 \,\int d^4 x\,\sqrt{- \bar \eta}\,
 \left(\frac14 \nabla_\alpha B_{\mu\nu}
 \nabla^\alpha B^{\mu\nu}-\frac{5}{64}\,
 \nabla_\alpha B
 \nabla^\alpha B
 \right)\,,
\end{equation}
with $\bar \eta$ and $\bar B$ indicating traces. The physical metric is then expanded for weak gravitational fields
$g_{\mu\nu}=\eta_{\mu\nu}+h_{\mu\nu}$, and the 
gravitational perturbation is related with $B_{\mu\nu}$ via $h_{\mu\nu}\,=\,B_{\mu\nu}-B\,\eta_{\mu\nu}/8$.
Unlike GR, the theory allows for preferred-frame effects and propagates, in general, six gravitational-wave polarizations. Historically, it played an important role as a consistent alternative model to explore deviations from GR in the PPN program.

The Lightman--Lee theory differs from Rosen's  bimetric theory discussed above, in that the background metric is not dynamical and the theory was explicitly constructed to be free of the pathologies affecting Rosen's model. Nevertheless, Lightman--Lee gravity is now excluded by high-precision tests of gravity: its fixed background structure induces preferred-frame PPN parameters that violate current bounds (notably on $\alpha_2$), and the predicted orbital decay in binary pulsar systems disagrees with observations. As a consequence, the theory serves today mainly as a useful theoretical benchmark for characterizing Lorentz-violating metric theories and their associated polarization content, rather than as a phenomenologically viable alternative to GR.
It is simple enough though that is worth
studying in our context.

We do so 
in this section, by mapping Lightman-Lee theory to the PPE parametrization we are implementing. 
In the Eardly's~\cite{Eardley:1973zuo}  $E(2)$ classification, Ligthman-Lee theory belongs to the $II_6$ class, making some of the six polarization observer-dependent.  

\subsubsection{Frequency domain waveforms}
The gravitational waveforms  for Lightman-Lee theory 
are reported in~\cite{Chatziioannou:2012rf}. In frequency domain, for the angular harmonic $\ell = |m|= 2$ these are given by
\begin{align}
    \tilde{h}^{(2,2)}_{T}(f) &=    \, A^{(2)}  e^{- i \Psi_{GR}^{(2)} } \left( - i \sqrt{\frac27} \right) \, e^{- i  \delta \Psi^{(2)}_{\text{LL}}}, \\
    \tilde{h}^{(2,2)}_{V}(f)  &=    \, A^{(2)}  e^{- i \Psi_{GR}^{(2)} } \left( \sqrt{\frac27} \right) \, e^{- i  \delta \Psi^{(2)}_{\text{LL}}},   \\
    \tilde{h}^{(2,2)}_{B} (f) &=   A^{(2)}  e^{- i \Psi_{GR}^{(2)} } \left( - i   \sqrt{\frac37} \right) \, e^{- i  \delta \Psi^{(2)}_{ \text{LL}}}, \\
    \tilde{h}^{(2,2)}_{L}(f) &=   A^{(2)}  e^{- i \Psi_{GR}^{(2)} } \left( 2 i   \sqrt{\frac{1}{21}} \right) \, e^{- i  \delta \Psi^{(2)}_{ \text{LL}}} ,
\end{align}
while for the $\ell = |m|=1$ angular harmonic we have
\begin{align}
    \tilde{h}^{(1,1)}_{V}(f)  &=       \, A^{(1)}  e^{- i \Psi_{GR}^{(1)} } \left( \frac{10}{3} \sqrt{\frac{10}{21}} {\cal G} \right) \, e^{- i  \delta \Psi^{(1)}_{\text{LL}}}, \\
    \tilde{h}^{(1,1)}_{B} (f) &=     \, A^{(1)}  e^{- i \Psi_{GR}^{(1)} } \left( - i\,\frac{25}{6} \sqrt{\frac{5}{21}} {\cal G} \right) \, e^{- i  \delta \Psi^{(1)}_{\text{LL}}}, \\
    \tilde{h}^{(1,1)}_{L}(f) &=      \, A^{(1)}  e^{- i \Psi_{GR}^{(1)} } \left(  - i\,\frac{10}{3} \sqrt{\frac{5}{21}} {\cal G} \right) \, e^{- i  \delta \Psi^{(1)}_{\text{LL}}},
\end{align}
with 
\begin{equation}
    {\cal G} \equiv \frac{s_1}{m_1} - \frac{s_2 }{m_2 } \,, 
\end{equation}
describing the difference  of the self-gravitating binding energy per unit mass of the binary constituents, where $s_i$ are the sensitivities of the two bodies. Its value in General Relativity is ${\cal G} = 0$.

The modified phase is defined as
\begin{equation}
    \delta \Psi_{\text{LL}}^{(m)} \,=
       \frac{3 |m|}{224 u_m^5}  - \frac{625 |m|}{16464} \, \frac{{\cal G}^2 \eta^{2/5}}{u_m^7} + \frac{3 |m| }{256 u^5_m}\,,
\end{equation}
As  for the case of Rosen's theory,   the  limit ${\cal G} \to 0$ of Lightman-Lee theory does not reproduce General Relativity: while the dipole radiation is eliminated, the $\ell = 2$ harmonics still contain non-vanishing vector and scalar contributions. 

\subsubsection{Mapping to PPE parametrization} 
Now we want to match these expressions to the PPE parameterization in  Eqs.~\eqref{Eq:PPE_Tlm},~\eqref{Eq:PPE_p22} and~\eqref{Eq:PPE_p11}. 
Comparing the expressions we find that $a = a_V = a_B =a_L=0$, and 
\begin{align}
    \alpha_T &= -1 - i \sqrt{\frac27}  ,
    \\
    \alpha_{V2} &= \sqrt{\frac27} \,, \qquad \alpha_{V1} = \frac{10}{3} \sqrt{\frac{10}{21}} {\cal G},
    \\
    \alpha_{B2} &= -i\sqrt{ \frac{ 3 }{7}} \,, \qquad \alpha_{B1} =  -i\,\frac{25}{6} \sqrt{\frac{5}{21}} {\cal G},
    \\
    \alpha_{L2} &= \frac{2 i }{\sqrt{21}} \,,\qquad \alpha_{L1} = -i\,\frac{10}{3} \sqrt{\frac{5}{21}} {\cal G}.
\end{align}
The phase factor is matched to $ 2\beta_P u_2^b= - \delta \Psi^{(2)}_{\text{LL}} $ , while the one of tensor modes can be found from $\beta=2\beta_P$. To extract the specific values of the PPE parameters, we have to consider separately the cases where the dynamics of the binary is driven by the dipole or the quadrupole contribution. 

\paragraph{Dipole}
As before, in this case we consider the $-1$PN contributions. This gives
\begin{equation}
    b = -5\,, \qquad \beta_P = - \frac{45}{1792}  \,,
\end{equation}

\paragraph{Quadrupole}
In this case we consider the $0$PN order,
while the phase factors are  
\begin{equation}
    b = -7\, \,, \qquad \beta_P = \frac{625 }{16464} {\cal G}^2 \eta^{2/5}\,.
\end{equation}
We point out that this second case is second order in the modified gravity parameters, since it depends on ${\cal G}^2$, hence in the standard PPE formalism, where only leading corrections are kept, this second case would not be considered.

\section{\label{sec:results_for_extra_pol}GW polarization forecast}

In this section, we present quantitative results on LISA’s ability to constrain the presence of extra polarizations with  MBHBs.
As explained  in Secs.~\ref{subsec:Extra_Polarizations} and~\ref{subsec:mapping_to_gravity_theories}, we adopt the PPE formalism to model both the amplitude and phase modifications in the scalar, vector and tensor polarizations (see Eqs.~\eqref{Eq:PPE_Tlm}-\eqref{Eq:PPE_p11}). This formalism  introduces a total of 13 free parameters in addition to the 11 standard parameters in GR (assuming quasi-circular orbits and anti/aligned spins). Exploring a 24-dimensional parameter space is unfeasible, and strong degeneracies are to be expected. Therefore, we restrict our analysis considering only one PPE parameter at the time  (see Secs.~\ref{subsec:Extra_Polarizations} and~\ref{subsec:mapping_to_gravity_theories} for a full discussion), and only in one particular case we will allow two to vary (see Section~\ref{subsec:amp-phase-modifications}). In particular, we will focus on the following PPE parameters:
\begin{itemize}
    \item $\alpha_T$: amplitude parameter for the modification in the tensor polarization;
    \item $(\alpha_{B1}, \alpha_{B2})$: amplitude parameters for modifications in the breathing polarization in the $(\ell = |m|=1)$ and $(\ell = |m|=2)$ harmonics, respectively;
    \item $(\alpha_{L1}, \alpha_{L2})$: amplitude parameters for modifications in the longitudinal polarization in the $(\ell = |m|=1)$ and $(\ell = |m|=2)$ harmonics, respectively;
    \item $(\alpha_{V1}, \alpha_{V2})$: amplitude parameters for modifications in the  vector polarizations in the $(\ell = |m|=1)$ and $(\ell = |m|=2)$ harmonics, respectively;
    \item $\beta$: phase parameter for modifications in the scalar, vector and tensor polarizations. 
\end{itemize}
This section organizes the Fisher forecast results on constraining these quantities as follows:
\begin{enumerate}
    \item  In Section \ref{subsec:SNR_plots} we present  SNR fractional variations and horizon-redshift plots for the amplitude PPE parameters. More specifically, we show: (i) the modifications to the average SNR when $\alpha_T \neq 0$, and (ii) the maximum redshift at which non-zero 
    $\alpha_{B1}, \alpha_{B2}, $ $\alpha_{L1}, \alpha_{L2},\alpha_{V1}, \alpha_{V2}$ can be observed with $\rm SNR=10$ when only the corresponding scalar or vector polarization is present. 
    \item In Section \ref{subsec:gr_injections} we show the results of performing GR injections for all the PPE parameters, namely the expected 68\% confidence intervals with Fisher matrix forecasts, spanning the parameter space of MBHBs.  
    \item In Section \ref{sec:constrain-specific-theories}, we translate the constraints obtained on the PPE parameters into expected bounds on the fundamental parameters of the beyond-GR theories considered, using the mapping established in  Tabs.~\ref{table:PPE_mapping_beta}-\ref{table:PPE_mapping_alpha1}. For Rosen's and Lightman-Lee theory, which are not smoothly connected to GR, we perform tailored non-GR injections before translating the bounds on these theories parameters.
    \item In Section \ref{sec:inspiral-only-injections} we explore how the constraints on the PPE parameters change when cutting the signal at merger, rather than tapering the modified gravity effects at merger and ringdown, as assumed in Section \ref{subsec:gr_injections}.
    \item In Section \ref{subsec:amp-phase-modifications} we report an analysis probing {\it two} PPE parameters simultaneously: $(\alpha_T,\beta)$, the two tensor polarization parameters. 
    \item In Section \ref{subsec:complex_alpha} we present a  generalization of the previous results by considering the possibility that $\alpha_T$ is a complex quantity.
\end{enumerate}

In order to obtain representative estimates, we perform a Montecarlo analysis to randomize over most binary parameters, and present the median results over 100 realizations. 
Unless otherwise specified, the mass ratio is drawn uniformly from $[1,20]$ \footnote{This is the case in which  \texttt{IMRPhenomXHM} has been mostly calibrated with numerical relativity simulations.}, while spin magnitudes are sampled uniformly in $[-1,1]$. The sky position and orbital angular momentum are distributed uniformly over the sky. The coalescence time is fixed at $t_c = 1 \,{\rm yr}$ before merger. This choice is motivated by the expectation that the strongest constraints arise from the loudest systems, which are those observed for the longest duration.  However,  MBHBs are, in the best cases, detected only $\sim 1$ month before merger (see Figure~2 in \cite{2020PhRvD.102h4056M}).
The coalescence phase $\phi_c$ is randomized in $[-\pi, \pi]$.
For the total mass and redshift, we explore a range between $z
\in [0.5,4]$ and $M\in [3\times 10^4,10^7] \msun$. This mass and redshift range captures the bulk of MBHB populations expected to contribute to LISA: cosmological models of MBH formation predict mergers originating from both light- and heavy-seed scenarios, leading to characteristic binary masses from a few $10^{4} \msun$ up to $\sim 10^{7} \msun$ (see Section 2.3 in \cite{LISA:2022yao} or \cite{Volonteri:2025iit} for more details). We restrict to $z \in [0.5,4]$ because this interval encompasses many of the predicted MBHB merger distribution, where hierarchical galaxy growth drives a high event rate, and aligns with the redshift range in which LISA achieves its highest SNR for $10^{4}$–$10^{7}\msun$ systems, providing also reliable Fisher  estimates.

\subsection{\label{subsec:SNR_plots}SNR fractional deviations and detectability horizon}

In this section we present the median SNR fractional variation results for the tensor polarization parameter, as well as horizon-redshift results for the scalar and vector polarization parameters. 
Since we  present results based on SNR calculations, we do not consider the cases of phase modifications. In preliminary stages of this work, we evaluated the impact of a phase modification in the tensor polarization on the SNR and found relative differences smaller than $10^{-4}$ in the most favorable cases, and even smaller otherwise. This behavior is expected, as the parameter $\beta$ induces a phase modification and therefore has a negligible effect on the SNR. For this reason, in Section~\ref{subsec:gr_injections} we explore the detectability of $\beta$ only through GR injections.

We emphasize that the SNR should be regarded only as a proxy for detection; the main results for LISA's ability to constrain amplitude and phase modifications are obtained from the Fisher estimates, as presented in the next Section. 

\subsubsection{SNR fractional variation}
\begin{figure}[h!]
\centering
\includegraphics[width = 0.49\textwidth]{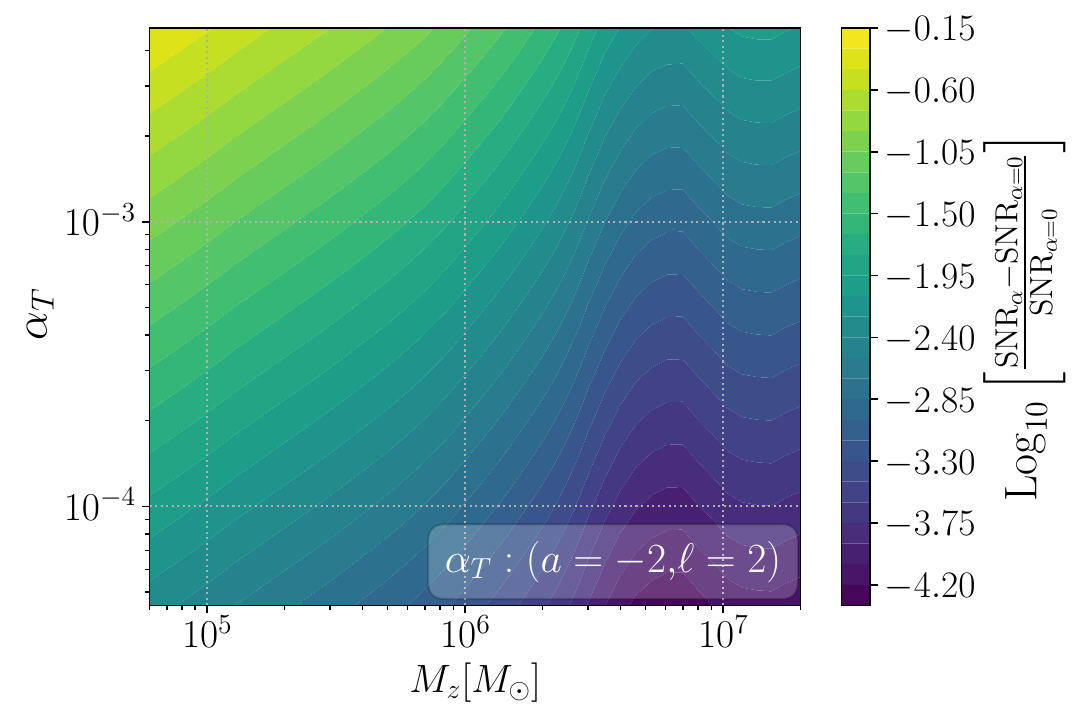}
\includegraphics[width = 0.49\textwidth]{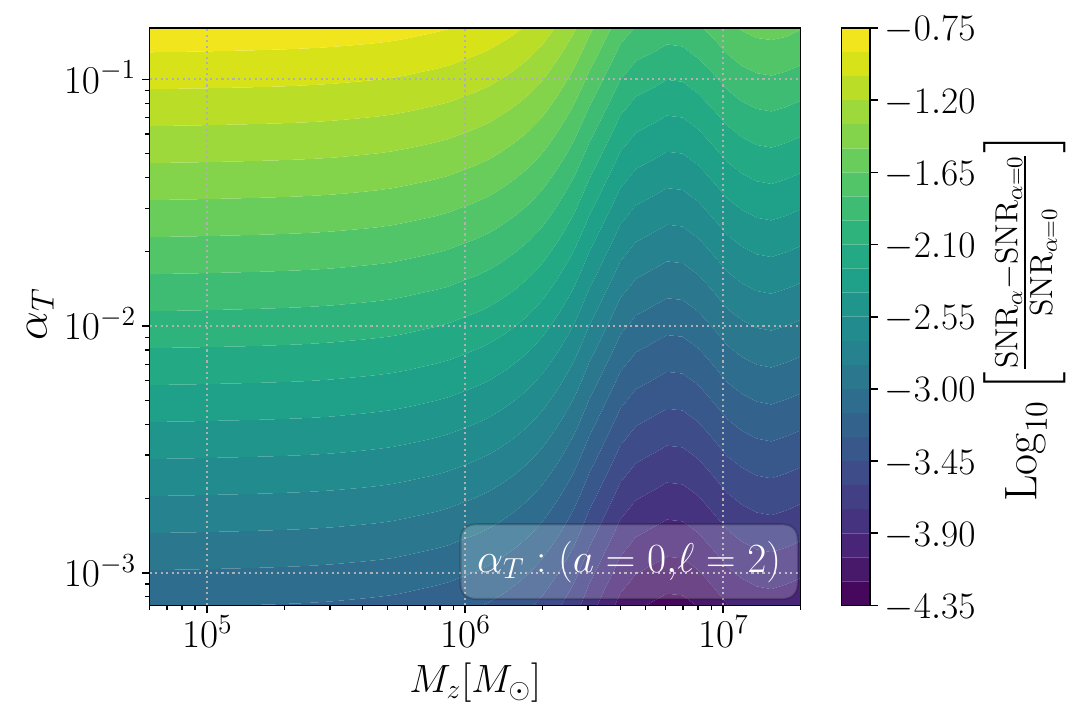}
 \caption{Median SNR fractional variations due to $\alpha_T\not=0$, as a function of redshifted total mass $M_z$ and $\alpha_T$. Brighter colors indicate larger variations. In the left (right) panel we report the results for the case  $a=-2$ ($a=0$). Note that the $y$-axis range and the color bars are different in the two panels. The values of $\alpha_T$ are chosen to cover the smallest and largest value of $\alpha_T$ LISA can constrain over all redshifted masses. Note the value of $\ell=2$ here is a reminder that we only modify the amplitude of the $(\ell = |m|=2)$ tensor angular harmonic.}
 \label{fig:SNRalphaT} 
\end{figure}

In Figure~\ref{fig:SNRalphaT} we show the median fractional SNR variation as a function of the redshifted total mass $M_z$ and the amplitude parameter $\alpha_T$ at $z=1$, with all other PPE parameters, including $\beta$, set to zero. Following Table~\ref{table:PPE_mapping_ab}, we consider two cases: $a=-2$ (left panel) and $a=0$ (right panel). The minimum and maximum values of $\alpha_T$ on the $y$-axis correspond to the minimum and maximum values of $\alpha_T$ that LISA can constrain across the redshifted mass range considered (see Figure~\ref{fig:gr_injection_alphaT}). For each value of $M_z$ and $\alpha_T$, we simulated 100 events, randomizing all the binaries parameters, we computed the SNR for these 100 events, and we report here the median.  

From Figure~\ref{fig:SNRalphaT} we observe that systems with larger $M_z$ exhibit smaller SNR variations. This is consistent with the fact that, for heavier systems, the SNR is dominated by the merger and ringdown phases, where the amplitude modification is effectively suppressed. For instance, for $a=0$ and $\alpha_T=10^{-2}$, the fractional SNR variation is $\sim 1\% ,\, 0.7\%$ and $0.2\%$ at $M_z = 10^5,\, 10^6$ and $10^7 \msun$, respectively. In contrast, at fixed $M_z$, the SNR variation increases
with $\alpha_T$. For example, at $M_z = 10^6 \msun$ and $a=0$, setting $\alpha_T = 10^{-3}$ ($\alpha_T = 10^{-1}$) yields a fractional SNR variation of $0.07\%$ ($\sim 8\%$). Similar trends hold for $a=-2$. 

At fixed value of $M_z$ and $\alpha_T$, we observe that $a=-2$ leads to larger SNR fractional variations compared to $a=0$. For instance, for $M_z = 10^5 \msun$ and $\alpha_T=10^{-3}$, the SNR changes by $\sim 0.1\%$ if $a=0$, but increases to $\sim 10\%$ for $a=-2$. This behavior originates from the typical values of $u_m$ in Eq.~\eqref{eq:ul-PPE-extra} for the redshifted masses and frequencies considered.

\subsubsection{Horizon-redshift}
\begin{figure}[H]
\centering
\includegraphics[width = 0.49\textwidth]{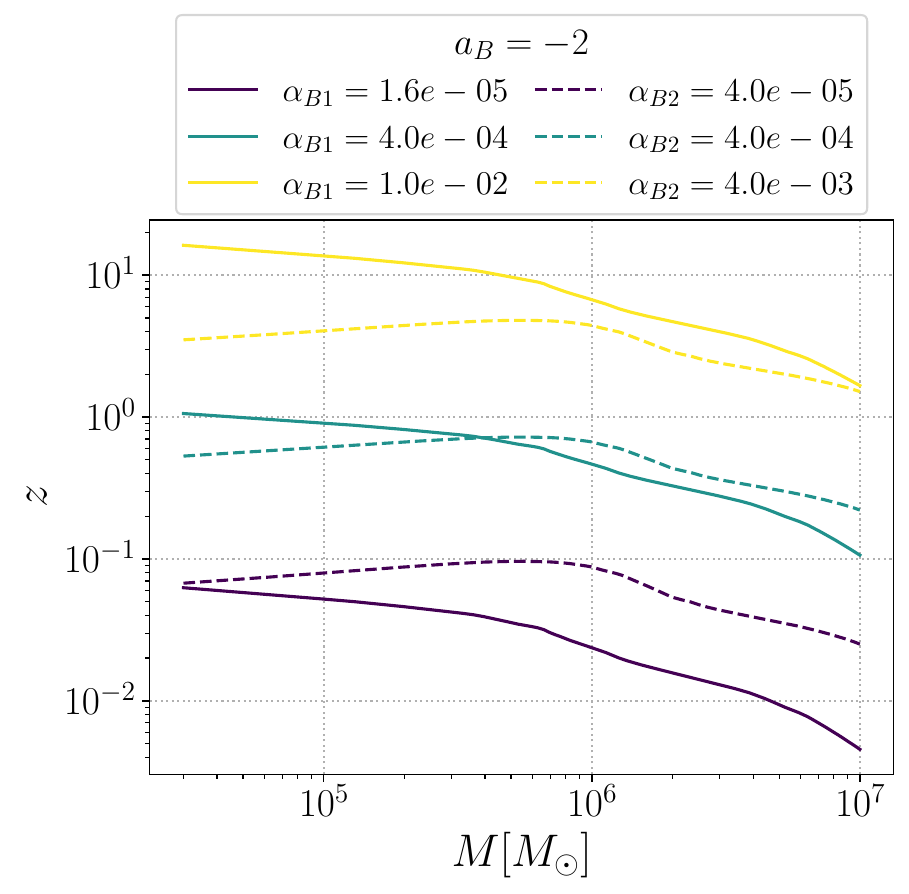}
\includegraphics[width = 0.49\textwidth]{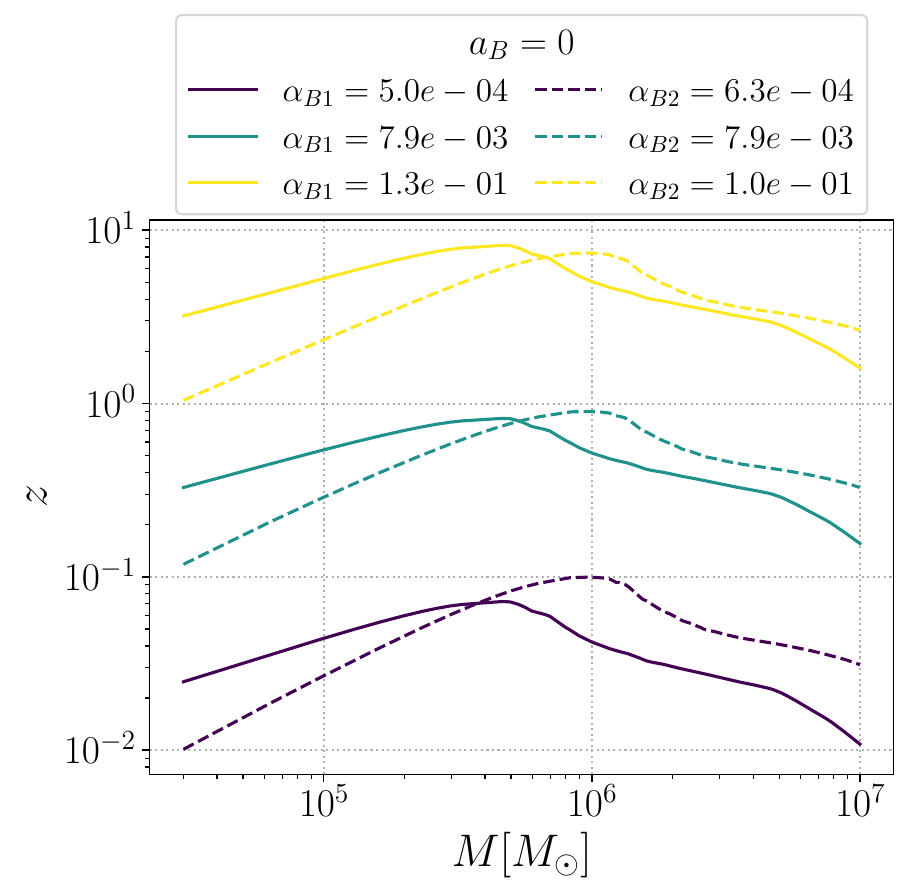} \\
 \caption{Average redshift horizon ($\rm SNR=10$) for different values of $\alpha_{B1}$ and $\alpha_{B2}$, as indicated in the legend of each plot. The left (right) panel shows the case $a_B=-2$ ($a_B=0$) and solid (dashed) lines correspond to $\ell = 1$ ($\ell = 2$). Note that the values of $\alpha_{B1}$ and $\alpha_{B2}$ reported in the left and right panels are different.
}  
 \label{fig:horizon_breathing} 
\end{figure}

\begin{figure}[h!]
\centering
\includegraphics[width = 0.49\textwidth]{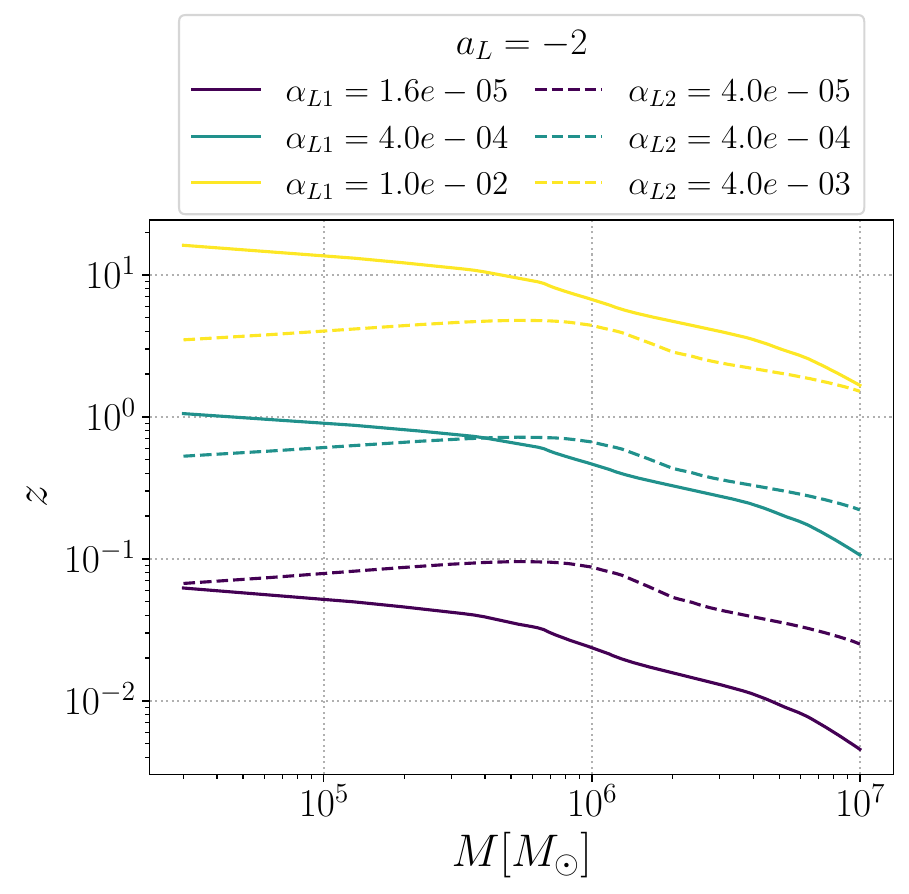}
\includegraphics[width = 0.49\textwidth]{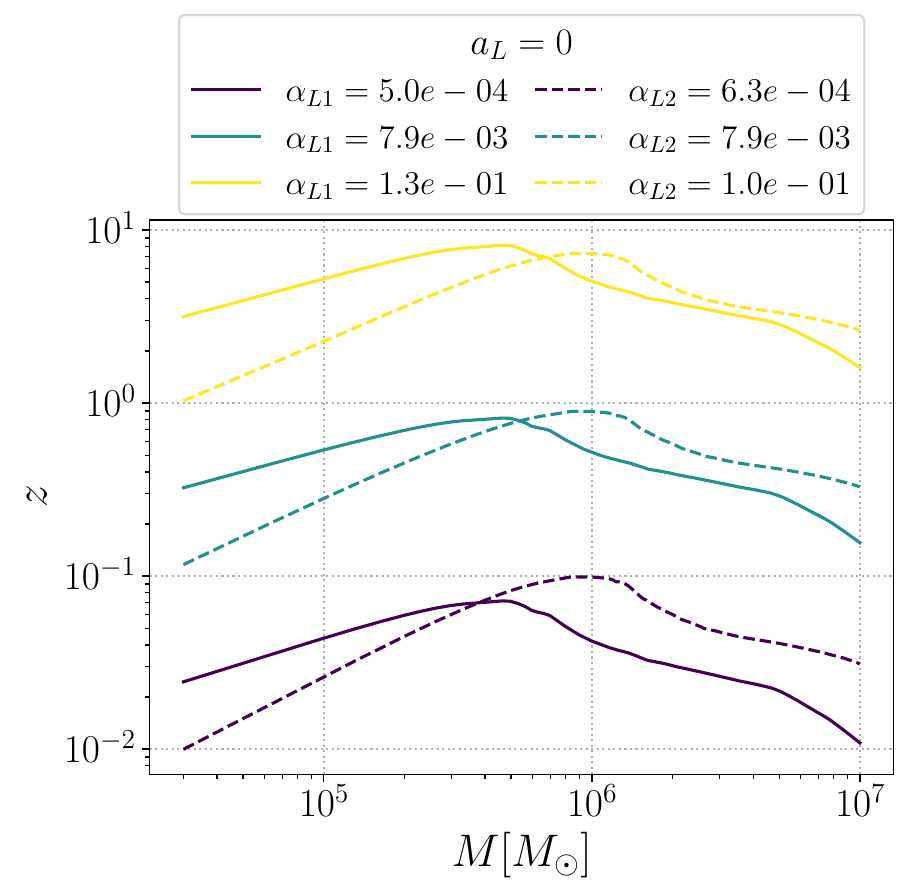} \\
\includegraphics[width = 0.49\textwidth]{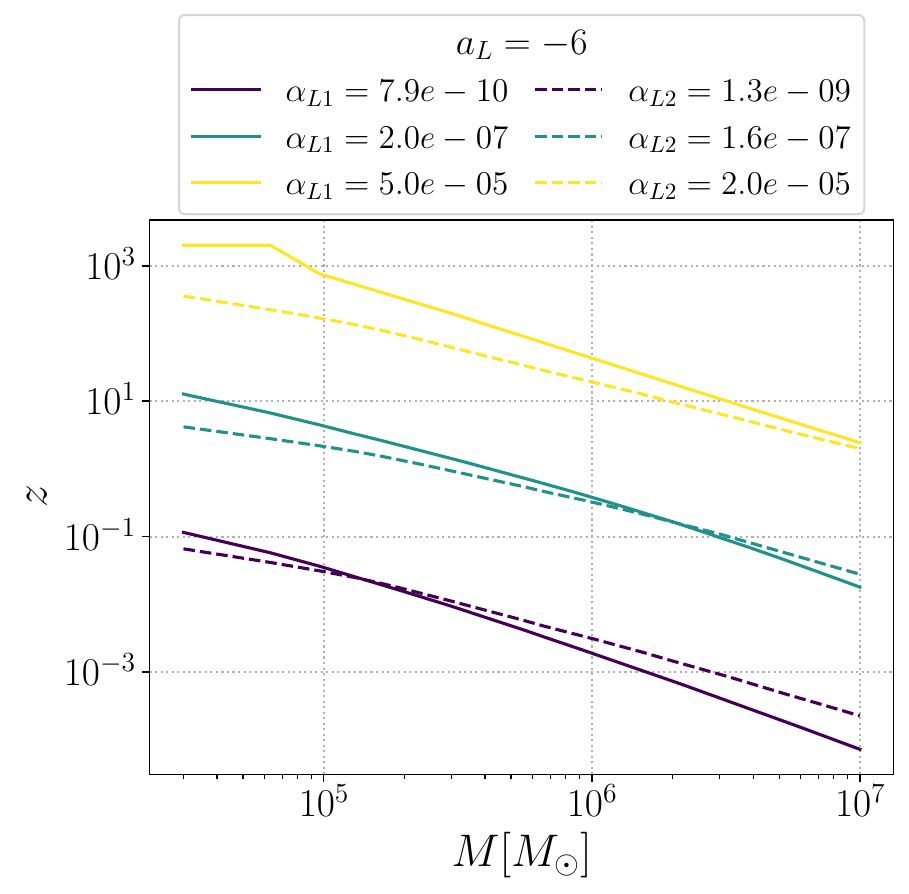}
 \caption{Average redshift horizon ($\rm SNR=10$) for different values of the longitudinal polarization parameters $\alpha_{L1}$ and $\alpha_{L2}$, for three frequency scalings $a_L=-2$ (top left), $a_L=0$ (top right), and $a_L=-6$ (bottom). Solid (dashed) lines correspond to $\ell = 1$ ($\ell = 2$) harmonics. Note that the values of $\alpha_{L1}$ and $\alpha_{L2}$ reported in the three panels are different, and the $y$-axis is different in the bottom panel. }  
 \label{fig:horizon_longitudinal} 
\end{figure}

\begin{figure}[h!]
\centering
\includegraphics[width = 0.49\textwidth]{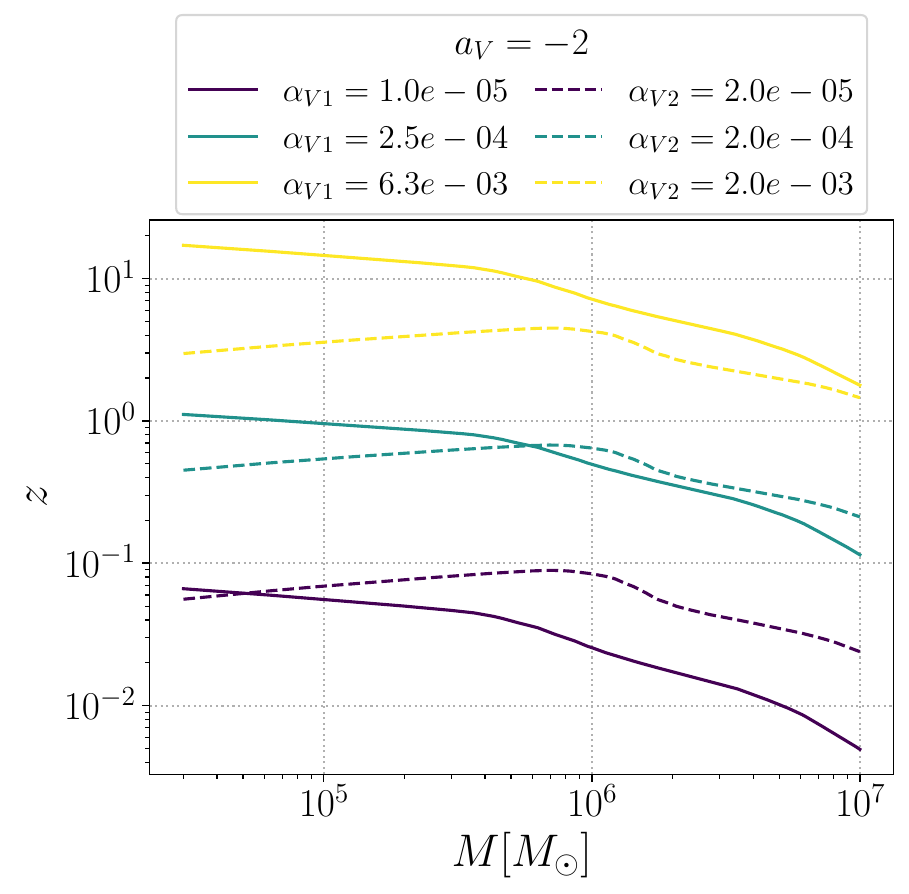}
\includegraphics[width = 0.49\textwidth]{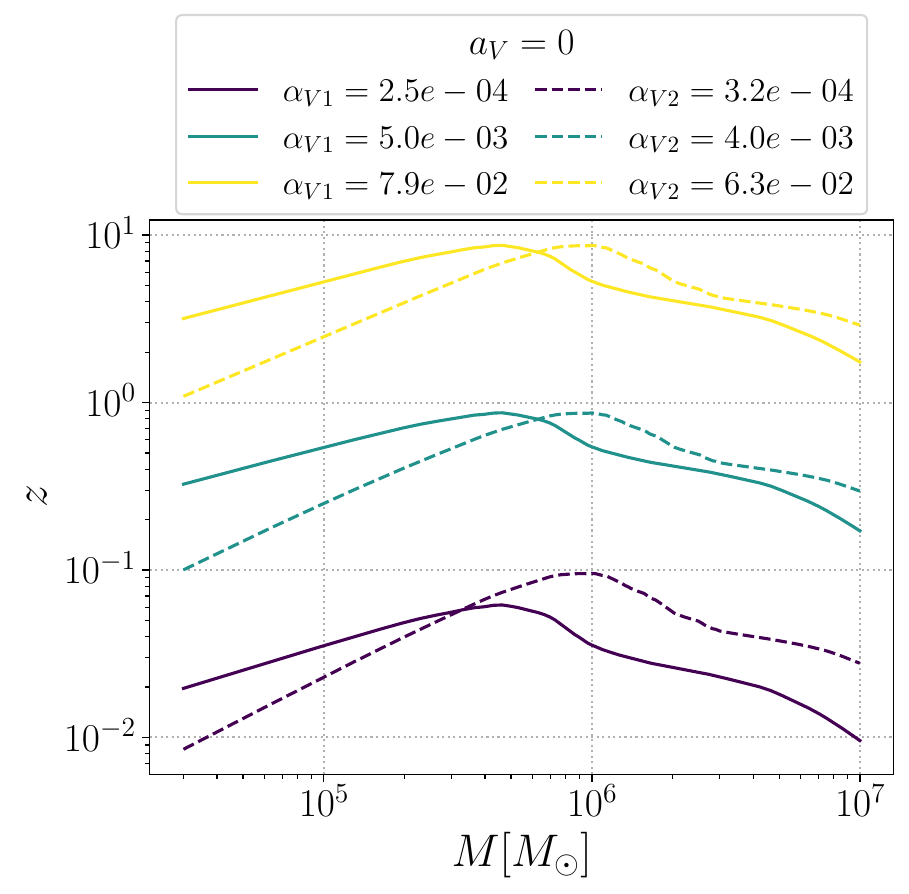} \\
 \caption{Average redshift horizon ($\rm SNR=10$) for different values of the vector polarization parameters $\alpha_{V1}$ and $\alpha_{V2}$, for two frequency scalings $a_V=-2$ (left panel), and $a_V=0$ (right panel).  Solid (dashed) lines correspond to $\ell = 1$ ($\ell = 2$) harmonics. Note that the values of $\alpha_{V1}$ and $\alpha_{V2}$ reported in the left and right panels are different.}  
 \label{fig:horizon_vector} 
\end{figure}

To analyze scalar and vector polarizations, we compute the SNR associated with each extra polarization individually, as a function of the total mass and redshift. From this, we construct average horizon-redshift plots for different values of $\{ \alpha_{B1}, \alpha_{B2} \} $ , $\{ \alpha_{L1}, \alpha_{L2} \} $  and $\{ \alpha_{V1}, \alpha_{V2} \} $. In the following, we present results corresponding to the minimum, maximum, and median values of these parameters that LISA can constrain. Note that the SNR scales linearly with the amplitude of these parameters (see Eqs.~\eqref{Eq:PPE_p22}--\eqref{Eq:PPE_p11}), so the redshift horizon for any other choice of parameter values can be readily obtained by rescaling the curves in Figures~\ref{fig:horizon_breathing}--\ref{fig:horizon_vector}.

\textbf{Breathing mode:} In Figure~\ref{fig:horizon_breathing} we report the average horizon-redshift for the breathing mode parameters $\alpha_{B1}$ and $\alpha_{B2}$, in the cases of frequency scalings $a_B=-2$ and $a_B=0$. The curves indicate the redshift and total mass at which the average SNR obtained \emph{only} from the breathing polarization is equal to 10, for a given value of $\alpha_{B1}$ or $\alpha_{B2}$. The reported values of $\alpha_{B1}$ and $\alpha_{B2}$ have been chosen as the minimum, maximum and median constrainable values by LISA, across all redshifted masses (as it will be shown in Figure\ \ref{fig:gr_injection_alphab}). 

We learn
from Figure~\ref{fig:horizon_breathing}  that larger values of $ \{\alpha_{B1}, \alpha_{B2}\}$ can be observed up to higher redshifts. For $a_B=-2$ and $\ell=1$, we find that the average horizon redshift decreases as a function of the total mass. This behavior can be explained recalling the definition of $u_m$ in Eq.~\eqref{eq:ul-PPE-extra}: $a_B=-2$ produces higher amplitude modifications at low frequencies and for smaller total masses \footnote{Remember that the total mass scales linearly with the chirp mass, cf.\ Table~\ref{tab:Notations}.}, leading to a larger horizon at lower total masses. If we compare the horizons for the case $\alpha_{B1} = \alpha_{B2}$ (green lines) we can see that the $\ell=1$ mode has larger horizon at low masses, while the $\ell=2$ can be detected further at higher masses.

For the case  $a_B=0$, we see the horizon curves are non-monotonic and they peak at $M \sim 5\times 10^5 \msun$ and $M \sim  10^6 \msun$ for $\ell=1$ and $\ell=2$, respectively. In fact, when $a_B=0$, the dependence on the redshifted chirp mass and frequency is analogous to GR, and hence the resulting average horizon redshifts have an analogous behavior to GR SNR horizons plots for LISA (see, e.g.\ the left panel of Figure 3.5 in \cite{LISA:2024hlh}). The slight shift toward lower masses in the case of $\ell=1$ is due to the fact that this mode has half the frequency of the $\ell=2$ mode.

\textbf{Longitudinal mode:} In Figure~\ref{fig:horizon_longitudinal} we report the average redshift horizon for the longitudinal polarization parameters $\{ \alpha_{L1}, \alpha_{L2} \}$. Notice that the values of $\{ \alpha_{L1}, \alpha_{L2} \}$ chosen for  $a_L = \{  -2,0 \}$ are those of $\{ \alpha_{B1}, \alpha_{B2} \}$ in Figure~\ref{fig:horizon_breathing}.
This is not coincidental: these values are fixed as the maximum and minimum values constrainable by LISA and the detector response to the two scalar modes coincides for most of the LISA sensitivity range, as shown in Figure~ \ref{fig:responsefcomp}. 
However, as reported in Table~\ref{table:PPE_mapping_ab}, the longitudinal mode also includes one option with $a_L=-6$, which we report in the bottom panel of  Figure~\ref{fig:horizon_longitudinal}. This scaling amplifies even more low frequencies and small total masses: for example, the horizon for $\alpha_{L1}=2\times 10^{-7}$ at $M=10^5 \msun$ is $z \simeq 10$ but this value reduces to $z \simeq 0.02$ for $M=10^7 \msun$.

\textbf{Vector modes: } In Figure~\ref{fig:horizon_vector} we show the average redshift horizon for the vector polarization parameters $\{\alpha_{V1}, \alpha_{V2}\}$, for two frequency scalings $a_V=-2$ (left panel), and $a_V=0$ (right panel). 
While the overall trend of these horizon plots is similar to the one of the two scalar modes, we notice that the values of $\{\alpha_{V1}, \alpha_{V2}\}$ chosen in  Figure~\ref{fig:horizon_vector} are smaller than those of $\{\alpha_{B1}, \alpha_{B2}\}$ and $\{\alpha_{L1}, \alpha_{L2}\}$ in Figures~\ref{fig:horizon_breathing} and~\ref{fig:horizon_longitudinal} respectively.
This is once again due to the LISA response to vector polarization, which is higher compared to the one for scalar polarizations (see Figure~\ref{fig:responsefcomp}) for most of the sensitivity frequency range. This means that LISA will have the ability to probe smaller deviations from GR in vector than in scalar polarizations.

\subsection{\label{subsec:gr_injections}GR injections}
In this section, we report the results from \emph{GR injections}. 
In the case of the amplitude PPE parameters ---namely $\alpha_T,\, \alpha_{B1},\, \alpha_{B2},\, \alpha_{L1},\, \alpha_{L2},\,\alpha_{V1},\, \alpha_{V2}$--- we set these to zero (the default value in GR) and we report the median $68\%$ confidence interval on these quantities, using the Fisher forecast formalism described in Section~\ref{subsec:numerical_implementation}.
Special attention is required for the phase modification $\beta$ when considering extra polarizations: in order for its presence to be relevant, the amplitude of the corresponding polarization must also be non-vanishing. Therefore, when constraining $\beta$, we will assume $\alpha_P \neq 0$, as explained in more detail later.

We present results as function of redshift and intrinsic total mass $M$. The total mass range is sampled log-uniformly at 43 values and the redshift range linearly at 15 values, yielding a grid of 645 points. These values were chosen to provide sufficiently fine coverage of the parameter space while keeping the computational cost manageable. For each grid point $(z,M)$ we report the median $\sigma$ error on a given PPE parameter, obtained from 100 realizations. Note that for all the realizations performed, we do not apply a detectability SNR threshold when computing the Fisher  matrix since, due to our chosen ranges in total mass and redshift, most of the MBHBs are detected with  $\rm SNR>10$.

\subsubsection{Tensor modes amplitude}

\begin{figure}[h!]
\centering
\includegraphics[scale=0.44]{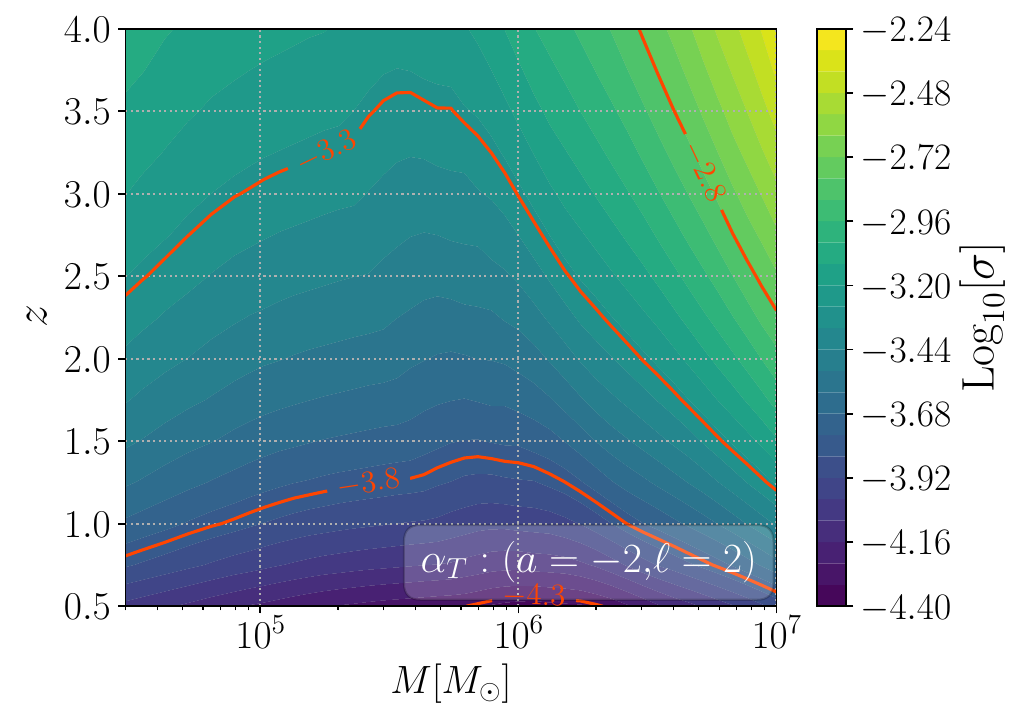} 
\includegraphics[scale=0.44]{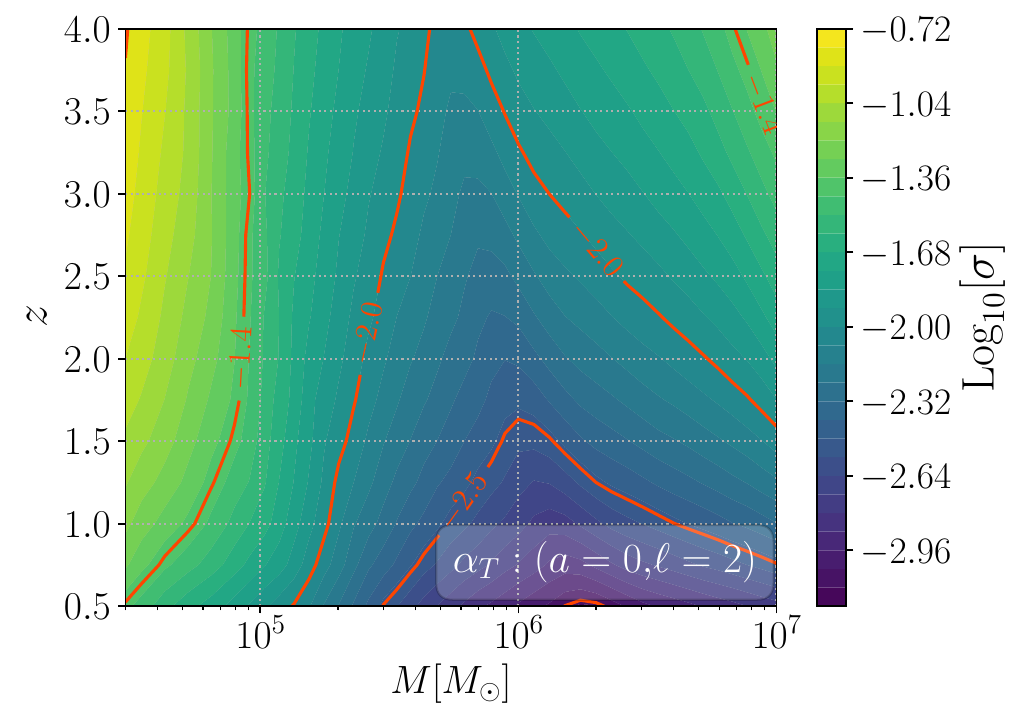} 
\caption{Median absolute errors on $\alpha_T$ as a function of the intrinsic total mass $M$ and redshift $z$ from GR injections in the case of $a=-2$ (left plot) and $a=0$ (right plot). Darker colors represent better constraints. The red lines indicate representative constraints and their values are set to guide the eyes of the readers. The choice of $a$ strongly affects the region where we can put the better constraints on $\alpha_T$.}
 \label{fig:gr_injection_alphaT} 
\end{figure}

In Figure~\ref{fig:gr_injection_alphaT} we report the constraints from the GR injections on $\alpha_T$ for the cases with $a=-2$ (left panel) and $a=0$ (right panel). 

\paragraph{$\mathbf{a = -2}$:} In this case, $\alpha_T$ is constrained on average with an absolute precision of $[ 1.7\times 10^{-4},\, 1.2\times 10^{-4} ,\, 4 \times 10^{-4}] $ for total masses of $ [10^5,\, 10^6 ,\, 10^7 ] \msun $ at $z=1$, respectively.
The constraints deteriorate at higher redshifts, as expected. We also note that the constraints worsen quicker when increasing redshift for heavy-mass systems (i.e.\ $M\sim 10^7 \msun$) with respect to  light-mass systems (i.e.\ $M\sim 10^5 \msun$). This is due to the frequency scaling ($a = -2$) of the amplitude's modification, which gives more weight to the low frequency part of the signal.

\paragraph{$\mathbf{a = 0}$:}
In this case, $\alpha_T$ can be constrained on average with an absolute precision of $ [ 2.7\times 10^{-2} ,\, 2.5\times 10^{-3} ,\, 7 \times 10^{-3} ] $ for total masses of $ [10^5,\, 10^6 ,\, 10^7 ] \msun $  at $z=1$, respectively.
This frequency scaling ($a=0$) affects the signal in the same way across all the frequencies, effectively leading to constraint plots which resembles closely typical SNR plots in LISA (e.g.\ see left panel in Figure 3.5 in \cite{LISA:2024hlh}). 
Therefore, the constraints are in general weaker compared to the $a = -2$ case. For instance, constraints at $M= [10^5,\, 10^7] \msun$ and $z = 1$ are respectively 2 and 1 orders of magnitude tighter in the $a=-2$, compared to $a=0$.

\subsubsection{Breathing  mode amplitude}
\begin{figure}[h!]
\centering
\includegraphics[scale=0.44]{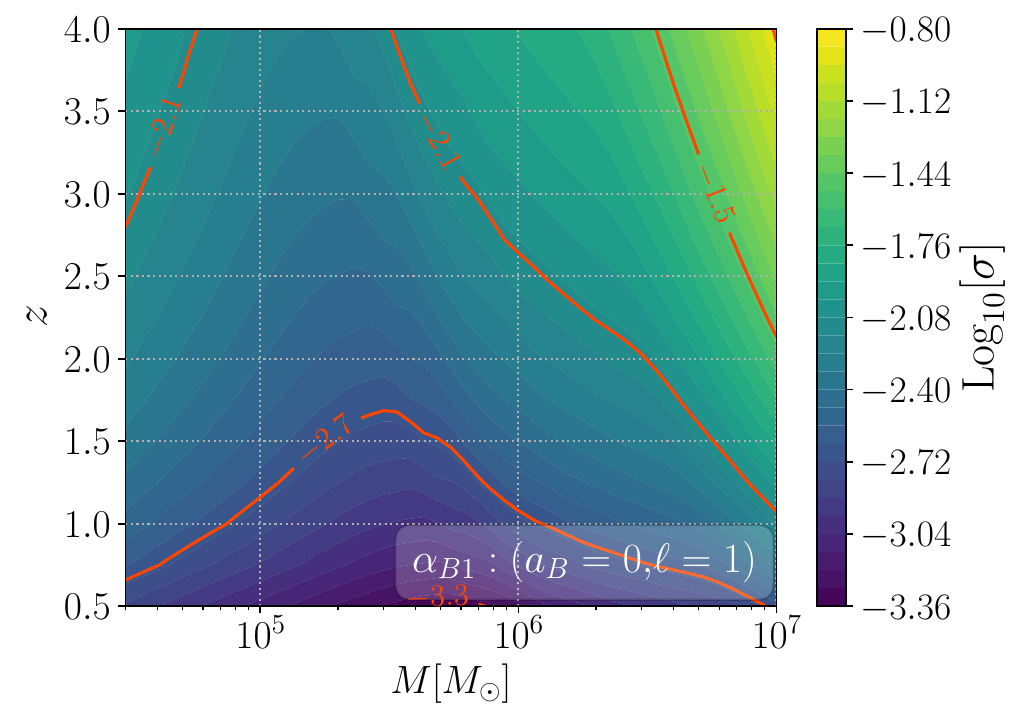}  
\includegraphics[scale=0.44]{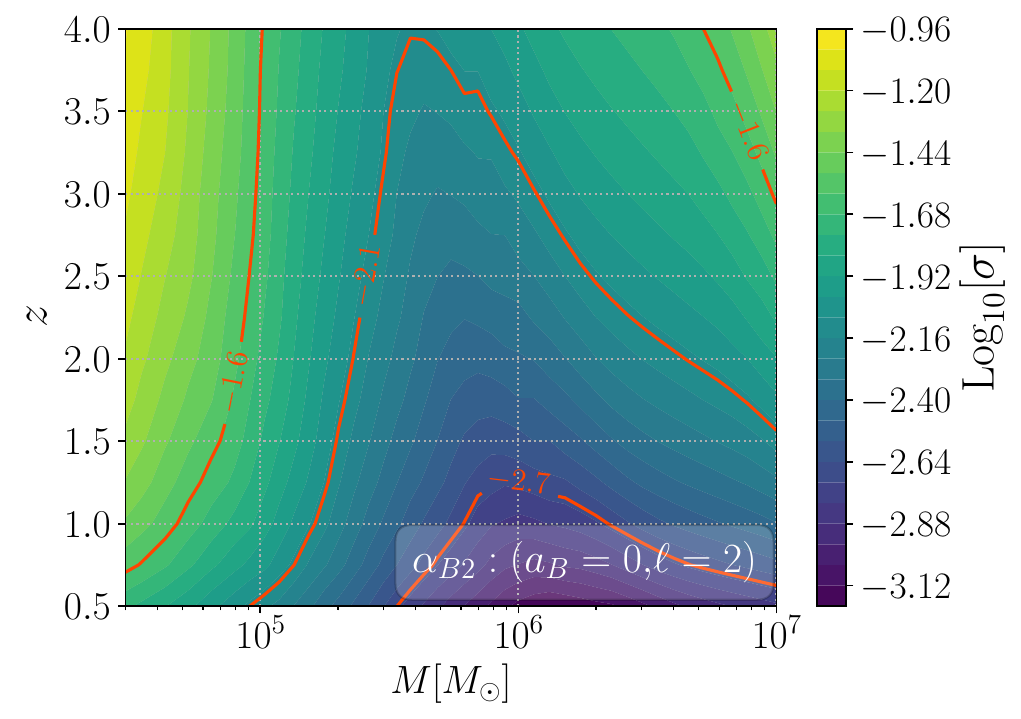}  \\
\includegraphics[scale=0.44]{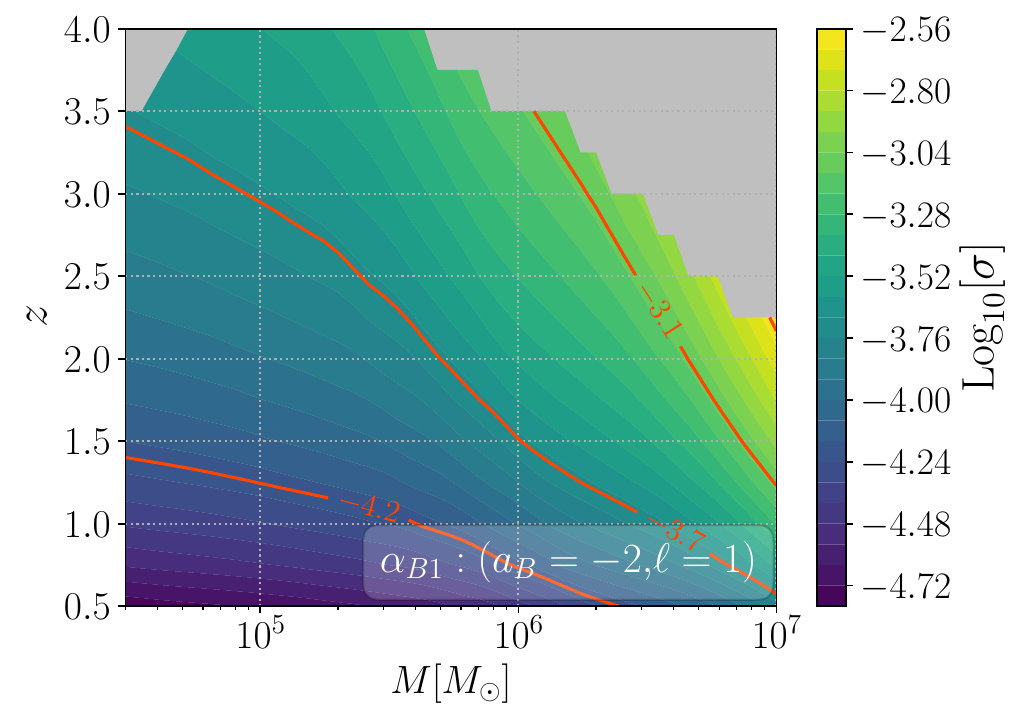}  
\includegraphics[scale=0.44]{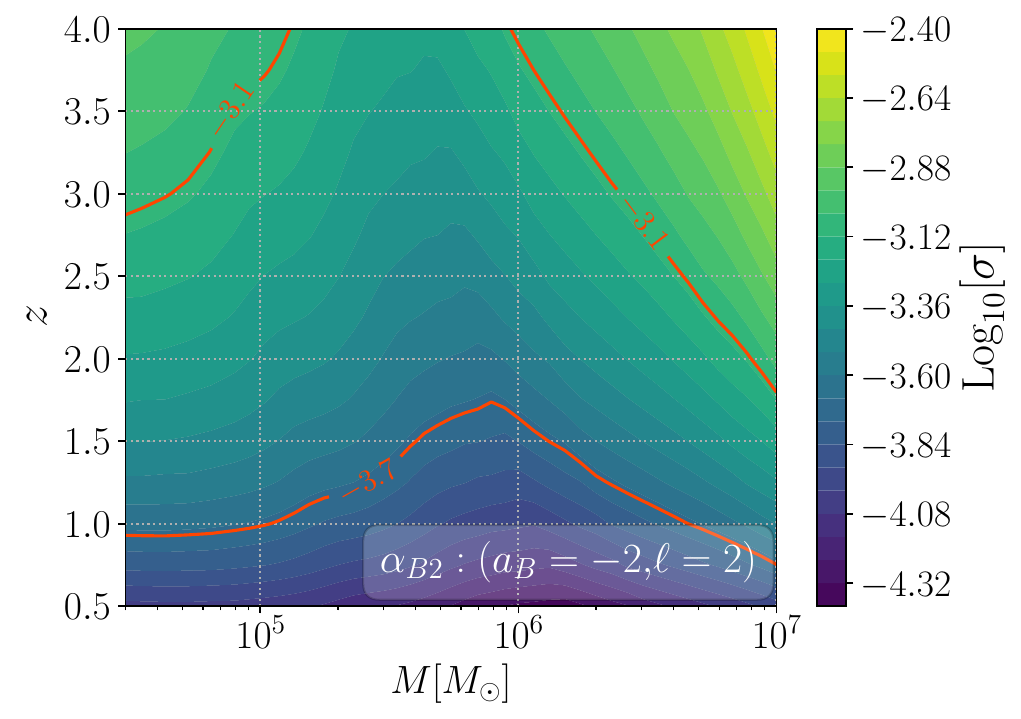} 
\caption{Median absolute errors on $\alpha_{B1}$ (left panels) and $\alpha_{B2}$ (right panels). The specific combination of $a_B$ and $\ell$ are reported in the legend of each plot. The gray area in the case ($a_B=-2$, $\ell=1$) corresponds to masked regions where the amplitude of the breathing mode is larger than that of the tensor mode (see Eq.~\eqref{Eq:limit_alpha_extrapol}).
}
\label{fig:gr_injection_alphab} 
\end{figure}

In Figure~\ref{fig:gr_injection_alphab} we report the constraints on $\{\alpha_{B1}, \alpha_{B2} \}$ from GR injections, for the cases with $a_B=0$ (top panels) and $a_B=-2$ (bottom panels), for both dipole and quadrupole harmonics.

Note that in this case, we require that the amplitude of the breathing emission is not larger than  the amplitude of the $(\ell=|m|=2)$ tensor mode for consistency of the PPE parametrized formalism (see Eq.~\eqref{Eq:limit_alpha_extrapol}).
Since we are performing GR injections and the fiducial value for  $\alpha_{B1}$ and $\alpha_{B2} $ are zero, we replace $\alpha_{Pm}$ in Eq.~\eqref{Eq:limit_alpha_extrapol} with the $68\%$ confidence interval obtained from GR injections, and
 in Figure~\ref{fig:gr_injection_alphab} 
we mask with gray areas the region where Eq.~\eqref{Eq:limit_alpha_extrapol} is not satisfied. The motivation is that, even though we were able to assess the precision of the constraints, the uncertainty would be so large that would go beyond the consistency conditions of our formalism, which assume a perturbative approach beyond GR. In a Bayesian parameter estimation, this would mean that Eq.~\eqref{Eq:limit_alpha_extrapol} can be considered as a condition that provides a prior on $\alpha_{Pm}$, and in our masked regions the expected constraints would be dominated by such a prior.

\paragraph{$\mathbf{\alpha_{B1}}$ vs $\mathbf{\alpha_{B2}}$:}

Due to the scaling of the extra-polarization amplitude with $\ell$ (see Eq.~\eqref{Eq:h_PN_aell}), we are able to put considerably better constraints on $\alpha_{B1}$ than $\alpha_{B2}$ at smaller masses ($M \lesssim 10^6 \msun$), while for $M \gtrsim 10^6 \msun$  the constraints on the two are comparable. For example, at $M=10^5 \msun$ and $z=1$ with $a_B=0$, $\alpha_{B1}$ can be constrained to $\simeq 1.6\times 10^{-3}$, whereas $\alpha_{B2}$ is constrained only to $\simeq 1.7\times 10^{-2}$.
When comparing the behavior for $a_B=-2$, we notice that $a_B=-2$ allows for tighter constraints due to the frequency dependence of the modifications of the waveform, as we found for the tensor polarizations previously. This is particularly important for the low mass systems.
Overall, the best constraints are placed on $\alpha_{B1}$ in the case $a_B=-2$, while the worst constraints are on $\alpha_{B2}$ in the case $a_B=0$.

\paragraph{Comparison with horizon-redshift plot:}
Comparing the GR injection  with the corresponding average horizon-redshift plots we find that, at a given combination of total mass and redshift, LISA can constrain values of $\alpha_{B2}$ smaller than those that would produce a SNR of 10 for the same parameters. Specifically, from the left panel of Figure~\ref{fig:horizon_breathing}, a system with $M=10^5 \msun$ and $z \simeq 0.6$ requires $\alpha_{B2} = 4\times 10^{-4}$ to reach SNR = 10. However, at the same mass and redshift for $a_B = -2$, the smallest constrainable $\alpha_{B2} \simeq 1.2\times 10^{-4}$ (see  lower-right panel of Figure~\ref{fig:gr_injection_alphab}). 
This consideration implies that, judging the detectability of extra polarizations solely based on SNR would yield a pessimistic estimate of LISA's true capabilities.

\subsubsection{\label{subsubsec:long_mode}Longitudinal  mode amplitude}
\begin{figure}[h!]
\centering
\includegraphics[scale=0.44]{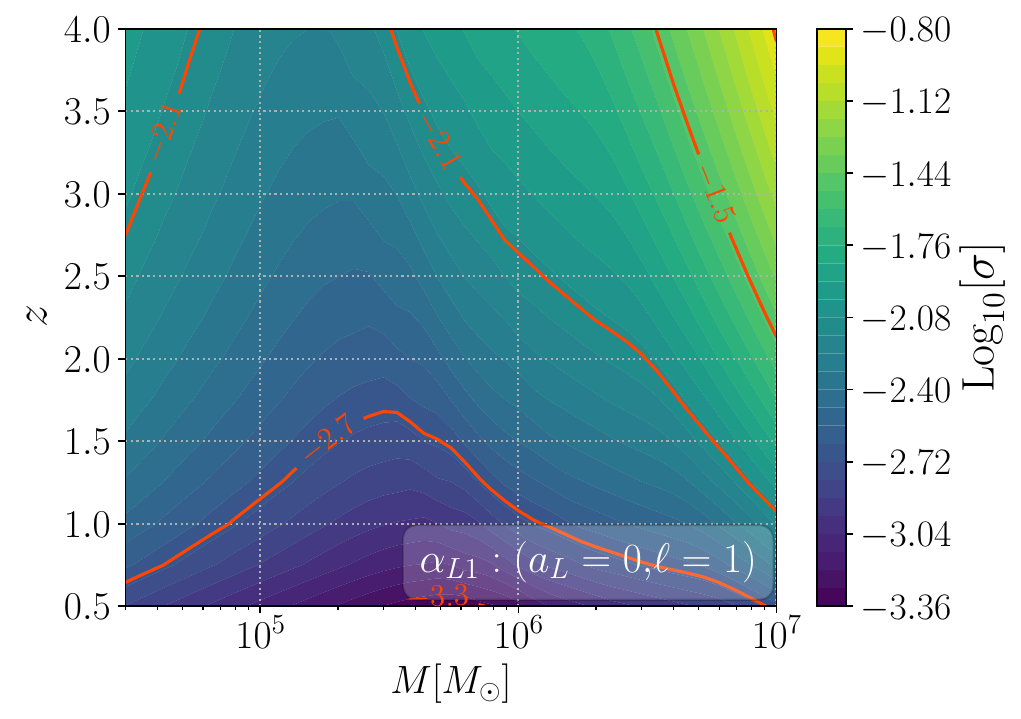}  
\includegraphics[scale=0.44]{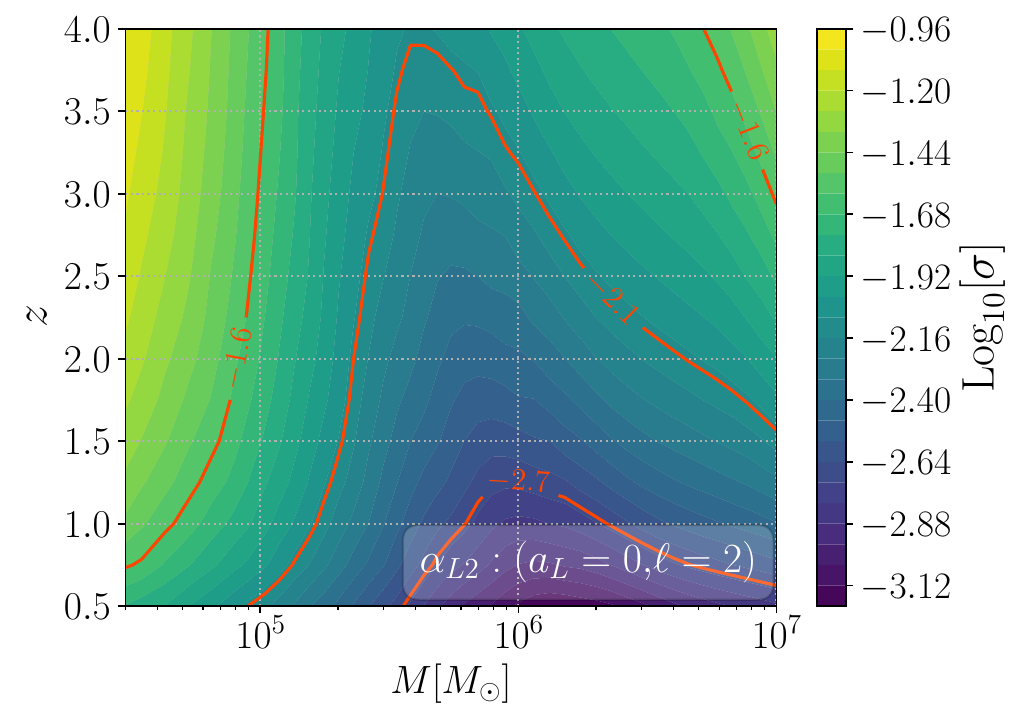}  \\
\includegraphics[scale=0.44]{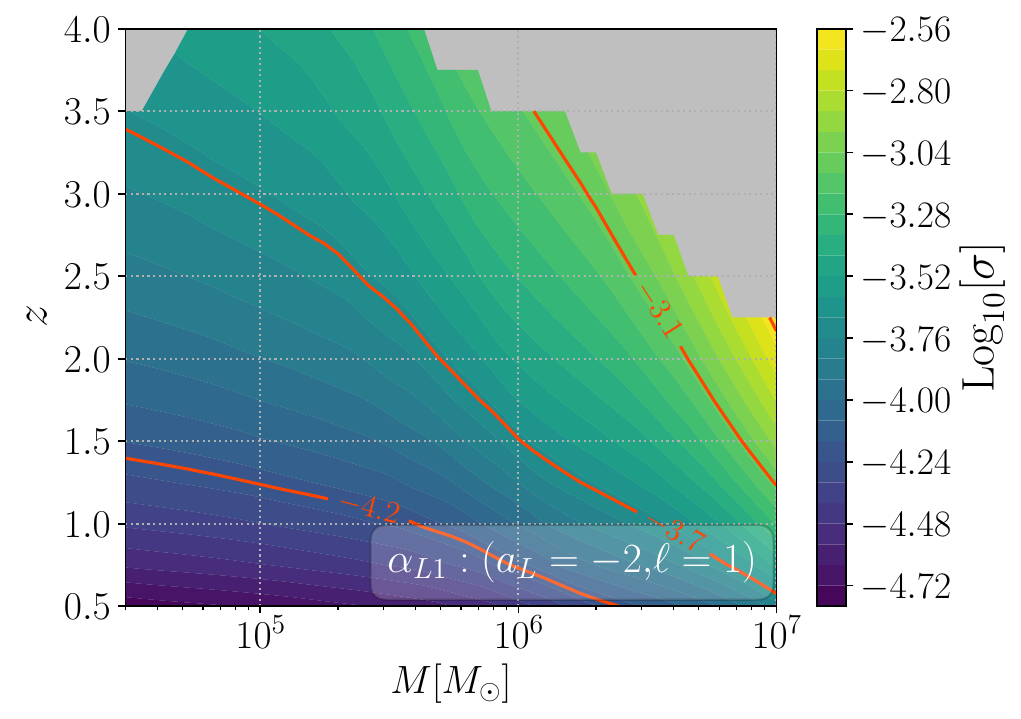}  
\includegraphics[scale=0.44]{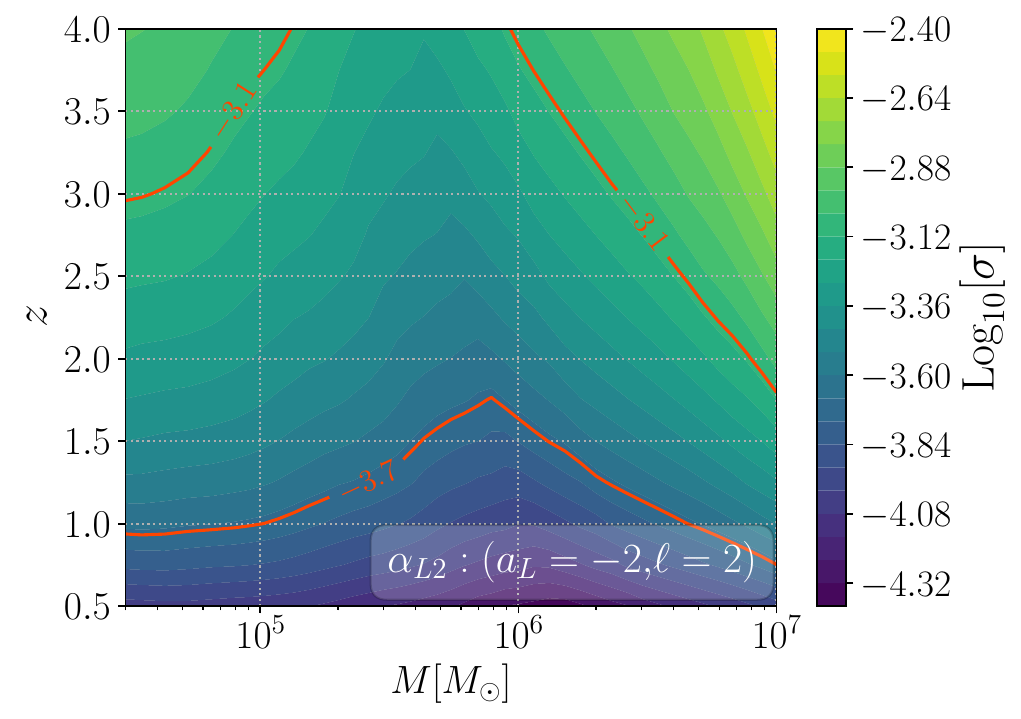} \\
\includegraphics[scale=0.44]{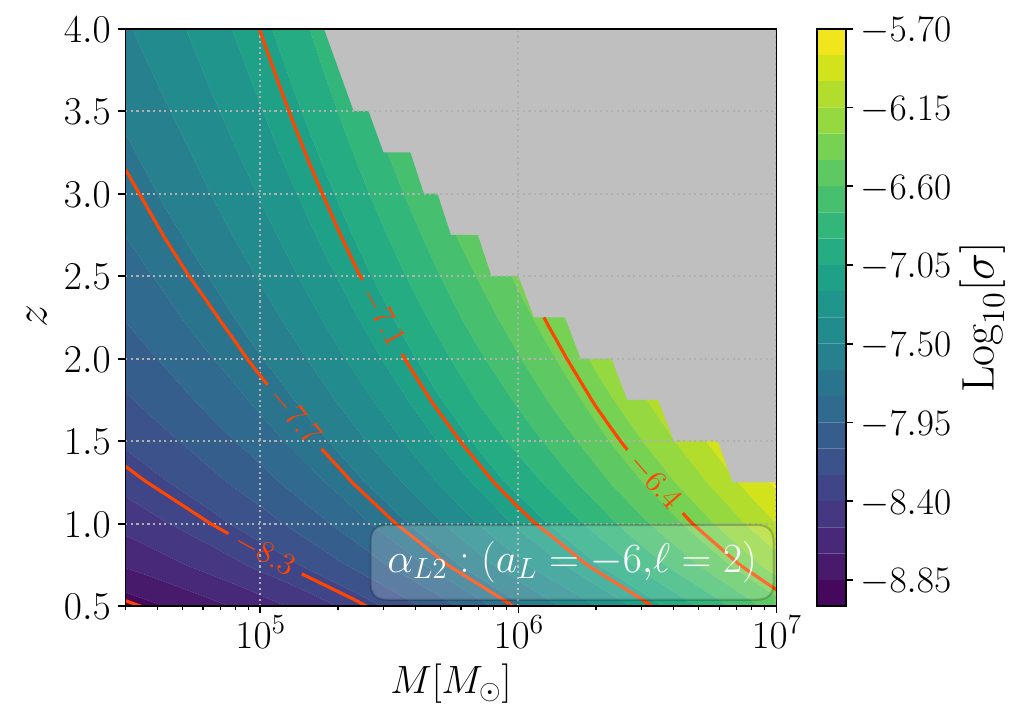}
\caption{Median absolute errors on $\alpha_{L1}$ (left plots) and $\alpha_{L2}$ (right plots). The specific combination of $a_L$ and $\ell$ are reported in the legend of each plot. We do not show the case for $a_L=-6$ and $\ell=1$ because that entire plot would be masked out.}
 \label{fig:gr_injection_alphal} 
\end{figure}

In Figure~\ref{fig:gr_injection_alphal}, we report the constraints for $\alpha_{L1}$ and $\alpha_{L2}$  from GR injections for $a_L=0$, $a_L=-2$ and $a_L=-6$.

Similarly to before, gray areas corresponds to regions of the parameter space where the uncertainties in $\{\alpha_{L1}, \alpha_{L2} \}$ are larger than the amplitude of the $(\ell=|m|=2)$ tensor mode (see Eq.~\eqref{Eq:limit_alpha_extrapol}), which we mask out. Indeed, we do not show the case for $a_L=-6$ and $\ell=1$ because that entire plot would be masked out.

\paragraph{$\mathbf{a_L = \{-6, -2 , 0\} }$:} For the cases $a_L=-2$ and $a_L=0$, the  constraints are identical to those derived for the breathing mode. Indeed, we have verified that the differences between the two are too small to be visible at the scale of the plots. This fact is a consequence of the LISA response to these two polarizations, which is very similar in the range of the parameter space considered (see Figure~\ref{fig:responsefcomp}). Interestingly, though, the longitudinal mode admits a case where $a_L=-6$, where the constraining power is improved for low-mass systems compared to the other case, once again, due to the strong frequency dependence of the modification to the GR waveform.

\paragraph{Distinguishing breathing and longitudinal modes:}
\begin{figure}
\centering
\includegraphics[width = 0.5\textwidth]{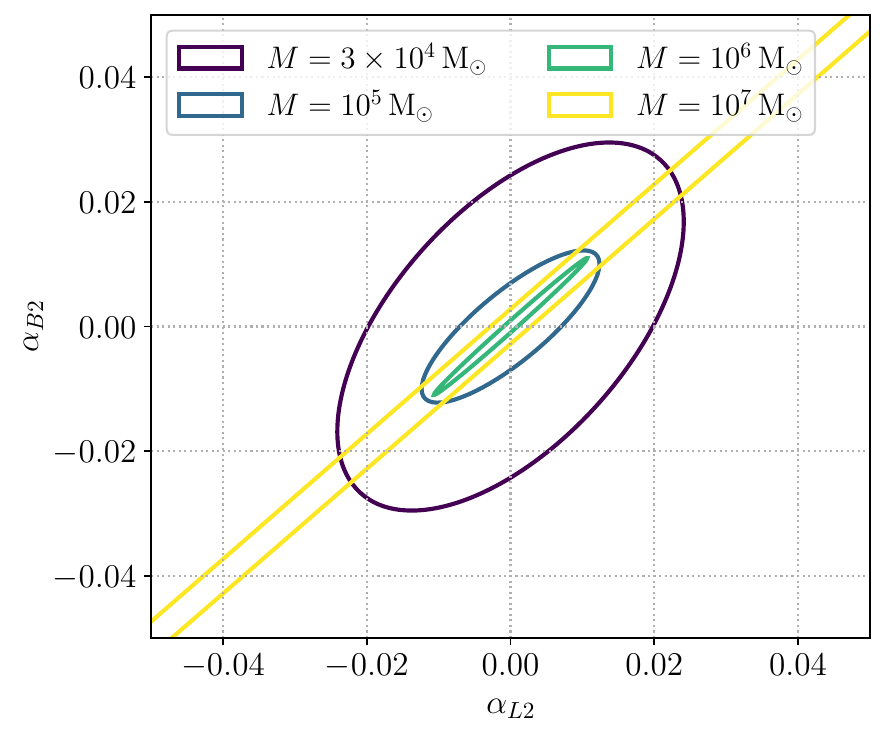}
 \caption{Example of correlation between $\alpha_{B2}$ and $\alpha_{L2}$ for four MBHB systems at $z=1$ and with total masses reported in the legend, with other parameters randomized.}  
 \label{fig:correlation_hb_hl} 
\end{figure}

From the previous results, it is clear that LISA will be able to provide comparable constraints on breathing and longitudinal polarizations and, as shown in Figure~\ref{fig:responsefcomp}, the LISA response to these two modes is identical at low frequencies. To further test if LISA can distinguish between breathing and longitudinal polarizations, we run an additional set of simulations where we include both $\alpha_{B2}$ and $\alpha_{L2}$ (for a total of 13 binary parameters and fixing $a_B=a_L=0$)
to check the correlation between these two quantities for different total masses. 
Similarly to the previous cases, we perform a Monte Carlo over 100 realizations for MBHBs with intrinsic total mass of $M = 3\times 10^4,\, 10^5, \, 10^6$ and $10^7 \msun$ at redshift $z=1$ \footnote{We obtain similar  results also for $z=3$ and $z=5$.}.
In Figure~\ref{fig:correlation_hb_hl} we plot the correlation between $\alpha_{B2}$ and $\alpha_{L2}$ for four representative systems. Two features appear clearly: (i) the constraints on $\alpha_{B2}$ and $\alpha_{L2}$ are the best for the system with $M = 10^6 \msun$ since these are the best observed systems in LISA; (ii) the correlation strongly increases as we move from low-mass to high-mass systems, since we see the shape of the joint posterior becoming more elliptical. This latter feature happens because the LISA response to scalar polarizations differs at high frequencies (see  Figure~\ref{fig:responsefcomp}), and hence the detector can distinguish them for systems that merge at frequencies $f\gtrsim 3\times 10^{-2}$Hz. For the lowest masses we test, $M=3\times 10^4 \msun $, we find a median correlation factor of $\sim 0.64_{-0.10}^{+0.11}$, which suggests that for light mass events the correlation factor can be smaller than 0.5.

We emphasize that performing independent measurements of the longitudinal and breathing mode can provide important constraints on specific modified gravity theories, when each scalar polarization depends on different theory parameters. One example is Horndeski theory, where the longitudinal polarization depends on the mass of the scalar field (see Tables \ref{table:PPE_mapping_alphaX2}-\ref{table:PPE_mapping_alpha1}), and thus an independent measurement of this polarization would bound the scalar mass. In the case of Einstein-\ae ther theory, we also find the longitudinal and breathing modes to depend on different parameters, and an independent measurement of them would allow to constrain the parameter $a_{bL}^{\text{\ae}}$. In the cases of Rosen and Lightman-Lee theories we find both scalar modes to depend on the same theory parameters, and thus a measurement of a combination of the scalar polarizations would be enough.

Altogether, our numerical results show that observations of MBHB mergers not always allow us to distinguish between breathing and longitudinal polarizations, with the most promising systems corresponding to light masses with $M\lesssim 10^4 \msun$. Such light systems are expected to be common in light-seed MBH formation scenarios but rare in heavy-seed MBH formation scenarios, and hence whether LISA will indeed detect events that distinguish these scalar polarizations is still unknown.
Nonetheless, there is the possibility that breathing and longitudinal modes could be distinguished with observations of different sources, such as stellar-mass BHBs or, potentially, EMRIs and/or intermediate MRIs. These scenarios are under active investigation and will be addressed in the future.

We emphasize that the conclusions presented here should be regarded as strong evidence rather than a definitive statement. A dedicated Bayesian analysis assessing the Bayes factor for breathing versus longitudinal polarizations is currently ongoing and will be presented in a future publication.

\subsubsection{Vector modes amplitude}
\begin{figure}[h!]
\centering
\includegraphics[scale=0.44]{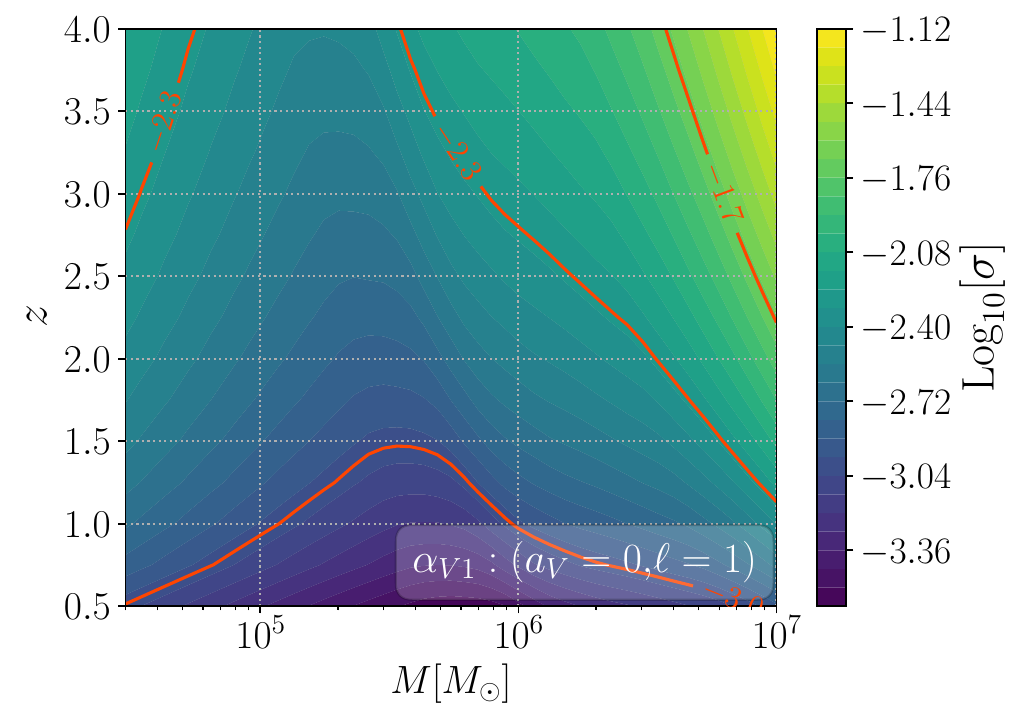}  
\includegraphics[scale=0.44]{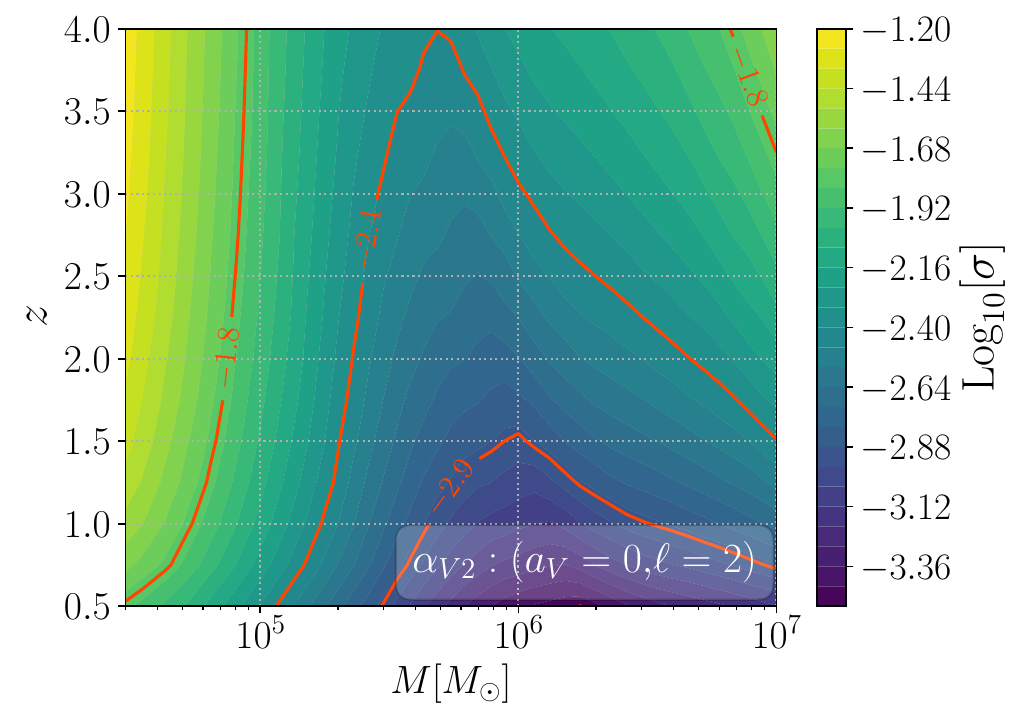}  \\
\includegraphics[scale=0.44]{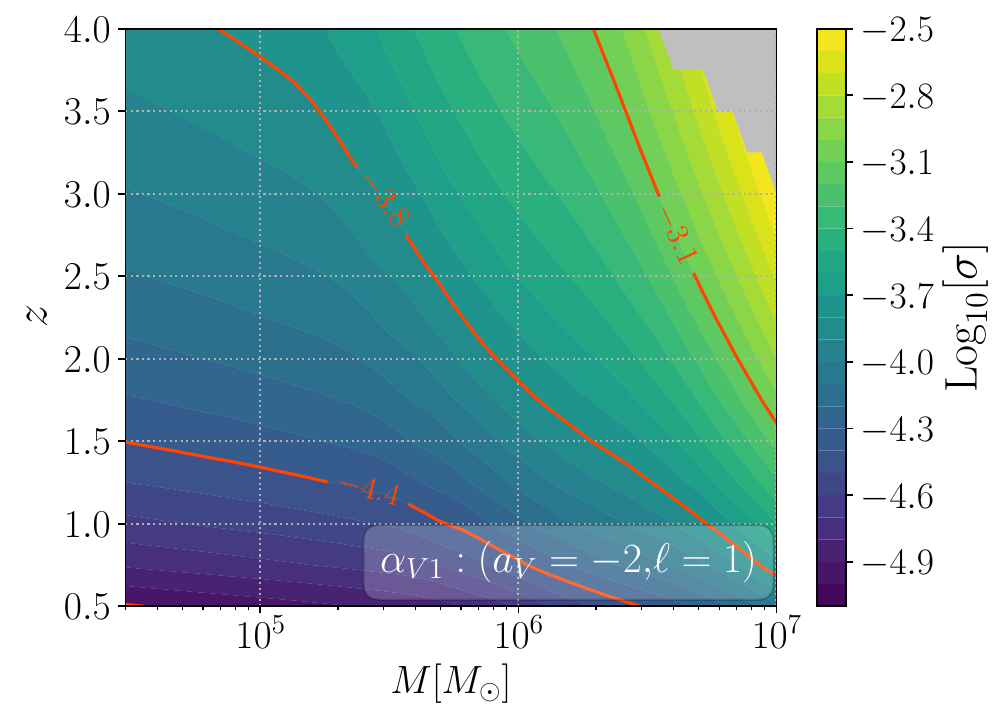}  
\includegraphics[scale=0.44]{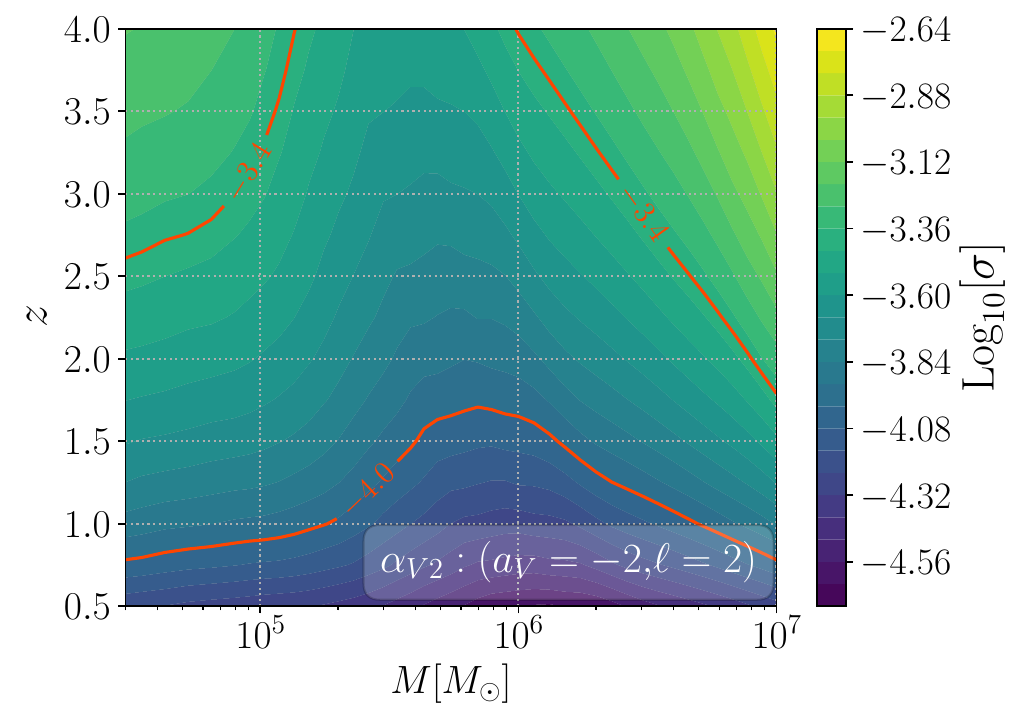} 
\caption{Same as Figure~\ref{fig:gr_injection_alphab} but for $\alpha_{V1}$ (left plots) and $\alpha_{V2}$ (right plots). The specific combination of $a_V$ and $\ell$ are reported in the legend of each plot. 
}
 \label{fig:gr_injection_alphaV} 
\end{figure}

In Figure~\ref{fig:gr_injection_alphaV} we report the constraints for $\{\alpha_{V1}, \alpha_{V2} \}$ from GR injections, for the cases with $a_V=0$ (top panels) and $a_V=-2$ (bottom panels), for both dipole and quadrupole harmonics.

The results can be interpreted analogously to the ones for the breathing and longitudinal modes. For example, for the case $a_V=0$, we can constrain $\alpha_{V1}$ ($\alpha_{V2}$) at $\sim 10^{-3}$ ($7.59 \times 10^{-3}$) for $M=10^5 \msun$ at $z=1$.

Comparing the constraints on vector modes with those on the two scalar modes, we find that the former are consistently tighter than the latter by a factor of 2-3 for a given frequency scaling, as can be seen in Figures~\ref{fig:gr_injection_alphaV}, \ref{fig:gr_injection_alphab}, and \ref{fig:gr_injection_alphal}. In particular, the red curves in these figures trace similar trajectories in the $(M,\, z)$ parameter space, yet the bounds associated with the vector modes are systematically more stringent. For example, for $a_V = -2$ at redshifts $z < 1.5$ the parameter $\alpha_{V1}$ ($\alpha_{V2}$) can be constrained down to $\sim 4 \times 10^{-5}$ ($\sim 10^{-4}$), whereas the corresponding curves in Figs.~\ref{fig:gr_injection_alphab} and \ref{fig:gr_injection_alphal} yield weaker constraints of order $\sim 6\times 10^{-5}$ ($\sim 2 \times 10^{-4}$) on both $\alpha_{B1}$ ($\alpha_{B2}$) and $\alpha_{L1}$ ($\alpha_{L2}$). We argue that this fact is once again related to the response function of LISA to these modes, and the fact that this is better for vector modes, compared to scalars (cf.\ with Figure~\ref{fig:responsefcomp}). 

Notice that this statement of stronger constraints on vector  than scalar modes is valid when assuming the same frequency scaling for these extra polarizations. Indeed, the longitudinal mode also is allowed to have a scaling $a_L = -6$ which is not present for the vector modes. In this case, the frequency scaling of the modification to the longitudinal amplitude is strong enough to produce better constraints than the vector amplitude parameters (when considering $a_V=0,-2$), regardless of the poorer detector response. For example, for $\ell =2$, $M = 10^5 \msun$ and $z = 1$, the constraints on $\alpha_{L2}$ reach a precision of order $\sim 6\times 10^{-9}$ when $a_L=-6$. 

\subsubsection{Phase modification}
\begin{figure}[h!]
\centering
\includegraphics[scale=0.44]{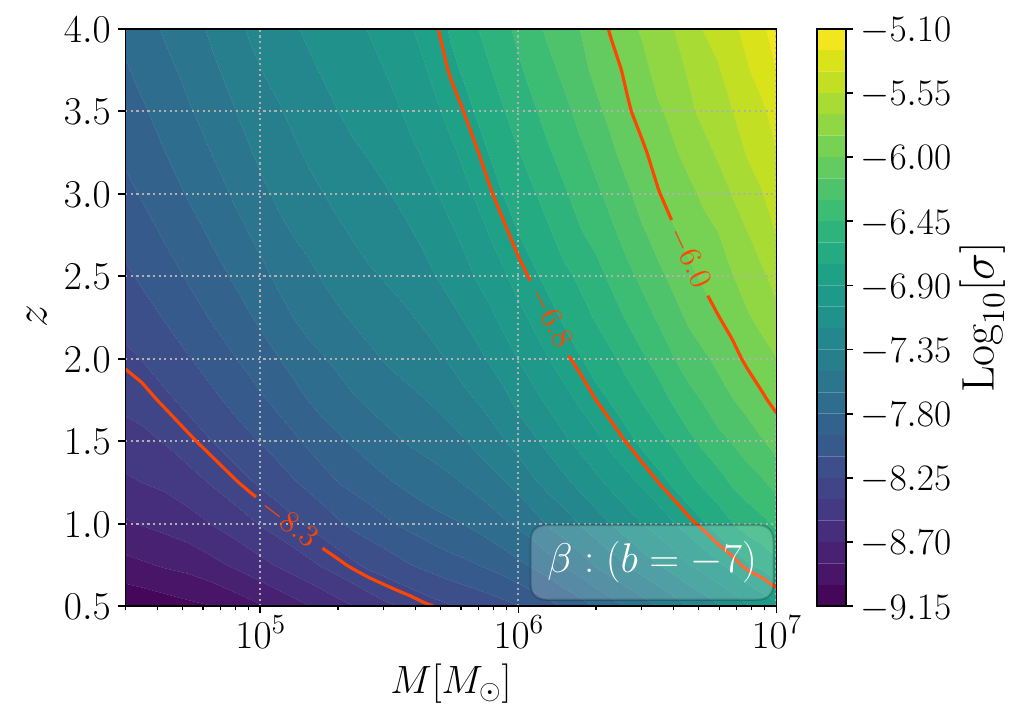} 
\includegraphics[scale=0.44]{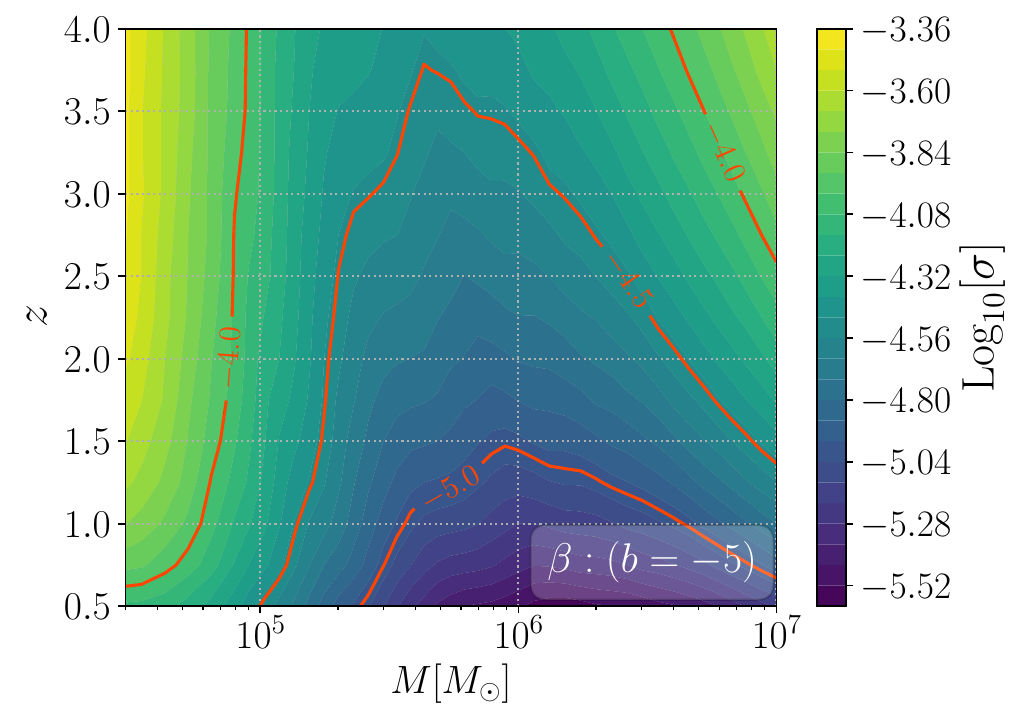}  
\caption{Same as Figure~\ref{fig:gr_injection_alphaT} but for $\beta$. The specific choice of $b$ is reported in the legend of each plot.
}
 \label{fig:gr_injection_betaT} 
\end{figure}

Figure~\ref{fig:gr_injection_betaT} shows the forecast constraints on $\beta$, namely the parameter describing the phase modifications of all polarizations. 
We include both $b = -7$ and $b=-5$ in line with Table~\ref{table:PPE_mapping_ab} and, for concreteness, consider that the modification induced by $\beta$ affects only the tensor modes. That is, we will assume all amplitude parameters to be vanishing $\alpha_T = \alpha_{P1} = \alpha_{P2}  = 0$. As expected, we get the best constraints for the case $b=-7$ and for low-mass systems, due to the enhancement of the modifications at low frequencies and this unique PN term in the phase. Similar to $\alpha_T$, the constraints on $\beta$ for $b=-5$ follows the typical SNR waterfall plots for LISA. In this case, $\beta$ is highly degenerate with the chirp mass, as discussed in Section~\ref{subsec:numerical_implementation}, which is why we use the transformation in Eq.~\eqref{eq:beta_prime} in order to obtain these results.

In Appendix \ref{sec:fti_comparison} we compare our $\beta$ constraints to those using the FTI formalism for LISA \cite{Piarulli:2025rvr}, and find them to be in agreement. We notice that these constraints obtained for LISA can be 2-3 orders of magnitude more precise than current LVK constraints \cite{LIGOScientific:2021sio,LIGOScientific:2025obp}. The work in \cite{Perkins:2020tra} also performed a population forecast for LISA and 3G ground-based detectors for tensor phase modifications, where it is found that LISA is expected to provide comparable or better constraints than 3G detectors, depending on the expected observed population of events and details of the 3G networks\footnote{We emphasize that this comparison assumes the PPE parameters to be constant across different binary systems but, depending on the modified gravity predictions, they may depend on other parameters such as total mass or mass ratio (see e.g.\ Tables \ref{table:PPE_mapping_beta}-\ref{table:PPE_mapping_alpha1}), in which case a proper comparison between the constraining power of stellar-mass objects observed by ground-based detectors versus massive black holes observed by LISA would need to be revisited.}. Our constraints on the tensor phase are comparable to those obtained in \cite{Perkins:2020tra}.

\bigskip
Similar results for the phase modification $\beta_P$ in the breathing, longitudinal and vector sectors are reported in Appendix~\ref{app:betap_injections}. However, we caution the reader that these constraints are not pure GR injections since, in order to perform them, we  choose non-zero values of $\alpha_{P1}, \alpha_{P2}$: otherwise the polarization mode would be vanishing, rendering  $\beta_P$ not constrainable  (see Eqs.~\eqref{Eq:PPE_p22}-\eqref{Eq:PPE_p11}). 
In this sense, the results for $\beta_P$ should more appropriately be referred to as ``almost-GR'' injections. Additionally, since a $\beta \neq 0$ affects both the extra polarization mode {\it and} the tensor mode, we modify also the phase of the tensor modes when performing these tests  (see Eq.~\eqref{Eq:PPE_Tlm},~\eqref{Eq:PPE_p22} and~\eqref{Eq:PPE_p11} with $2\beta_P  =\beta$). In order to produce such constraints, we take the median errors from GR injections for each $(M, z)$ combination as values for $\alpha_{P1}, \alpha_{P2}$, and set $\beta_P = 0$.  We find that the corresponding constraints are mostly dominated by the tensor polarization (due to their modified phase), leading the ``almost-GR'' injection $\beta_P$ plots to closely resemble those shown in Figure~\ref{fig:gr_injection_betaT}.

\subsection{\label{sec:constrain-specific-theories}Constraints on specific theories}

In this section, we map the constraints obtained from GR injections in Section~\ref{subsec:gr_injections} to those of specific theories. To perform this mapping, we adopt the expressions connecting $\alpha_T$, $\alpha_{B1}$, $\alpha_{B2}$, $\alpha_{L1}$, $\alpha_{L2}$, $\alpha_{V1}$, $\alpha_{V2}$ and $\beta$ to the theory-specific parameters listed in Tables~\ref{table:PPE_mapping_beta}–\ref{table:PPE_mapping_alpha1}. 
For each event, we use the 12-dimensional covariance matrix derived from the GR injections to generate $10^5$ samples of all binary parameters, including the PPE coefficients. We then apply the relations in Tables~\ref{table:PPE_mapping_beta}–\ref{table:PPE_mapping_alpha1} to map these samples from the PPE parameters (and, when necessary, additional binary parameters) to constraints on the parameters of the specific theories. Similarly to the GR injections, we also report the 68\% confidence interval.  
For the theories listed in Tables~\ref{table:PPE_mapping_beta}–\ref{table:PPE_mapping_alphaX2} that do not admit a GR limit, we instead perform a set of dedicated non-GR injections to obtain the corresponding constraints.

\textbf{Horndeski theory:} In Table~\ref{table:constrain_hornsdeski}, we report the constraints on the parameters of Horndeski theory. Apart from the considerations already discussed in Section~\ref{subsec:gr_injections}, namely, that the results depend sensitively on the adopted values of $a$, $a_B$, $a_L$, and $a_V$, as well as on the choice between $\ell=1$ and $\ell=2$, we also highlight a few noteworthy cases. For instance, for $a=-2$, the parameters $\alpha_T$ and $\beta$ map onto a similar combination of Horndeski parameters; the difference is in the power of $G_*/G_N$. Therefore, a joint analysis of amplitude and phase modifications for this case could allow us to constrain $G_*/G_N$ and $\kappa_4(\Delta\hat\alpha)^2$ independently.
By contrast, for tensor emission with $a = 0$, as well as for scalar polarizations, the corresponding PPE coefficients map onto different combinations involving four or five theory parameters. This makes it challenging to place meaningful constraints on individual parameters. This observation highlights the need for higher-order expressions within this theory, where these combinations may enter differently, potentially enabling independent constraints. 

Moreover, we emphasize that the constraints on additional polarization modes obtained here apply only to a specific subclass of Horndeski theories. In particular, the mere presence of an additional scalar degree of freedom in the gravitational radiation does not guarantee the existence of measurable scalar polarizations in the detector response. While the existence of an extra degree of freedom is a fundamental property of the underlying theory, the appearance of additional polarizations depends sensitively on the form of the non-minimal coupling between the scalar field and the physical metric, which ultimately determines how matter responds to the radiation field.
As shown explicitly in Appendix~\ref{sApp:HorndeskiPolarizations}, a massless scalar field generically does not support a longitudinal polarization, while theories lacking a non-trivial non-minimal coupling to the Ricci scalar, specifically those for which $\sigma \propto g_4 = 0$, where $\sigma$ is defined in Eq.~\eqref{eq:defsigma}, do not admit any scalar polarizations at all. These statements are clearly reflected in Table~\ref{table:constrain_hornsdeski}. Conversely, a positive detection of an additional polarization mode would strongly suggest new physics beyond GR.

\begin{table}[h]
\centering
\begin{tabular}{|c|c|c|c|c|}
\hline
 \multicolumn{5}{|c|}{Horndeski Theory (Section \ref{sec:ST_theory}) } \\
\hline
PPE & Theory  & \multicolumn{3}{c|}{Median $1\sigma$ error} \\ \cline{3-5}
Parameter & Parameter & $M = 3 \times 10^5$ M$_{\odot}$ & $M = 10^6$ M$_{\odot}$ & $M = 10^7$ M$_{\odot}$ \\
\hline
\multirow{2}{*}{$\alpha_T$} 
    & \cellcolor{blue!5} $\kappa_4\biggl(\frac{G_*}{G_N}\biggr)^{\frac16}(\Delta\hat{\alpha})^2$ 
    & $3.05 \times 10^{-3}$ & $3.04 \times 10^{-3}$ & $8.68 \times 10^{-3}$ \\ \cline{2-5}
    & \cellcolor{green!5} $\left(\frac{G_*}{G_N}\right)^{5/6} \left(1 + \frac{1}{3}\delta_0 - \frac{\kappa_4}{12}\Gamma^2\right)$ 
    & $5.39 \times 10^{-3}$ & $2.02 \times 10^{-3}$ & $4.69 \times 10^{-3}$ \\ 
\hline
\multirow{2}{*}{$\beta$} 
    &  \cellcolor{blue!5}  $\kappa_4\biggl(\frac{G_*}{G_N}\biggr)^{-\frac73}(\Delta\hat{\alpha})^2$ 
    & $9.17 \times 10^{-6}$ & $9.17 \times 10^{-5}$ & $4.00 \times 10^{-4}$ \\ \cline{2-5}
    &\cellcolor{green!5}  $\left(\frac{G_*}{G_N}\right)^{-5/3} \left(1 - \frac{2}{3}\delta_0 - \frac{\kappa_4}{6}\Gamma^2\right)$ 
    & $5.38 \times 10^{-4}$ & $2.19 \times 10^{-4}$ & $2.76 \times 10^{-4}$ \\
\hline
$\alpha_{B2}$ 
    & $\kappa_4 g_4 \Gamma \left(\frac{G_*}{G_N}\right)^{5/6}$ 
    & $9.50 \times 10^{-3}$ & $3.63 \times 10^{-3}$ & $9.60 \times 10^{-3}$ \\
    \hline
$\alpha_{B1}$ 
    & $\kappa_4 g_4 (\Delta\hat{\alpha}) \left(\frac{G_*}{G_N}\right)^{1/2}$ 
    & $1.21 \times 10^{-3}$ & $1.94 \times 10^{-3}$ & $7.70 \times 10^{-3}$ \\
\hline
$\alpha_{L2}$ 
    &  $\alpha_{B2} m_s^2G_N/(c\hbar) $ 
    & $2.97 \times 10^{-93}$ & $9.21 \times 10^{-94}$ & $1.25 \times 10^{-94}$ \\
\hline
\end{tabular}
\caption{Median $1\sigma$ uncertainties obtained from Fisher  forecasts on specific Horndeski theory parameters. We quote the results for three masses of MBHBs, all at $z=1$. The first column indicates from which PPE parameter we get the constraints while the second column reports the specific Horndeski parameters over which we are providing the constrains.
For the case $\alpha_{L2}$, we factor out $ \mathcal{M}_z^2 G_N/(4c \hbar)$ in order to constrain a \emph{dimensionless} unknown theory parameter (we also recover explicitly $c$ and $\hbar$ for clarity). The extreme values of $\sim 10^{-93}$ can be explained by the fact that the quantity $ \mathcal{M}_z^2 G_N/(4c \hbar)$ 
 has a typical value of $\sim 10^{87}$ for $\mathcal{M}_z\sim 10^6 \msun$.
We don't perform the mapping for $\alpha_{L1}$ because the case $\alpha_L=-6$ and $\ell=1$ was masked out in Fig.~\ref{fig:gr_injection_alphal}. }
\label{table:constrain_hornsdeski}
\end{table}

\begin{table*}[htb]
\centering
\begin{tabular}{|c|c|c|c|c|}
\hline
 \multicolumn{5}{|c|}{Einstein-\ae ther  (Section \ref{sec:EA_theory}) } \\
\hline
PPE & Theory  & \multicolumn{3}{c|}{Median $1\sigma$ error} \\ \cline{3-5}
Parameter & Parameter & $M = 3 \times 10^5$ M$_{\odot}$ & $M = 10^6$ M$_{\odot}$ & $M = 10^7$ M$_{\odot}$ \\
\hline
\multirow{2}{*}{$\alpha_T$} 
    & \cellcolor{blue!5} $\frac{ \epsilon_x}{  [(1-s_1)(1-s_2)]^{2/3}  }$ 
    & $6.35\times10^{-4}$ & $6.33\times10^{-4}$ & $1.81\times10^{-3}$ \\ \cline{2-5}
    & \cellcolor{green!5} $\left[ \sqrt{\frac{(2-c_a)}{ \kappa_3}} [(1-s_1)(1-s_2)]^{\frac13} \right]^{(1)}$ 
    & $7.62\times10^{-3}$ & $2.86\times10^{-3}$ & $6.64\times10^{-3}$ \\ \hline
\multirow{2}{*}{$\beta$} 
    & \cellcolor{blue!5} $ \epsilon_x [(1-s_1)(1-s_2)]^\frac23$ 
    & $1.91\times10^{-6}$ & $5.09\times10^{-6}$ & $8.32\times10^{-5}$ \\ \cline{2-5}
    & \cellcolor{green!5} $\left[ \frac{[(1-s_1)(1-s_2)]^{-\frac23}}{(2-c_a)\kappa_3}\right]^{(1)}$ 
    & $2.69\times10^{-4}$ & $2.19\times10^{-4}$ & $1.38\times10^{-4}$ \\    \hline
\multirow{2}{*}{$\alpha_{B2}$} & \cellcolor{blue!5} $-2\sqrt{6}\eta^{-2/5} \alpha_{B2}$ 
    & $2.11\times10^{-3}$ & $1.41\times10^{-3}$ & $4.25\times10^{-3}$  \\ \cline{2-5}
    & \cellcolor{green!5} $\sqrt{12}\alpha_{B2}$
    & $1.34\times10^{-2}$ & $5.13\times10^{-3}$ & $1.36\times10^{-2}$  \\  \hline
\multirow{2}{*}{$\alpha_{B1}$} &  \cellcolor{blue!5} $-2\sqrt{\frac{96}{5}} \eta^{-2/5}\alpha_{B1}$ & $1.35\times10^{-3}$ & $2.60\times10^{-3}$ & $1.28\times10^{-2}$  \\ \cline{2-5}
    & \cellcolor{green!5} $\sqrt{\frac{96}{5}}\alpha_{B1}$
    & $4.83\times10^{-3}$ & $7.78\times10^{-3}$ & $3.08\times10^{-2}$   \\  \hline
\multirow{2}{*}{$\alpha_{L2}$} &  \cellcolor{blue!5} $-2\sqrt{6}\eta^{-2/5} \alpha_{L2}$ & $2.14\times10^{-3}$  &  $1.42\times10^{-3}$ &  $4.22\times10^{-3}$  \\ \cline{2-5}
    & \cellcolor{green!5} $\sqrt{12}\alpha_{L2}$
    & $1.39\times10^{-2}$ & $5.12\times10^{-3}$  & $1.36\times10^{-2}$    \\  \hline
\multirow{2}{*}{$\alpha_{L1}$} &  \cellcolor{blue!5} $-2\sqrt{\frac{96}{5}} \eta^{-2/5}\alpha_{L1}$ & $1.34\times10^{-3}$  &  $2.58\times10^{-3}$ &  $1.28\times10^{-2}$  \\ \cline{2-5}
    & \cellcolor{green!5} $\sqrt{\frac{96}{5}}\alpha_{L1}$
    & $4.85\times10^{-3}$  & $7.86\times10^{-3}$  & $3.07\times10^{-2}$   \\ \hline
\multirow{2}{*}{$\alpha_{V2}$} &  \cellcolor{blue!5}  $-4\eta^{-2/5} \alpha_{V2}$ & $8.71\times10^{-4}$  &  $5.98\times10^{-4}$ &  $1.63\times10^{-3}$   \\ \cline{2-5}
    & \cellcolor{green!5} $2\sqrt{2}\alpha_{V2}$
    & $5.61\times10^{-3}$  & $2.19\times10^{-3}$  & $5.74\times10^{-3}$   \\ \hline
\multirow{2}{*}{$\alpha_{V1}$} &  \cellcolor{blue!5}  $-2\sqrt{\frac{192}{5}} \eta^{-2/5}\alpha_{V1}$ & $1.18\times10^{-3}$  &  $2.13\times10^{-3}$ &  $1.06\times10^{-2}$   \\ \cline{2-5}
 & \cellcolor{green!5} $\sqrt{\frac{192}{5}}\alpha_{V1}$
    & $4.13\times10^{-3}$  & $6.50\times10^{-3}$  & $2.49\times10^{-2}$   \\ \hline
\end{tabular}
\caption{Same as Table~\ref{table:constrain_hornsdeski} but for Einstein-\ae ther theory. Whenever the expressions for the theory parameters were too long to fit the table, we report only the name of the PPE parameter together with the factors we constrained. For example, in the case of $\alpha_{B2}$ and $a_b=-2$, we start from Eq.~\eqref{eq:EAalphaB2a2} and we move to the left-hand side the numerical constant that can be isolated ($-2\sqrt{6}$) and $\eta$. }
\label{table:constrain_ae}
\end{table*}

\textbf{Einstein-\ae ther theory:} 
Table~\ref{table:constrain_ae} reports the mapping of the constraints from the PPE parameters to the Einstein-\ae ther theory fundamental parameters.
Interestingly, for $a = -2$, $\alpha_T$ maps onto the ratio between $\epsilon_x$ and $[(1 - s_1)(1 - s_2)]^{2/3}$, while $\beta$ maps onto their product. If both amplitude and phase modifications were allowed simultaneously, this different functional dependence could, in principle, enable independent constraints on $\epsilon_x$ and $[(1 - s_1)(1 - s_2)]^{2/3}$. We discuss this possibility in more detail in Section~\ref{subsec:amp-phase-modifications}.

\begin{table*}[htb]
\centering
\begin{tabular}{|c|c|c|c|c|}
\hline
 \multicolumn{5}{|c|}{Rosen's theory  (Section \ref{sec:Rosen_theory})} \\
\hline
PPE & Theory  & \multicolumn{3}{c|}{Median $1\sigma$ error} \\ \cline{3-5}
Parameter & Parameter & $M = 3 \times 10^5$ M$_{\odot}$ & $M = 10^6$ M$_{\odot}$ & $M = 10^7$ M$_{\odot}$ \\
\hline
$\alpha_{T}$ &  \fcolorbox{red}{white}{$\sqrt{1 + \frac{2}{7}}$} & $1.03\times 10^{-2}$  &  $3.14\times 10^{-3}$ &  $6.96\times 10^{-3}$   \\ \hline
\multirow{2}{*}{$\beta$} &  \cellcolor{blue!5}   $\kappa_{\rm R}^{-\frac23} {\cal G}^2 $ & $8.42\times10^{-6}$   &  $5.34\times10^{-6}$ &  $1.30\times10^{-5}$   \\ \cline{2-5}
& \cellcolor{green!5} \fcolorbox{red}{green!5}{$ -\frac{3}{256} \left[ 1 + \frac87  \right] $}
    & $6.86\times 10^{-6}$  & $1.03\times 10^{-5}$  & $5.51\times 10^{-5}$   \\ \hline
$\alpha_{B2}$ &  \fcolorbox{red}{white}{$\frac{1}{\sqrt{21}}$} & $1.03\times 10^{-2}$   &  $5.51\times 10^{-3}$ &  $1.28\times 10^{-2}$   \\ \hline
$\alpha_{B1}$ &  $\mathcal{G} \kappa_{\rm R}^{-7/12}$ & $1.69\times10^{-3}$   &  $2.73\times10^{-3}$  &  $1.08\times10^{-2}$   \\ \hline
$\alpha_{L2}$ & \fcolorbox{red}{white}{$\frac{2}{\sqrt{21}}$} & $1.52\times 10^{-2}$  & $7.66\times 10^{-3}$  & $1.62\times 10^{-2}$   \\ \hline
$\alpha_{L1}$ & $\mathcal{G} \kappa_{\rm R}^{-7/12}$ & $1.70\times10^{-3}$  & $2.76\times10^{-3}$  & $1.08\times10^{-2}$   \\ \hline
 $\alpha_{V2}$ & \fcolorbox{red}{white}{$\frac{2}{\sqrt{7}}$} & $7.07\times 10^{-3}$   &  $3.44\times 10^{-3}$ &  $7.32\times 10^{-3}$   \\ \hline
$\alpha_{V1}$ & $\mathcal{G} \kappa_{\rm R}^{-7/12}$ & $7.25\times10^{-4}$   &  $1.14\times10^{-3}$ &  $4.36\times10^{-3}$   \\ \hline
\end{tabular}
\caption{Same as Table~\ref{table:constrain_hornsdeski} but for Rosen's theory. Red boxes represent cases that do not admit GR limits.}
\label{table:constrain_rosen}
\end{table*}

\begin{table*}[htb]
\centering
\begin{tabular}{|c|c|c|c|c|}
\hline
 \multicolumn{5}{|c|}{Lightman-Lee Theory  (Section \ref{sec:LL_theory})} \\
\hline
PPE & Theory  & \multicolumn{3}{c|}{Median $1\sigma$ error} \\ \cline{3-5}
Parameter & Parameter & $M = 3 \times 10^5$ M$_{\odot}$ & $M = 10^6$ M$_{\odot}$ & $M = 10^7$ M$_{\odot}$ \\
\hline
$\alpha_{T}$ &   \fcolorbox{red}{white}{$\sqrt{1 + \frac{2}{7}}$} & $1.03\times 10^{-2}$  &  $3.14\times 10^{-3}$ &  $6.96\times 10^{-3}$   \\ \hline
\multirow{2}{*}{$\beta$} &  \cellcolor{blue!5}   ${\cal G}^2 $ & $6.73\times10^{-7}$   &  $4.27\times10^{-7}$ &   $1.04\times10^{-6}$    \\ \cline{2-5}
& \cellcolor{green!5} \fcolorbox{red}{green!5}{$-\frac{3}{224} - \frac{3}{256} $}
    &  $6.86\times 10^{-6}$  & $1.03\times 10^{-5}$  & $5.51\times 10^{-5}$   \\ \hline
$\alpha_{B2}$ &  \fcolorbox{red}{white}{$\frac{3}{\sqrt{7}}$} & $1.82 \times 10^{-2}$   &  $9.99 \times 10^{-3}$ &  $2.40 \times 10^{-2}$   \\ \hline
$\alpha_{B1}$ &  $\mathcal{G} $ & $5.42\times10^{-4}$   &  $8.73\times10^{-4}$  &   $3.46\times10^{-3}$    \\ \hline
$\alpha_{L2}$ & \fcolorbox{red}{white}{$\frac{2}{\sqrt{21}}$} & $1.52\times 10^{-2}$  & $7.66\times 10^{-3}$  & $1.62\times 10^{-2}$   \\ \hline
$\alpha_{L1}$ & $\mathcal{G}$ & $6.80\times10^{-4}$  & $1.10\times10^{-3}$  & $4.31\times10^{-3}$   \\ \hline
 $\alpha_{V2}$ & \fcolorbox{red}{white}{$\frac{2}{\sqrt{7}}$} & $7.07\times 10^{-3}$   &  $3.44\times 10^{-3}$ &  $7.32\times 10^{-3}$  \\ \hline
$\alpha_{V1}$ & $\mathcal{G}$ &  $2.90\times10^{-4}$   &  $4.56\times10^{-4}$  &  $1.74\times10^{-3}$   \\ \hline
\end{tabular}
\caption{Same as Table~\ref{table:constrain_hornsdeski} but for Lightman-Lee theory. Red boxes represent cases that do not admit GR limits.}
\label{table:constrain_LL}
\end{table*}

\textbf{Rosen's and Lightman-Lee theories:} 
Tables~\ref{table:constrain_rosen} and~\ref{table:constrain_LL} report the mapping of the constraints from the PPE parameters to the fundamental parameters of Rosen and Lightman-Lee theory, respectively.
We recall that Rosen's bimetric theory does not admit black hole solutions (see Section~\ref{sec:Rosen_theory}), which means that the GR merger and ringdown waveform model is not self-consistent for this theory. The most robust constraints for this theory are therefore those derived from the inspiral phase alone (Section~\ref{sec:inspiral-only-injections}). For low-mass systems, where the inspiral dominates the in-band SNR, the inspiral-only and IMR constraints are expected to be of comparable order of magnitude; for high-mass systems, where the merger and ringdown contribute significantly, the inspiral-only constraints are degraded relative to the IMR results. In any case, Rosen's theory is observationally ruled out and the constraints derived here are therefore illustrative of the PPE methodology and the role of extra polarizations, rather than physically meaningful bounds on a viable theory of gravity.

\subsection{\label{sec:inspiral-only-injections}Inspiral-only GR injections}

\begin{figure}
\centering
\includegraphics[scale=0.44]{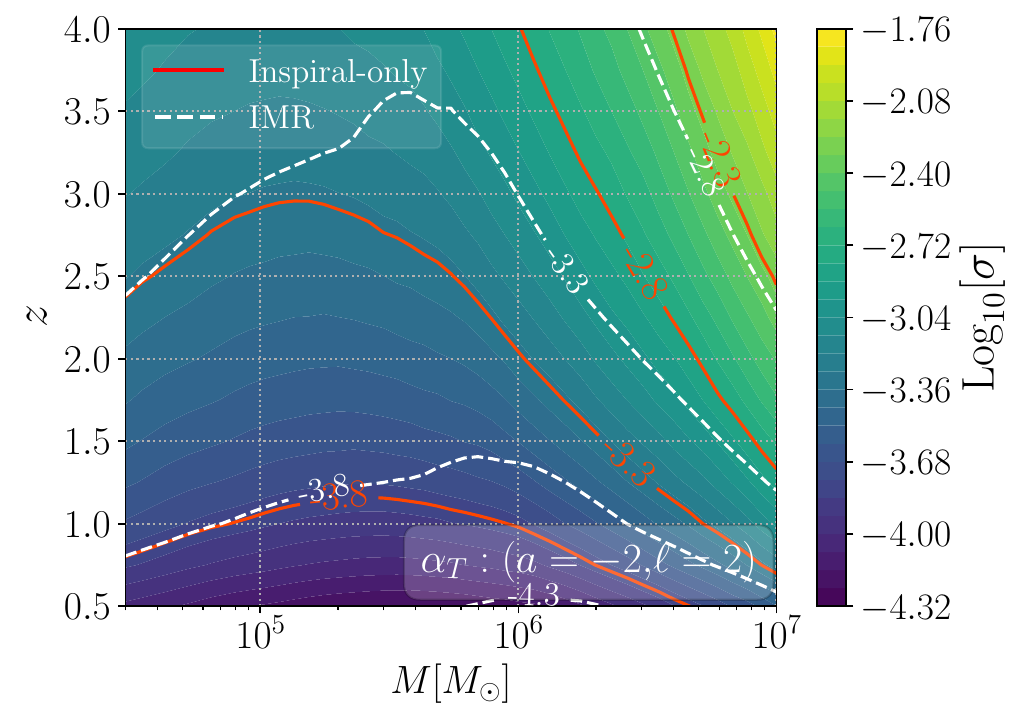}  
\includegraphics[scale=0.44]{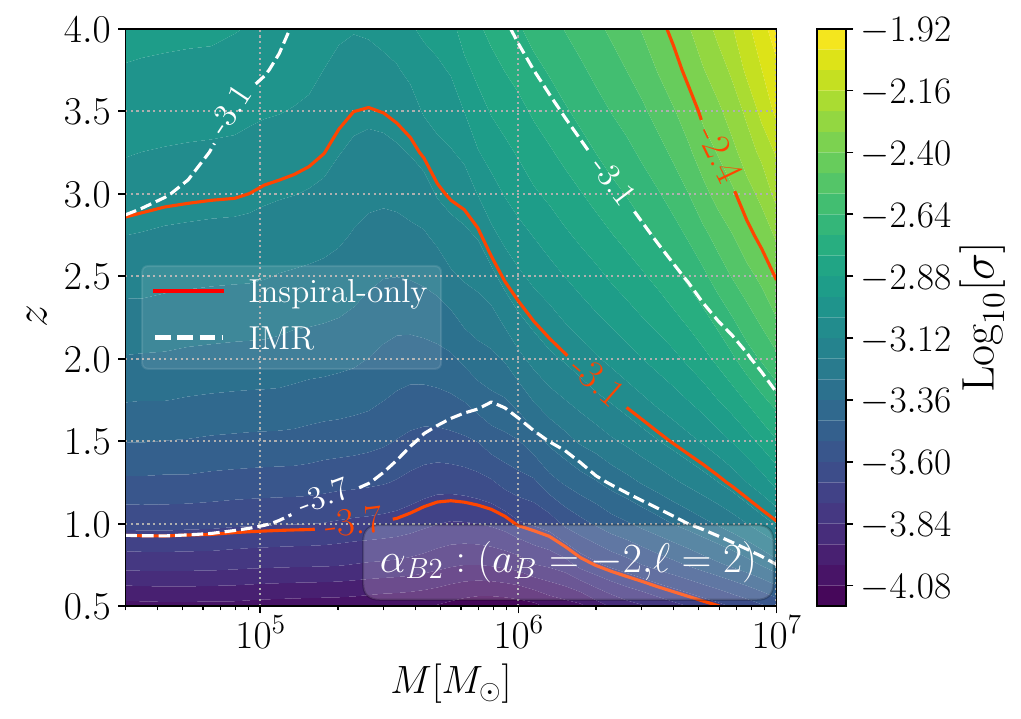}  \\
\includegraphics[scale=0.44]{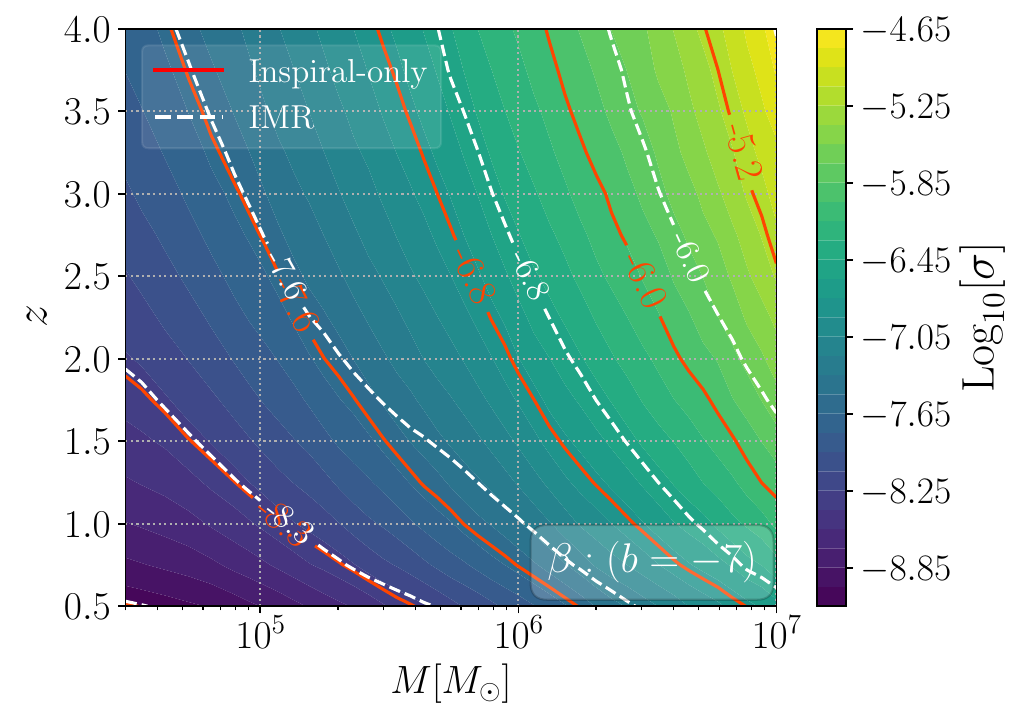}
\includegraphics[scale=0.44]{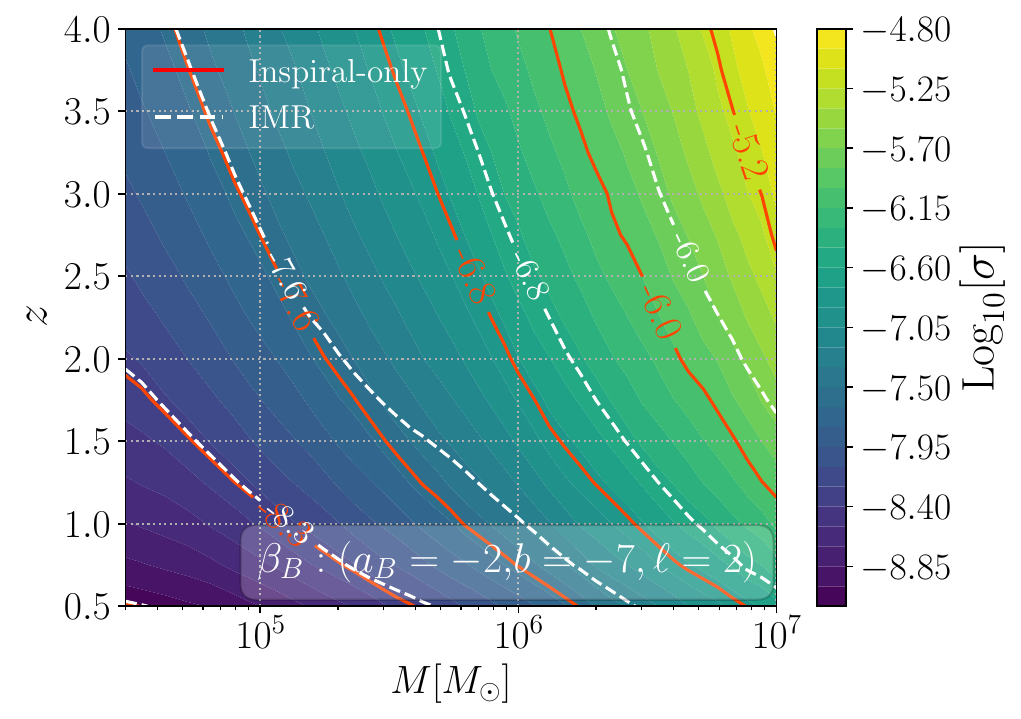}
\caption{Same as Figure~\ref{fig:gr_injection_alphaT}-\ref{fig:gr_injection_betaT} but for the case where we consider only the inspiral portion of the signal. Red lines and colormaps refer to inspiral-only GR injections. Dashed white lines represent the constraints from inspiral–merger–ringdown (IMR) GR injections, i.e. those presented in Section~\ref{subsec:gr_injections}.}
\label{fig:gr_injection_inspiral_only} 
\end{figure}

In Section~\ref{subsec:numerical_implementation} we described the tapering applied to the non-GR part of the signal, in order to include also the merger and ringdown. A more conservative approach would be to consider only the inspiral as it would not require the introduction of an arbitrary window function, nor the assumption of vanishing modifications after the inspiral.

In this section we present the constraints we obtain from GR injections in the case we consider only the inspiral portion of the signal and we compare them with the full inspiral-merger-ringdown (IMR) results presented in Section~\ref{subsec:gr_injections}. We follow the same procedure described in Section~\ref{subsec:gr_injections} with the only modification that the maximum frequency of the GW signal is set at the peak of the 22 mode, in order to include only the inspiral. For the extra polarizations in the $\ell=|m|=1$ angular harmonic, the frequency is cut at half the peak of the 22 mode. This ensures all harmonics are cut at a fixed time of observation near the merger.

In Figure~\ref{fig:gr_injection_inspiral_only} we present the results from inspiral-only GR injections for a few representative cases: $\alpha_T$ and $\alpha_{B2}$ with $a=a_B=-2$, as well as $\beta$ in the tensor and breathing modes with $b=-7$. To facilitate the comparison, we also show, with dashed white lines, the results from the corresponding IMR GR injections we presented in Section~\ref{subsec:gr_injections}. Overall, we find that the inspiral-only and IMR approaches produce similar results for MBHBs with $M < 10^5 \msun$, while the largest differences arise at higher masses. This behavior is expected since low-mass systems are inspiral-dominated, so removing the merger and ringdown portions of the signal does not affect the available signal significantly. By contrast, most of the information gained for heavy systems comes from the merger and ringdown and thus neglecting these contributions leads to larger uncertainties on the PPE parameters.
For example, for $M = 10^6 \msun$ and $\alpha_T$, we have that $\sigma = 1.58\times 10^{-4}$ ($\sigma = 5.01 \times 10^{-4}$) at $z \simeq 1$ ($z \simeq 2$) in the inspiral-only case. However, in the case of IMR GR injections we were able to achieve the same level of constraints up to $z\simeq 1.4$ ($z\simeq 3$).  Similar trends are observed for both $\alpha_{B2}$ and $\beta$. 
These results naturally raise the question of which of the two approaches should be considered more reliable: the inspiral-only or the IMR analysis. As  discussed at the end of Section~\ref{subsec:numerical_implementation} and at the beginning of this section, the more conservative choice is to rely on the inspiral-only results, as they are not biased by the ad hoc tapering applied near merger, and do not assume modified gravity to approach GR in the merger and ringdown.

For the case of a phase modification in the breathing mode (which also includes phase modifications in the tensor mode due to the relation $\beta=2\beta_B$) shown in the bottom-right panel, we again recover the same behavior as in the tensor-only phase modification shown in the bottom-left panel, consistent with the discussion in Section~\ref{subsec:gr_injections}.

\subsection{\label{subsec:amp-phase-modifications}Joint analysis of amplitude and phase modifications}

Tests of GR such as the PPE or FTI parameterizations typically allow only one beyond-GR parameter to vary at a time. This choice is largely dictated by feasibility: the full PPE framework contains 13 parameters (see Section \ref{subsec:numerical_implementation}), and varying all of them simultaneously would make the analysis intractable.  
However, most modified gravity theories predict simultaneous deviations in both the amplitude and the phase of each multipole mode, potentially affecting multiple polarization modes at once. In the PPE framework, this situation is naturally accommodated by allowing several parameters to be nonzero simultaneously.  
In this Section, we take a first step in this direction and explore a representative example in which both an amplitude and a phase modification are activated in the tensor sector, i.e.\ $\alpha_T, \beta \neq 0$. This should be regarded as a proof-of-concept test: extending the full analysis to all possible two-parameter combinations would be computationally intensive and is left for future work.

\begin{figure}[h!]
\centering
\includegraphics[width = 1\textwidth]{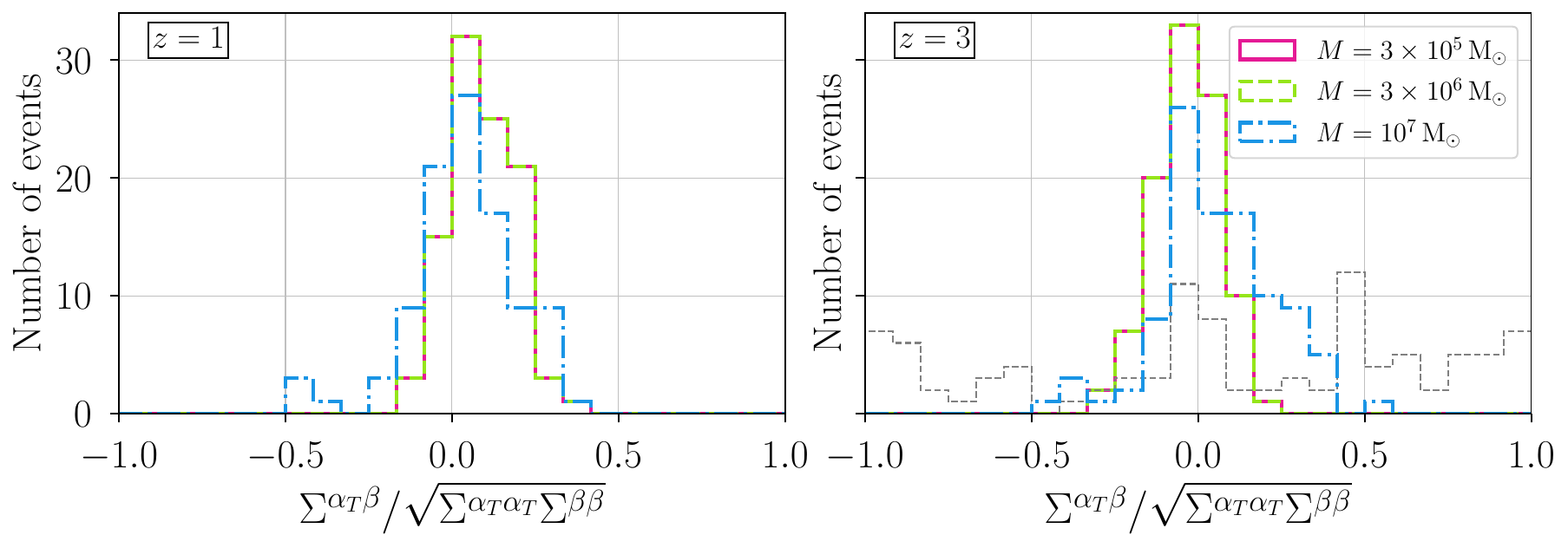}
 \caption{Correlation distributions between $\alpha_T$ and $\beta$ normalized over the corresponding errors for three total masses as reported in the legend. Left (right) panel is for $z=1$ ($z=3$). The thin gray dashed line shows the correlation distribution between $d_L$ and $\iota$ for the case $M=3\times 10^6 \msun$ and it is included here to illustrate the typical level of correlation between two quantities that are degenerate in GW analyses.}
 \label{fig:corre_alphaT_beta} 
\end{figure}

In Figure~\ref{fig:corre_alphaT_beta} we present the correlation between $\alpha_T$ and $\beta$. As it can be appreciated, none of the cases show a strong correlation between the two parameters, as the correlation is centered around $0$.
This can be understood by noting that parameters entering the amplitude are typically not degenerate with those entering the phase, as they modify distinct parts of the GW signal.
Consequently, we are able to recover $\alpha_T$ and $\beta$ at the same level of precision as in Figures~\ref{fig:gr_injection_alphaT}-\ref{fig:gr_injection_betaT}. Indeed, comparing the median errors on $\alpha_T$ and $\beta$, we found a relative difference $\lesssim 6.7\%$.

The fact that parameters modifying the phase and those modifying the amplitude are effectively uncorrelated can have an important impact when placing constraints on specific modified gravity theories, as opposed to working within a purely agnostic parametrized framework. In concrete theories, any modification to observables is ultimately related to the same underlying fundamental parameters. Consequently, measuring both types of modifications with good precision can help break degeneracies that would otherwise remain hidden when considering phase or amplitude effects independently. 
As an example, we consider Einstein-\ae ther theory, where the PPE coefficients $\beta$ and $\alpha_T$ are mapped to the fundamental theory parameters as shown in Tables~\ref{table:PPE_mapping_beta} and~\ref{table:PPE_mapping_alpha2}. Simultaneous measurements of $\beta$ and $\alpha_T$ would allow us to constrain the sensitivities of the two bodies through the relation $\beta / \alpha_T = 3 [(1-s_1)(1-s_2)]^{4/3} / 112$ if $a_T = -2$, while for $a_T = 0$ one could access directly the parameter $\kappa_3$ of the theory (see Table~\ref{tab:AEparameters}) through $\beta \alpha^2_T = 3/(64 \kappa_3)$. 
Thus, joint constraints on amplitude and phase corrections provide a powerful way to disentangle different combinations of fundamental parameters in concrete modified gravity theories.

\subsection{What if $\alpha_T$ is complex?}\label{subsec:complex_alpha}

\begin{figure}
\centering
\includegraphics[width = 1.\textwidth]{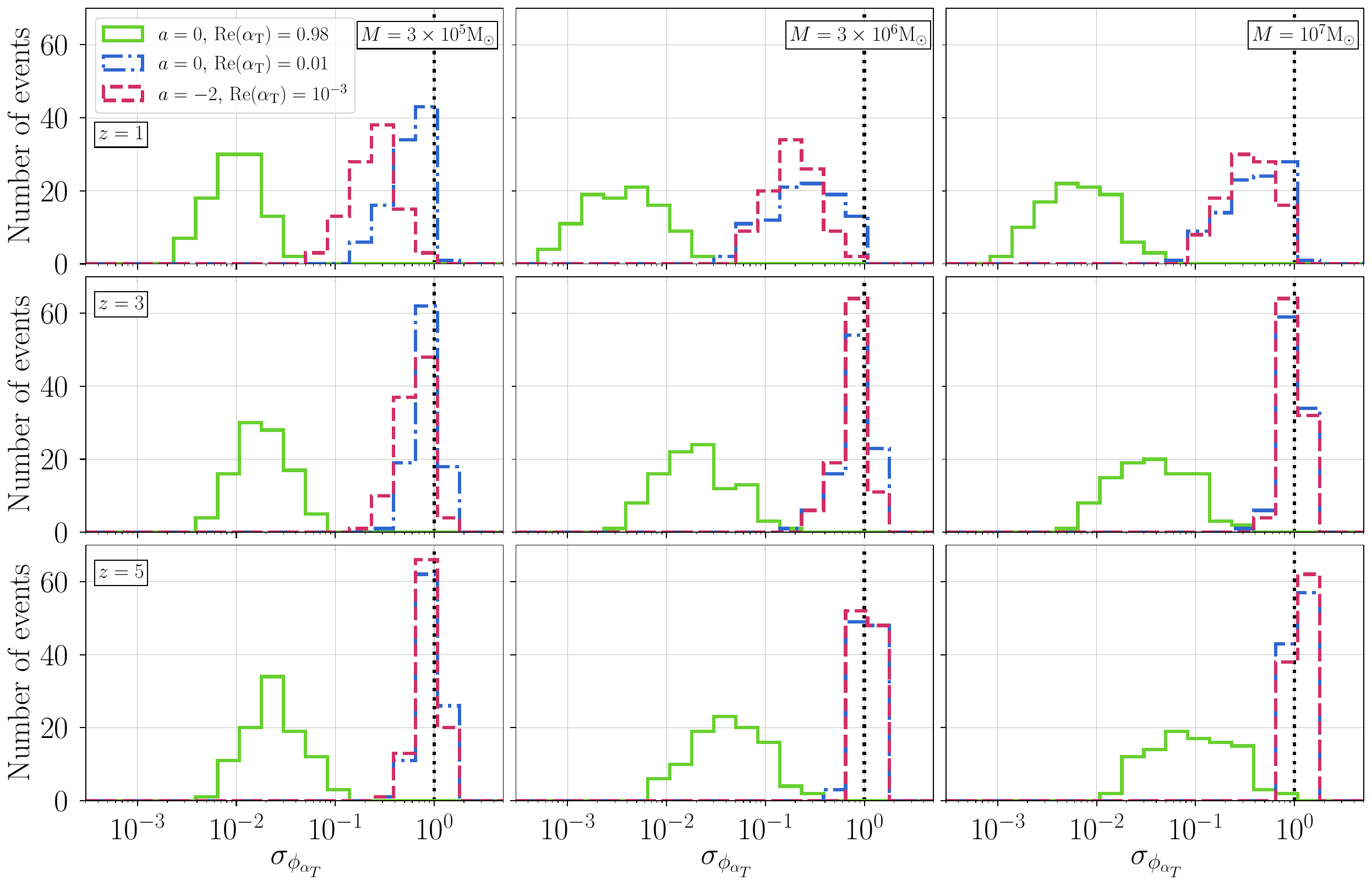}
 \caption{Absolute error distributions on the phase of $\alpha_T$ from GR-injections (i.e. ${\rm Im}(\alpha_T) = \phi_{\alpha_T} = 0$) in the case where $\alpha_T$ is complex for different values of $a$ and $\rm Re(\alpha_T$) as reported in the legend and for different combinations of intrinsic total masses and redshifts. The vertical dotted lines correspond to an error of 100\%. Overall the constraints are strongly affected by the value of the real part of $\alpha_T$ as well as the value of $a$. }
 \label{fig:alpha_complex} 
\end{figure}

As described in Section~\ref{subsec:Extra_Polarizations}, in our analysis we have assumed $\alpha_T \in \mathbb{R}$. However, a natural generalization is to consider $\alpha_T \in \mathbb{C}$, i.e.\ a  complex quantity. In this case, we can write $\alpha_T = |\alpha_T| \exp{(i\phi_{\alpha_T})}$ where $|\alpha_T| $ represents the modulus and $\phi_{\alpha_T}$ the phase\footnote{Note that we also assume the PPE amplitude parameters of extra polarizations to be complex, but their phase can be reabsorbed into the definition of their PPE phase parameter. This is not the case of $\alpha_T$, where the phase $\phi_{\alpha_T}$ cannot be reabsorbed into the phase parameter $\beta$.}. 

In \cite{Islam:2019dmk} the authors also perform GR injections in the modulus and phase of $\alpha_T$, finding that the phase cannot be constrained for the systems studied. This is expected because, if the modulus is very small, the phase can take any arbitrary value. Therefore, to test the case of a complex $\alpha_T$, we partially relax the assumptions of GR injections and, in particular, assume that the real part of the amplitude modification, ${\rm Re}(\alpha_T)$, is nonzero, while we perform GR injections in the imaginary part, i.e.\ ${\rm Im}(\alpha_T) = 0$. We then derive results for $[{\rm Re}(\alpha_T), {\rm Im}(\alpha_T)]$ and convert the constraints to $[|\alpha_T|,\phi_{\alpha_T}]$.

In Figure~\ref{fig:alpha_complex}
we report the absolute uncertainties of $\phi_{\alpha_T}$ from GR injections of MBHBs with intrinsic total mass of $M = 3\times 10^5, \, 3\times 10^6$ and $10^7 \msun$ at redshift $z=[1,3,5]$. In the case of $a=0$ we explored two cases: the most optimistic case in which ${\rm Re}(\alpha_T) = 0.98$ that correspond to the maximum value we could apply (see Eq.~\eqref{Eq:limit_alpha_tensor}) and a case with a smaller value of ${\rm Re}(\alpha_T)=10^{-2}$ to see how our estimates change if the modification is weaker. In addition we also consider the case $a=-2$.

As expected, $\phi_{\alpha_T}$ is best constrained in the first case with $a=0$ and ${\rm Re}(\alpha_T) = 0.98$. For systems at $z=1$, the phase can be on average constrained with a precision of 0.9\%, 0.4\% and 0.6\% for $M = 3\times 10^5, \, 3\times 10^6$ and $10^7 \msun$, respectively. However in the case where $a$ is kept fixed but ${\rm Re}(\alpha_T) = 0.01$, the uncertainties deteriorate by more than one order of magnitude, leading to median absolute uncertainties of $65\%$, $25\%$ and $42\%$ still for $z=1$. In the case of $a=-2$ and ${\rm Re}(\alpha_T) = 10^{-3}$ we obtain similar forecasts. The results worsen also if we keep the total mass fixed and increase the redshift.

\section{\label{sec:discu_and_concl}Discussion and conclusions}
This paper presents a comprehensive Fisher forecast analysis of LISA's ability to constrain gravitational wave (GW) polarizations with massive black hole binary mergers. We employ the Parametrized Post-Einsteinian (PPE) formalism to model frequency-dependent deviations from General Relativity (GR) in both the amplitude and phase of GWs across all six polarization modes predicted by metric theories of gravity: two tensor, two vector, and two scalar (breathing and longitudinal) modes. Our PPE parametrization is motivated by four previously studied modified gravity theories (Horndeski gravity, Einstein-\ae ther theory, Rosen's bimetric theory, and Lightman-Lee theory) hence providing  concrete examples of how extra polarizations manifest in GW signals.

An important distinction highlighted by our work is between modified gravity theories that possess additional dynamical degrees of freedom, versus those that actually excite additional GW polarizations. Some modified gravity theories may modify the tensor polarization waveform without generating scalar or vector modes, while others produce unmistakable signatures through the presence of extra polarizations. The detection of non-tensorial polarizations would therefore provide a smoking gun for certain classes of modified gravity, complementing common searches for modifications within the tensor polarizations alone.

For all polarizations, we explored two different frequency evolutions during the inspiral, motivated by the leading and sub-leading modifications to GR in a Post-Newtonian (PN) approximation in the four theories considered here. In addition, for all non-tensorial polarizations we probe their presence in the $\ell=|m|=1$ and $\ell=|m|=2$ angular harmonics. In all these cases, we perform Fisher  forecasts with GR injections and provide the median uncertainty expected on the amplitude and phase parameters of GW polarizations, as a function of redshift and total mass, after randomizing other parameters such as sky location, inclination, etc. 

For the amplitude of tensor polarizations, we find that in the case of frequency scaling $a = -2$, constraints on the amplitude parameter $\alpha_T$ reach an uncertainty of order $\sigma\sim [10^{-4}, 10^{-3}]$ for events at $z = [1,4]$, respectively. These constraints are typically one to two orders of magnitude tighter than those obtained with frequency scaling $a = 0$, highlighting the importance of the frequency dependence of the modifications. Phase modifications of the tensor polarizations are characterized by the parameter $\beta$:   we find that LISA can constrain deviations from GR with remarkable precision. We expect precisions of order $\sigma\sim [10^{-7},10^{-5}]$rad for events at $z = [1,4]$  respectively, with $M\sim 10^6 \msun$ when the frequency evolution is $b=-7$ (-1 PN order). These results are in agreement with previous LISA studies \cite{Piarulli:2025rvr}. These constraints can be 2-3 orders of magnitude more precise than current LVK constraints \cite{LIGOScientific:2021sio,LIGOScientific:2025obp}.  LISA is also expected to provide comparable or better constraints than 3G ground-based detectors \cite{Perkins:2020tra}, depending on the expected observed population of events and details of the 3G networks.  

Regarding extra polarizations, LISA is expected to reach amplitude constraints on scalar polarizations with precision of order $\sigma\sim [10^{-2},10^{-3},10^{-7}]$ for the frequency scalings $a=[0,-2,-6]$ considered here. For vector polarizations, we predict a precision of order $\sigma\sim [10^{-2},10^{-4}]$ for the frequency scalings $a=[0,-2]$. Most tests of extra polarizations using current observations have ruled out signals to consist of purely scalar or purely vector polarizations, but have not been able to constrain the possibility of tensor polarizations occurring simultaneously with scalar or vector polarizations (as assumed here) \cite{LIGOScientific:2021sio}. Forecasts for LVK at design sensitivity indicate that constraints on the scalar and vector amplitude parameters will have uncertainties between $\sigma\sim 10^{-1}-1$ for the frequency scaling $a=0$ \cite{Takeda:2018uai}, based on low-mass events with long inspirals. This represents one to two orders of magnitude worse sensitivity than our predictions for LISA.

In addition, for scalar polarizations we demonstrate that LISA is potentially able to distinguish between breathing and longitudinal modes---an important capability that current ground-based detectors lack. However, this discrimination is only effective for relatively low-mass systems (see Section \ref{subsec:gr_injections}). Our analysis shows that for systems with total mass $M \lesssim 10^4 \msun$, the detector responses to breathing and longitudinal modes differ sufficiently to allow independent constraints on their amplitudes (see Section \ref{subsec:aet_channels}). By measuring both polarizations independently, LISA would be able to set constraints on specific parameters of modified gravity theories, such as the mass of the scalar field in Horndeski theory, which would otherwise remain degenerate with other theory parameters. Nonetheless, for heavier systems, the two scalar modes become largely degenerate in LISA's response, limiting its ability to separate their contributions. Moreover, constraints on the scalar mass should also include associated cosmological dispersion effects not taken into account in this work. 

Importantly, our results reveal that LISA is more sensitive to vector polarizations than to scalar polarizations across most of the parameter space we explored. The vector amplitude values that LISA can constrain are about $\sim 2-3$ times better than those for scalar modes (for the same given inspiral frequency scaling for vectors and scalars), owing to the enhanced detector response to vector modes compared to scalar modes throughout LISA's sensitivity band. This suggests that if modified gravity theories predict the presence of vector polarizations, LISA observations could place more stringent bounds on such theories.  
Indeed, we find that in the case of Lightman-Lee theory the modified gravity coupling parameter is slightly better constrained with vector rather than scalar modes. In more complicated theories like Einstein-\ae ther (a model with various modified gravity coupling parameters) the scalar and vector polarizations depend on different theory parameters and hence scalar and vector observations would be complementary.

A key technical contribution of this work is our implementation of a complete inspiral-merger-ringdown analysis with proper waveform alignment and tapering procedures. While many previous PPE studies have focused on inspiral-only waveforms \cite{Cornish:2011ys, Sampson:2013lpa, Mishra:2010tp, Li:2011cg, Chatziioannou:2012rf, Mezzasoma:2022pjb, Xie:2025voe, Xie:2024ubm} or sometimes do not detail their alignment and tapering methodology \cite{Perkins:2020tra, Schumacher:2023cxh}, in this work we have developed a consistent framework that applies modifications throughout the entire coalescence, from inspiral through merger and into ringdown, motivated by numerical relativity simulations that currently show modified gravity effects becoming negligible after inspiral. This is particularly important for LISA observations of massive black hole binaries, where the merger and ringdown phases contribute significantly to the total signal-to-noise ratio, especially for systems with $M \gtrsim 10^6 \msun$.

Several caveats and limitations of our analysis should be noted. First, our results are based on single-event forecasts using Fisher matrix methods, with a limited range both in redshift and total mass. In the GR injections performed in Section \ref{subsec:gr_injections}, we considered 
systems with $M \in [3\times 10^4, 10^7]  \msun$ and redshift $z \in [0.5, 4]$. If we examine the three astrophysically motivated MBHB populations adopted in \cite{2022PhRvD.106j3017M} (see also \cite{2020ApJ...904...16B} and references therein), we find that only $\sim 45\%$, $\sim 1.6\%$, and $\sim 14\%$ of the events fall within this mass–redshift range for the Q3d, Pop3, and Q3nd populations, respectively. In reality, if GR is modified, such modifications should be consistent across all events from the same astrophysical population. A hierarchical Bayesian analysis combining multiple detections would improve constraints on modified gravity parameters, though such an analysis would require careful treatment of the broader mass and redshift range where many MBHBs lie outside the regime of validity of the Fisher approximation. A follow-up study presenting Bayesian parameter estimation will address these questions more thoroughly.

Second, our analysis assumes that only one massive black hole binary is present at a time. In practice, LISA will observe multiple overlapping signals from various source types including MBHBs, extreme-mass-ratio inspirals, galactic binaries, and stellar-mass black hole binaries \cite{LISA:2024hlh}. A proper global-fit analysis accounting for these overlapping signals is necessary to realistically assess LISA's ability to test General Relativity (see \cite{2023PhRvD.107f3004L,2025PhRvD.111j3014D,2025PhRvD.111b4060K} for three different global-fit pipelines). Whether modified gravity effects should be searched for after producing the global fit or incorporated directly into the fitting pipeline remains an open question that requires further investigation.

Third, in the Fisher forecasts we assumed that all gravitational polarizations arrive at the detector simultaneously. This effectively implies, up to error margins, that all polarization modes propagate at the same speed. Given that, in a number of theories, the amplitude of extra polarization modes may vanish when imposing luminal propagation speed, allowing for distinct propagation speeds (and hence arrival times) would constitute a valuable extension of this work. 
Likewise, incorporating cosmological effects from modified dispersion relations is expected to provide additional constraining power. 
In the present analysis, we have neglected cosmological propagation effects beyond GR, accounting only for standard Hubble dilution and redshift through a rescaling of the mass parameters and frequency and the use of the luminosity distance. This assumption is justified insofar as the modified gravity theories considered here are not intended to model the dark sector of the cosmological model, but rather to parametrize potential deviations from GR in the strong-field regime of gravity around binary black holes.

Finally, while our PPE model captures the phase and amplitude modifications at specific post-Newtonian orders motivated by the four theories we studied, exploring corrections at different PN orders could reveal additional structure in the parameter space of modified gravity theories. However, establishing the connection between such corrections and specific theoretical models would require additional theoretical development.

Despite these limitations, our work establishes LISA as a powerful instrument for testing the polarization content of GWs and constraining deviations from GR in the strong-field, dynamical regime. The combination of long observation times, high signal-to-noise ratios, and time-varying detector response due to LISA's orbital motion will enable complementary tests of gravity to those obtained from ground-based detectors. As LISA approaches its planned launch, refined analyses incorporating Bayesian inference, population studies, and global-fit approaches will further sharpen our understanding of what LISA can teach us about the fundamental nature of gravity.

\begin{acknowledgments}
We wish to thank Jonathan Gair, Chunshan Lin, Andrea Maselli, Elise M. S$\rm \ddot{a}$nger, Jan Steinhoff, Kazufumi Takahashi and Nicol\'{a}s Yunes for fruitful discussions. We also thank Giorgio Orlando for contributions in the first part of this work. We thank the chairs of the LISA Cosmology working group for their support during this project. 
AM acknowledges support from the postdoctoral fellowships of IN2P3 (CNRS). This project has received funding from the European Union’s Horizon 2020 research and innovation program under the Marie Skłodowska-Curie grant agreement No. 101066346 (MASSIVEBAYES). GT is partially funded by the STFC
grants ST/T000813/1 and ST/X000648/1. SA, RT and MZ were supported by the grant No.\ UMO-2021/42/E/ST9/00260 from the National Science Centre, Poland. SA was also supported by MEXT-JSPS Grant-in-Aid for Transformative Research Areas (A) “Extreme
Universe”, No.\ JP21H05189. PCMD is supported by the Czech Science Foundation (GAČR) project PreCOG (Grant No.\ 24-10780S). JZ is supported by funding from the Swiss National Science Foundation
(Grant No.\ 222346) and the Janggen-Pöhn-Foundation. The Center of Gravity is a Center of Excellence funded by the Danish National Research Foundation under Grant No.\ 184.
NAN was financed by the Institute for Basic Science under the project code IBS-R018-D3 and acknowledges support from PSL/Observatoire de Paris.
AG is supported by funds provided by the Center for Particle
Cosmology at the University of Pennsylvania.
YX acknowledges the support of the Natural Sciences and Engineering Research Council of Canada (NSERC) (funding reference number 513671).
KSA would like to acknowledge the National Science Foundation (NSF) Graduate Research Fellowship Program under Grant No. DGE – 1746047 and the NSF Postdoctoral Fellowship: MPS-Ascend under Grant No. 2503256. GGL is supported by the postdoctoral fellowship program of the
University of Lleida.

The code adopted for this study is publicly available in the \texttt{extra\_polar} branch at \url{https://gitlab.in2p3.fr/marsat/lisabeta/-/tree/extra_polar?ref_type=heads}. The data underlying this article, together with the post-processing scripts, are publicly available at
\url{https://gitlab.in2p3.fr/mcorman/testing_polarizations_lisabeta}.

\end{acknowledgments}

\paragraph{Authors' Contributions.}
SA: contributed to the theoretical formulation of the work, including identifying the two sub-cases of waveform modifications in Tables \ref{table:PPE_mapping_ab}-\ref{table:PPE_mapping_alpha1} and helping with Appendix~\ref{app:rotations}. He also wrote the Horndeski work in Section~\ref{sec:ST_theory} and Appendix \ref{sec:app-deerivation-ST}. He generally assisted in writing the manuscript, contributing to improvements in consistency and notation throughout. 
MC: extended and run \texttt{lisabeta} code, including phase and amplitude waveform modifications, phase alignment, comparison to FTI, and re-parameterization of the 0PN phase parameter $\beta$. She also performed GR MCMC parameter estimation to validate Fisher analyses, ran non-GR injections to test GR Fisher estimates, and produced various plots. She investigated the degeneracy between the breathing and longitudinal modes, and contributed to Sections~\ref{sec:ppe_formalism} and \ref{sec:results_for_extra_pol} as well as Appendices \ref{sec:alignment}, \ref{sec:fti_comparison}, \ref{app:edgb_case} and \ref{app:betap_injections} of the manuscript. She also assisted in writing and reviewing the manuscript. 
PCMD: extended and run \texttt{lisabeta} code, including the analysis of the appropriate Fisher steps for the extra polarization parameters, GR and almost-GR injections fisher forecasts for the inspiral-only and inspiral-merger-ringdown cases, performed the non-GR injections for the specific theories that do not posses a GR limit, run Fisher analysis with modifications in both amplitude and phase to investigate correlations, contributed to the initial stages of SNR variation investigation, and to Appendices \ref{sec:Fisher steps}, \ref{sec:fti_comparison}, and \ref{app:betap_injections}. She also assisted in writing and reviewing the manuscript. AG: developed the results presented in Sections~\ref{subsec:mapping_to_gravity_theories},~\ref{sec:EA_theory},~\ref{sec:Rosen_theory},~\ref{sec:LL_theory}, as well as contributed to the theoretical formulation of the work, and assisted in writing the entire manuscript. 
ML: proposed and coordinated the project with AM, supervising the theoretical work on modified gravity theories and PPE parametrization. She contributed to formulating the theoretical work in Section \ref{subsec:spin_weight}, \ref{subsec:spin_weighted_spherical_decomposition}, \ref{subsec:waveform_frequency_domain}, \ref{subsec:tensor_polarizations}, \ref{subsec:Extra_Polarizations}, \ref{subsec:mapping_to_gravity_theories} and Appendix~\ref{app:rotations}. She also assisted in writing and reviewing all sections of the manuscript. 
AM: proposed and coordinated the project with ML and coordinated the data analysis part, supervising the implementation of the code and the Fisher results. He contributed to the results presented in Section~\ref{sec:results_for_extra_pol}, in particular the mapping to specific theories in Section~\ref{sec:constrain-specific-theories}, as well as to the results shown in Appendices~\ref{sec:Fisher steps}, \ref{sec:alignment}, and \ref{app:edgb_case}. He also contributed to the writing of Sections~\ref{sec:ppe_formalism} and~\ref{sec:results_for_extra_pol}, Appendices~\ref{sec:Fisher steps}--\ref{app:betap_injections}, and participated in the review of the entire manuscript. 
SM: implemented in \texttt{lisabeta} the response to extra polarizations and provided support for their integration in the code, as well as assistance in the formulation of the corresponding results. 
MP: adapted and implemented the FTI framework in \texttt{lisabeta}, which served as the baseline for the PPE implementation presented in this work, as well as for the re-parameterization of the 0PN order. He also performed the FTI Fisher analyses to facilitate the direct comparison with the PPE results. 
GT: developed the results presented in Sections~\ref{subsec:aet_channels} and~\ref{sec:LL_theory}, contributed to the theoretical entire formulation of the work, and assisted in writing the manuscript. 
JZ: contributed the theoretical foundation of the project summarized in Sections \ref{sec:revpol}, \ref{sec:Detecting_extra_polarizations}, \ref{sec:ppe_formalism} and wrote Appendix \ref{App:Horndeski plus}. Helped at resolving conceptual questions in comparison to  previous works in Sections~\ref{sec:Detecting_extra_polarizations}, \ref{sec:ModGrav} and Appendix \ref{app:rotations}, and assisted in writing the manuscript. 
GGL: contributed to Section \ref{subsec:geodesic_equation}, \ref{subsec:Diffeomorphism_Invariance}, \ref{subsec:Polarizations}, and \ref{subsec:spin_weight}. He also assisted in writing the manuscript. 
NAN: contributed to the theoretical calculation of waveforms in specific modified gravity theories. LP: contributed to the writing of Section~\ref{sec:EA_theory}.
KSA: contributed to the formulation of the PPE model, the writing of Section \ref{subsec:tensor_polarizations} and \ref{subsec:Extra_Polarizations}, and discussions of \AE{}theory for Section \ref{sec:EA_theory}. 
BS: implemented \texttt{lisabeta} to perform Fisher forecast analyses on stellar-mass BHs, which served as benchmark tests to validate the code in GR. 
RT: contributed the analysis of the LISA antenna response and the writing of Section \ref{subsec:LISA_Antenna_Response}. She also implemented \texttt{lisabeta} to perform Fisher-forecast analyses for MBHBs, which served as benchmark tests to validate the code in GR. 
AV: implemented \texttt{lisabeta} to perform Fisher-forecast analyses on SBHB injections. These analyses served as benchmark tests, providing a systematic way to validate and verify the code’s performance and accuracy within the GR framework. 
YX: contributed to the formulation of the PPE model and the writing of Section \ref{subsec:tensor_polarizations} and \ref{subsec:Extra_Polarizations}.
MZ: contributed to the theoretical formulation of the work and writing of Section \ref{subsec:mapping_to_gravity_theories} and \ref{sec:ST_theory}. He also assisted in writing the manuscript and reviewed the entire manuscript.

\appendix

\section{Horndeski gravity: Dynamical degrees of freedom vs Polarizations}\label{App:Horndeski plus}

In this Appendix, we wish to clarify the conceptual difference between dynamical degree of freedom (DoF) and polarization mode. We will do so using Horndeski gravity as an example to illustrate such concept.  Throughout this appendix, we closely follow Ref.~\cite{Heisenberg:2024cjk}.

The full action of Horndeski gravity can be written in the following way \cite{Kobayashi:2019hrl}
\begin{eqnarray}
    S = \frac{1}{2\kappa}\int d^4x \sqrt{-g} \left ( \sum_{i=2}^{5} \mathcal{L}_i[g,\phi]\right ) +S^\text{min}_\text{m}[g,\Psi_\text{m}]\,,
    \label{eqn:genprocaaction}
\end{eqnarray}
where
\begin{align}
    \mathcal{L}_2  =& \text{  } G_2(\phi, X)  \,, \nonumber \\
    \mathcal{L}_3  =& \text{  } -G_3(\phi, X) \Box \phi \,, \nonumber \\
    \mathcal{L}_4  =& \text{  } G_4(\phi, X)R + G_{4,X}\left [ (\Box \phi)^2 - \nabla_\rho \nabla_\sigma \phi \nabla^\rho \nabla^\sigma \phi \right ]  \,, \nonumber \\
    \mathcal{L}_5  =& \text{  } G_5(\phi, X)G_{\mu\nu} \nabla^\mu \nabla^\nu \phi  - \frac{1}{6}G_{5,X} \left [ (\Box \phi)^3 
    + 2 (\nabla_\rho \nabla^\sigma \phi) (\nabla_\gamma \nabla^\rho \phi) (\nabla_\sigma \nabla^\gamma \phi)  \right. \nonumber \\
    & \left. - 3(\Box \phi) (\nabla_\rho \nabla_\sigma \phi) (\nabla^\sigma \nabla^\rho \phi) \right ]  \,.
\end{align}
The functions $G_i$'s are arbitrary scalar functions that depend the scalar field $\phi$ and $X \equiv -\nabla_\mu \phi \nabla^\mu \phi$. Moreover, we write $G_{i,Z} \equiv \partial G_i / \partial Z$ for partial derivatives of these scalar functions and define $\Box \phi \equiv \nabla_\mu \nabla^\mu \phi$. Finally, following the general definition in Eq.~\eqref{eq:ActionMetricTheory}, $S^\text{min}_\text{m}[g,\Psi_\text{m}]$ is the matter action that depends on minimally coupled matter fields $\Psi_\text{m}$. 
Horndeski theory encompasses many specific gravity theories beyond GR. In fact, by construction any scalar-tensor theory with second-order equations of motion is captured by the Horndeski class. Well-known examples are BD theory \cite{Brans:1961sx,Dicke:1961gz}, $f(R)$ gravity \cite{Bergmann:1968aj,Ruzma:1969JETP,Buchdahl:1970MN,Sotiriou:2008rp,DeFelice:2010aj} and sGB gravity \cite{Zwiebach:1985uq,Gross:1986mw}. These theories are recovered by choosing specific forms of the general functionals $G_i$. Explicitly, BD theory is recovered through the specific choices
\begin{equation}
    G_2 = \frac{\omega X}{\phi}\,,\;\;G_4 = \phi\,,\;\;G_3 = G_5 = 0\,, \label{eq:BD}
\end{equation}
while $f(R)$ gravity with $f''\neq 0$ arises from the combination 
\begin{equation}
    G_2 = f(\phi)-\phi f'(\phi)\,,\;\;G_4 = f'(\phi)\,,\;\;G_3 = G_5 = 0\,, \label{eq:f(R)}
\end{equation}
and sGB gravity is equivalent to choosing \cite{Kobayashi:2011nu,Kobayashi:2019hrl}
\begin{align}
    G_2 &= X +8f^{(4)}(\phi)X^2(3-\ln X) ,  \nonumber \\
    G_3 &= 4f^{(3)}(\phi)X(7-3\ln X) , \nonumber \\
    G_4 &= 1 +4f^{(2)}(\phi)X(2-\ln X) ,  \nonumber \\
    G_5 &= -f^{(1)}(\phi)\ln X ,  \label{eq:sGB}
\end{align}
where $f^{(n)} \equiv \partial^n f/ \partial \phi^n$ is the derivative of some scalar function $f(\phi)$.

\subsection{SVT Decomposition and Dynamical Degrees of Freedom.} 

Let us now assume the presence of a radiative localized source within an asymptotically flat spacetime, as we do throughout this work. Hence, assume that the fields of Horndeski theory asymptote towards a constant background value with the physical metric $g_{\mu\nu}$ reducing to the Minkowski metric $\eta_{\mu\nu}$ on top of which we consider perturbations $h_{\mu\nu}$ at $\mathcal{O}(1/r)$. Similarly, for the scalar field this implies the asymptotic decomposition
\begin{equation}
    \phi=\bar{\phi}+\varphi\,,
\end{equation}
where $\bar{\phi}$ is a constant and $\varphi$ an $\mathcal{O}(1/r)$ perturbation. 

As in Section~\ref{sec:revpol}, we want to identify the gauge invariant variables of the perturbed Horndeski theory. The SVT decomposition of the physical metric of any metric theory was already given in Eq.~\eqref{eq:SVTmetricDecomp}, together with the corresponding six manifestly gauge invariant variables in Eq.~\eqref{eqn:GaugeInvQuantity}. On the other hand, the SVT decomposition of the scalar field $\phi$ is of course trivial, while its perturbation is already gauge invariant on the constant background
\begin{equation}\label{eq:gaugeTransfScalar}
    \varphi \rightarrow \varphi - \mathcal{L}_\xi\,\bar{\phi}= \varphi-\xi^\alpha \partial_{\alpha} \bar{\phi}=\varphi\,.
\end{equation}

Starting from the Horndeski action~\eqref{eqn:genprocaaction}, one can find the linearized equations of motion of the gauge invariant quantities and determine the dynamical DoFs of the theory

As one could have expected from the definition of Horndeski theory as a scalar-tensor theory, only the scalar field perturbation $\varphi$ and the TT part of the metric perturbation satisfy dynamical equations of motion that in the asymptotically flat limit can be written as
\begin{align}
    \Box E^\text{TT}_{ij} = 0 \,, \quad (\Box-m_s^2)\varphi = 0 \,. \label{eqn:HorndeskiWaveEq}
\end{align}
where the effective mass of the scalar field perturbation is given by
\begin{align}
    m_s^2 \equiv \frac{\bar{G}_{2,\phi\phi}}{\bar{G}_{2,X}-2\bar{G}_{3,\phi}+3(\bar{G}_{4,\phi})^2 / \bar{G_4} } \,. \label{eq:defeffmass}
\end{align}
All other gauge invariant variables do not propagate and rather admit constraint equations that can be put into the following form
\begin{align}\label{eqn:HorndeskiConstraintEq}
    \Theta &= - \sigma \varphi \,, &
    \Phi &= \frac{1}{2} \sigma \varphi\,, &
    \Xi^i & = 0 \,,
\end{align}
where
\begin{equation}\label{eq:defsigma}
    \sigma \equiv \bar{G}_{4,\phi}/\bar{G}_4 \,.
\end{equation}
 We have introduced here the short hand notation $\bar{G}_i\equiv G_{i}(\bar{\phi},0)$ evaluated on the background and assumed the condition $\bar G_4\neq 0$, which ensures that the solutions are physical and tensor waves propagate. 
 
 Thus, Horndeski theory has $2+1=3$ dynamical DoFs characterized by the three independent and gauge invariant modes that satisfy a dynamical propagation equation in Eqs.~\eqref{eqn:HorndeskiWaveEq}. The two TT tensor modes remain massless as in GR, while the scalar DoF potentially comes with a mass that alters its propagation speed. Concretely, Eq.~\eqref{eqn:HorndeskiWaveEq} implies a dispersion relation of the scalar mode of the form
\begin{equation}
    \omega(k)=\sqrt{k^2+m_s^2}\,,
\end{equation}
with corresponding group velocity
\begin{equation}\label{eq:SpeedMassiveDOF}
    v_S\equiv\frac{d\omega}{dk}=\frac{k}{\sqrt{k^2+m_s^2}}=\sqrt{1-\frac{m_s^2}{\omega^2}}\,.
\end{equation}
Observe that this group velocity, which corresponds to the physical propagation speed, naturally preserves the Lorentz symmetric constraint $|v_S|<1$ in units of $c$.
It is related to the phase velocity introduced in Eq.~\eqref{eq:PhaseVelocityDef} through
\begin{equation}\label{eq:phasevelocityHorndeski}
    V_S\equiv \frac{\omega}{k}=\frac{1}{v_S}\,.
\end{equation}
Therefore, in terms of the group velocity of a massive mode, the asymptotic relation Eq.~\eqref{eq:identity derivatives} becomes
\begin{equation}\label{eq:identity derivatives Horndeski}
    \partial_{i} \varphi = -v_S\, n_i\, \partial_0\varphi\,.
\end{equation}

\subsection{Gravitational Polarizations in Horndeski Theory.}\label{sApp:HorndeskiPolarizations}

Based on the general treatment in Section~\ref{subsec:Polarizations}, it is now straightforward to determine the polarization content in Horndeski gravity. Namely, plugin the constraints in Eqs.~\eqref{eqn:HorndeskiConstraintEq} into the general relation of GW polarizations and gauge invariant metric perturbations in Eq.~\eqref{eq:GWgenPoldef}, we obtain the following result in terms of the three dynamical DoFs $E^{TT}_{ij}$ and $\varphi$ 
\begin{subequations}\label{eq:GWgenPolHorndeski}
\begin{align}
S_+&=\frac{1}{2}e_+^{ij}\,E^{TT}_{ij}\,,& S_\times&=\frac{1}{2}e_\times^{ij}\,E^{TT}_{ij}\,,& S_b&=-\sigma \varphi\,,\\
S_{v1}&= 0\,,& S_{v2}&= 0\,,& S_l&= \,\sigma \varphi (-1 +v_S^2)\,.
\end{align}
\end{subequations}
In words, Horndeski theory has no additional vector polarization modes as could have been expected. Moreover, the theory admits two tensor polarizaitons $S_+$ and $S_\times$, which are directly related to the two TT tensor DoFs, just as it is the case in GR. On top of these tensor polarizations, Horndeski gravity includes however the possibility for the presence of additional scalar breathing $S_b$ and longitudinal $S_l$ polarizations. 

Observe that within this framework, it is explicit that these polarizations are excited through the additional dynamical scalar degree of freedom $\varphi$ in the theory. However, such an excitation depends on the values of the background coefficient $\sigma$ defined in Eq.~\eqref{eq:defsigma} and the combination $(-1 +v_S^2)$. Hence, the excitation of the additional scalar GW polarizations depends on the specific form of non-minimal coupling between the additional scalar degree of freedom and the physical metric, as well as on the propagation speed of the scalar mode. More precisely, for a non-minimal coupling of the scalar field with the Ricci scalar that admits $\bar G_{4,\phi}\neq 0$, and a non-zero mass of the scalar DoF, hence $v_S\neq 1$, the scalar DoF excites a linear combination of both the breathing and the longitudinal polarization. If the scalar polarization is luminal, however, only the breathing mode is present, whereas if $\bar G_{4,\phi}=0$, no additional scalar polarizaiton is excited. Therefore, while Horndeski theory always propagates three dynamical DoFs, it can either admit two, three or four different gravitational polarizations within the gravitational detector response, depending on the exact form of the theory. This nicely exemplifies the difference between the concepts of gravitational polarizations, as the polarizations within the GW detector response of a metric theory, and the propagating degrees of freedom that are a priori independent from the couplings within the theory.

These different possibilities are particularly well illustrated through the different sub-theories within Horndeski gravity introduced above. BD theory defined in Eq.~\eqref{eq:BD} for instance satisfies 
\begin{equation}
    \sigma_\text{BD}=\frac{1}{\bar\phi}\neq 0\,.
\end{equation}
Hence, BD gravity does in general excite an additional scalar gravitational polarization. However, since the speed of propagation of the scalar wave on top of the asymptotically flat background is luminal due to the absence of a mass 
\begin{equation}
    m_\text{BD}=0\,,
\end{equation}
hence $v_S=1$, only the breathing mode is present.

On the other hand from Eq.~\eqref{eq:f(R)}, the background equations in $f(R)$ gravity require that $\bar\phi=f(\bar\phi)=0$, while $f'(\phi)\neq 0$ and $f''(\phi)\neq 0$ implying that the theory admits parameters that read
\begin{equation}
    \sigma_{f(R)}=\frac{f''(\bar\phi)}{f'(\bar\phi)}\neq 0\,,
\end{equation}
and 
\begin{equation}
    m_{f(R)}^2=\frac{f'(\bar\phi)}{3f''(\bar\phi)}\neq 0\,.
\end{equation}
Therefore the additional scalar DoF within $f(R)$ gravity excites both the $S_b$ and $S_l$ scalar polarizations.

Finally, the $\sigma$ parameter of sGB is computed through Eq.~\eqref{eq:sGB} to be 
\begin{equation}
    \sigma_\text{sGB}=0\,.
\end{equation}
Thus, sGB gravity, although propagating an additional scalar DoF, does not admit an additional scalar GW polarization.


\section{Rotation of Angular harmonics\label{app:rotations}}
In order to compare to the literature it is useful to make the arbitrary choice in polarization angle $\psi$ explicit. This choice corresponds to a rotation about the axis of propagation and therefore rotates the definitions of polarizations in the tensor and vector sectors among each other as:
\begin{subequations}\label{eq:TransformationsSpinWeightMetricPerturbations}
\begin{align}
h'_+&=\cos 2\psi \, h_+-\sin 2\psi \, h_\times\;,& h'_\times&=\cos 2\psi \, h_\times +\sin 2\psi \, h_+\,, 
\label{eq:T_rotation}\\
h'_{v1}&=\cos\psi \, h_{v1}-\sin\psi \,  h_{v2}\;,& h'_{v2}&=\cos\psi \,  h_{v2}+\sin\psi \,  h_{v1}\,,\\
h'_b&=h_b\;,& h'_l&=h_l\,.
\end{align}
\end{subequations}
In this sense, the definitions of the polarizations in each sector depends on an arbitrary choice that is usually not uniform across the literature. 

Notice that the choice of $\psi$ is arbitrary and it should not change the physics. Indeed, by changing the choice of $\psi$ one also changes the antenna pattern  appropriately such that the response stays invariant. In the TDI formalism, one can show that the one-way Doppler response obtained in Eq.\ \eqref{eq:DopplerResponse1} stays invariant by verifying that 
\begin{align}\label{freq_inv}
   \ell^{i'}\ell^{j'}h_{ij}'=\ell^i\ell^jh_{ij},
\end{align}
and $\hat{k'}\cdot \hat{\ell'}=\hat{k}\cdot \hat{\ell}$. In order to do this, we can, without loss of generality, consider the GW propagation direction to be $\hat{z}$. In this case, under a rotation around the $z$-axis (i.e.\ around the propagation direction), the vector pointing from spacecraft A and B changes as
\begin{align}
\hat\ell_{\rho}'=\Lambda_{\rho\sigma}\hat\ell_\sigma,
\end{align}
where $\Lambda_{\rho\sigma}$ is a three-dimensional rotation matrix given by
\begin{align}
(\Lambda)_{i j}=
\left(\begin{array}{ccc}
\cos\psi & -\sin\psi & 0 \\
\sin\psi & \cos\psi & 0 \\
0 & 0 & 1
\end{array}\right).
\end{align}
It is straightforward to notice that $\hat{k}\cdot \hat{\ell}=\ell_z$ is invariant under rotations around the $z$ axis. 

By using this rotation matrix, one can also express the transformed GW tensor $h'_{ij}$ as
\begin{align}
h'_{ij}=\Lambda_{ik}\Lambda_{jl}h_{kl},
\end{align}
where
\begin{align}
h'_{ij} &= \begin{pmatrix}
h_b'+h_+' & h_\times' & h_{v1}'\\
h_\times' & h_b'-h_+' & h_{v2}'\\
h_{v1}' & h_{v2}' & h_l'\\
\end{pmatrix},
\end{align}
with $h_p'$ being defined in Eq.~\eqref{eq:TransformationsSpinWeightMetricPerturbations}.
We can then straightforwardly show
\begin{align}
\hat\ell_{i}'\hat\ell_{j}'h'_{ij}=\hat\ell_{i}\hat\ell_{j}h_{ij},
\end{align}
where we used the orthogonality of the rotation matrix, i.e.\ $\Lambda_{ij}\Lambda_{ik}=\delta_{jk}$, confirming that observables will be independent of the choice $\psi$.

In terms of SWSH defined in Eq.~\eqref{eq:SWSHPols}, this freedom can be made explicit by defining the functions of spinweight $S$
\begin{align}
    &{}_{p_1}\mathcal{Y}_{\text{S}}^{(\ell,m)}(\iota,\phi_c,\psi)\equiv\frac{1}{2}\left(Y_{\text{S}}^{(\ell,|m|)}e^{i\text{S}\psi}+(-1)^{(\ell+\text{S})} Y_{\text{S}}^{(\ell,-m)*}e^{-i\text{S}\psi} \right),\\
    &{}_{p_2}\mathcal{Y}_{\text{S}}^{(\ell,m)}(\iota,\phi_c,\psi)\equiv\frac{i}{2}\left(Y_{\text{S}}^{(\ell,|m|)}e^{i\text{S}\psi}-(-1)^{(\ell+\text{S})} Y_{\text{S}}^{(\ell,-|m|)*}e^{-i\text{S}\psi} \right),
\end{align}
for $p_1=+,v1$ and $p_2=\times,v2$, depending on the value of the spin-weight S. Note that for the scalar modes, which are invariant under such rotations, the inclusion of the polarization angle has no effect.
As an example, for the $(\ell,|m|)=(2,2)$ and $(\ell,|m|)=(1,1)$ harmonics, the relevant angular functions for tensor and vector modes take the following form:
\begin{subequations}
\begin{align}
   {}_{+}\mathcal{Y}_{\text{-2}}^{(2,2)}(\iota,\phi_c,\psi)&=  \frac{1}{16}\sqrt{\frac{5}{\pi}} \left[(3+\cos 2\iota)\cos 2\psi-4i\cos\iota\sin 2\psi\right]e^{2i\phi_c},\\
    {}_{\times}\mathcal{Y}_{\text{-2}}^{(2,2)}(\iota,\phi_c,\psi)&=  \frac{i}{16}\sqrt{\frac{5}{\pi}} \left[-i(3+\cos 2\iota)\sin 2\psi+4\cos\iota\cos 2\psi\right]e^{2i\phi_c},\\
 {}_{v1}\mathcal{Y}_{\text{-1}}^{(2,2)}(\iota,\phi_c,\psi)&=  \frac{1}{4}\sqrt{\frac{5}{\pi}} \sin\iota(\cos\psi-i\cos\iota\sin\psi)e^{2i\phi_c},\\
    {}_{v2}\mathcal{Y}_{\text{-1}}^{(2,2)}(\iota,\phi_c,\psi)&=  \frac{-i}{4}\sqrt{\frac{5}{\pi}}\sin\iota(i\sin\psi-\cos\iota\cos\psi)e^{2i\phi_c},\\
    {}_{v1}\mathcal{Y}_{\text{-1}}^{(1,1)}(\iota,\phi_c,\psi)&= -\frac{1}{4}\sqrt{\frac{3}{\pi}}(\cos\psi-i\cos\iota\sin\psi)e^{i\phi_c},\\
    {}_{v2}\mathcal{Y}_{\text{-1}}^{(1,1)}(\iota,\phi_c,\psi)&=  -\frac{1}{4}\sqrt{\frac{3}{\pi}}(\sin\psi+i\cos\iota\cos\psi)e^{i\phi_c}.
\end{align}
\end{subequations}
In this form, it becomes apparent that the definitions of the SWSH associated to the specific polarizations in each sector depend on the arbitrary choice of $\psi$ and rotate into each other for different values of the polarization angle.
As an example, for the canonical choice of $\psi=0$, implemented in the \texttt{lisabeta} code used in this paper,  one recovers the definitions in Eq.~\eqref{eq:SWSHPols}. For the $(\ell,|m|)=(2,2)$ and $(\ell,|m|)=(1,1)$ harmonics they are explicitly given by:
\begin{subequations}
\begin{align}
    {}_{+}\mathcal{Y}_{\text{-2}}^{(2,2)}(\iota,\phi_c,0) &=Y_+^{(2,2)}(\iota, \phi_c) =  \frac{1}{8}\sqrt{\frac{5}{\pi}} (1+\cos^2\iota)e^{2i\phi_c},\\
   {}_{\times}\mathcal{Y}_{\text{-2}}^{(2,2)}(\iota,\phi_c,0)&=Y_\times^{(2,2)}(\iota, \phi_c) =  \frac{i}{4}\sqrt{\frac{5}{\pi}} \cos\iota e^{2i\phi_c},\\
   {}_{v1}\mathcal{Y}_{\text{-1}}^{(2,2)}(\iota,\phi_c,0)&=Y_{v1}^{(2,2)}(\iota, \phi_c) = \frac{1}{4}\sqrt{\frac{5}{\pi}} \sin \iota e^{2i\phi_c},\\
   {}_{v2}\mathcal{Y}_{\text{-1}}^{(2,2)}(\iota,\phi_c,0)&= Y_{v2}^{(2,2)} (\iota, \phi_c)= \frac{i}{8}\sqrt{\frac{5}{\pi}} \sin 2\iota e^{2i\phi_c},\\
   {}_{v1}\mathcal{Y}_{\text{-1}}^{(1,1)}(\iota,\phi_c,0)&= Y_{v1}^{(1,1)} (\iota, \phi_c)=-\frac{1}{4}\sqrt{\frac{3}{\pi}}e^{i\phi_c},\\
   {}_{v2}\mathcal{Y}_{\text{-1}}^{(1,1)}(\iota,\phi_c,0)&= Y_{v2}^{(1,1)} (\iota, \phi_c)=-\frac{i}{4}\sqrt{\frac{3}{\pi}}\cos\iota e^{i\phi_c},\\
    Y_{s}^{(2,2)}(\iota, \phi_c) &= \frac{1}{4}\sqrt{\frac{15}{2\pi}} \sin^2 \iota e^{2i\phi_c},\\
    Y_{s}^{(1,1)}(\iota, \phi_c) &= -\frac{1}{2}\sqrt{\frac{3}{2\pi}} \sin \iota e^{i\phi_c}.
\end{align}
\end{subequations}

In the literature, another common choice is $\psi=\pi/2$, in which case the angular polarization functions become 
\begin{subequations}
\begin{align}
    {}_{+}\mathcal{Y}_{\text{-2}}^{(2,2)}(\iota,\phi_c,\pi/2) &=-Y_+^{(2,2)}(\iota, \phi_c) =  -\frac{1}{8}\sqrt{\frac{5}{\pi}} (1+\cos^2\iota)e^{2i\phi_c},\\
   {}_{\times}\mathcal{Y}_{\text{-2}}^{(2,2)}(\iota,\phi_c,\pi/2)&=-Y_\times^{(2,2)}(\iota, \phi_c) =  -\frac{i}{4}\sqrt{\frac{5}{\pi}} \cos\iota e^{2i\phi_c},\\
   {}_{v1}\mathcal{Y}_{\text{-1}}^{(2,2)}(\iota,\phi_c,\pi/2)&=-Y_{v2}^{(2,2)}(\iota, \phi_c) = 
   -\frac{i}{8}\sqrt{\frac{5}{\pi}} \sin 2\iota e^{2i\phi_c},\\
   {}_{v2}\mathcal{Y}_{\text{-1}}^{(2,2)}(\iota,\phi_c,\pi/2)&= Y_{v1}^{(2,2)} (\iota, \phi_c)= \frac{1}{4}\sqrt{\frac{5}{\pi}} \sin \iota e^{2i\phi_c},\\
   {}_{v1}\mathcal{Y}_{\text{-1}}^{(1,1)}(\iota,\phi_c,\pi/2)&= -Y_{v2}^{(1,1)} (\iota, \phi_c)=\frac{i}{4}\sqrt{\frac{3}{\pi}}\cos\iota e^{i\phi_c},\\
   {}_{v2}\mathcal{Y}_{\text{-1}}^{(1,1)}(\iota,\phi_c,\pi/2)&= Y_{v1}^{(1,1)} (\iota, \phi_c)=-\frac{1}{4}\sqrt{\frac{3}{\pi}}e^{i\phi_c}.
\end{align}
\end{subequations}
These scalings agree with the results in \cite{Chatziioannou:2012rf,Schumacher:2023cxh}, which are used to obtain the waveforms for Einstein-\ae ther, Lightman-Lee theory, and Rosen's theory. In order to use the convention of \texttt{lisabeta} in this paper, $\psi=0$, we take the waveforms in \cite{Chatziioannou:2012rf,Schumacher:2023cxh} and perform a rotation of $\delta \psi=-\pi/2$, following Eq.\ \eqref{eq:TransformationsSpinWeightMetricPerturbations}. The tensor and vector polarizations quoted in Section \ref{sec:ModGrav} are thus obtained from those in  \cite{Chatziioannou:2012rf,Schumacher:2023cxh} as:
\begin{align}
    h'_+&=- h_+,& h'_\times&=- \, h_\times ,\\
h'_{v1}&=+  h_{v2}\;,& h'_{v2}&=-  h_{v1}\,.
\end{align}
Notice that the scalar polarizations do not change and thus this  paper will match exactly the scalars quoted in \cite{Chatziioannou:2012rf,Schumacher:2023cxh}.

\section{Waveforms of extra polarizations in Horndeski gravity}\label{sec:app-deerivation-ST}
For the subclass of Hordeski gravity defined in Eq.~\eqref{eq:Horndeskisubclass} in the main text, the waveforms in the time domain have been studied in Ref.~\cite{Higashino:2022izi}. However, the waveforms in the frequency domain have been studied only for the tensor modes, while Ref.~\cite{Liu:2018sia} studied the frequency-waveform in the screened modified gravity. Below, we briefly summarize the derivation of waveforms of breathing and longitudinal modes in the frequency domain in the subclass of Horndeski gravity. 
The waveforms in the time domain are of the form~\cite{Higashino:2022izi}:
\begin{align}
    h_b & \nonumber = \frac{4\kappa_4g_4\mu}{D}G_* \Bigg\{ (\Delta\hat\alpha) v \sin \gamma \cos \Phi - \frac{\Gamma}{2} v^2 \sin^2 \gamma \cos 2\Phi \\
    &\quad\ - \int_0^{\infty} dz J_1(z) \left[ (\Delta\hat\alpha) \frac{v}{w^2} \sin \gamma \cos \Phi - \frac{\Gamma}{2w^3} v^2 \sin^2 \gamma \cos 2\Phi \right] \Bigg\} ~,\\
    h_l & \nonumber = \frac{4\kappa_4g_4\mu}{D}G_* \int_0^{\infty} dz J_1(z) \left( \frac{1}{w^2} - 1 \right)\left[ (\Delta\hat\alpha) \frac{v}{w^2} \sin \gamma \cos \Phi - \frac{\Gamma}{2} \frac{v^2}{w^3} \sin^2 \gamma \cos 2\Phi \right] ~,
\end{align}
where $\mu=\mathcal{M}^{5/3}M^{-2/3}$ and $w:=\sqrt{1+z^2/(m_s^2D^2)}$, $D$ denotes the distance from the source to the observer, and the relative velocity $v$ satisfies $v^2=\tilde G M/r$ with $\tilde G:=G_*[1+4\delta_0\hat\alpha_1\hat\alpha_2(1+m_s r)e^{-m_s r}]$ and the relative distance $r$. Here, the difference between $G_*$ and $\tilde G$ appears in the waveforms as the cubic order in the scalar hairs $\hat\alpha_{(1,2)}$. However, since we assume that these scalar hairs are much smaller than unity, we neglect any term of cubic or higher-order in the hairs in the waveforms and approximate $\tilde G\simeq G_*$.
The angles $\theta$ and $\varphi$ are defined through the propagation direction vector $\hat n=(0,\sin\gamma,\cos\gamma)$ and the relative vector $\hat r=(\cos\Phi,\sin\Phi,0)$, and $J_1$ denotes a Bessel function of the first kind. The other parameters are summarized in Table~(\ref{tab:Notations-ST}).

In general, the above $z$ integrals cannot be evaluated analytically, but approximate results can be obtained in the limit $D\to\infty$, as has been shown in Ref.~\cite{Liu:2018sia,Liu:2020moh}. See also Ref.~\cite{Alsing:2011er} for the method for calculating these integrals. In light of $v\propto\omega^{1/3}$ where $\omega$ is the orbital frequency, we need to evaluate
\begin{align}
I_1&=\int_0^\infty {\rm d}z J_1(z)\frac{1}{w^2}[\omega(t-Dw)]^{1/3}\cos[\Phi(t-Dw)],\\
I_2&=\int_0^\infty {\rm d}z J_1(z)\frac{1}{w^3}[\omega(t-Dw)]^{2/3}\cos[2\Phi(t-Dw)],\\
I_3&=\int_0^\infty {\rm d}z J_1(z)\biggl(\frac{1}{w^2}-1\biggr)\frac{1}{w^2}[\omega(t-Dw)]^{1/3}\cos[\Phi(t-Dw)],\\
I_4&=\int_0^\infty {\rm d}z J_1(z)\biggl(\frac{1}{w^2}-1\biggr)\frac{1}{w^3}[\omega(t-Dw)]^{2/3}\cos[2\Phi(t-Dw)].
\end{align}
Here, as has been shown in Refs.~\cite{Liu:2018sia,Liu:2020moh}, there is no scalar polarization for $m_s\gg\omega$.
Furthermore, similarly to Ref.~\cite{Liu:2018sia}, one may impose $m_s\ll\omega$ in order to avoid an over production of vacuum gravi-Cerenkov radiation.
We thus focus on the case $m_s\ll\omega$.
In this case, the above integrals in the limit $D\to\infty$ have been calculated in Ref.~\cite{Liu:2018sia,Liu:2020moh} as
\begin{align}
I_1&\simeq [\omega(t-D)]^{1/3}\cos[\Phi(t-D)]-[\omega(t-Dw_1)]^{-2/3}\sqrt{[\omega(t-Dw_1)]^2-m_s^2}\notag\\
&\quad \times \cos\biggl[\frac{m_s^2D}{\sqrt{[\omega(t-Dw_1)]^2-m_s^2}}+\Phi(t-Dw_1)\biggr],\\
I_2&\simeq [\omega(t-D)]^{2/3}\cos[2\Phi(t-D)]-[\omega(t-Dw_2)]^{2/3}\biggl\{1-\frac{m_s^2}{4[\omega(t-Dw_2)]^2}\biggr\}\notag\\
&\quad \times \cos\biggl[\frac{m_s^2D}{\sqrt{4[\omega(t-Dw_2)]^2-m_s^2}}+2\Phi(t-Dw_2)\biggr],\\
I_3&\simeq \frac{m_s^2}{\omega^{8/3}}\sqrt{\omega^2-m_s^2}\cos\biggl(\frac{m_s^2D}{\sqrt{\omega^2-m_s^2}}+\Phi\biggr)\biggr|_{t-Dw_1},\\
I_4&\simeq \frac{m_s^2}{4\omega^2}\biggl(1-\frac{m_s^2}{4\omega^2}\biggr)\omega^{2/3}\cos\biggl(\frac{m_s^2D}{\sqrt{4\omega^2-m_s^2}}+2\Phi\biggr)\biggr|_{t-Dw_2},
\end{align}
where $w_n:=n\omega(t-D)/\sqrt{n^2[\omega(t-D)]^2-m_s^2}$.
In this limit, $h_b$ and $h_l$ can be written as
\begin{align}
h_b&=h_{b1}+h_{b2},\\
h_l&=h_{l1}+h_{l2},
\end{align}
where
\begin{align}
h_{b1}&:=\frac{4\kappa_4g_4\mu}{D}G_*\biggl\{(G_* M)^{1/3}(\Delta\hat\alpha)\sin\gamma[\omega(t-Dw_1)]^{-2/3}\sqrt{[\omega(t-Dw_1)]^2-m_s^2}\notag\\
&\quad \times \cos\biggl\{\frac{m_s^2D}{\sqrt{[\omega(t-Dw_1)]^2-m_s^2}}+\Phi(t-Dw_1)\biggr\}\biggr\},\\
h_{b2}&:=-\frac{4\kappa_4g_4\mu}{D}G_*\cdot\frac{1}{2}\Gamma(G_* M)^{2/3}\sin^2\gamma[\omega(t-Dw_2)]^{2/3}\biggl\{1-\frac{m_s^2}{4[\omega(t-Dw_2)]^2}\biggr\}\notag\\
&\quad \times \cos\biggl[\frac{m_s^2D}{\sqrt{4[\omega(t-Dw_2)]^2-m_s^2}}+2\Phi(t-Dw_2)\biggr],\\
h_{l1}&=\frac{m_s^2}{\omega^2}h_{b1},\\
h_{l2}&=\frac{m_s^2}{4\omega^2}h_{b2}.
\end{align}
The time variation of the orbital frequency for $m_s\ll\omega$ has been obtained as~\cite{Higashino:2022izi}:
\begin{align}
\dot\omega&\simeq\frac{96}{5}(G_*\mathcal{M})^{5/3}\omega^{11/3}\biggl[1+\frac{2}{3}\delta_0+\frac{\kappa_4}{6}\Gamma^2+\frac{5\kappa_4}{24}\frac{(\Delta\hat\alpha)^2}{(G_*M\omega)^{2/3}}\biggr],
\end{align}
where we assumed $|\delta_0|\ll1$ and $|\kappa_4\Gamma^2|\ll1$.
The remaining steps in the derivation of frequency-domain waveforms follow those of Ref.~\cite{Liu:2018sia}. Following Ref.~\cite{Liu:2018sia}, the waveform of the breathing mode in the frequency domain is obtained as
\begin{align}
\tilde h_b(f)=\tilde h_{b1}(f)+\tilde h_{b2}(f),
\end{align}
where
\begin{align}
\tilde h_{b1}(f)&=\biggl(\frac{5\pi}{48}\biggr)^{1/2}\frac{2\kappa_4g_4}{D}G_*\mu(\Delta\hat\alpha)\eta^{-1/5}(G_*\mathcal{M})^{-1/2}(2\pi f)^{-3/2}\sin\gamma e^{i\Psi_{b1}}, \label{eq:hb1-frequencydomain}\\
\tilde h_{b2}(f)&=-\frac{1}{4}\biggl(\frac{5\pi}{24}\biggr)^{1/2}\frac{2\kappa_4g_4}{D}G_*\mathcal{M}\Gamma(G_*\mathcal{M})^{-1/6}(\pi f)^{-7/6}\sin^2\gamma e^{i\Psi_{b2}},
\end{align}
with
\begin{align}
\Psi_{b1}&:=2\pi ft_c-\frac{m_s^2D}{4\pi f}-\frac{\pi}{4}-\Phi_c\notag\\
&\quad +\frac{3}{256}(2\pi fG_*\mathcal{M})^{-5/3}\biggl[1-\frac{2}{3}\delta_0-\frac{\kappa_4}{6}\Gamma^2-\frac{5}{42}\kappa_4\eta^{2/5}(\Delta\hat\alpha)^2(G_*\mathcal{M}2\pi f)^{-2/3}\biggr],\\
\Psi_{b2}&:=2\pi ft_c-\frac{m_s^2D}{4\pi f}-\frac{\pi}{4}-2\Phi_c\notag\\
&\quad +\frac{3}{128}(\pi fG_*\mathcal{M})^{-5/3}\biggl[1-\frac{2}{3}\delta_0-\frac{\kappa_4}{6}\Gamma^2-\frac{5}{42}\kappa_4\eta^{2/5}(\Delta\hat\alpha)^2(G_*\mathcal{M}\pi f)^{-2/3}\biggr],
\end{align} 
and $\Phi_c=-\phi_c$.
In these expressions, we ignored the subleading-order terms in the amplitudes, while we picked up the next-to-leading-order terms in the phases for the purpose of mapping into PPE parametrization adopted in the main text.
Similarly, the longitudinal mode $h_l$ in the frequency domain is obtained as
\begin{align}
\tilde h_{l1}(f)&=\biggl(\frac{5\pi}{48}\biggr)^{1/2}\frac{2\kappa_4g_4}{D}G_*\mu(\Delta\hat\alpha)\eta^{-1/5}m_s^2(G_*\mathcal{M})^{-1/2}(2\pi f)^{-7/2}\sin\gamma e^{i\Psi_{l1}},\\
\tilde h_{l2}(f)&=-\frac{1}{16}\biggl(\frac{5\pi}{24}\biggr)^{1/2}\frac{2\kappa_4g_4}{D}G_*\mathcal{M}\Gamma m_s^2(G_*\mathcal{M})^{-1/6}(\pi f)^{-19/6}\sin^2\gamma e^{i\Psi_{l2}},
\end{align}
where $\Psi_{l1}=\Psi_{b1}$ and $\Psi_{l2}=\Psi_{b2}$. Here, the ratio of the longitudinal mode to the breathing one is given by $\tilde h_l/\tilde h_b\sim m_s^2/f^2$. Since we assumed $m_s\ll f$, the amplitude of the longitudinal mode is much smaller than that of the breathing one. 

In this paper, we retain the frequency-independent terms in the approximate expression for $\dot\omega$ and neglect subleading corrections to the amplitudes of the extra polarization modes, in contrast to the waveform analysis of screened modified gravity in Ref.~\cite{Liu:2018sia}.
Aside from these differences, our results are found to be consistent with those obtained in Ref.~\cite{Liu:2018sia} under an appropriate reparametrization of parameters, since the two theories are conformally related.

After taking into account the cosmological effects (e.g., $D\to d_L$), the results in Section~\ref{sec:ST_theory} are obtained.

\section{Validation of the Fisher  analysis}\label{sec:Fisher steps}

\begin{figure}[h!]
\centering
\includegraphics[width = 0.7
\textwidth]{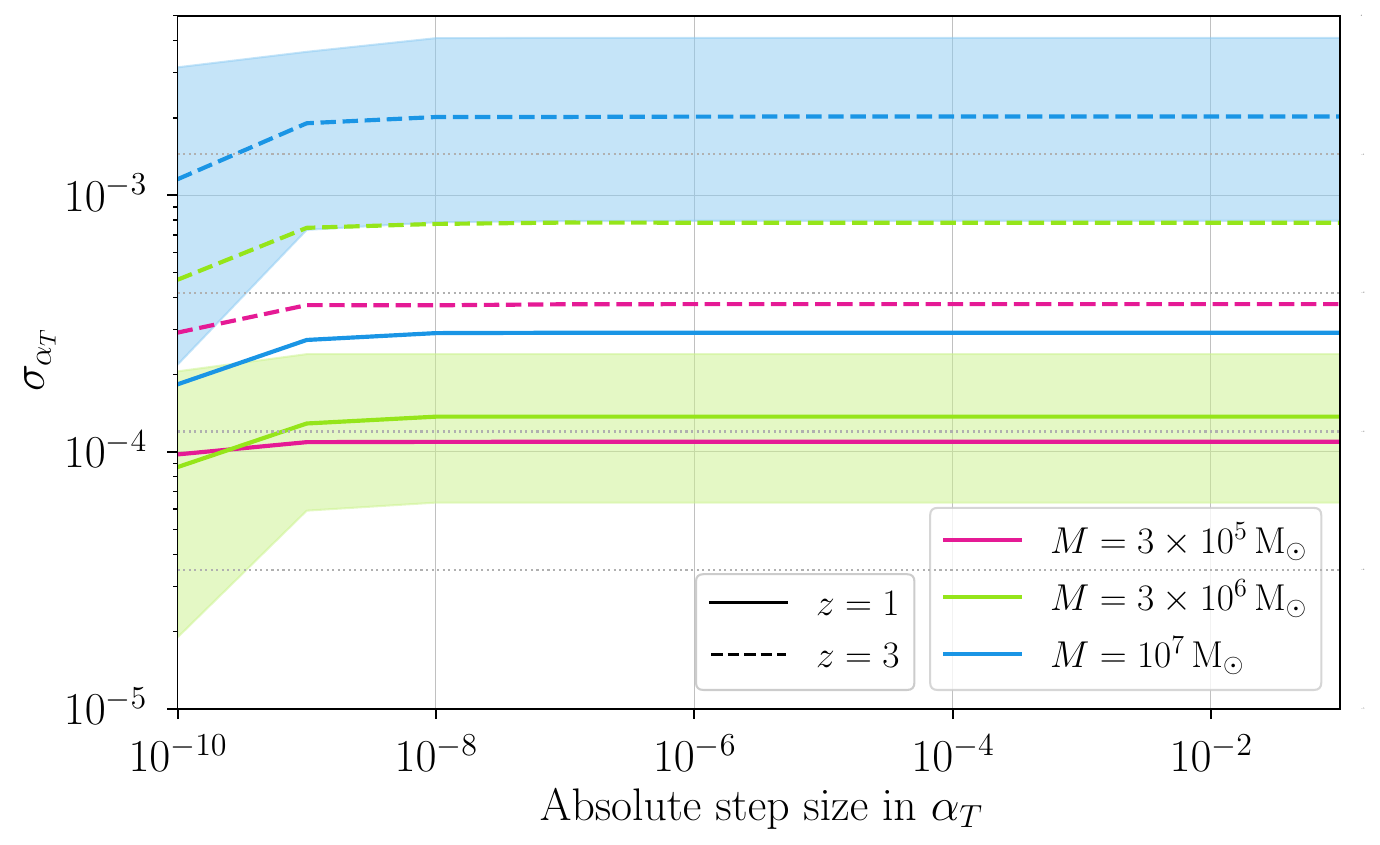}
 \caption{Absolute error on $\alpha_T$ as a function of the absolute step size in $\alpha_T$ for different masses and redshifts, as reported in the legend. For two cases, we show in shaded regions the $\pm 1 \sigma $ dispersion regions obtained averaging over a Montecarlo of 100 events,  randomly sampling the other binary parameters.}
 \label{fig:stepsT} 
\end{figure}

\begin{figure}
\centering
\includegraphics[width = 0.49\textwidth]{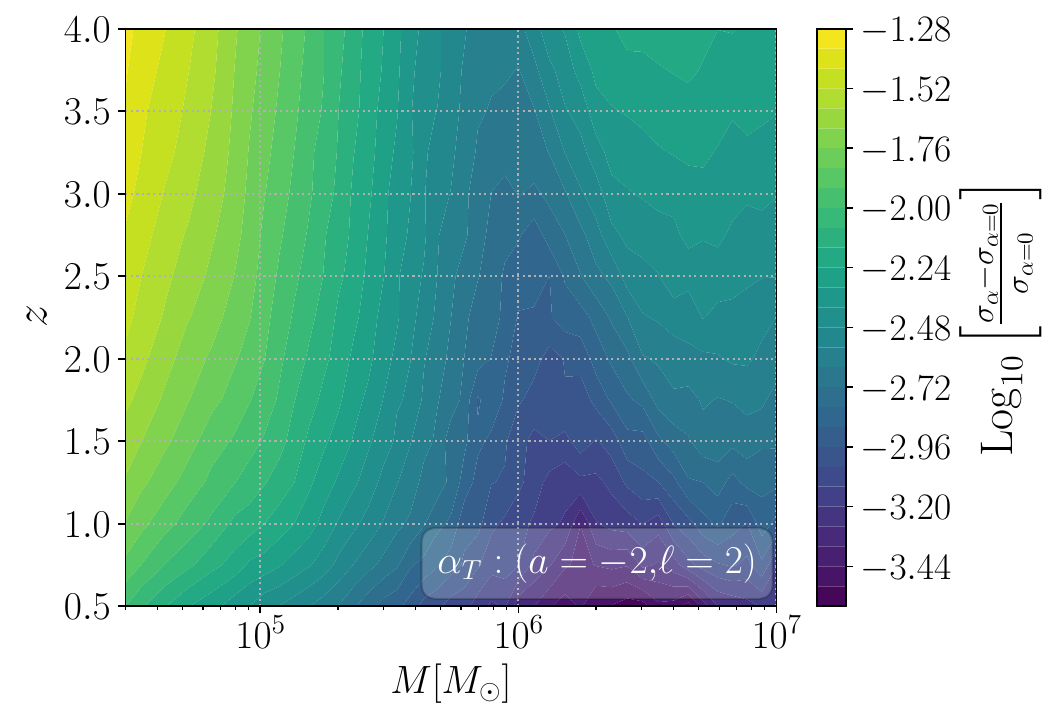}  
\includegraphics[width = 0.49\textwidth]{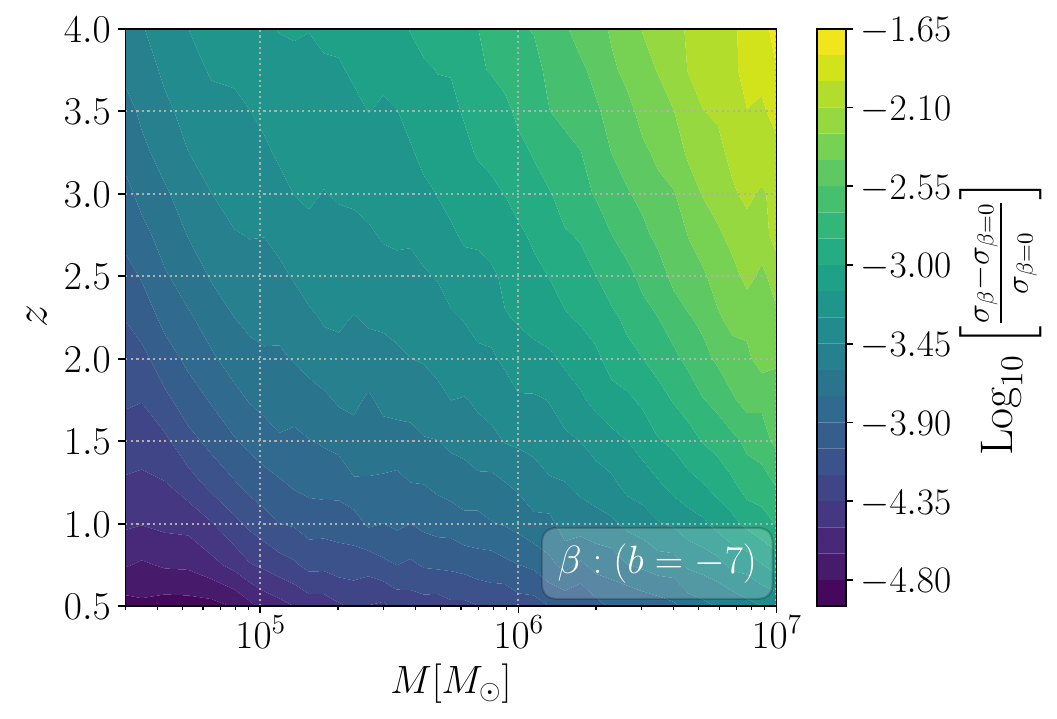}  \\
\caption{Relative difference between the $68\%$ confidence intervals derived from GR injections (Section~\ref{subsec:gr_injections}) and those obtained from non-GR injections performed using the GR-derived $68\%$ interval boundaries as injected PPE parameter values as a function of total mass and redshift. Left (right) plot reports the results for $\alpha_T$ and $a=-2$ ($\beta$ and $b=-7$). }
 \label{fig:rel_error_plot} 
\end{figure}

In this Appendix we describe some of the numerical checks we performed to assess the validity of our Fisher  results. 

As described in Section~\ref{subsec:numerical_implementation}, we compute the derivatives of the GW signals numerically. To determine the optimal step size for each PPE parameter, we examined the stability of the expected errors from the Fisher matrix, as a function of the chosen step size, looking for a plateau region. The results for $\alpha_T$ with $a=-2$ are shown in Figure~\ref{fig:stepsT} for three different total masses and redshifts. 

We find that step sizes $\lesssim 10^{-8}$ lead to strong fluctuations in the recovered error on $\alpha_T$, while values $\gtrsim 10^{-8}$ yield stable results. For two representative combinations of masses and redshifts, we also report the $\pm 1 \sigma$ error regions, demonstrating that the chosen step size is not determined by a single event. Based on this behavior, we set the step size for $\alpha_T$ to $10^{-3}$. The same value is adopted for
$\alpha_{B1}, \alpha_{B2},\alpha_{L1}, \alpha_{L2},\alpha_{V1}, \alpha_{V2}$,  which exhibit plateau regions similar to those shown in Figure~\ref{fig:stepsT}. For all GR injections of the amplitude PPE parameters presented in Section \ref{subsec:gr_injections}, we use an absolute step size of  $10^{-3}$, while for the non-GR injections presented in Section~\ref{sec:constrain-specific-theories}, we instead adopt $10^{-3}$ as the fractional step size. For the angle parameter $\beta$ we perform a similar test and set the step size to $10^{-9}$. We use these absolute steps in both GR and non-GR injections.

We also verify the stability of the Fisher matrix inversion for a subset of representative systems in 
the case of $\alpha_T$ with $a=-2$ and $a=0$. Specifically, we consider binaries with 
$M = 3\times 10^4 \msun$ and $M = 10^7 \msun$ at $z=0.5$ and $z=4$, i.e., the edges of the 
grid adopted in Section~\ref{sec:results_for_extra_pol}. For each combination of total mass and 
redshift, we perform a Montecarlo  of 100 realizations, randomizing over all binary parameters. 
For every event, we compare the product of the Fisher matrix and the covariance matrix (its 
inverse) with the identity matrix, verifying that the deviation from identity remains sufficiently 
small. In practice, we check that
\begin{equation}
    \max_{i,j}\left| \Gamma^{ij}\Sigma^{ij} - I^{ij} \right| < \epsilon_{\rm min},
\end{equation}
where $I^{ij}$ is the identity matrix and we set $\epsilon_{\rm min} = 10^{-3}$. In all cases tested, 
all realizations satisfied this condition, ensuring that the matrix inversion was performed correctly.

As an additional validation step, we compare the PE results obtained for GR injections with those from a corresponding set of non-GR injections performed at the constraint levels inferred in the former case. Specifically, we take the $68\%$ confidence intervals on the PPE parameters derived from the GR injections in Section~\ref{subsec:gr_injections} and use the boundaries of these intervals as injected values for new non-GR signals. This procedure tests the robustness of our results in the vicinity of the GR limit: if a parameter is injected at a $1\sigma$ deviation from zero, it should be recoverable as non-zero at the same level of significance (i.e. at $1\sigma$). In this way we check that the Fisher estimates behave well near the GR limit, showing no pathological non-linearities or systematic biases. By comparing the estimated errors between the GR and non-GR cases, we verify that deviations of the magnitude allowed by the GR analysis would indeed be detectable with the expected accuracy.

The results are reported in Figure~\ref{fig:rel_error_plot}
for $\alpha_T$ ($\beta$) in the case of $a=2$ ($b=-7$). 
As it can be appreciated in both cases the maximum relative  difference between the GR and non-GR constrain is $\lesssim 5\%$, supporting the results of our analysis.

\section{\label{sec:alignment}Alignment of the waveform}

As mentioned at the end of Section~\ref{subsec:numerical_implementation}, a phase modification in the tensor sector introduces a misalignment in the time-to-frequency correspondence. 
In the FTI approach, the re-alignment is performed fixing the two constants that come out of the integration of the second frequency derivative of the phase multiplied by the window function. 
At an early stage of this work, however, we implemented a slightly different alignment procedure. As detailed in this Appendix, there were two main differences: (i) the window function was applied directly to the phase modification rather than to its second derivative, and (ii) we introduced two additional parameters (later referred to as $a_1$ and $a_0$) accounting for extra time and phase shifts, which we fixed by requiring the phase modification to vanish at the reference frequency. We first describe this earlier implementation and then compare the resulting constraints with those obtained using the FTI approach. Our goal is to highlight the subtleties associated with waveform alignment, as different choices may lead to discrepancies in the inferred constraints of order $\sim10\%$, and in a few extreme cases, up to $\sim100\%$.

The relation between the orbital energy $E(x)$ and the energy flux emitted in GWs $F(x)$ can be written as 
\begin{equation}
    F(x) = -\frac{dE(x)}{dt},
    \label{eq:flux_energy_eq}
\end{equation}
where $ x = \left( G_NM\omega_{\rm orb}/c^3 \right)^{2/3}$.
Inverting Eq.~\eqref{eq:flux_energy_eq} we can write 
\begin{equation}
    dt = - \frac{dE(x)}{F(x)} = - \frac{dE(x)}{dx}\frac{dx}{F(x)}= -\frac{E'(x)}{F(x)} dx,
     \label{eq:dt_expression}
\end{equation}
and inverting the expression of $x$ we can write the phase derivative $\dot{\phi}=d\phi/dt$ as 
\begin{equation}
    d\phi = \frac{c^3}{G_NM} x^{3/2} dt = - \frac{c^3}{G_NM} x^{3/2} \frac{E'(x)}{F(x)} dx.
     \label{eq:dphi_expression}
\end{equation}
In the SPA approach, the phase for the $(\ell, m)$ harmonic can be written as \cite{Marsat:2018oam}
\begin{equation}
    \psi_{SPA}^{lm} = [m\phi(\omega) - m \omega t(\omega)]_{\omega = \frac{2 \pi f }{m}} + \frac{\pi}{4} ,
\end{equation}
and we also define 
\begin{align}
  \mathcal{F}(\omega) & \equiv  2 [\phi(\omega) - \omega t(\omega)],
  \label{eq:f_of_omega} \\
  \Delta \mathcal{F}(\omega) &\equiv  2 \Delta \phi(\omega) -2 \omega \Delta t(\omega), 
\end{align}
for later convenience. For the alignment, we require that $\phi(\omega_{ref}) = \phi_{ref}$ and $t(\omega_{ref}) = t_{ref}$, i.e. that the phase and time are equal to the phase and time reference, $\phi_{ref}$ and $t_{ref}$, at the frequency reference $\omega_{ref}$. Therefore any phase and time modifications must disappear at $\omega_{ref}$, i.e. we must impose that $\Delta \phi(\omega_{ref}) = 0$ and $\Delta t(\omega_{ref}) = 0$. We can show that these latter conditions are equivalent to $\Delta \mathcal{F}(\omega_{ref}) = 0$ and $\Delta \mathcal{F}'(\omega_{ref}) =  0$.
Indeed, using Eqs.~\eqref{eq:dt_expression}-\eqref{eq:dphi_expression}, we can write that
\begin{equation}
\begin{cases}
 \phi(x ) = &  \frac{c^3}{G_NM} \int dx x^{3/2} G(x), \\
 t(x) = & \int dx G(x) ,
\end{cases}
\label{eq:phi_and_t}
\end{equation}
where $ G(x) \equiv - E'(x)/F(x)$. If we differentiate $ G(x)$ as $\Delta G = - (E' + \Delta E')/(F + \Delta F) - \left( -E'/F\right)$, Eq.~\eqref{eq:phi_and_t} becomes
\begin{equation}
\begin{cases}
 \Delta \phi(x ) = &  \frac{c^3}{G_NM} \int dx x^{3/2} \Delta G(x), \\
 \Delta t(x) = & \int dx \Delta G(x) ,
\end{cases}
\end{equation}
or, alternatively,
\begin{equation}
\begin{cases}
 \frac{d}{dx}\Delta \phi(x ) = &  \frac{c^3}{G_NM}  x^{3/2} \Delta G(x) = \omega  \Delta G(x),\\
 \frac{d}{dx} \Delta t(x) = &  \Delta G(x) .
\end{cases}
\end{equation}
Differentiating $\mathcal{F}(\omega)$ with respect to $\omega$
\begin{align}
\frac{d}{d\omega} \Delta \mathcal{F}(\omega) & = 2 \frac{d}{d\omega} \Delta \phi(\omega) - 2 \Delta t(\omega) -2 \omega \frac{d}{d\omega} \Delta t(\omega) \\
& = 2 \frac{dx}{d\omega} \frac{d}{dx} \Delta \phi(x) -2\Delta t(\omega) -2 \omega \frac{dx}{d\omega} \frac{d}{dx} \Delta t(x) \\
&= -2 \Delta t(\omega) + 2 \frac{dx}{d\omega} \left( \frac{d}{dx}\Delta \phi(x) - \omega \frac{d}{dx}\Delta t(x) \right) \\
&= -2 \Delta t(\omega) + 2 \frac{dx}{d\omega} \left( \omega \Delta G(x) - \omega \Delta G(x)  \right) \\
&= -2 \Delta t(\omega).
\end{align}
Therefore, imposing $\Delta \mathcal{F}'(\omega) = 0$ implies that $\Delta t(\omega)=0$. If we add the condition $\Delta \mathcal{F}(\omega) = 0 $ then, recalling Eq.~\eqref{eq:f_of_omega}, we have also $\Delta \phi(\omega)=0$ so the two conditions are equivalent.

In our case, for the $\ell=m=2$ harmonic, the total phase modification was defined as 
\begin{equation}
    \Delta \mathcal{F}(\omega) = W(\omega)\beta u^b(\omega) + 2(\omega-\omega_{ref})a_1 +a_0,
\end{equation}
where $W(\omega)$ is the window function from Eq.~\eqref{eq:window_func} expressed as a function of $\omega$, $\omega_{ref}= \pi f^{\rm ref}_{\ell m}$, $a_1$ and $a_0$ are the two parameters that we wanted to determine and $f^{\rm ref}_{\ell m}$ is defined in Eq.~\eqref{eq:fref_lm}.
From the condition $\Delta \mathcal{F}(\omega_{ref}) =0$, we obtained
\begin{equation}
    a_0= - W(\omega_{ref})\beta u^b(\omega_{ref}),
\end{equation}
and from $\Delta \mathcal{F}'(\omega_{ref}) =0$, we got
\begin{equation}
    -2 a_1 = \beta  W'(\omega)|_{\omega=\omega_{ref}} u^b(\omega_{ref}) + \beta   W(\omega_{ref}) \frac{b}{3}\mathcal{M}_z u^{b-3}(\omega_{ref}),
\end{equation}
where we use that $u = (\mathcal{M}_z \omega)^{1/3}$ from Eq.~\eqref{eq:ul-PPE-extra} 

This procedure allowed us to compute the values of $a_0$ and $a_1$ such that the total phase modification as $\omega_{ref}$ is zero and recovering, in this way, the original time-to-frequency correspondence defined as $\phi(\omega_{ref}) = \phi_{ref}$ and $t(\omega_{ref}) = t_{ref}$. 

\begin{figure}
\centering
\includegraphics[scale=0.43]{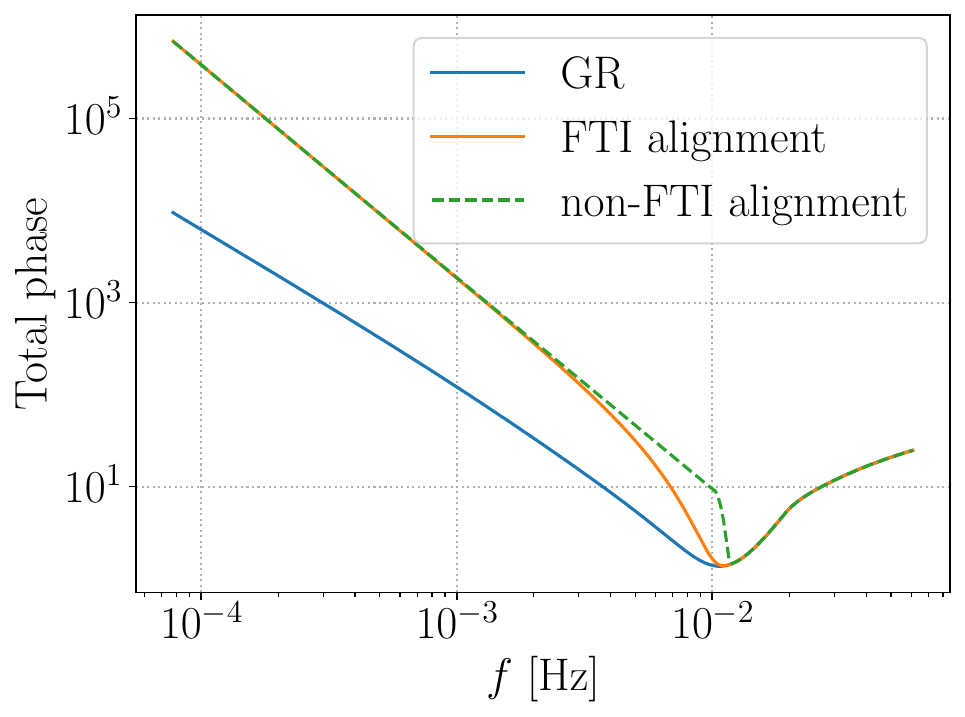}  
\includegraphics[scale=0.46]{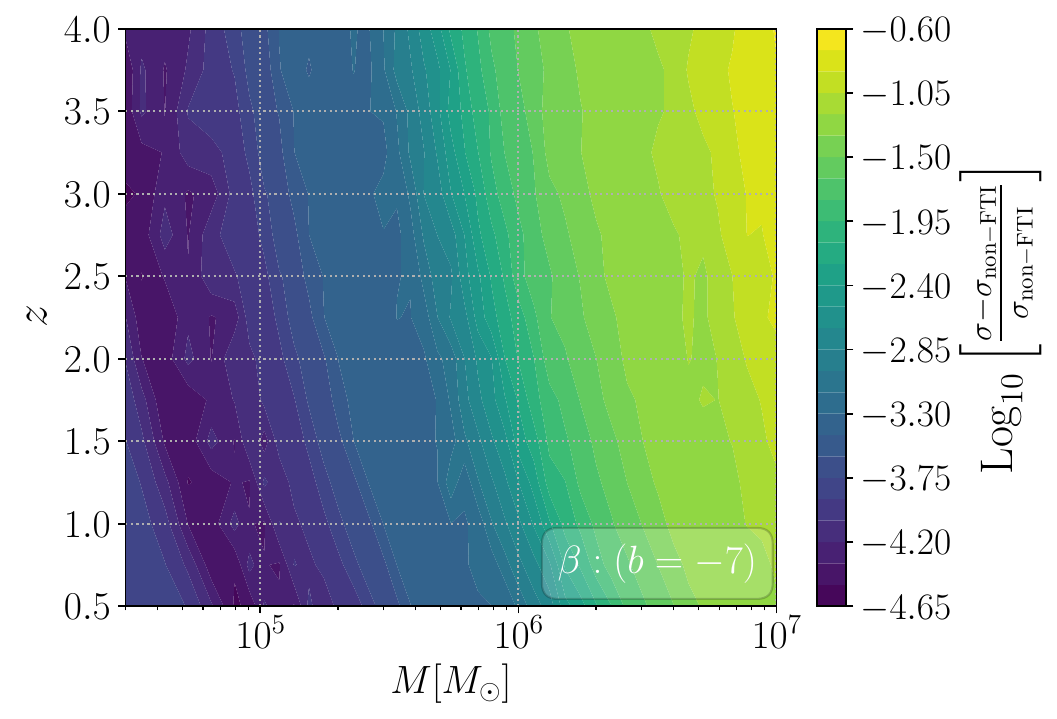}  \\
\caption{Left panel: Total phase as a function of frequency for three cases: (i) GR only (solid blue line), (ii) a phase modification aligned using the FTI method (solid orange line) and (iii) a phase modification aligned following the methodology described in this Appendix (dashed green line). The latter approach produces a noticeably sharper transition at the reference frequency.
Right panel: Relative difference between the $68\%$ confidence intervals obtained using the FTI alignment and those derived using the alternative alignment procedure described in this Appendix. For larger masses, the relative difference is typically $\gtrsim 10\%$.}
\label{fig:alignment_comparison} 
\end{figure}

In Figure~\ref{fig:alignment_comparison}, we compare the phases and the corresponding $68\%$ confidence intervals obtained when performing the alignment using either the FTI formalism or the alternative procedure described in this Appendix. From the left panel, it is apparent that, although both methods align the waveform at the same reference frequency, the FTI approach leads to a smoother transition than our original implementation. The right panel shows that the choice of alignment procedure can significantly affect the inferred constraints. While for MBHBs with total masses below $10^6,\msun$ the impact is negligible, for more massive systems the resulting differences can exceed $\gtrsim 10\%$. In some extreme cases, we found differences of up to $\sim 100\%$, highlighting the importance of the description of the analysis setup.

We stress once again that, for the main results of the paper, we adopted the alignment as prescribed in the FTI formalism and the alternative alignment procedure described in this Appendix is adopted only to produce the results in Figure~\ref{fig:alignment_comparison}.

\section{Comparison with FTI \label{sec:fti_comparison}}
\begin{figure}
\centering
\includegraphics[scale=0.48]{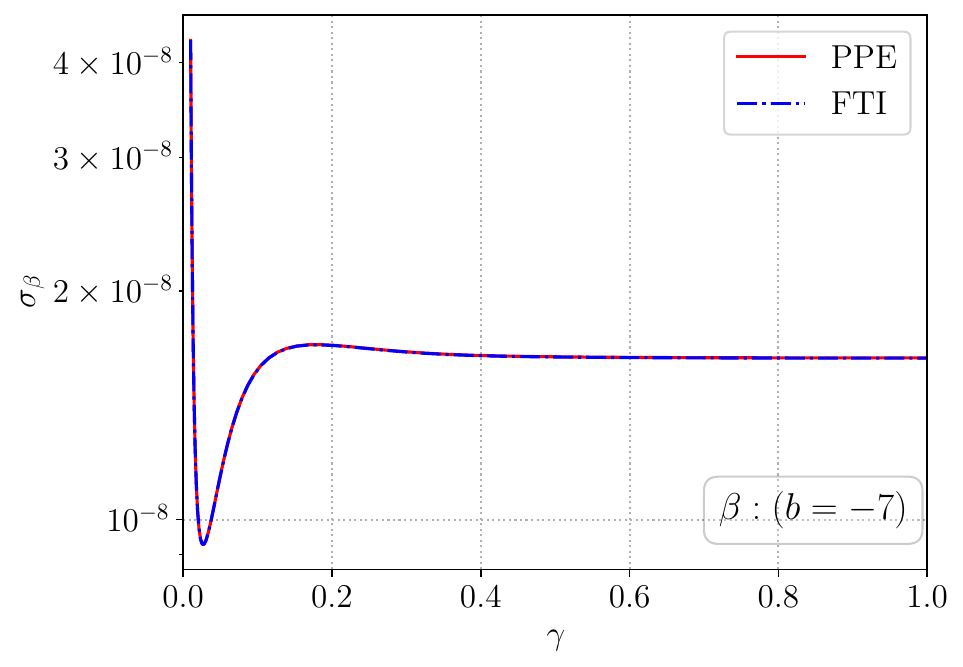}  
\includegraphics[scale=0.48]{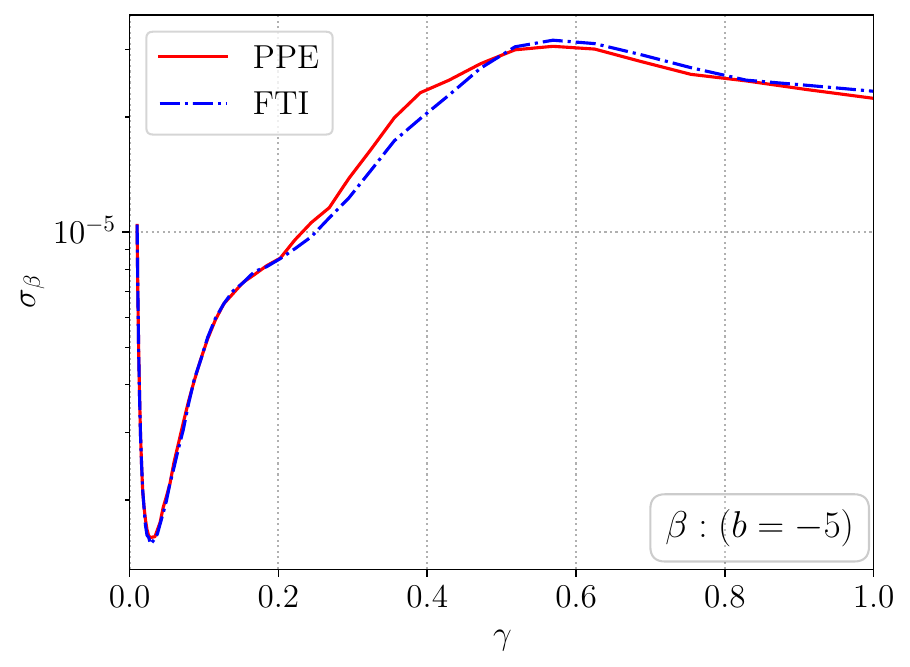}  \\
\caption{Absolute error for $\beta$ as a function of the $\gamma$, obtained in the PPE and FTI approach as described in the legend. Left (right) plot is for $b=-7$ ($b=-5$). }
 \label{fig:ppe_fti_comparison} 
\end{figure}

Let us compare the results obtained in the scope of FTI \cite{Piarulli:2025rvr} to the GR injection result of the present work for $\beta$ with $b=-7$. In FTI, the (2,2)-mode of the frequency-domain waveform is written as
\begin{align}
    \tilde{h}^{(2,2)}_T(f)=A^{(2)}e^{-i(\Psi_{GR}^{(2)}+\delta\psi_{22})},
\end{align}
where the amplitude and GR phase are given in Eqs.\ \eqref{Eq:h_PN_aell}-\eqref{Eq:Psi_GR}, while $\delta\psi_{22}$ describes modifications of GR in all of the PN coefficients as:
\begin{align}\label{eq:FTI_mods}
    \delta\psi_{22}=\frac{3}{128 \bar{q}\nu^5} \sum_{n=-2}^{7} \delta\psi_n \nu^n,
\end{align}
with $\nu=(\pi f M)^{1/3}$ and $\bar{q}=q/(1+q)^2$, and $\delta\psi_n$ assumed to be constant deviation coefficients. For the PPE phase modification that scales with $b=-7$, the corresponding FTI term is the $-1$PN order modification (i.e.\ $n=-2$ in Eq.\ \eqref{eq:FTI_mods}). Therefore, we care about the term
\begin{align}
    \delta\psi_{22}=\frac{3}{128 \bar{q}\nu^7}\delta\psi_{-2}.
\end{align}
A comparison between the definitions of the FTI and PPE approaches allows us to find the correspondence between our parameter $\beta$ and the coefficient $\delta\psi_{-2}$:
\begin{align}
    \beta=-\frac{3}{128 \bar{q}}\left(\frac{M}{\mathcal{M}_z}\right)^{-7/3}\delta\psi_{-2}.
\end{align}
A similar analysis can be made for the phase modification that scales with $b=-5$.

In Figure \ref{fig:ppe_fti_comparison} we compare the constraints on $\beta$ obtained with the PPE and FTI approach, showing very good agreement. The slight discrepancy in the case $b=-5$ at $\gamma\sim 0.4$ is due to the reparametrization we perform on $\beta$ to reduce the degeneracy with the chirp mass (see Eq.~\eqref{eq:beta_prime}).

\section{\label{app:edgb_case} Constraints on the Einstein-dilaton Gauss Bonnet theory}

As a further example of a theory that has recently attracted significant interest in the literature, we also present constraints on the coupling parameter $\alpha_{\mathrm{EdGB}}$ of the Einstein-dilaton Gauss Bonnet (EdGB) theory. Briefly, EdGB gravity is a well-known extension of GR, which emerges naturally in the framework of low-energy effective string theories \cite{Kanti:1995vq,Gross:1986mw,Mignemi:1992nt} and gives one of the simplest viable high-energy modifications to GR. The corresponding lagrangian density is given by: 
\begin{equation}
\mathcal{L}_{\mathrm{EdGB}}=\alpha_{\rm EdGB}\; \phi \left[ R_{\mu\nu\delta\sigma}R^{\mu\nu\delta\sigma}-4R_{\mu\nu}R^{\mu\nu}+R^2\right] -\frac{1}{2}\nabla_{\mu}\phi \nabla^{\mu}\phi,
\end{equation}
with $\alpha_{\mathrm{EdGB}}$ being the coupling constant between the scalar field $\phi$ and the metric. In this theory, the effect of the scalar field only appears indirectly through modifications to the tensor polarizations, with no direct excitation of extra polarizations. 
At leading order (which appears at -1 PN order in amplitude) one can model the tensor modifications as:
\begin{align}
    \alpha_T&=-\frac{5}{192}\zeta_{\rm EdGB}\frac{(m_1^2s_2-m_2^2s_1)^2}{M^4\eta^{18/5}},\\
    a&=-2,\\
    \beta&=-\frac{5}{7168}\zeta_{\rm EdGB}\frac{(m_1^2s_2-m_2^2s_1)^2}{M^4\eta^{18/5}},\\
    b&=-7,
\end{align}
where $\zeta_{\mathrm{EdGB}}=16\pi c^8 \alpha^2_{\mathrm{EdGB}}/(G_N^4 M^4)\ll 1$ is the EdGB coupling to be constrained. Notice that $\zeta_{\mathrm{EdGB}}$ is dimensionless and we have partially recovered the $c$ and $G$ factors such that the mass $M$ is in physical units, yet $\alpha_{\mathrm{EdGB}}$ is in geometric units (as usually quoted in the literature) with dimensions of $\text{length}^2$. Also, we have defined the spin-dependent dimensionless factors $s_A$ related to the BH scalar charges:
\begin{eqnarray}
    s_A= 2\left(  \sqrt{1-\chi_A^2}-1+\chi_A^2\right)/\chi_A^2,
\end{eqnarray}
with $\chi_A=|\vec{S}_A|/m_A^2$ being the magnitude of the spin angular momentum of the Ath body normalized by its mass squared. The scalar charge associated to each BH is given by $\alpha_{EdGB}s_A/m_A$.

In Table~\ref{table:constrain_edgb}, we report the constraints on $\sqrt{\alpha_{\mathrm{EdGB}}}$. 
\begin{table}[h!]
\centering
\begin{tabular}{|c|c|c|c|c|}
\hline
 \multicolumn{5}{|c|}{ Einstein-dilaton Gauss Bonnet (EdGB)} \\
\hline
PPE & Theory  & \multicolumn{3}{c|}{Median $1\sigma$ error} \\ \cline{3-5}
Parameter & Parameter & $M = 3 \times 10^5$ M$_{\odot}$ & $M = 10^6$ M$_{\odot}$ & $M = 10^7$ M$_{\odot}$ \\
\hline
$\alpha_{T}$ &  $\multirow{2}{*}{$\sqrt{\alpha_{EdGB}}$ [km]}$ & $4.23\times 10^{3}$  &  $1.41\times 10^{4}$ &  $1.95\times 10^{5}$   \\ \hhline{-~---} 
$\beta$ &  & $9.70\times 10^{2}$  &  $4.01\times 10^{3}$ &  $8.10\times 10^{4}$   \\ \hline
\end{tabular}
\caption{Same as Table~\ref{table:constrain_hornsdeski} but for EdGB case.}
\label{table:constrain_edgb}
\end{table}

Similar to the previous cases, the strongest constraints are derived from the phase modification. However, since both $\alpha_T$ and $\beta$ scale with the total mass as $\sim M^{-4}$, MBHBs are not optimal sources for constraining $\sqrt{\alpha_{\mathrm{EdGB}}}$. As a result, the corresponding bounds are not competitive with those from LVK observations (see \cite{Julie:2024fwy}), with forecasts from EMRIs (see Figure~3 in \cite{Speri:2024qak}), or forecasts from 3G ground-based detectors (see Figure 20 in \cite{Perkins:2020tra}).  Our results are also comparable to the ones reported in Figure~20 of  \cite{Perkins:2020tra} (note that, in that case, the authors report cumulative constraints, i.e.\ combining multiple events).

\section{\label{app:betap_injections} ``Almost-GR'' injections for the extra polarization phase}

As mentioned in Section~\ref{subsec:Extra_Polarizations}, we assume that the phase modification in the scalar and vector sectors is related to that of the tensor polarization as $2\beta_P = \beta$. Since $b$ is also shared across polarizations, any phase modification in the scalar and vector sectors will necessarily be accompanied by a corresponding phase modification in the tensor one.

\begin{figure}[!ht]
\centering
\includegraphics[scale=0.44]{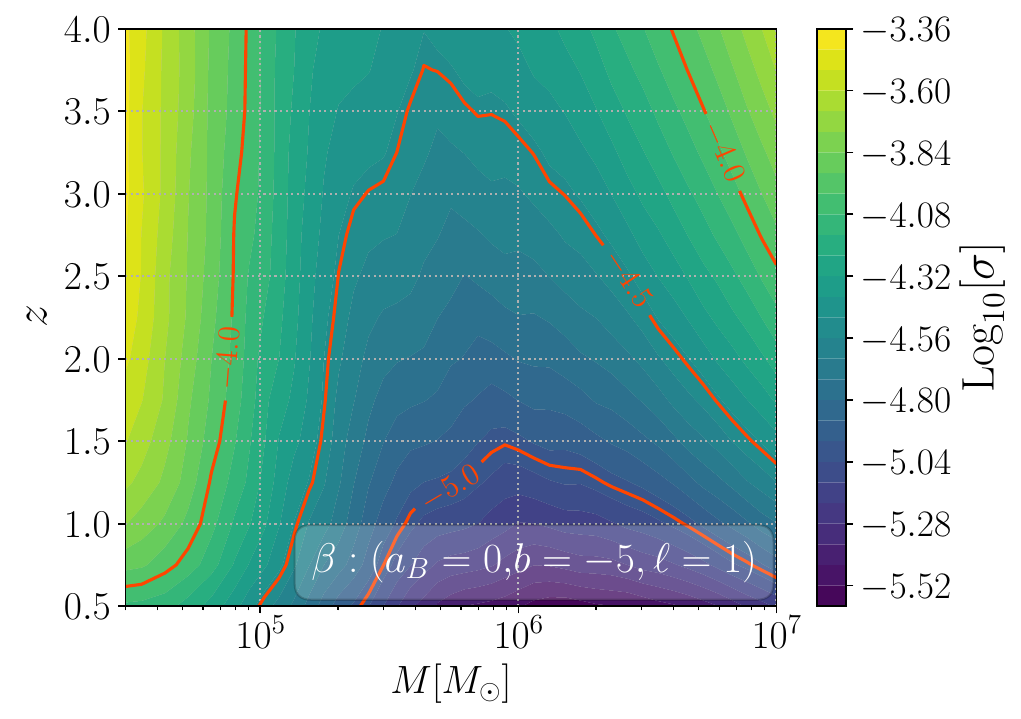}  
\includegraphics[scale=0.44]{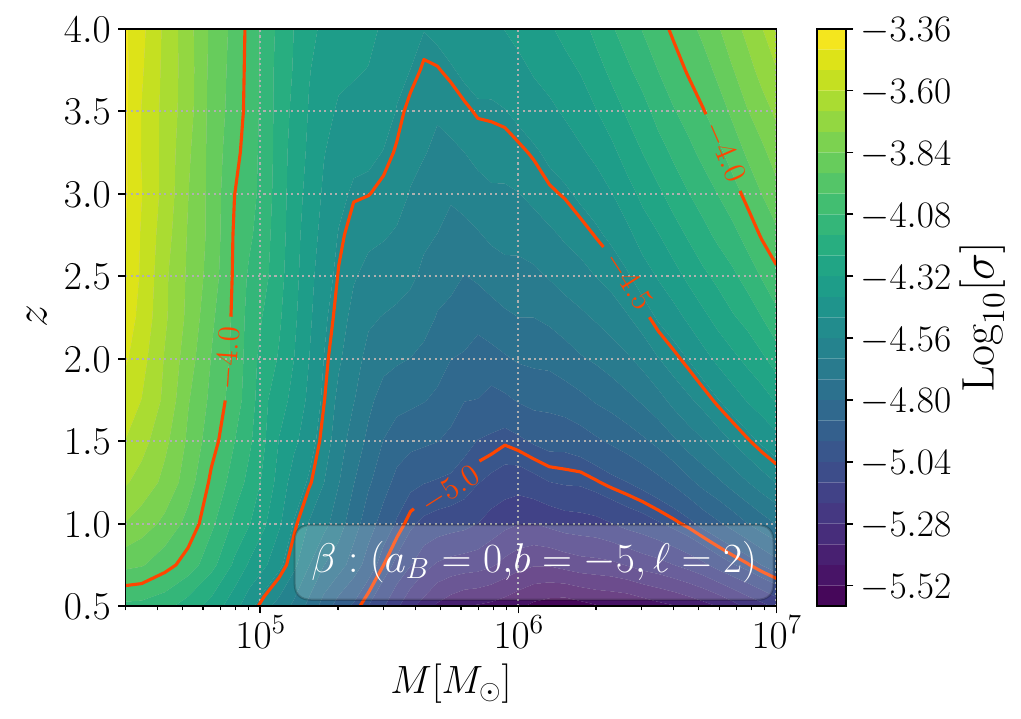}  \\
\includegraphics[scale=0.44]{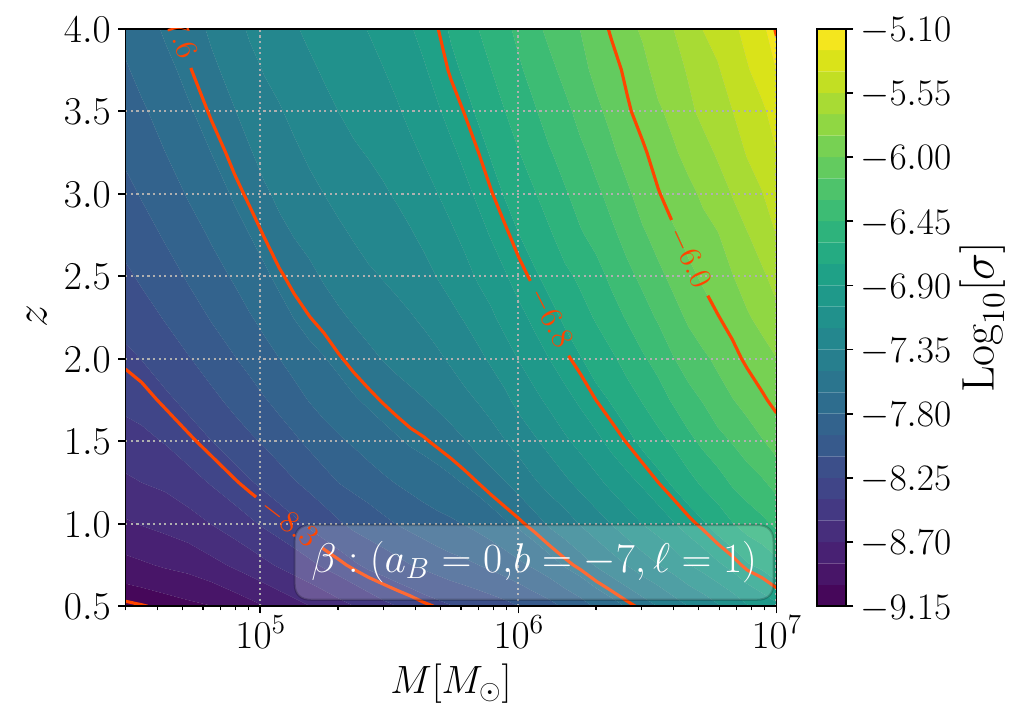}  
\includegraphics[scale=0.44]{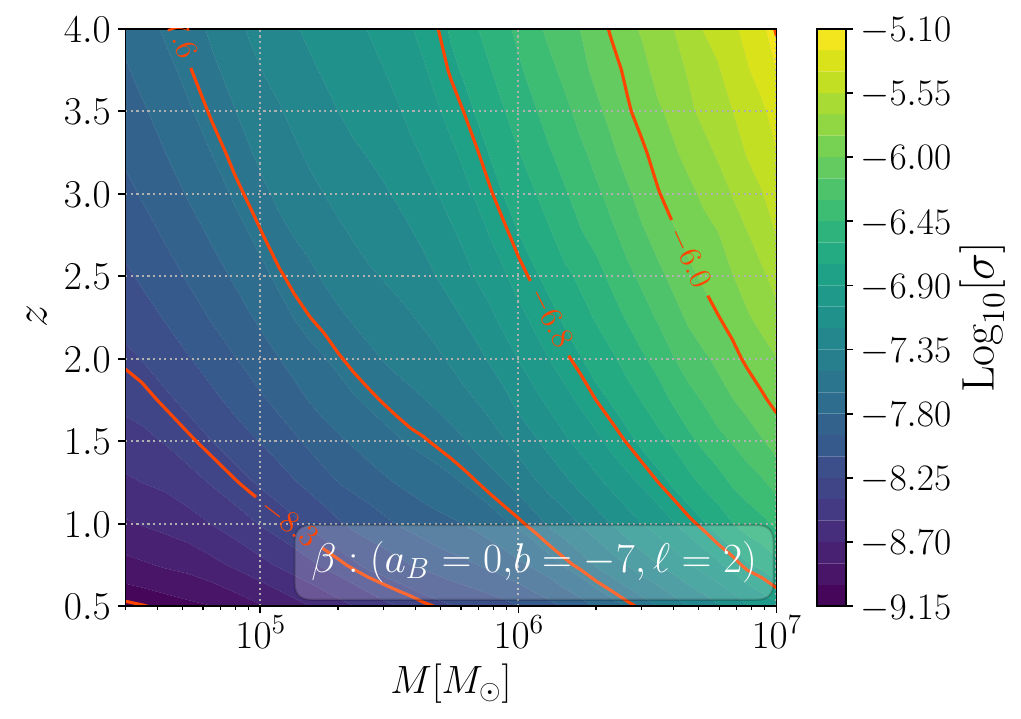} \\
\includegraphics[scale=0.44]{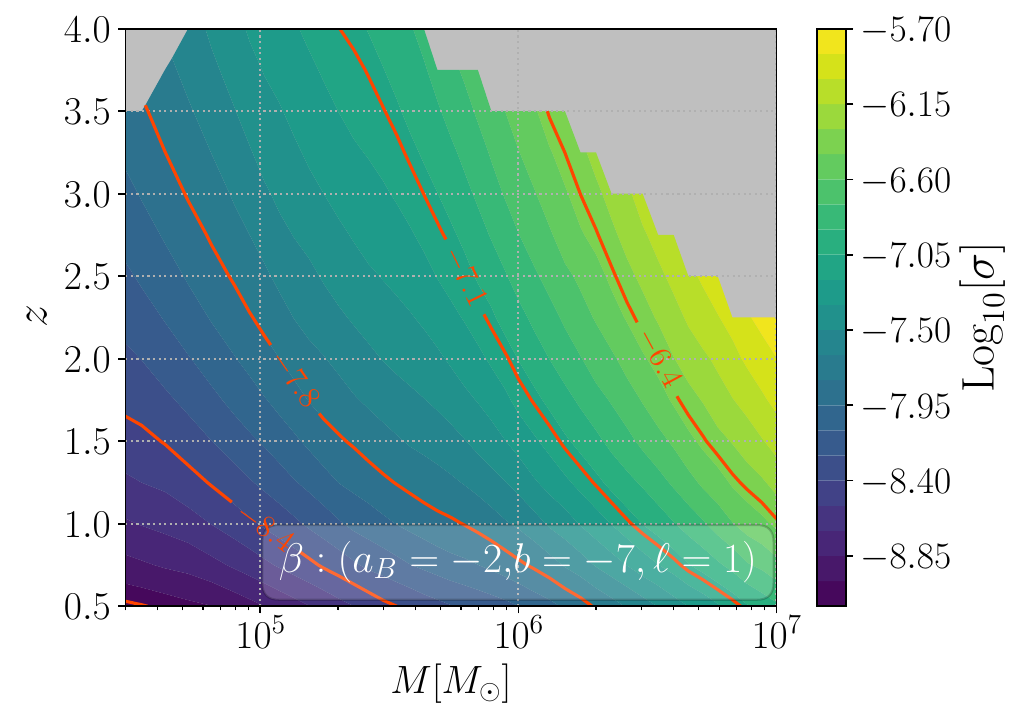}  
\includegraphics[scale=0.44]{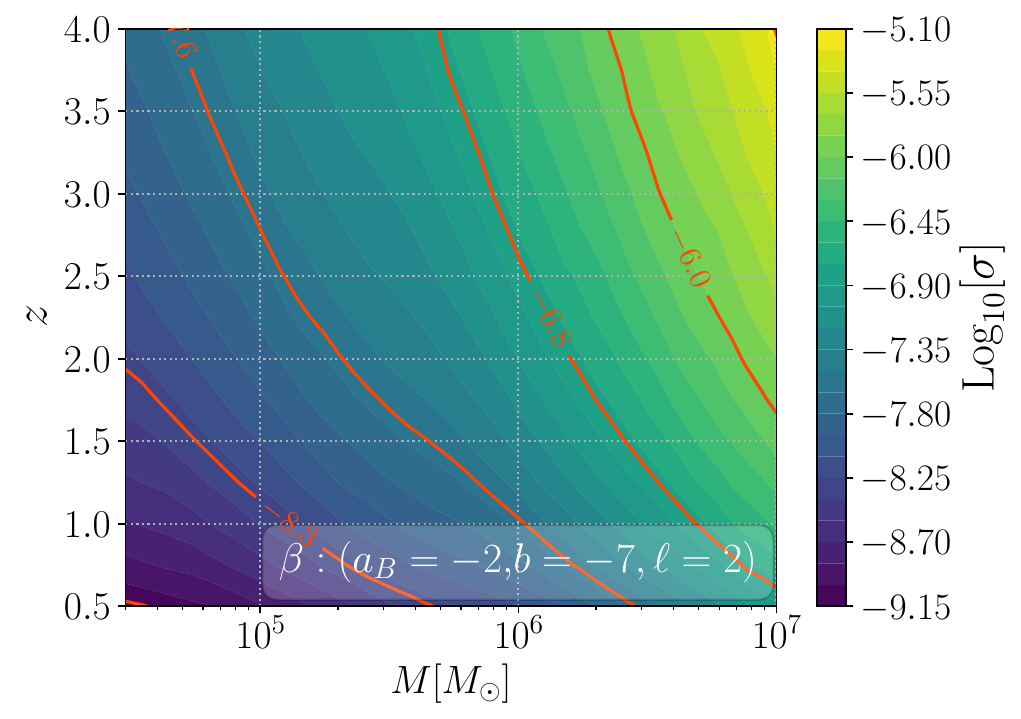}
\caption{``Almost-GR" injections for phase modification in the breathing sector for different combination of $a_B$, $b$ and $\ell$ as reported in the legend of each panel. }
 \label{fig:gr_injection_betab} 
\end{figure}

\begin{figure}[!ht]
\centering
\includegraphics[scale=0.44]{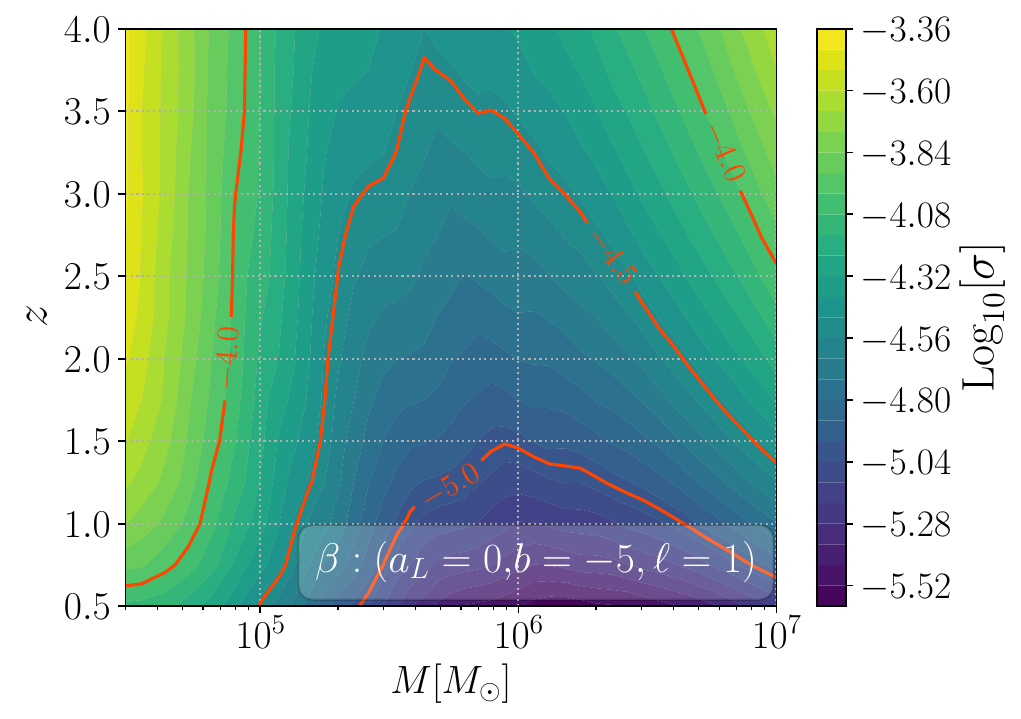}  
\includegraphics[scale=0.44]{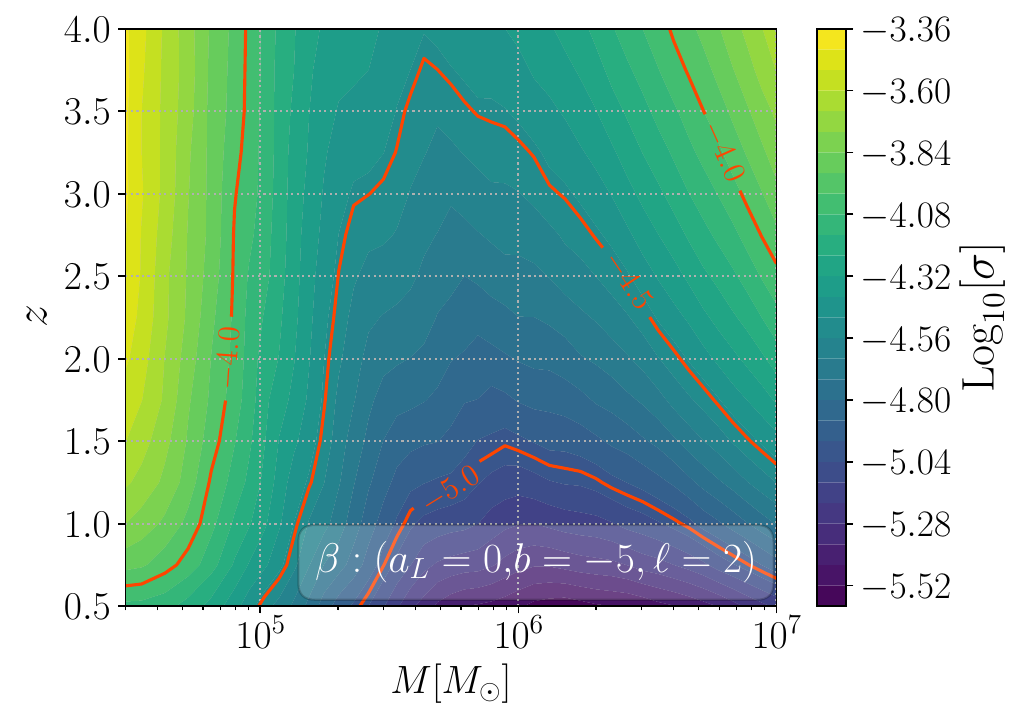}  \\
\includegraphics[scale=0.44]{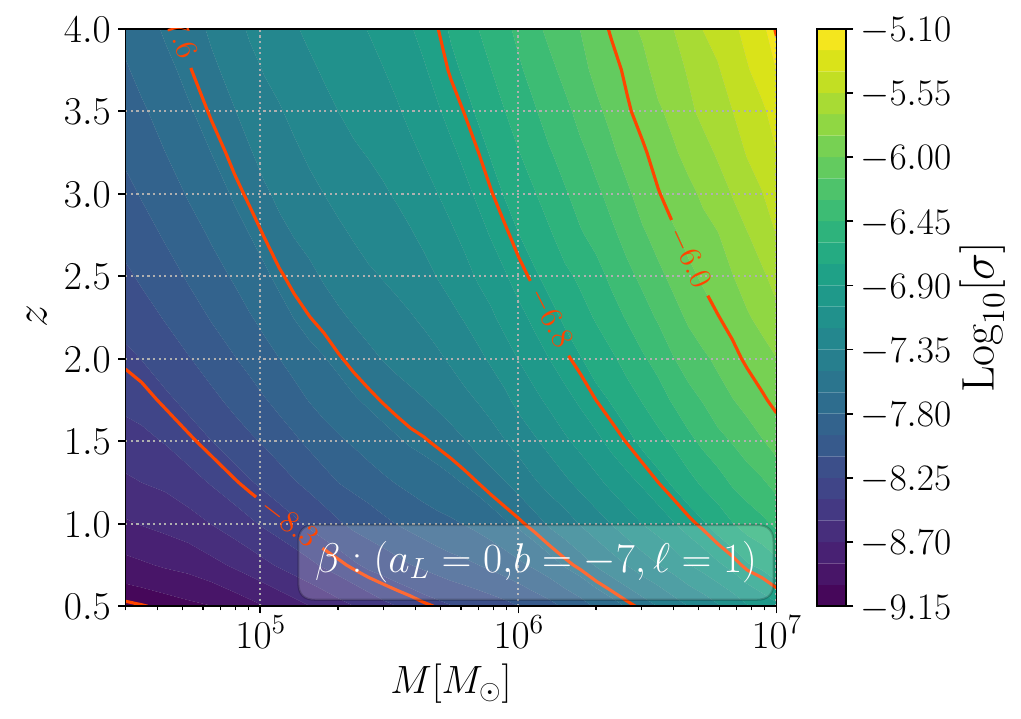}  
\includegraphics[scale=0.44]{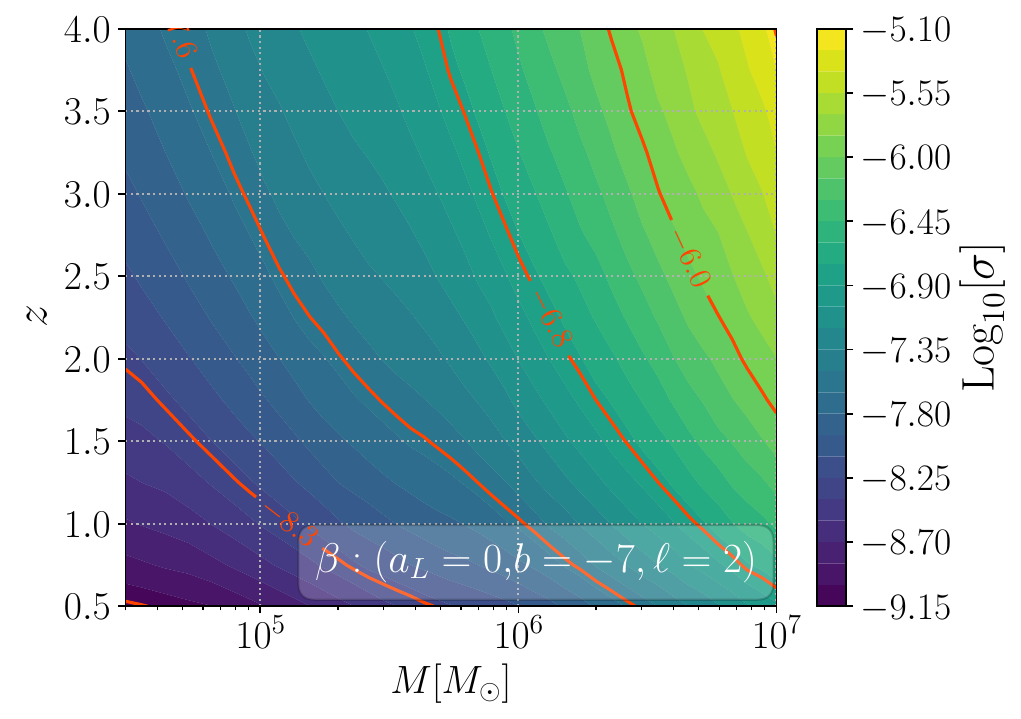} \\
\includegraphics[scale=0.44]{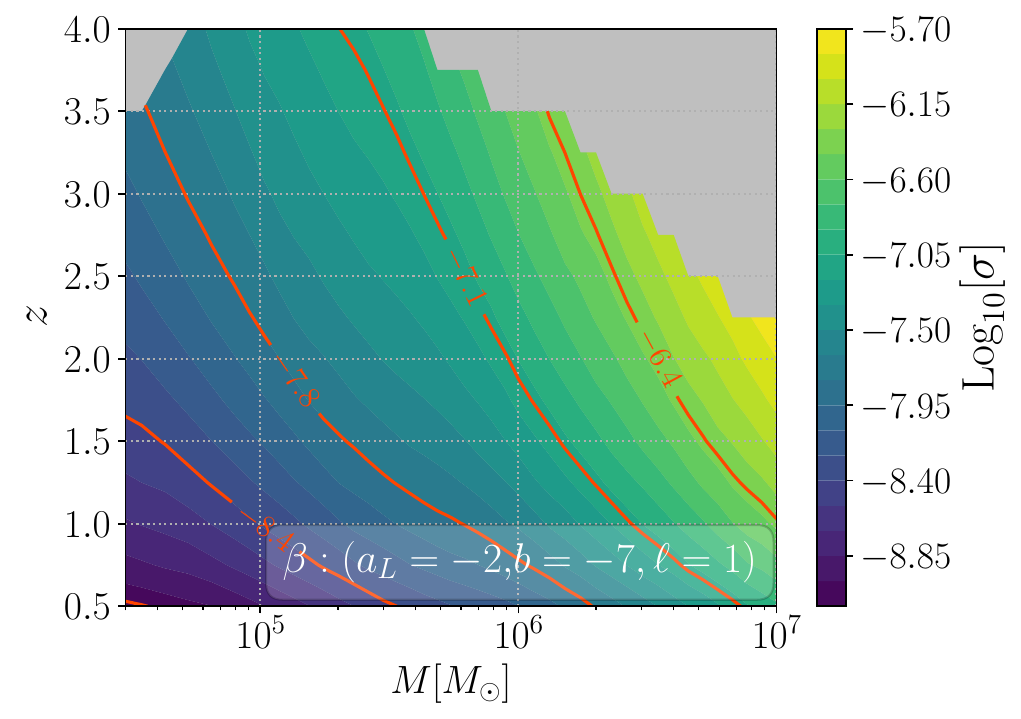}  
\includegraphics[scale=0.44]{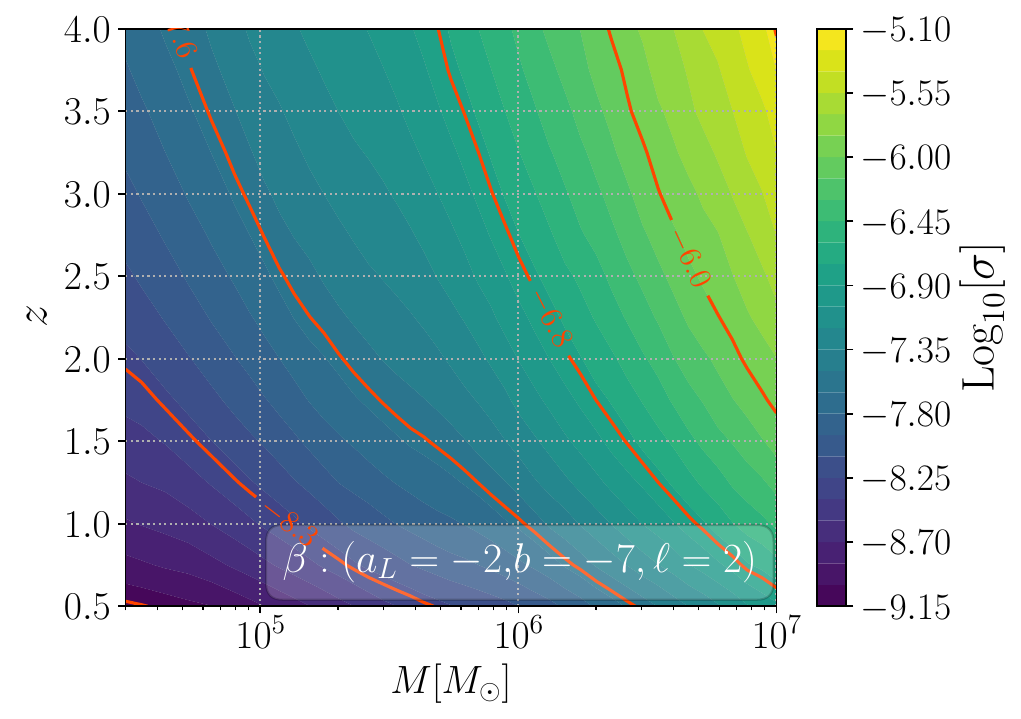} \\
\includegraphics[scale=0.44]{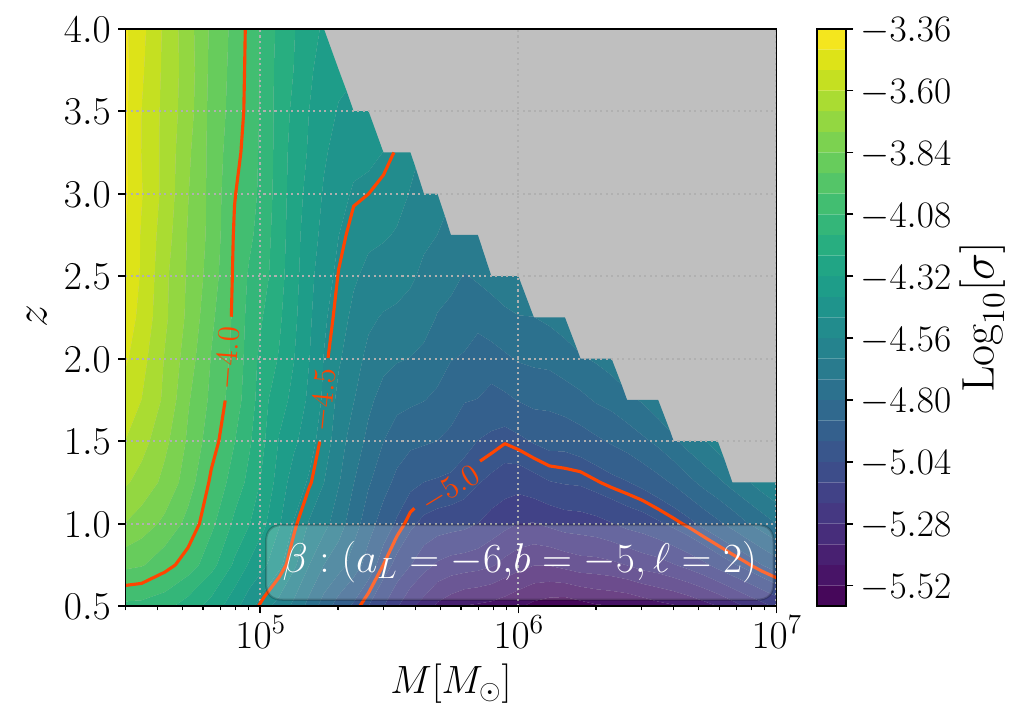}  
\includegraphics[scale=0.44]{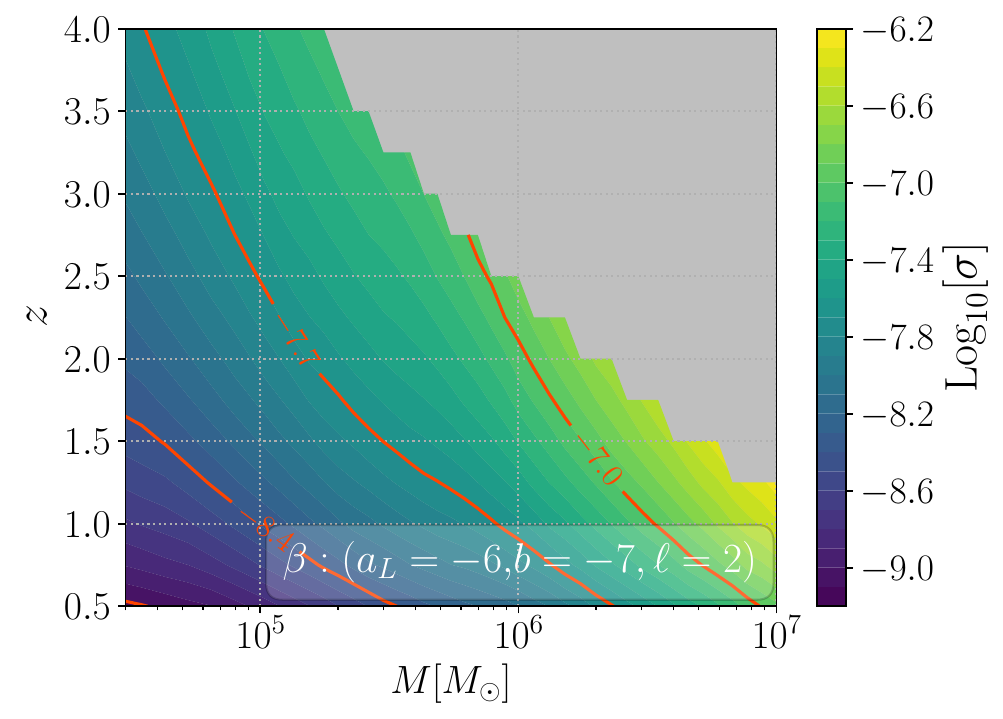} \\
\caption{Same as Figure~\ref{fig:gr_injection_betab} but for longitudinal polarization. The cases $(a_L=-6,b=-5,\ell=1)$ and  $(a_L=-6,b=-7,\ell=1)$ are not shown, as the whole parameter space is unconstrainable. }
 \label{fig:gr_injection_betal} 
\end{figure}

\begin{figure}[!ht]
\centering
\includegraphics[scale=0.44]{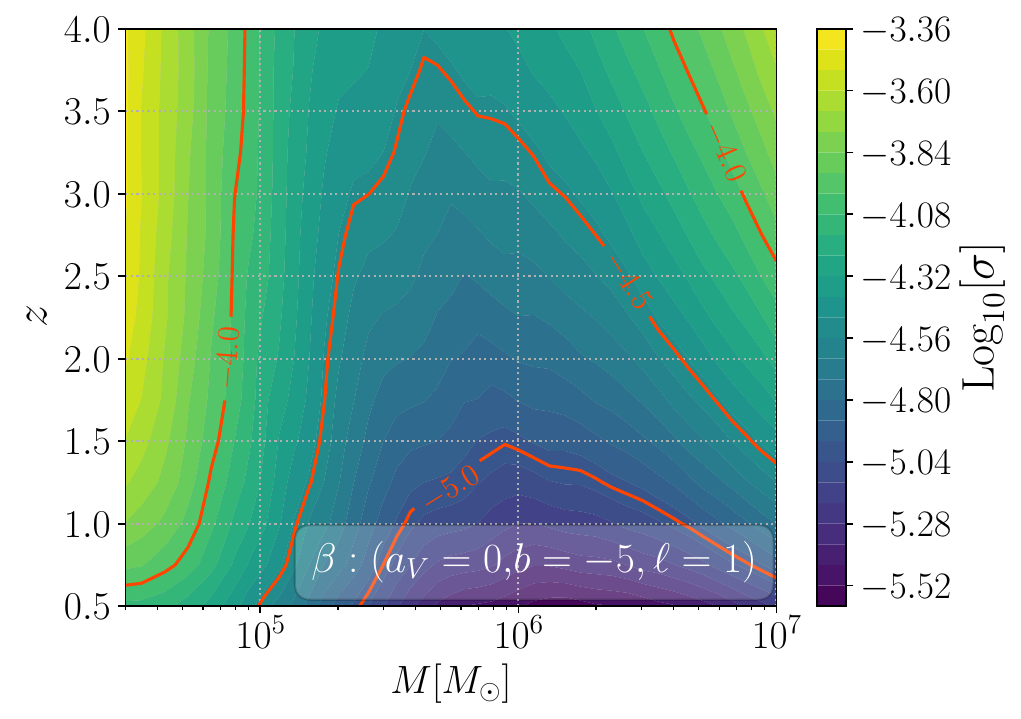}  
\includegraphics[scale=0.44]{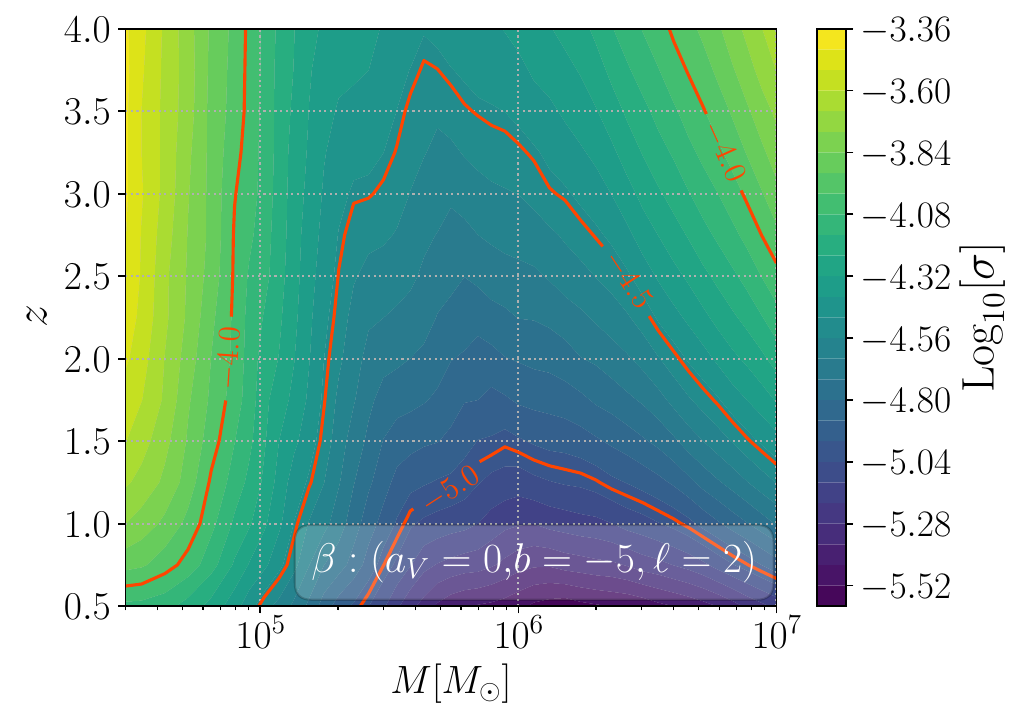}  \\
\includegraphics[scale=0.44]{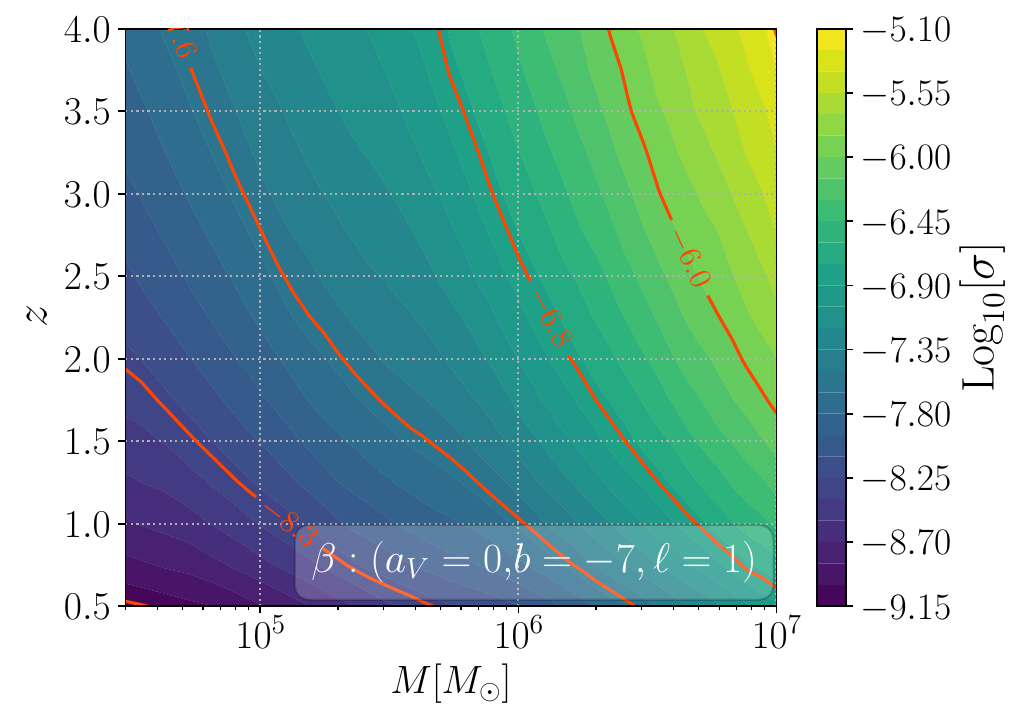}  
\includegraphics[scale=0.44]{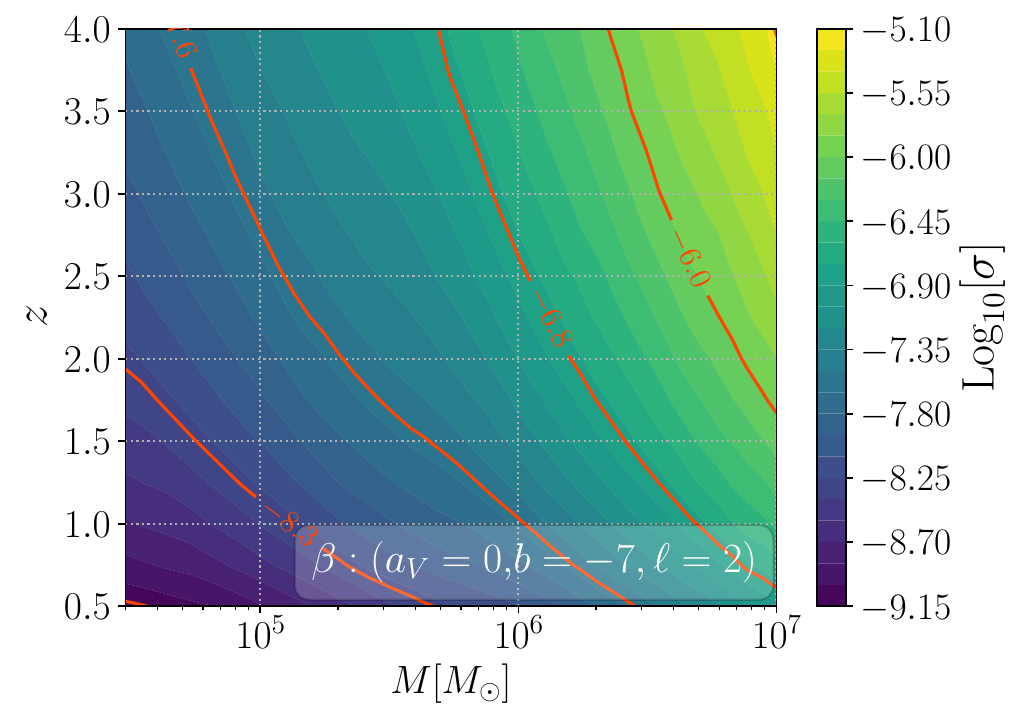} \\
\includegraphics[scale=0.44]{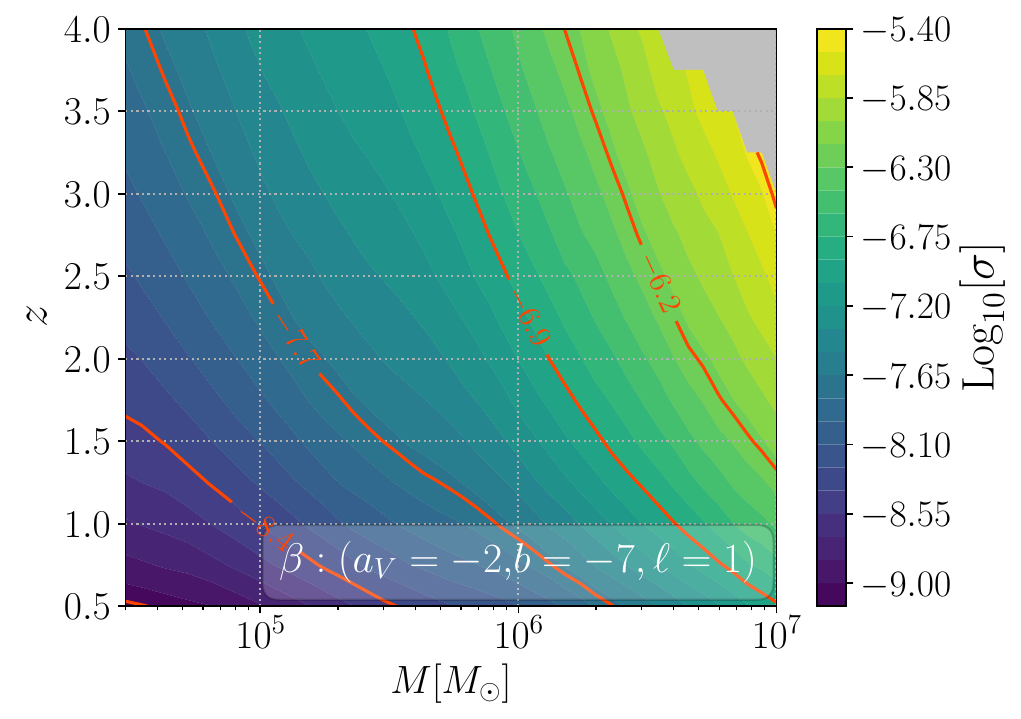}  
\includegraphics[scale=0.44]{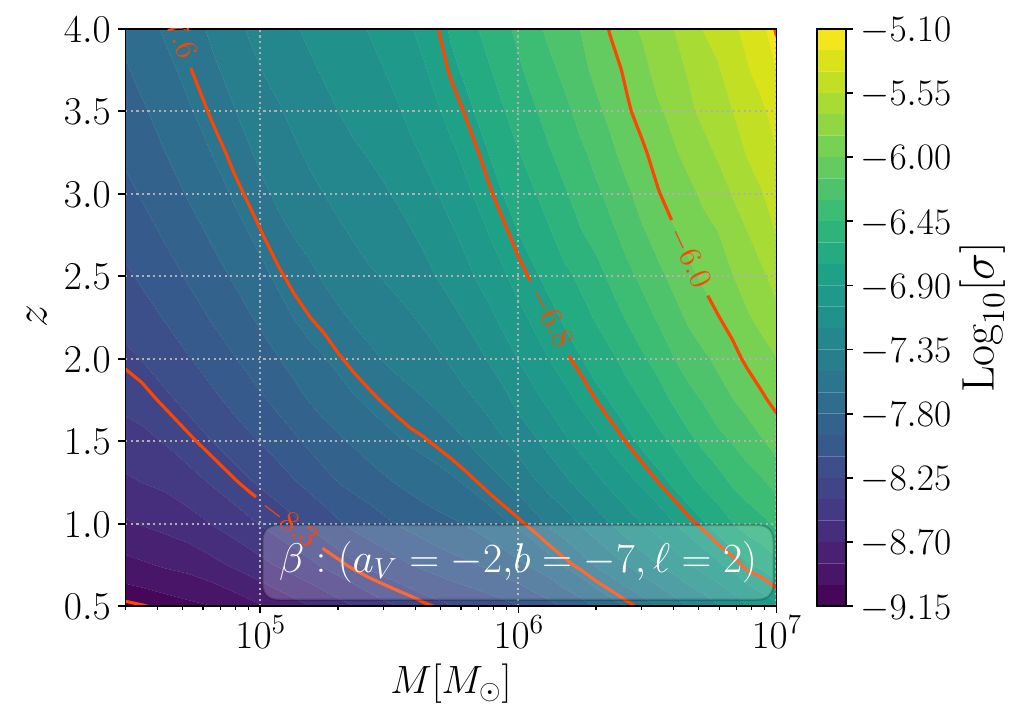}
\caption{Same as Figure~\ref{fig:gr_injection_betab} but for vector polarization.}
 \label{fig:gr_injection_betaV} 
\end{figure}

During the preparation of the manuscript, we found that the tensor polarization alone provides most of the constraining power on phase modifications. As a result, the GR injections for phase modifications in the scalar and vector sectors are practically identical to the constraints already shown in Figure~\ref{fig:gr_injection_betaT}. However, for completeness, we also report in Figures~\ref{fig:gr_injection_betab}, \ref{fig:gr_injection_betal} and \ref{fig:gr_injection_betaV} the results of ``almost-GR" injections for the phase of breathing, longitudinal, and vector polarizations, respectively. We recall that the almost-GR injections are performed by setting the uncertainties on $\alpha_P$ from the GR injections obtained in Section~\ref{subsec:gr_injections} as the amplitude of the polarization $P$, while the phase is set to $\beta=0$. This approach is necessary in the case of phase modifications of extra polarizations, as $\alpha_P=0$ implies a vanishing waveform for the polarization $P$, making a fully-GR injection not possible. Additionally, we recall that the gray areas represent parameter space regions where the amplitude parameter $\alpha_P$ could not be constrained.

\FloatBarrier
\bibliographystyle{JHEP}
\bibliography{refs}

\end{document}